 \documentclass[12pt,a4paper,twoside,openright]{report}
 \let\openright=\cleardoublepage

\usepackage{setspace}
\let\openright=\clearpage
\usepackage[utf8]{inputenc}

 \usepackage{wrapfig}

\usepackage{enumitem}
\usepackage{graphicx}
\usepackage{amsthm}
\usepackage[usenames,dvipsnames]{xcolor}
\usepackage{amsmath, amsthm, amssymb,gensymb}
\usepackage{upquote,textcomp}
\usepackage{color}
\usepackage{fancyvrb}
\usepackage{marginnote}
\VerbatimFootnotes

\usepackage{setspace}
\usepackage{tikz}
\usepackage{parskip}

\newlength{\RoundedBoxWidth}
\newsavebox{\GrayRoundedBox}
\newenvironment{prolbox}[1][\dimexpr\textwidth-1.5ex]%
   {\setlength{\RoundedBoxWidth}{\dimexpr#1}
    \begin{lrbox}{\GrayRoundedBox}
       \begin{minipage}{\RoundedBoxWidth}}%
   {   \end{minipage}
    \end{lrbox}
    \begin{center}
    \begin{tikzpicture}%
       \draw node[draw=yellow!5,fill=yellow!5,%
             inner sep=0ex,text width=\RoundedBoxWidth]%
             {\usebox{\GrayRoundedBox}};
    \end{tikzpicture}
    \end{center}}
\newenvironment{grambox}[1][\dimexpr\textwidth-1.5ex]%
   {\setlength{\RoundedBoxWidth}{\dimexpr#1}
    \begin{lrbox}{\GrayRoundedBox}
       \begin{minipage}{\RoundedBoxWidth}}%
   {   \end{minipage}
    \end{lrbox}
    \begin{center}
    \begin{tikzpicture}%
       \draw node[draw=blue!5,fill=blue!5,%
             inner sep=0ex,text width=\RoundedBoxWidth]%
             {\usebox{\GrayRoundedBox}};
    \end{tikzpicture}
    \end{center}}
\newenvironment{exambox}[1][\dimexpr\textwidth-1.5ex]%
   {\setlength{\RoundedBoxWidth}{\dimexpr#1}
    \begin{lrbox}{\GrayRoundedBox}
       \begin{minipage}{\RoundedBoxWidth}}%
   {   \end{minipage}
    \end{lrbox}
    \begin{center}
    \begin{tikzpicture}%
       \draw node[draw=green!5,fill=green!5,%
             inner sep=0ex,text width=\RoundedBoxWidth]%
             {\usebox{\GrayRoundedBox}};
    \end{tikzpicture}
    \end{center}}
\def\regreg{{$\mathrm{REG^{REG}}$}}
\def\peg{$\mathrm{PEG}$}
\def\extext#1{{\bf\it ``#1''}}

\usepackage[ps2pdf,unicode]{hyperref}   
\hypersetup{pdftitle=Pattern matching in compilers}
\hypersetup{pdfauthor=Ondřej Bílka}

\makeatletter
\def\@makechapterhead#1{
  {\parindent \z@ \raggedright \normalfont
   \Huge\bfseries \thechapter. #1
   \par\nobreak
   \vskip 20\p@
}}
\def\@makeschapterhead#1{
  {\parindent \z@ \raggedright \normalfont
   \Huge\bfseries #1
   \par\nobreak
   \vskip 20\p@
}}
\makeatother

\begin{document}
\VerbatimFootnotes
\def \floor#1{\lfloor #1 \rfloor}
\def \ceil#1{\lceil #1 \rceil }
\def \tilde{{\raise-.80ex\hbox{\~{}}}}
\def \tw#1{{\texttt{#1}}}
\def \col#1{\multicolumn{2}{|l|}{#1}}
\input epsf
\newtheorem{thm}[section]{Theorem}
\newtheorem{qst}[section]{Question}
\newtheorem{obs}[section]{Observation}
\newtheorem{cor}[section]{Corollary}
\newtheorem{claim}[section]{Claim}
\newtheorem{lem}[section]{Lemma}

\lefthyphenmin=2
\righthyphenmin=2


\pagestyle{empty}
\begin{center}

\large

Charles University in Prague

\medskip

Faculty of Mathematics and Physics

\vfill

{\bf\Large MASTER THESIS}

\vfill

\centerline{\mbox{\includegraphics[width=60mm]{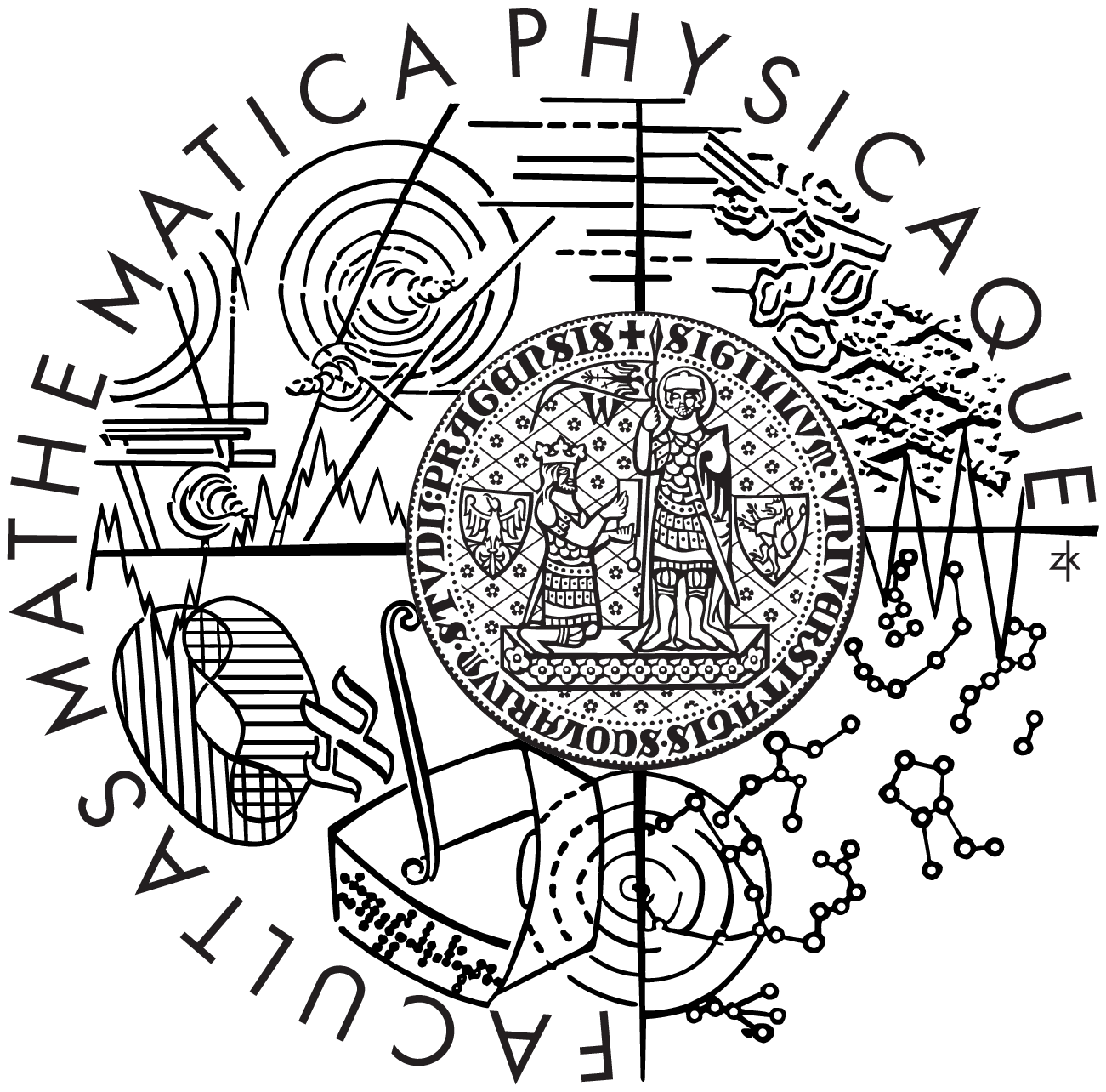}}}

\vfill
\vspace{5mm}

{\LARGE Ondřej Bílka}

\vspace{15mm}

{\LARGE\bfseries Pattern matching in compilers}

\vfill

Department of applied mathematics

\vfill

\begin{tabular}{rl}

Supervisor of the master thesis: & Jan Hubička \\
\noalign{\vspace{2mm}}
Study programme: & Diskrétní modely a algoritmy \\
\noalign{\vspace{2mm}}
Specialization: & Diskrétní modely a algoritmy\\
\end{tabular}

\vfill

Prague 2012

\end{center}

\newpage



\openright

\noindent
\section*{Acknowledgements}
I would like to thank to everybody at Charles University who made this possible. 

I would like to thank Pavel Klav\'ik for help with the typography. 

I would like to thank Andrew Goodall for a help with traps of the English grammar.

I would like to thank my advisor Honza Hubička for help, time, endurance, valuable comments and insights how compilers work in real world.

I would like to thank to many authors on whose work is this thesis built.

I would like to thank my family for their love.

\newpage


\vglue 0pt plus 1fill

\noindent
I declare that I carried out this master thesis independently, and only with the cited
sources, literature and other professional sources.

\medskip\noindent
I understand that my work relates to the rights and obligations under the Act No.
121/2000 Coll., the Copyright Act, as amended, in particular the fact that the Charles
University in Prague has the right to conclude a license agreement on the use of this
work as a school work pursuant to Section 60 paragraph 1 of the Copyright Act.

\vspace{10mm}

\hbox{\hbox to 0.5\hsize{%
In ........ date ............
\hss}\hbox to 0.5\hsize{%
signature of the author
\hss}}

\vspace{20mm}
\newpage


\vbox to 0.5\vsize{
\setlength\parindent{0mm}
\setlength\parskip{5mm}

Název práce:
 Pattern matching in compilers

Autor:
Ondřej Bílka

Katedra:  
Katedra Aplikovan\'e Matematiky

Vedoucí diplomové práce:
Jan Hubička, Katedra Aplikovan\'e Matematiky

Abstrakt:
V t\'eto pr\'aci vyvineme n\'astroje na efektivn\'i a flexibiln\'i pattern matching. P\v{r}edstav\'ime specializovan\'y programovac\'i jazyk amethyst. 
Jedna z funkc\'i amethystu je generatov\'an\'i parser\r{u}. Tak\'e m\r{u}\v{z}e slou\v{z}it jako alternativa k regul\'arn\'im v\'yrazum. Na\v{s} syst\'em um\'i generovat dynamick\'e parsery. Jejich hlavn\'i uplatn\v{e}n\'i je tvorba n\'astroju do IDE jako nap\v{r}. interaktivn\'i zv\'yraz\v{n}ova\v{c} syntaxe nebo detektor chyb.
Amethyst um\'i zpracov\'avat i obecn\'e datov\'e struktury. Pl\'anovan\'e vyu\v{z}it\'i je implementace kompil\'atorov\'ych optimalizac\'i jako nap\v{r}iklad propagace konstant \v{c}i rozvrhov\'an\'i instrukc\'i a jin\'e optimalizace zalo\v{z}en\'e na dataflow analyze. 

Generovan\'e parsery jsou v\'icem\'en\v{e} top-down parsery. P\v{r}edstav\'ime nov\'y algoritmus pro parsovan\'i strukturovan\'ych gramatik v linearn\'im \v{c}ase.
Amethyst pou\v{z}\'iv\'a techniky z kompilator\r{u} pro optimalizovan\'i generovan\'ych parser\r{u}.

Klíčová slova:
packrat parsov\'an\'i, dynamick\'e parsov\'an\'i, strukturovan\'e gramatiky, funkcion\'aln\'i programov\'an\'i

\vss}\nobreak\vbox to 0.49\vsize{
\setlength\parindent{0mm}
\setlength\parskip{5mm}

Title:
 Pattern matching in compilers


Author:
Ondřej Bílka

Department:
Department of Applied Mathematics

Supervisor:
Jan Hubička, Department of Applied Mathematics

Abstract:

In this thesis we develop tools for effective and flexible pattern matching. We introduce a new pattern matching system called amethyst. 
Amethyst is not only a generator of parsers of programming languages, but can also serve as an alternative to tools for matching regular expressions. 
 Our framework also  produces dynamic parsers. Its intended use is in the context of IDE (accurate syntax highlighting and error detection on the fly).
Amethyst offers pattern matching of general data structures. This makes it a useful tool for implementing compiler optimizations such as constant folding, instruction scheduling, and dataflow analysis in general. 

The parsers produced are essentially top-down parsers. Linear time complexity is obtained by introducing the novel notion of structured grammars and regularized regular expressions.
Amethyst uses techniques known from compiler optimizations to produce effective parsers.


Keywords:
Packrat parsing, dynamic parsing, structured grammars, functional programming

\vss}

\newpage


\openright
\pagestyle{plain}
\setcounter{page}{1}
\tableofcontents


\chapter*{A bigger picture}

This thesis forms the basis of an overarching project: We wish to experiment with language design and optimizations.
We seek  a language with flexibility beyond Lisp and with good syntax. Our base language is Ruby, which is often described as Lisp with syntax.
We strive for high performance. One of our goals is to make a dynamic optimizations and optimizations made on the data structure level possible.

In this work we solve one aspect of this project:
A major part of every compiler implementation consists of various forms of pattern matching often written in ad-hoc way. For example tokenization, parsing, expression simplification, dataflow analysis or instruction selection can be seen as instances of pattern matching.
We introduce amethyst which is a tool for pattern matching of arbitrary data. 

To reach our goals of high effectivity and flexibility we use top-down parsers. For a long time bottom-up parsers were viewed as the only alternative to handle reasonably rich class of languages.
However top-down parsers have received a lot of attention recently. 

The new formalism of boolean grammars \cite{boolean} extends context free grammars (introduced by Chomsky in 1956 \cite{contextfree})  to a wider family of languages that includes most of programming languages. 
A variant of the top-down parser that archives linear time by memoization was introduced by Ford \cite{ford}. These parsers can be generated from description in \peg{} format.
We extend this research by introducing notion structured grammars that overcomes several limitations of \peg{}. We provide a linear time algorithm for parsers of structured grammars that gives exactly the same output as a backtracking top-down parser.
The class of languages recognized by \peg{} is equivalent to the class \regreg{} recognized by amethyst. 

Amethyst takes inspiration from an OMeta (2007) \cite{ometa} which extended parsing expression grammars (2002), which extended regular expressions (1956), \cite{regexp} which were introduced as a way to describe finite state machines (1943) \cite{finstatemachine}. 

One of the extensions made in OMeta is pattern matching of tree-like data structures. We further extend this work in several respects. One is extending pattern matching to arbitrary data structures. In OMeta there are hints of functional language.
Amethyst provides several high-level constructs known from functional languages (lambdas, trackable state). A goal is to make grammars made in amethyst more maintainable.

We introduce a framework to  make  parsing dynamic (Chapter \ref{dynparsing}), probably the first time this has been done. An editor can add or remove characters and obtains updates to a syntax tree. A dynamic parser recomputes only rules it needs to recompute. For typical workloads a change takes only $O(\log n)$ time. 
One application will be to make syntax highlighting  and other tools easier to write and more accurate. 

As a step in experimenting with language design we created a simple dynamic programming language called Peridot.

\chapter{Amethyst language}

Amethyst is a pattern matching system.

The purpose of this chapter is to introduce amethyst language and to teach how to use it effectively. 
We describe amethyst as an evolution of concepts from various pattern matching systems. We will progressively see more hints of functional programming style. In fact amethyst turned out to be full functional language.

Our starting point are regular expressions and we will visit several different settings, each more general than the previous one.

Then we move to the problem of parsing with the focus on top-down parsing. We introduce  \peg{} parser and generalize it to more flexible \regreg{} parser. (In fact, \regreg{} stands for relativized regular expressions.)
Next we can move into pattern matching of tree-like structures.  
Finally we generalize our pattern matching to objects that can form arbitrary graphs.

\section{Notation}

For better readability our examples use syntax highlighting. We also use the following notational conventions:

\begin{exambox}
An example code is enclosed in a box like this.\\
In examples an result of expression is written with the following syntax:\\
\verb$2+2 #-> 4$
\end{exambox}

\begin{prolbox}
Most of the amethyst functionality is implemented by normal amethyst code in standard prologue file. We also show portions of standard prologue in boxes with color like this.
\end{prolbox}

\section{Technical prerequisites}

We assume that the reader knows the basic syntax of Ruby language (we give a brief overview in the next section.) and is familiar with the basics of formal language theory.
From Section \ref{sec:parameter} onward we expect an understanding of basic functional programming techniques. 

\newpage
\subsection{Basics of Ruby} \label{basruby}

We show several examples of Ruby expressions that we will use in later sections.

Arithmetic offers no surprise:
\vskip -1.8em\begin{spacing}{0.8}
{\small
\begin{exambox}
\Verb$#$\Verb$ $\Verb$T$\Verb$h$\Verb$i$\Verb$s$\Verb$ $\Verb$i$\Verb$s$\Verb$ $\Verb$a$\Verb$ $\Verb$c$\Verb$o$\Verb$m$\Verb$m$\Verb$e$\Verb$n$\Verb$t$\Verb$.$\\\Verb$#$\Verb$ $\Verb$E$\Verb$x$\Verb$p$\Verb$e$\Verb$c$\Verb$t$\Verb$e$\Verb$d$\Verb$ $\Verb$r$\Verb$e$\Verb$s$\Verb$u$\Verb$l$\Verb$t$\Verb$s$\Verb$ $\Verb$o$\Verb$f$\Verb$ $\Verb$e$\Verb$x$\Verb$p$\Verb$r$\Verb$e$\Verb$s$\Verb$s$\Verb$i$\Verb$o$\Verb$n$\Verb$s$\Verb$ $\Verb$a$\Verb$r$\Verb$e$\Verb$ $\Verb$d$\Verb$e$\Verb$n$\Verb$o$\Verb$t$\Verb$e$\Verb$d$\Verb$ $\Verb$b$\Verb$y$\\\Verb$#$\Verb$ $\Verb$c$\Verb$o$\Verb$m$\Verb$m$\Verb$e$\Verb$n$\Verb$t$\Verb$ $\Verb$#$\Verb$-$\Verb$>$\Verb$ $\Verb$r$\Verb$e$\Verb$s$\Verb$u$\Verb$l$\Verb$t$\\\Verb$x$\Verb$=$\Verb$2$\\\Verb$y$\Verb$=$\Verb$3$\\\Verb$x$\Verb$+$\Verb$y$\Verb$*$\Verb$y$\Verb$ $\Verb$#$\Verb$-$\Verb$>$\Verb$ $\Verb$1$\Verb$1$
\end{exambox}
}
\end{spacing}\vskip -0.4em

Function definition and function call are written as follows:
\vskip -1.8em\begin{spacing}{0.8}
{\small
\begin{exambox}
\Verb$d$\Verb$e$\Verb$f$\Verb$ $\Verb$p$\Verb$y$\Verb$t$\Verb$h$\Verb$($\Verb$x$\Verb$,$\Verb$y$\Verb$)$\\\Verb$ $\Verb$x$\Verb$*$\Verb$x$\Verb$+$\Verb$y$\Verb$*$\Verb$y$\\\Verb$e$\Verb$n$\Verb$d$\\\Verb$p$\Verb$y$\Verb$t$\Verb$h$\Verb$($\Verb$3$\Verb$,$\Verb$4$\Verb$)$\Verb$ $\Verb$#$\Verb$-$\Verb$>$\Verb$ $\Verb$2$\Verb$5$
\end{exambox}
}
\end{spacing}\vskip -0.4em

We use the following operations with arrays:

\vskip -1.8em\begin{spacing}{0.8}
{\small
\begin{exambox}
\Verb$[$\Verb$1$\Verb$,$\Verb$2$\Verb$]$\Verb$+$\Verb$[$\Verb$3$\Verb$,$\Verb$4$\Verb$]$\Verb$ $\Verb$#$\Verb$-$\Verb$>$\Verb$ $\Verb$[$\Verb$1$\Verb$,$\Verb$2$\Verb$,$\Verb$3$\Verb$,$\Verb$4$\Verb$]$\\\Verb$x$\Verb$ $\Verb$=$\Verb$ $\Verb$[$\Verb$1$\Verb$,$\Verb$2$\Verb$]$\\\Verb$#$\Verb$ $\Verb$W$\Verb$e$\Verb$ $\Verb$u$\Verb$s$\Verb$e$\Verb$ $\Verb$s$\Verb$p$\Verb$l$\Verb$a$\Verb$t$\Verb$ $\Verb$o$\Verb$p$\Verb$e$\Verb$r$\Verb$a$\Verb$t$\Verb$o$\Verb$r$\Verb$ $\Verb$t$\Verb$o$\Verb$ $\Verb$e$\Verb$x$\Verb$p$\Verb$a$\Verb$n$\Verb$d$\Verb$ $\Verb$a$\Verb$r$\Verb$r$\Verb$a$\Verb$y$\Verb$s$\\\Verb$[$\Verb$ $\Verb$x$\Verb$,$\Verb$ $\Verb$*$\Verb$x$\Verb$]$\Verb$ $\Verb$#$\Verb$-$\Verb$>$\Verb$ $\Verb$[$\Verb$[$\Verb$1$\Verb$,$\Verb$2$\Verb$]$\Verb$,$\Verb$1$\Verb$,$\Verb$2$\Verb$]$\\\Verb$#$\Verb$ $\Verb$S$\Verb$p$\Verb$l$\Verb$a$\Verb$t$\Verb$ $\Verb$c$\Verb$a$\Verb$n$\Verb$ $\Verb$b$\Verb$e$\Verb$ $\Verb$u$\Verb$s$\Verb$e$\Verb$d$\Verb$ $\Verb$i$\Verb$n$\Verb$ $\Verb$f$\Verb$u$\Verb$n$\Verb$c$\Verb$t$\Verb$i$\Verb$o$\Verb$n$\Verb$ $\Verb$c$\Verb$a$\Verb$l$\Verb$l$\Verb$s$\\\Verb$p$\Verb$y$\Verb$t$\Verb$h$\Verb$($\Verb$*$\Verb$x$\Verb$)$\Verb$ $\Verb$#$\Verb$-$\Verb$>$\Verb$ $\Verb$5$
\end{exambox}
}
\end{spacing}\vskip -0.4em

A {\it closure} is an important concept from functional programing \cite{steele}. Both Ruby and amethyst use closures. An example in Ruby follows:

\vskip -1.8em\begin{spacing}{0.8}
{\small
\begin{exambox}
\Verb$d$\Verb$e$\Verb$f$\Verb$ $\Verb$f$\Verb$o$\Verb$o$\Verb$($\Verb$x$\Verb$,$\Verb$y$\Verb$)$\\\Verb$ $\Verb$ $\Verb$p$\Verb$r$\Verb$o$\Verb$c$\Verb${$\\\Verb$ $\Verb$ $\Verb$ $\Verb$ $\Verb$x$\Verb$=$\Verb$x$\Verb$+$\Verb$1$\\\Verb$ $\Verb$ $\Verb$ $\Verb$ $\Verb$x$\Verb$+$\Verb$y$\\\Verb$ $\Verb$ $\Verb$}$\\\Verb$e$\Verb$n$\Verb$d$\\\Verb$x$\Verb$=$\Verb$1$\\\Verb$z$\Verb$=$\Verb$f$\Verb$o$\Verb$o$\Verb$($\Verb$x$\Verb$,$\Verb$2$\Verb$)$\Verb$ $\\\Verb$z$\Verb$.$\Verb$c$\Verb$a$\Verb$l$\Verb$l$\Verb$ $\Verb$ $\Verb$ $\Verb$ $\Verb$ $\Verb$#$\Verb$-$\Verb$>$\Verb$ $\Verb$4$\\\Verb$x$\Verb$=$\Verb$0$\\\Verb$z$\Verb$.$\Verb$c$\Verb$a$\Verb$l$\Verb$l$\Verb$ $\Verb$ $\Verb$ $\Verb$ $\Verb$ $\Verb$#$\Verb$-$\Verb$>$\Verb$ $\Verb$5$
\end{exambox}
}
\end{spacing}\vskip -0.4em

\subsection{Getting sources}

A source of amethyst can be obtained from git repository by the following command:

\begin{verbatim}
git clone git://github.com/neleai/mthyst.git
\end{verbatim}

This thesis refers to a version of amethyst that can be obtained by running the following command:

\begin{verbatim}
git checkout thesis
\end{verbatim}

Examples used in this thesis are also in the \verb$doc$ directory of the amethyst.

The peridot language can be obtained by:

\begin{verbatim}
git clone git://github.com/neleai/peridot.git
\end{verbatim}

Installation and running instructions are in \verb$README$ files.

\newpage
\subsection{Using amethyst}

To use amethyst in a Ruby program you need to load it first:
\vskip -1.8em\begin{spacing}{0.8}
{\small
\begin{exambox}
\Verb$r$\Verb$e$\Verb$q$\Verb$u$\Verb$i$\Verb$r$\Verb$e$\Verb$ $\Verb$'$\Verb$a$\Verb$m$\Verb$e$\Verb$t$\Verb$h$\Verb$y$\Verb$s$\Verb$t$\Verb$'$
\end{exambox}
}
\end{spacing}\vskip -0.4em

Then you can load amethyst source files as in the following example:

\vskip -1.8em\begin{spacing}{0.8}
{\small
\begin{exambox}
\Verb$A$\Verb$m$\Verb$e$\Verb$t$\Verb$h$\Verb$y$\Verb$s$\Verb$t$\Verb$:$\Verb$:$\Verb$f$\Verb$i$\Verb$l$\Verb$e$\Verb$ $\Verb$'$\Verb$e$\Verb$x$\Verb$a$\Verb$m$\Verb$p$\Verb$l$\Verb$e$\Verb$1$\Verb$.$\Verb$a$\Verb$m$\Verb$e$\Verb$'$
\end{exambox}
}
\end{spacing}\vskip -0.4em

An amethyst source file is a Ruby source file except for grammar definitions with the following syntax:

\vskip -1.8em\begin{spacing}{0.8}
{\small
\begin{exambox}
\Verb$a$\Verb$m$\Verb$e$\Verb$t$\Verb$h$\Verb$y$\Verb$s$\Verb$t$\Verb$ $\Verb$G$\Verb$r$\Verb$a$\Verb$m$\Verb$m$\Verb$a$\Verb$r$\Verb${$\\\Verb$ $\Verb$ ${\color{red}\Verb$r$}{\color{red}\Verb$u$}{\color{red}\Verb$l$}{\color{red}\Verb$e$}{\color{red}\Verb$s$}\\\Verb$}$
\end{exambox}
}
\end{spacing}\vskip -0.4em

Hello world program in parser generators is a simple calculator. We follow this tradition too. Constructions used will be the topic of the following chapters.

The source consist amethyst source file \verb$calculator.ame$:

\vskip -1.8em\begin{spacing}{0.8}
{\small
\begin{exambox}
\Verb$a$\Verb$m$\Verb$e$\Verb$t$\Verb$h$\Verb$y$\Verb$s$\Verb$t$\Verb$ $\Verb$C$\Verb$a$\Verb$l$\Verb$c$\Verb$u$\Verb$l$\Verb$a$\Verb$t$\Verb$o$\Verb$r$\Verb$ $\Verb${$\\\Verb$ $\Verb$ ${\color{red}\Verb$c$}{\color{red}\Verb$a$}{\color{red}\Verb$l$}{\color{red}\Verb$c$}{\color{red}\Verb$u$}{\color{red}\Verb$l$}{\color{red}\Verb$a$}{\color{red}\Verb$t$}{\color{red}\Verb$e$}\Verb$ $\Verb$=$\Verb$ ${\color{red}\Verb$a$}{\color{red}\Verb$d$}{\color{red}\Verb$d$}{\color{red}\Verb$_$}{\color{red}\Verb$e$}{\color{red}\Verb$x$}{\color{red}\Verb$p$}{\color{red}\Verb$r$}\Verb$ $\Verb$ $\\\\\Verb$ $\Verb$ ${\color{red}\Verb$a$}{\color{red}\Verb$d$}{\color{red}\Verb$d$}{\color{red}\Verb$_$}{\color{red}\Verb$e$}{\color{red}\Verb$x$}{\color{red}\Verb$p$}{\color{red}\Verb$r$}\Verb$ $\Verb$ $\Verb$=$\Verb$ ${\color{red}\Verb$a$}{\color{red}\Verb$d$}{\color{red}\Verb$d$}{\color{red}\Verb$_$}{\color{red}\Verb$e$}{\color{red}\Verb$x$}{\color{red}\Verb$p$}{\color{red}\Verb$r$}{\color{Aquamarine}\Verb$:$}{\color{Aquamarine}\Verb$x$}\Verb$ ${\color{black}\Verb$"$}{\color{black}\Verb$+$}{\color{black}\Verb$"$}\Verb$ ${\color{red}\Verb$m$}{\color{red}\Verb$u$}{\color{red}\Verb$l$}{\color{red}\Verb$_$}{\color{red}\Verb$e$}{\color{red}\Verb$x$}{\color{red}\Verb$p$}{\color{red}\Verb$r$}{\color{Aquamarine}\Verb$:$}{\color{Aquamarine}\Verb$y$}\Verb$ $\Verb$ $\Verb$ ${\color{Tan}\Verb$-$}{\color{Tan}\Verb$>$}{\color{Tan}\Verb$ $}{\color{Aquamarine}\Verb$x$}{\color{Tan}\Verb$+$}{\color{Aquamarine}\Verb$y$}{\color{Tan}\\}{\color{Tan}\Verb$ $}\Verb$ $\Verb$ $\Verb$ $\Verb$ $\Verb$ $\Verb$ $\Verb$ $\Verb$ $\Verb$ $\Verb$ $\Verb$ $\Verb$|$\Verb$ ${\color{red}\Verb$a$}{\color{red}\Verb$d$}{\color{red}\Verb$d$}{\color{red}\Verb$_$}{\color{red}\Verb$e$}{\color{red}\Verb$x$}{\color{red}\Verb$p$}{\color{red}\Verb$r$}{\color{Aquamarine}\Verb$:$}{\color{Aquamarine}\Verb$x$}\Verb$ ${\color{black}\Verb$"$}{\color{black}\Verb$-$}{\color{black}\Verb$"$}\Verb$ ${\color{red}\Verb$m$}{\color{red}\Verb$u$}{\color{red}\Verb$l$}{\color{red}\Verb$_$}{\color{red}\Verb$e$}{\color{red}\Verb$x$}{\color{red}\Verb$p$}{\color{red}\Verb$r$}{\color{Aquamarine}\Verb$:$}{\color{Aquamarine}\Verb$y$}\Verb$ $\Verb$ $\Verb$ ${\color{Tan}\Verb$-$}{\color{Tan}\Verb$>$}{\color{Tan}\Verb$ $}{\color{Aquamarine}\Verb$x$}{\color{Tan}\Verb$-$}{\color{Aquamarine}\Verb$y$}{\color{Tan}\\}{\color{Tan}\Verb$ $}\Verb$ $\Verb$ $\Verb$ $\Verb$ $\Verb$ $\Verb$ $\Verb$ $\Verb$ $\Verb$ $\Verb$ $\Verb$ $\Verb$|$\Verb$ ${\color{red}\Verb$m$}{\color{red}\Verb$u$}{\color{red}\Verb$l$}{\color{red}\Verb$_$}{\color{red}\Verb$e$}{\color{red}\Verb$x$}{\color{red}\Verb$p$}{\color{red}\Verb$r$}\\\\\Verb$ $\Verb$ ${\color{red}\Verb$m$}{\color{red}\Verb$u$}{\color{red}\Verb$l$}{\color{red}\Verb$_$}{\color{red}\Verb$e$}{\color{red}\Verb$x$}{\color{red}\Verb$p$}{\color{red}\Verb$r$}\Verb$ $\Verb$ $\Verb$=$\Verb$ ${\color{red}\Verb$m$}{\color{red}\Verb$u$}{\color{red}\Verb$l$}{\color{red}\Verb$_$}{\color{red}\Verb$e$}{\color{red}\Verb$x$}{\color{red}\Verb$p$}{\color{red}\Verb$r$}{\color{Aquamarine}\Verb$:$}{\color{Aquamarine}\Verb$x$}\Verb$ ${\color{black}\Verb$"$}{\color{black}\Verb$*$}{\color{black}\Verb$"$}\Verb$ ${\color{red}\Verb$a$}{\color{red}\Verb$t$}{\color{red}\Verb$o$}{\color{red}\Verb$m$}{\color{red}\Verb$_$}{\color{red}\Verb$e$}{\color{red}\Verb$x$}{\color{red}\Verb$p$}{\color{red}\Verb$r$}{\color{Aquamarine}\Verb$:$}{\color{Aquamarine}\Verb$y$}\Verb$ $\Verb$ ${\color{Tan}\Verb$-$}{\color{Tan}\Verb$>$}{\color{Tan}\Verb$ $}{\color{Aquamarine}\Verb$x$}{\color{Tan}\Verb$*$}{\color{Aquamarine}\Verb$y$}{\color{Tan}\\}{\color{Tan}\Verb$ $}\Verb$ $\Verb$ $\Verb$ $\Verb$ $\Verb$ $\Verb$ $\Verb$ $\Verb$ $\Verb$ $\Verb$ $\Verb$ $\Verb$|$\Verb$ ${\color{red}\Verb$m$}{\color{red}\Verb$u$}{\color{red}\Verb$l$}{\color{red}\Verb$_$}{\color{red}\Verb$e$}{\color{red}\Verb$x$}{\color{red}\Verb$p$}{\color{red}\Verb$r$}{\color{Aquamarine}\Verb$:$}{\color{Aquamarine}\Verb$x$}\Verb$ ${\color{black}\Verb$"$}{\color{black}\Verb$/$}{\color{black}\Verb$"$}\Verb$ ${\color{red}\Verb$a$}{\color{red}\Verb$t$}{\color{red}\Verb$o$}{\color{red}\Verb$m$}{\color{red}\Verb$_$}{\color{red}\Verb$e$}{\color{red}\Verb$x$}{\color{red}\Verb$p$}{\color{red}\Verb$r$}{\color{Aquamarine}\Verb$:$}{\color{Aquamarine}\Verb$y$}\Verb$ $\Verb$ ${\color{Tan}\Verb$-$}{\color{Tan}\Verb$>$}{\color{Tan}\Verb$ $}{\color{Aquamarine}\Verb$x$}{\color{Tan}\Verb$/$}{\color{Aquamarine}\Verb$y$}{\color{Tan}\\}{\color{Tan}\Verb$ $}\Verb$ $\Verb$ $\Verb$ $\Verb$ $\Verb$ $\Verb$ $\Verb$ $\Verb$ $\Verb$ $\Verb$ $\Verb$ $\Verb$|$\Verb$ ${\color{red}\Verb$a$}{\color{red}\Verb$t$}{\color{red}\Verb$o$}{\color{red}\Verb$m$}{\color{red}\Verb$_$}{\color{red}\Verb$e$}{\color{red}\Verb$x$}{\color{red}\Verb$p$}{\color{red}\Verb$r$}\Verb$ $\Verb$ $\Verb$ $\Verb$ $\Verb$ $\Verb$ $\Verb$ $\Verb$ $\Verb$ $\Verb$ $\Verb$ $\\\Verb$ $\\\Verb$ $\Verb$ ${\color{red}\Verb$a$}{\color{red}\Verb$t$}{\color{red}\Verb$o$}{\color{red}\Verb$m$}{\color{red}\Verb$_$}{\color{red}\Verb$e$}{\color{red}\Verb$x$}{\color{red}\Verb$p$}{\color{red}\Verb$r$}\Verb$ $\Verb$=$\Verb$ ${\color{black}\Verb$"$}{\color{black}\Verb$($}{\color{black}\Verb$"$}\Verb$ ${\color{red}\Verb$a$}{\color{red}\Verb$d$}{\color{red}\Verb$d$}{\color{red}\Verb$_$}{\color{red}\Verb$e$}{\color{red}\Verb$x$}{\color{red}\Verb$p$}{\color{red}\Verb$r$}{\color{Aquamarine}\Verb$:$}{\color{Aquamarine}\Verb$x$}\Verb$ ${\color{black}\Verb$"$}{\color{black}\Verb$)$}{\color{black}\Verb$"$}\Verb$ $\Verb$ $\Verb$ $\Verb$ $\Verb$ $\Verb$ $\Verb$ $\Verb$ $\Verb$ $\Verb$ ${\color{Tan}\Verb$-$}{\color{Tan}\Verb$>$}{\color{Tan}\Verb$ $}{\color{Tan}\Verb$ $}{\color{Aquamarine}\Verb$x$}{\color{Tan}\\}{\color{Tan}\Verb$ $}\Verb$ $\Verb$ $\Verb$ $\Verb$ $\Verb$ $\Verb$ $\Verb$ $\Verb$ $\Verb$ $\Verb$ $\Verb$ $\Verb$|$\Verb$ ${\color{red}\Verb$f$}{\color{red}\Verb$l$}{\color{red}\Verb$o$}{\color{red}\Verb$a$}{\color{red}\Verb$t$}\\\Verb$}$\\\Verb$p$\Verb$u$\Verb$t$\Verb$s$\Verb$ $\Verb$C$\Verb$a$\Verb$l$\Verb$c$\Verb$u$\Verb$l$\Verb$a$\Verb$t$\Verb$o$\Verb$r$\Verb$.$\Verb$c$\Verb$a$\Verb$l$\Verb$c$\Verb$u$\Verb$l$\Verb$a$\Verb$t$\Verb$e$\Verb$($\Verb$"$\Verb$2$\Verb$-$\Verb$4$\Verb$+$\Verb$2$\Verb$*$\Verb$2$\Verb$-$\Verb$-$\Verb$2$\Verb$"$\Verb$)$\Verb$ $\Verb$#$\Verb$-$\Verb$>$\Verb$4$
\end{exambox}
}
\end{spacing}\vskip -0.4em

and Ruby source file \verb$amethyst.rb$:

\vskip -1.8em\begin{spacing}{0.8}
{\small
\begin{exambox}
\Verb$r$\Verb$e$\Verb$q$\Verb$u$\Verb$i$\Verb$r$\Verb$e$\Verb$ $\Verb$'$\Verb$a$\Verb$m$\Verb$e$\Verb$t$\Verb$h$\Verb$y$\Verb$s$\Verb$t$\Verb$'$\\\Verb$A$\Verb$m$\Verb$e$\Verb$t$\Verb$h$\Verb$y$\Verb$s$\Verb$t$\Verb$:$\Verb$:$\Verb$f$\Verb$i$\Verb$l$\Verb$e$\Verb$ $\Verb$'$\Verb$c$\Verb$a$\Verb$l$\Verb$c$\Verb$u$\Verb$l$\Verb$a$\Verb$t$\Verb$o$\Verb$r$\Verb$.$\Verb$a$\Verb$m$\Verb$e$\Verb$'$\\\Verb$w$\Verb$h$\Verb$i$\Verb$l$\Verb$e$\Verb$ $\Verb$t$\Verb$r$\Verb$u$\Verb$e$\\\Verb$ $\Verb$ $\Verb$i$\Verb$n$\Verb$p$\Verb$u$\Verb$t$\Verb$ $\Verb$=$\Verb$ $\Verb$g$\Verb$e$\Verb$t$\Verb$s$\\\Verb$ $\Verb$ $\Verb$p$\Verb$u$\Verb$t$\Verb$s$\Verb$ $\Verb$C$\Verb$a$\Verb$l$\Verb$c$\Verb$u$\Verb$l$\Verb$a$\Verb$t$\Verb$o$\Verb$r$\Verb$.$\Verb$c$\Verb$a$\Verb$l$\Verb$c$\Verb$u$\Verb$l$\Verb$a$\Verb$t$\Verb$e$\Verb$($\Verb$i$\Verb$n$\Verb$p$\Verb$u$\Verb$t$\Verb$)$\\\Verb$e$\Verb$n$\Verb$d$
\end{exambox}
}
\end{spacing}\vskip -0.4em

The file \verb$calculator.rb$ is run by the command:

\begin{verbatim}
ruby calculator.rb
\end{verbatim}

For tasks where a simple expression suffices, defining full grammar is not necessary. 
We can enclose arbitrary amethyst expression {\color{blue}\Verb$e$} in the following construction: \Verb$($\Verb$|$\Verb$ ${\color{blue}\Verb$e$}\Verb$ $\Verb$|$\Verb$)$. This creates an object that can be handled in a similar way as a regular expression. So instead of writing:

\vskip -1.8em\begin{spacing}{0.8}
{\small
\begin{exambox}
\Verb$a$\Verb$m$\Verb$e$\Verb$t$\Verb$h$\Verb$y$\Verb$s$\Verb$t$\Verb$ $\Verb$H$\Verb$e$\Verb$l$\Verb$l$\Verb$o$\Verb$_$\Verb$W$\Verb$o$\Verb$r$\Verb$l$\Verb$d$\Verb$ $\Verb${$\\\Verb$ $\Verb$ ${\color{red}\Verb$h$}{\color{red}\Verb$e$}{\color{red}\Verb$l$}{\color{red}\Verb$l$}{\color{red}\Verb$o$}\Verb$ $\Verb$=$\Verb$ ${\color{black}\Verb$'$}{\color{black}\Verb$h$}{\color{black}\Verb$e$}{\color{black}\Verb$l$}{\color{black}\Verb$l$}{\color{black}\Verb$o$}{\color{black}\Verb$'$}\\\Verb$ $\Verb$ ${\color{red}\Verb$w$}{\color{red}\Verb$o$}{\color{red}\Verb$r$}{\color{red}\Verb$l$}{\color{red}\Verb$d$}\Verb$ $\Verb$=$\Verb$ ${\color{black}\Verb$'$}{\color{black}\Verb$w$}{\color{black}\Verb$o$}{\color{black}\Verb$r$}{\color{black}\Verb$l$}{\color{black}\Verb$d$}{\color{black}\Verb$'$}\\\Verb$ $\Verb$ ${\color{red}\Verb$h$}{\color{red}\Verb$e$}{\color{red}\Verb$l$}{\color{red}\Verb$l$}{\color{red}\Verb$o$}{\color{red}\Verb$_$}{\color{red}\Verb$w$}{\color{red}\Verb$o$}{\color{red}\Verb$r$}{\color{red}\Verb$l$}{\color{red}\Verb$d$}\Verb$ $\Verb$=$\Verb$ ${\color{red}\Verb$h$}{\color{red}\Verb$e$}{\color{red}\Verb$l$}{\color{red}\Verb$l$}{\color{red}\Verb$o$}\Verb$ ${\color{black}\Verb$'$}{\color{black}\Verb$ $}{\color{black}\Verb$'$}\Verb$ ${\color{red}\Verb$w$}{\color{red}\Verb$o$}{\color{red}\Verb$r$}{\color{red}\Verb$l$}{\color{red}\Verb$d$}\\\Verb$}$\\\Verb$H$\Verb$e$\Verb$l$\Verb$l$\Verb$o$\Verb$_$\Verb$W$\Verb$o$\Verb$r$\Verb$l$\Verb$d$\Verb$.$\Verb$h$\Verb$e$\Verb$l$\Verb$l$\Verb$o$\Verb$_$\Verb$w$\Verb$o$\Verb$r$\Verb$l$\Verb$d$\Verb$($\Verb$s$\Verb$)$
\end{exambox}
}
\end{spacing}\vskip -0.4em

one can write:

\vskip -1.8em\begin{spacing}{0.8}
{\small
\begin{exambox}
{\color{Violet}\Verb$($}{\color{Violet}\Verb$|$}\Verb$ ${\color{black}\Verb$'$}{\color{black}\Verb$h$}{\color{black}\Verb$e$}{\color{black}\Verb$l$}{\color{black}\Verb$l$}{\color{black}\Verb$o$}{\color{black}\Verb$ $}{\color{black}\Verb$w$}{\color{black}\Verb$o$}{\color{black}\Verb$r$}{\color{black}\Verb$l$}{\color{black}\Verb$d$}{\color{black}\Verb$'$}\Verb$ ${\color{Violet}\Verb$|$}{\color{Violet}\Verb$)$}\Verb$ $\Verb$=$\Verb$=$\Verb$=$\Verb$ $\Verb$s$
\end{exambox}
}
\end{spacing}\vskip -0.4em

\newpage
\section{Regular expressions}

Regular expressions provide a way to match strings of text and are widely supported by many languages and libraries.
They extend search and replace functionality of text editors like \verb$vi$.
Typically implementations add nonstandard extensions which we will not consider in this work.

Regular expressions can be formed recursively: An expression can be an atomic expression that can not be decomposed (and typically matches single character) or expressions composed from smaller expressions by some operator.

\subsection*{Atomic expressions}

\vskip 0.2em
\vskip 0.2em\noindent\begin{tabular}{| l | p{9.5cm} |}
\hline
Syntax & Description \\
\hline
\verb$c                   $& Match character c\footnotemark[1].\\
\verb$. $& Match arbitrary character.\\
\verb$[group]$ & Match character described in character group.\\
\hline
\end{tabular}\vskip 0.2em
\vskip 0.2em
\footnotetext[1]{Unless c has special meaning in which case you have to escape it.}
\subsection*{Operators}

\vskip 0.2em\noindent\begin{tabular}{| l | p{9.5cm} |}
\hline
{\color{blue}\Verb$e$}{\color{red}\Verb$1$}{\color{blue}\Verb$e$}{\color{blue}\Verb$2$}            &Sequencing\\
{\color{blue}\Verb$e$}{\color{blue}\Verb$1$}\Verb$|${\color{blue}\Verb$e$}{\color{blue}\Verb$2$}\verb$               $&Choice\\
\Verb$(${\color{blue}\Verb$e$}\Verb$)$                      &Grouping\\
{\color{blue}\Verb$e$}{\color{black}\Verb$*$} & Iteration: match {\color{blue}\Verb$e$} 0 or more times.\\
{\color{blue}\Verb$e$}{\color{black}\Verb$+$} & Iteration: match {\color{blue}\Verb$e$} 1 or more times.\\
{\color{blue}\Verb$e$}{\color{black}\Verb$?$} & Iteration: match {\color{blue}\Verb$e$} 0 or 1    times.\\
\hline
\end{tabular}\vskip 0.2em

For example, the expression \verb$[Hh]ello (world|worlds)$ matches the strings \\
\extext{Hello world}, \extext{hello world}, \extext{Hello worlds}, \extext{hello worlds}.

In Ruby regular expressions are enclosed by ``\verb$/$". We match the example above against the string \extext{hello world} by writing:

\vskip -1.8em\begin{spacing}{0.8}
{\small
\begin{exambox}
\Verb$/$\Verb$[$\Verb$H$\Verb$h$\Verb$]$\Verb$e$\Verb$l$\Verb$l$\Verb$o$\Verb$ $\Verb$($\Verb$w$\Verb$o$\Verb$r$\Verb$l$\Verb$d$\Verb$|$\Verb$w$\Verb$o$\Verb$r$\Verb$l$\Verb$d$\Verb$s$\Verb$)$\Verb$/$\Verb$ $\Verb$=$\Verb$=$\Verb$=$\Verb$ $\Verb$"$\Verb$h$\Verb$e$\Verb$l$\Verb$l$\Verb$o$\Verb$ $\Verb$w$\Verb$o$\Verb$r$\Verb$l$\Verb$d$\Verb$"$
\end{exambox}
}
\end{spacing}\vskip -0.4em


Note that the space is also matched literary. This becomes problematic for more complex expressions as they can not be reformatted.

\section{Amethyst grammars and expressions}

The syntax of a regular expression and its equivalent amethyst expression is similar. 

We embed amethyst {\color{blue} expressions} with \Verb$($\Verb$|$\Verb$ ${\color{blue}\Verb$e$}\Verb$ $\Verb$|$\Verb$)$ syntax. The example from previous section becomes:

\vskip -1.8em\begin{spacing}{0.8}
{\small
\begin{exambox}
{\color{Violet}\Verb$($}{\color{Violet}\Verb$|$}\Verb$ $\Verb$<$\Verb$H$\Verb$h$\Verb$>$\Verb$ ${\color{black}\Verb$'$}{\color{black}\Verb$e$}{\color{black}\Verb$l$}{\color{black}\Verb$l$}{\color{black}\Verb$o$}{\color{black}\Verb$ $}{\color{black}\Verb$'$}\Verb$ $\Verb$(${\color{black}\Verb$'$}{\color{black}\Verb$w$}{\color{black}\Verb$o$}{\color{black}\Verb$r$}{\color{black}\Verb$l$}{\color{black}\Verb$d$}{\color{black}\Verb$'$}\Verb$|${\color{black}\Verb$'$}{\color{black}\Verb$w$}{\color{black}\Verb$o$}{\color{black}\Verb$r$}{\color{black}\Verb$l$}{\color{black}\Verb$d$}{\color{black}\Verb$s$}{\color{black}\Verb$'$}\Verb$)$\Verb$ ${\color{Violet}\Verb$|$}{\color{Violet}\Verb$)$}\Verb$ $\Verb$=$\Verb$=$\Verb$=$\Verb$ $\Verb$"$\Verb$h$\Verb$e$\Verb$l$\Verb$l$\Verb$o$\Verb$ $\Verb$w$\Verb$o$\Verb$r$\Verb$l$\Verb$d$\Verb$"$
\end{exambox}
}
\end{spacing}\vskip -0.4em

Amethyst is whitespace insensitive. We need to enclose matched strings with single quotes. The reason why \verb$[]$ turns to \verb$<>$ will be explained in Section \ref{matchtree}.

\subsection*{Grammars}

Expressions are useful for simple tasks. More complicated tasks are described by {\it grammars}.
Amethyst source file consist of grammars that contain {\it rules}. Syntax of rule definition and calls is the following:

\noindent\begin{tabular}{| l  || p{11.5cm} |}
\hline
Pattern & Description \\
\hline
{\color{red}\Verb$n$}{\color{red}\Verb$a$}{\color{red}\Verb$m$}{\color{red}\Verb$e$}\Verb$ $\Verb$ $\Verb$=$\Verb$ ${\color{blue}\Verb$e$} &  Rule definition\\
{\color{red}\Verb$n$}{\color{red}\Verb$a$}{\color{red}\Verb$m$}{\color{red}\Verb$e$} & Rule call\\
\hline
\end{tabular}\vskip 0.2em

If we want hello world program to be whitespace insensitive we can write it as:

\vskip -1.8em\begin{spacing}{0.8}
{\small
\begin{exambox}
\Verb$a$\Verb$m$\Verb$e$\Verb$t$\Verb$h$\Verb$y$\Verb$s$\Verb$t$\Verb$ $\Verb$G$\Verb$r$\Verb$a$\Verb$m$\Verb$m$\Verb$a$\Verb$r$\Verb$ $\Verb${$\\\Verb$ $\Verb$ ${\color{red}\Verb$s$}{\color{red}\Verb$p$}{\color{red}\Verb$a$}{\color{red}\Verb$c$}{\color{red}\Verb$e$}\Verb$ $\Verb$ $\Verb$=$\Verb$ $\Verb$<$\Verb$ $\Verb$\$\Verb$t$\Verb$\$\Verb$r$\Verb$\$\Verb$n$\Verb$>$\\\Verb$ $\Verb$ $\\\Verb$ $\Verb$ ${\color{red}\Verb$h$}{\color{red}\Verb$e$}{\color{red}\Verb$l$}{\color{red}\Verb$l$}{\color{red}\Verb$o$}\Verb$ $\Verb$ $\Verb$=$\Verb$ ${\color{black}\Verb$'$}{\color{black}\Verb$h$}{\color{black}\Verb$e$}{\color{black}\Verb$l$}{\color{black}\Verb$l$}{\color{black}\Verb$o$}{\color{black}\Verb$'$}\Verb$ ${\color{red}\Verb$s$}{\color{red}\Verb$p$}{\color{red}\Verb$a$}{\color{red}\Verb$c$}{\color{red}\Verb$e$}{\color{black}\Verb$+$}\Verb$ ${\color{black}\Verb$'$}{\color{black}\Verb$w$}{\color{black}\Verb$o$}{\color{black}\Verb$r$}{\color{black}\Verb$l$}{\color{black}\Verb$d$}{\color{black}\Verb$'$}\\\Verb$}$
\end{exambox}
}
\end{spacing}\vskip -0.4em

As a less trivial example we show  rules recognizing integers:\footnote{If we do not care what this rule returns.}

\vskip -1.8em\begin{spacing}{0.8}
{\small
\begin{exambox}
\Verb$a$\Verb$m$\Verb$e$\Verb$t$\Verb$h$\Verb$y$\Verb$s$\Verb$t$\Verb$ $\Verb$G$\Verb$r$\Verb$a$\Verb$m$\Verb$m$\Verb$a$\Verb$r$\Verb$2$\Verb$ $\Verb${$\\\Verb$ $\Verb$ ${\color{red}\Verb$d$}{\color{red}\Verb$i$}{\color{red}\Verb$g$}{\color{red}\Verb$i$}{\color{red}\Verb$t$}\Verb$ $\Verb$ $\Verb$=$\Verb$ $\Verb$<$\Verb$0$\Verb$-$\Verb$9$\Verb$>$\\\Verb$ $\Verb$ ${\color{red}\Verb$i$}{\color{red}\Verb$n$}{\color{red}\Verb$t$}\Verb$ $\Verb$ $\Verb$ $\Verb$ $\Verb$=$\Verb$ ${\color{black}\Verb$'$}{\color{black}\Verb$-$}{\color{black}\Verb$'$}{\color{black}\Verb$?$}\Verb$ ${\color{red}\Verb$d$}{\color{red}\Verb$i$}{\color{red}\Verb$g$}{\color{red}\Verb$i$}{\color{red}\Verb$t$}{\color{black}\Verb$+$}\\\Verb$}$\\\\\Verb$#$\Verb$ $\Verb$A$\Verb$ $\Verb$r$\Verb$u$\Verb$l$\Verb$e$\Verb$ $\Verb$c$\Verb$a$\Verb$n$\Verb$ $\Verb$b$\Verb$e$\Verb$ $\Verb$i$\Verb$n$\Verb$v$\Verb$o$\Verb$k$\Verb$e$\Verb$d$\Verb$ $\Verb$i$\Verb$n$\Verb$ $\Verb$t$\Verb$h$\Verb$e$\Verb$ $\Verb$f$\Verb$o$\Verb$l$\Verb$l$\Verb$o$\Verb$w$\Verb$i$\Verb$n$\Verb$g$\Verb$ $\Verb$w$\Verb$a$\Verb$y$\Verb$:$\\\Verb$ $\Verb$ $\Verb$ $\Verb$G$\Verb$r$\Verb$a$\Verb$m$\Verb$m$\Verb$a$\Verb$r$\Verb$.$\Verb$i$\Verb$n$\Verb$t$\Verb$($\Verb$ $\Verb$ $\Verb$ $\Verb$ $\Verb$ $\Verb$ $\Verb$ $\Verb$"$\Verb$4$\Verb$2$\Verb$1$\Verb$"$\Verb$)$\Verb$ $\Verb$#$\Verb$-$\Verb$>$\Verb$ $\Verb$[$\Verb$'$\Verb$4$\Verb$'$\Verb$,$\Verb$'$\Verb$2$\Verb$'$\Verb$,$\Verb$'$\Verb$1$\Verb$'$\Verb$]$
\end{exambox}
}
\end{spacing}\vskip -0.4em

\subsection*{Character groups}
Character groups provide a concise way to match single character from given set of characters.

Following constructions are supported:

\vskip 0.2em\noindent\begin{tabular}{| l | l || p{7.8cm} |}
\hline
Regular expression & Amethyst   & Description\\
\hline
\verb$[a]$          & \Verb$<$\Verb$a$\Verb$>$        & Match character \extext{a}\\
\verb$[aeiou]$      & \Verb$<$\Verb$a$\Verb$e$\Verb$i$\Verb$o$\Verb$u$\Verb$>$    & Match any of characters \extext{aeiou}\\
\verb$[a-z]$        & \Verb$<$\Verb$a$\Verb$-$\Verb$z$\Verb$>$      & Match any of characters from \extext{a} to \extext{z}\\
\verb$[^abc]$       & \Verb$<$\Verb$^$\Verb$a$\Verb$b$\Verb$c$\Verb$>$    & Match any character except \extext{abc}\\
\verb$[[:digit:]]$  & \Verb$<$\Verb$<$\Verb$d$\Verb$i$\Verb$g$\Verb$i$\Verb$t$\Verb$>$\Verb$>$ & Match predefined class\\
\hline
\end{tabular}\vskip 0.2em

In definitions above characters ``\verb$<>\$'' have to be escaped.

\noindent There are several predefined rules to match POSIX character classes (\verb$alpha$, \verb$alnum$, \verb$digit$, \ldots).
User can also define custom character class, say vowels:

\begin{exambox}
\begin{spacing}{0.8}
{\small
{\color{red}\Verb$v$}{\color{red}\Verb$o$}{\color{red}\Verb$w$}{\color{red}\Verb$e$}{\color{red}\Verb$l$}{\color{red}\Verb$s$}\Verb$ $\Verb$=$\Verb$ $\Verb$<$\Verb$a$\Verb$e$\Verb$i$\Verb$o$\Verb$u$\Verb$>$
}\end{spacing}

\end{exambox}

And use it as character group class: \verb$<<${\color{red}\verb$vowels$}\verb$>0-9>$. 

\newpage
\section{Amethyst expressions} \label{opers}

Amethyst consist of a small set of core operators. The rest of amethyst syntax is a syntax sugar that is converted to core operators. 
Most of amethyst functionality is done by ordinary rules. These rules are contained in file called {\it standard prologue} in Appendix \ref{append_prolog}. 
We will show relevant parts of standard prologue as an example.

Explaining exact semantic of core operators will take some time. In this section we only briefly summarize core operator syntax. Various aspects of core operators will be covered later.

\subsection*{Basic operators}

Like most pattern matching systems amethyst supports following operators:

\vskip 0.2em\noindent\begin{tabular}{| l | p{11.5cm} |}
\hline
Name & Description \\
\hline
{\color{blue}\Verb$e$}{\color{blue}\Verb$1$}\Verb$ ${\color{blue}\Verb$e$}{\color{blue}\Verb$2$}&Sequencing  \\
{\color{blue}\Verb$e$}{\color{blue}\Verb$1$}\Verb$|${\color{blue}\Verb$e$}{\color{blue}\Verb$2$}&Priorized choice  \\
\Verb$(${\color{blue}\Verb$e$}\Verb$)$&Grouping   \\
{\color{blue}\Verb$e$}{\color{black}\Verb$*$}\Verb$,$\Verb$ ${\color{blue}\Verb$e$}{\color{black}\Verb$+$}\Verb$,$\Verb$ ${\color{blue}\Verb$e$}{\color{black}\Verb$?$} & Iteration\\
\hline
\end{tabular}\vskip 0.2em

\subsection*{Lookaheads}

When parsing programming languages the  decisions which alternative should be used often depend on the future input. This is done by means of {\it \color{violet} lookaheads}. Due to limited memory of computers in the 1970's  the lookaheads were limited to next token. 
A \peg{} parser relaxes this restriction by allowing unlimited lookahead. Amethyst also supports unlimited lookaheads but with slightly different syntax:

\vskip 0.2em\noindent\begin{tabular}{| l | p{11.5cm} |}
\hline
{\color{blue}\Verb$e$}{\color{blue}\Verb$1$}\Verb$ ${\color{Violet}\Verb$&$}{\color{Violet}\Verb$ $}{\color{blue}\Verb$e$}{\color{blue}\Verb$2$}\verb$  $   & Positive lookahead\\
{\color{Violet}\Verb$~$}{\color{blue}\Verb$e$}& Negative lookahead\\
\hline
\end{tabular}\vskip 0.2em

Positive lookahead is similar to intersection. If input can be matched by {\color{blue}\Verb$e$}{\color{blue}\Verb$1$} then lookahead matches input by {\color{blue}\Verb$e$}{\color{blue}\Verb$2$}, otherwise it fails. 
Negative lookahead succeeds if and only if {\color{blue}\Verb$e$}  fails and consumes no input.

In amethyst integers can be recognized by a rule {\color{red}\Verb$i$}{\color{red}\Verb$n$}{\color{red}\Verb$t$}. Based on first character one can decide if integer is positive or negative.
We can use positive lookaheads to match positive integers and negative lookaheads to match negative integers in the following way:

\vskip -1.8em\begin{spacing}{0.8}
{\small
\begin{exambox}
\Verb$a$\Verb$m$\Verb$e$\Verb$t$\Verb$h$\Verb$y$\Verb$s$\Verb$t$\Verb$ $\Verb$N$\Verb$u$\Verb$m$\Verb$b$\Verb$e$\Verb$r$\Verb$s$\Verb$ $\Verb${$\\\Verb$ $\Verb$ ${\color{red}\Verb$n$}{\color{red}\Verb$e$}{\color{red}\Verb$g$}{\color{red}\Verb$a$}{\color{red}\Verb$t$}{\color{red}\Verb$i$}{\color{red}\Verb$v$}{\color{red}\Verb$e$}{\color{red}\Verb$_$}{\color{red}\Verb$i$}{\color{red}\Verb$n$}{\color{red}\Verb$t$}\Verb$ $\Verb$=$\Verb$ ${\color{Violet}\Verb$~$}{\color{Violet}\Verb$<$}\Verb$0$\Verb$-$\Verb$9${\color{Violet}\Verb$>$}\Verb$ $\Verb$ $\Verb$ ${\color{red}\Verb$i$}{\color{red}\Verb$n$}{\color{red}\Verb$t$}\\\Verb$ $\Verb$ ${\color{red}\Verb$p$}{\color{red}\Verb$o$}{\color{red}\Verb$s$}{\color{red}\Verb$i$}{\color{red}\Verb$t$}{\color{red}\Verb$i$}{\color{red}\Verb$v$}{\color{red}\Verb$e$}{\color{red}\Verb$_$}{\color{red}\Verb$i$}{\color{red}\Verb$n$}{\color{red}\Verb$t$}\Verb$ $\Verb$=$\Verb$ $\Verb$ $\Verb$<$\Verb$1$\Verb$-$\Verb$9$\Verb$>$\Verb$ ${\color{Violet}\Verb$&$}{\color{Violet}\Verb$ $}\Verb$i$\Verb$n$\Verb$t$\\\Verb$}$
\end{exambox}
}
\end{spacing}\vskip -0.4em

\newpage
\subsection*{Atomic expressions}

We represent the following atomic expressions as the calling of a rule:

\vskip 0.2em\noindent\begin{tabular}{| l | l || p{7.2cm} |}
\hline
Atomic expression    & Rule call & Description\\
\hline
\Verb$.$& {\color{red}\Verb$a$}{\color{red}\Verb$n$}{\color{red}\Verb$y$}{\color{red}\Verb$t$}{\color{red}\Verb$h$}{\color{red}\Verb$i$}{\color{red}\Verb$n$}{\color{red}\Verb$g$}   & Match single object(character)\\
{\color{black}\Verb$'$}{\color{black}\Verb$s$}{\color{black}\Verb$t$}{\color{black}\Verb$r$}{\color{black}\Verb$'$}& {\color{red}\Verb$s$}{\color{red}\Verb$e$}{\color{red}\Verb$q$}{\color{green}\Verb$($}{\color{green}\Verb$"$}{\color{green}\Verb$s$}{\color{green}\Verb$t$}{\color{green}\Verb$r$}{\color{green}\Verb$"$}{\color{green}\Verb$)$}   & Match string \verb$str$\\
{\color{black}\Verb$"$}{\color{black}\Verb$s$}{\color{black}\Verb$t$}{\color{black}\Verb$r$}{\color{black}\Verb$"$}& {\color{red}\Verb$t$}{\color{red}\Verb$o$}{\color{red}\Verb$k$}{\color{red}\Verb$e$}{\color{red}\Verb$n$}{\color{green}\Verb$($}{\color{green}\Verb$"$}{\color{green}\Verb$s$}{\color{green}\Verb$t$}{\color{green}\Verb$r$}{\color{green}\Verb$"$}{\color{green}\Verb$)$}       & Match string preceded by whitespaces\\
\Verb$<$\Verb$s$\Verb$t$\Verb$r$\Verb$>$& {\color{red}\Verb$r$}{\color{red}\Verb$e$}{\color{red}\Verb$g$}{\color{red}\Verb$c$}{\color{red}\Verb$h$}{\color{green}\Verb$($}{\color{green}\Verb$"$}{\color{green}\Verb$s$}{\color{green}\Verb$t$}{\color{green}\Verb$r$}{\color{green}\Verb$"$}{\color{green}\Verb$)$} & Like \verb$[str]$ in regular expressions\\
\hline
\end{tabular}\vskip 0.2em

Rules {\color{red}\Verb$a$}{\color{red}\Verb$n$}{\color{red}\Verb$y$}{\color{red}\Verb$t$}{\color{red}\Verb$h$}{\color{red}\Verb$i$}{\color{red}\Verb$n$}{\color{red}\Verb$g$}, {\color{red}\Verb$s$}{\color{red}\Verb$e$}{\color{red}\Verb$q$} and {\color{red}\Verb$r$}{\color{red}\Verb$e$}{\color{red}\Verb$g$}{\color{red}\Verb$c$}{\color{red}\Verb$h$} are implemented as core functionality. Rule {\color{red}\Verb$t$}{\color{red}\Verb$o$}{\color{red}\Verb$k$}{\color{red}\Verb$e$}{\color{red}\Verb$n$} is derived and the relevant part of standard prologue follows:

\vskip -0.5em
{\small
\begin{spacing}{0.8}
\begin{prolbox}
{\color{red}\Verb$_$}\Verb$ $\Verb$ $\Verb$ $\Verb$ $\Verb$ $\Verb$ $\Verb$ $\Verb$ $\Verb$ $\Verb$=$\Verb$ $\Verb$<$\Verb$ $\Verb$\$\Verb$t$\Verb$\$\Verb$r$\Verb$\$\Verb$n$\Verb$>$\\{\color{red}\Verb$t$}{\color{red}\Verb$o$}{\color{red}\Verb$k$}{\color{red}\Verb$e$}{\color{red}\Verb$n$}{\color{green}\Verb$($}{\color{green}\Verb$x$}{\color{green}\Verb$)$}\Verb$ $\Verb$ $\Verb$=$\Verb$ ${\color{red}\Verb$_$}{\color{black}\Verb$*$}\Verb$ ${\color{red}\Verb$s$}{\color{red}\Verb$e$}{\color{red}\Verb$q$}{\color{green}\Verb$($}{\color{green}\Verb$x$}{\color{green}\Verb$)$}
\end{prolbox}\end{spacing}
}

\vskip -0.5em

We recommend reading Appendix \ref{append_prolog} containing standard prologue. It is expected that you will not completely understand some parts now\footnote{Explaining them is the topic of this chapter.}. Make a guess what the unknown parts do. This is the best way how to learn a new language and amethyst is no exception. 

\subsection*{Enter operator}

The operators covered so far deal mainly with matching strings. We need an additional operator {\color{blue} \verb$Enter$} to deal with general (possibly cyclic) data structures.
\verb$Enter$ operator is a powerful tool essential to sections \ref{matchtree} and \ref{objmatch}.

\vskip 0.2em\noindent\begin{tabular}{| l | p{11.7cm} |}
\hline
Name & Description\\
\hline
{\color{blue}\Verb$e$}{\color{blue}\Verb$1$}{\color{blue}\Verb$[$}\Verb$ ${\color{blue}\Verb$e$}{\color{blue}\Verb$2$}\Verb$ ${\color{blue}\Verb$]$}\verb$ $ &  Enter operator.\\
\hline
\end{tabular}\vskip 0.2em

An \verb$Enter$ operator matches {\color{blue}\Verb$e$}{\color{blue}\Verb$1$}. Then it recursively invokes parser to match {\color{blue}\Verb$e$}{\color{blue}\Verb$2$} on the result of {\color{blue}\Verb$e$}{\color{blue}\Verb$1$}.

\verb$Enter$ operator is one of the most important generalizations of amethyst. It allows us to do pattern matching of object with high level of abstraction which is the topic of Section \ref{matchtree} and Section \ref{objmatch}.

\newpage

\section{External interface}

So far we can only decide if an expression matches an input or not. 
To get useful work done we need integrate amethyst with a programming language. Amethyst is designed to be language independent and the particular language that is used is  called the {\it host language}.
 In this thesis we use Ruby as the host language.

In amethyst each expression yields a result. Results can be bound to rule-local variables using {\it \color{Aquamarine} variable binding} and processed by host language expressions which we call {\it \color{Tan} semantic actions}. In shortcuts  {\color{Aquamarine} a} denotes an anonymous variable which does not occur elsewhere.

Functional languages use the notion of referential transparency \cite{reftrans}. Amethyst requires a weaker condition:
execution is done in a persistent way. When an alternative fails we revert all changes it made and pretend they never happened.

\subsection*{Semantic actions and variable binding}

The syntax of  semantic actions and variable binding is the following:

\vskip 0.2em\noindent\begin{tabular}{| l | l || p{7.5cm} |}
\hline
Pattern     &Expansion&Description \\
\hline
{\color{Tan}\Verb${$}{\color{Tan}\Verb$c$}{\color{Tan}\Verb$}$}& core &Semantic action.\\
{\color{Tan}\Verb$-$}{\color{Tan}\Verb$>$}{\color{Tan}\Verb$ $}{\color{Tan}\Verb$c$}{\color{Tan}\Verb$ $}{\color{Tan}\Verb$n$}{\color{Tan}\Verb$e$}{\color{Tan}\Verb$w$}{\color{Tan}\Verb$l$}{\color{Tan}\Verb$i$}{\color{Tan}\Verb$n$}{\color{Tan}\Verb$e$}&{\color{Tan}\Verb${$}{\color{Tan}\Verb$c$}{\color{Tan}\Verb$}$}\verb$           $  & Alternative syntax. \\
{\color{blue}\Verb$e$}{\color{Aquamarine}\Verb$:$}{\color{Aquamarine}\Verb$v$}& core & Variable binding.\\
\hline
\end{tabular}\vskip 0.2em

We use Ruby closure support to capture scope as this example shows:

\vskip -1.8em\begin{spacing}{0.8}
{\small
\begin{exambox}
{\color{Violet}\Verb$($}{\color{Violet}\Verb$|$}\Verb$ ${\color{red}\Verb$i$}{\color{red}\Verb$n$}{\color{red}\Verb$t$}{\color{Aquamarine}\Verb$:$}{\color{Aquamarine}\Verb$x$}\Verb$ ${\color{black}\Verb$"$}{\color{black}\Verb$+$}{\color{black}\Verb$"$}\Verb$ ${\color{red}\Verb$i$}{\color{red}\Verb$n$}{\color{red}\Verb$t$}{\color{Aquamarine}\Verb$:$}{\color{Aquamarine}\Verb$y$}\Verb$ ${\color{Violet}\Verb$|$}{\color{Violet}\Verb$)$}\Verb$.$\Verb$m$\Verb$a$\Verb$t$\Verb$c$\Verb$h$\Verb$($\Verb$"$\Verb$2$\Verb$+$\Verb$2$\Verb$"$\Verb$)$\\\Verb$p$\Verb$u$\Verb$t$\Verb$s$\Verb$ $\Verb$x$\Verb$+$\Verb$y$\Verb$ $\Verb$#$\Verb$-$\Verb$>$\Verb$ $\Verb$4$
\end{exambox}
}
\end{spacing}\vskip -0.4em

\subsection*{Syntax sugar for variable binding}

It is common to collect results in an array or do simple conversions. First be expressed by the following syntax sugar:
\vskip 0.2em\noindent\begin{tabular}{| l | l || p{7.5cm} |}
\hline
{\color{blue}\Verb$e$}{\color{Aquamarine}\Verb$:$}{\color{Tan}\Verb${$}{\color{Tan}\Verb$ $}{\color{Tan}\Verb$c$}{\color{Tan}\Verb$}$}\verb$     $ &{\color{blue}\Verb$e$}{\color{Aquamarine}\Verb$:$}{\color{Aquamarine}\Verb$i$}{\color{Aquamarine}\Verb$t$}\Verb$ ${\color{Tan}\Verb${$}{\color{Tan}\Verb$c$}{\color{Tan}\Verb$}$}\verb$        $& Do conversion using variable {\color{Aquamarine}\verb$it$}.\\
\hline
\end{tabular}\vskip 0.2em

For example, imagine that a third party has written a {\color{red}\Verb$f$}{\color{red}\Verb$l$}{\color{red}\Verb$o$}{\color{red}\Verb$a$}{\color{red}\Verb$t$} rule. Their API however returns the result as a string. If we want to return a number instead we can write:

\noindent\begin{spacing}{0.8}
{\small
{\color{red}\Verb$f$}{\color{red}\Verb$l$}{\color{red}\Verb$o$}{\color{red}\Verb$a$}{\color{red}\Verb$t$}{\color{red}\Verb$2$}\Verb$ $\Verb$=$\Verb$ ${\color{red}\Verb$f$}{\color{red}\Verb$l$}{\color{red}\Verb$o$}{\color{red}\Verb$a$}{\color{red}\Verb$t$}{\color{Aquamarine}\Verb$:$}{\color{Tan}\Verb${$}{\color{Aquamarine}\Verb$i$}{\color{Aquamarine}\Verb$t$}{\color{Tan}\Verb$.$}{\color{Tan}\Verb$t$}{\color{Tan}\Verb$o$}{\color{Tan}\Verb$_$}{\color{Tan}\Verb$f$}{\color{Tan}\Verb$}$}
}\end{spacing}

When collecting results into array a parameter can be an arbitrary host language expression not just variable:

\vskip 0.2em\noindent\begin{tabular}{| l | l || p{7.5cm} |}
\hline
{\color{blue}\Verb$e$}{\color{Aquamarine}\Verb$:$}{\color{Aquamarine}\Verb$[$}{\color{Aquamarine}\Verb$ $}{\color{Aquamarine}\Verb$c$}{\color{Aquamarine}\Verb$]$}\verb$     $ &{\color{blue}\Verb$e$}{\color{Aquamarine}\Verb$:$}{\color{Tan}\Verb${$}{\color{Tan}\Verb$c$}{\color{Tan}\Verb$=$}{\color{Tan}\Verb$[$}{\color{Tan}\Verb$*$}{\color{Tan}\Verb$c$}{\color{Tan}\Verb$,$}{\color{Tan}\Verb$ $}{\color{Aquamarine}\Verb$i$}{\color{Aquamarine}\Verb$t$}{\color{Tan}\Verb$]$}{\color{Tan}\Verb$ $}{\color{Tan}\Verb$}$}&Append      result to array {\color{red}\Verb$c$}.\\
{\color{blue}\Verb$e$}{\color{Aquamarine}\Verb$:$}{\color{Aquamarine}\Verb$[$}{\color{Aquamarine}\Verb$*$}{\color{Aquamarine}\Verb$c$}{\color{Aquamarine}\Verb$]$}&{\color{blue}\Verb$e$}{\color{Aquamarine}\Verb$:$}{\color{Tan}\Verb${$}{\color{Tan}\Verb$c$}{\color{Tan}\Verb$=$}{\color{Tan}\Verb$[$}{\color{Tan}\Verb$*$}{\color{Tan}\Verb$c$}{\color{Tan}\Verb$,$}{\color{Tan}\Verb$*$}{\color{Aquamarine}\Verb$i$}{\color{Aquamarine}\Verb$t$}{\color{Tan}\Verb$]$}{\color{Tan}\Verb$ $}{\color{Tan}\Verb$}$}&Concatenate result to array {\color{red}\Verb$c$}.\\
\hline
\end{tabular}\vskip 0.2em

\subsection*{Semantic predicates}

It is possible to test arbitrary properties by {\it \color{violet} semantic predicates} which are implemented as:

\vskip 0.2em\noindent\begin{tabular}{| l | l || p{9.2cm} |}
\hline
Expression & Expansion & Description\\
\hline
{\color{Violet}\Verb$&$}{\color{Violet}\Verb${$}{\color{Tan}\Verb$c$}{\color{Violet}\Verb$}$} & {\color{red}\Verb$c$}{\color{red}\Verb$o$}{\color{red}\Verb$r$}{\color{red}\Verb$e$}\verb$     $ & Semantic predicate.\\
{\color{Violet}\Verb$~$}{\color{Violet}\Verb${$}{\color{Tan}\Verb$c$}{\color{Violet}\Verb$}$} & {\color{Violet}\Verb$&$}{\color{Violet}\Verb${$}{\color{Tan}\Verb$!$}{\color{Tan}\Verb$c$}{\color{Violet}\Verb$}$} & Negative semantic predicate.\\
\hline
\end{tabular}\vskip 0.2em

A semantic predicate expression accepts only if a predicate evaluates to true otherwise it rejects. Otherwise it behaves exactly as the semantic action.

For example, even integers are matched as follows:

\begin{exambox}
\begin{spacing}{0.8}
{\small
{\color{red}\Verb$e$}{\color{red}\Verb$v$}{\color{red}\Verb$e$}{\color{red}\Verb$n$}\Verb$ $\Verb$=$\Verb$ ${\color{red}\Verb$i$}{\color{red}\Verb$n$}{\color{red}\Verb$t$}{\color{Aquamarine}\Verb$:$}{\color{Aquamarine}\Verb$x$}\Verb$ ${\color{Violet}\Verb$&$}{\color{Violet}\Verb${$}{\color{Tan}\Verb$ $}{\color{Aquamarine}\Verb$x$}{\color{Tan}\Verb$
}\end{spacing}

\end{exambox}

Sometimes we need to represent an expression that always fails. We define the following rule in standard prologue:

{\small
\begin{spacing}{0.8}
\begin{prolbox}
{\color{red}\Verb$f$}{\color{red}\Verb$a$}{\color{red}\Verb$i$}{\color{red}\Verb$l$}{\color{red}\Verb$s$}\Verb$ $\Verb$=$\Verb$ ${\color{Violet}\Verb$&$}{\color{Violet}\Verb${$}{\color{Tan}\Verb$ $}{\color{Tan}\Verb$f$}{\color{Tan}\Verb$a$}{\color{Tan}\Verb$l$}{\color{Tan}\Verb$s$}{\color{Tan}\Verb$e$}{\color{Tan}\Verb$ $}{\color{Violet}\Verb$}$}
\end{prolbox}\end{spacing}
}

\subsection*{Results of operators}

The result of an atomic expressions is typically a string matched by that expression. 

Result of operators can be described by the following identities:

\vskip 0.2em\noindent\begin{tabular}{| l | l |}
\hline
Expression & Expansion \\
\hline
\Verb$(${\color{blue}\Verb$e$}{\color{blue}\Verb$1$}\Verb$ ${\color{blue}\Verb$e$}{\color{blue}\Verb$2$}\Verb$)${\color{Aquamarine}\Verb$:$}{\color{Aquamarine}\Verb$v$}\verb$ $& {\color{blue}\Verb$e$}{\color{blue}\Verb$1$}\Verb$ $\Verb$ $\Verb$ $\Verb$ $\Verb$ ${\color{blue}\Verb$e$}{\color{blue}\Verb$2$}{\color{Aquamarine}\Verb$:$}{\color{Aquamarine}\Verb$v$} \\
\Verb$(${\color{blue}\Verb$e$}{\color{blue}\Verb$1$}\Verb$|${\color{blue}\Verb$e$}{\color{blue}\Verb$2$}\Verb$)${\color{Aquamarine}\Verb$:$}{\color{Aquamarine}\Verb$v$}& {\color{blue}\Verb$e$}{\color{blue}\Verb$1$}{\color{Aquamarine}\Verb$:$}{\color{Aquamarine}\Verb$v$}\Verb$ $\Verb$|$\Verb$ ${\color{blue}\Verb$e$}{\color{blue}\Verb$2$}{\color{Aquamarine}\Verb$:$}{\color{Aquamarine}\Verb$v$} \\
\Verb$(${\color{blue}\Verb$e$}{\color{blue}\Verb$1$}{\color{Violet}\Verb$&$}{\color{blue}\Verb$e$}{\color{blue}\Verb$2$}\Verb$)${\color{Aquamarine}\Verb$:$}{\color{Aquamarine}\Verb$v$}& {\color{blue}\Verb$e$}{\color{blue}\Verb$1$}\Verb$ $\Verb$ $\Verb$ $\Verb$ $\Verb$ ${\color{blue}\Verb$e$}{\color{blue}\Verb$2$}{\color{Aquamarine}\Verb$:$}{\color{Aquamarine}\Verb$v$} \\
\Verb$(${\color{Violet}\Verb$~$}{\color{blue}\Verb$e$}\Verb$)${\color{Aquamarine}\Verb$:$}{\color{Aquamarine}\Verb$v$}& {\color{blue}\Verb$e$} \\
\Verb$(${\color{blue}\Verb$e$}\Verb$)${\color{Aquamarine}\Verb$:$}{\color{Aquamarine}\Verb$v$}& \Verb$(${\color{blue}\Verb$e$}{\color{Aquamarine}\Verb$:$}{\color{Aquamarine}\Verb$v$}\Verb$)$ \\
{\color{blue}\Verb$e$}{\color{black}\Verb$*$}{\color{Aquamarine}\Verb$:$}{\color{Aquamarine}\Verb$v$}& {\color{Tan}\Verb${$}{\color{Tan}\Verb$[$}{\color{Tan}\Verb$]$}{\color{Tan}\Verb$}$}{\color{Aquamarine}\Verb$:$}{\color{Aquamarine}\Verb$a$}\Verb$ $\Verb$($\Verb$ ${\color{blue}\Verb$e$}{\color{Aquamarine}\Verb$:$}{\color{Aquamarine}\Verb$[$}{\color{Aquamarine}\Verb$a$}{\color{Aquamarine}\Verb$]$}\Verb$ $\Verb$)${\color{black}\Verb$*$}\Verb$ ${\color{Tan}\Verb${$}{\color{Aquamarine}\Verb$a$}{\color{Tan}\Verb$}$}{\color{Aquamarine}\Verb$:$}{\color{Aquamarine}\Verb$v$} \\
{\color{blue}\Verb$e$}{\color{black}\Verb$+$}{\color{Aquamarine}\Verb$:$}{\color{Aquamarine}\Verb$v$}& {\color{Tan}\Verb${$}{\color{Tan}\Verb$[$}{\color{Tan}\Verb$]$}{\color{Tan}\Verb$}$}{\color{Aquamarine}\Verb$:$}{\color{Aquamarine}\Verb$a$}\Verb$ $\Verb$($\Verb$ ${\color{blue}\Verb$e$}{\color{Aquamarine}\Verb$:$}{\color{Aquamarine}\Verb$[$}{\color{Aquamarine}\Verb$a$}{\color{Aquamarine}\Verb$]$}\Verb$ $\Verb$)${\color{black}\Verb$+$}\Verb$ ${\color{Tan}\Verb${$}{\color{Aquamarine}\Verb$a$}{\color{Tan}\Verb$}$}{\color{Aquamarine}\Verb$:$}{\color{Aquamarine}\Verb$v$}\verb$                             $ \\
\hline
\end{tabular}\vskip 0.2em

Note that lookaheads are always reverted. The main reason is maintainability as lookaheads often cause a rule to be called more times than expected.

\subsection*{Results of rules}

Results of rules are passed by an instance variable of parser named \verb$@@_result$.

\vskip 0.2em\noindent\begin{tabular}{| l | l |}
\hline
Expression & Expansion \\
\hline
{\color{red}\Verb$n$}{\color{red}\Verb$a$}{\color{red}\Verb$m$}{\color{red}\Verb$e$}\Verb$ $\Verb$=$\Verb$ ${\color{red}\Verb$e$}{\color{red}\Verb$x$}{\color{red}\Verb$p$}& {\color{red}\Verb$n$}{\color{red}\Verb$a$}{\color{red}\Verb$m$}{\color{red}\Verb$e$}\Verb$=${\color{red}\Verb$e$}{\color{red}\Verb$x$}{\color{red}\Verb$p$}{\color{Aquamarine}\Verb$:$}{\color{Aquamarine}\Verb$@$}{\color{Aquamarine}\Verb$@$}{\color{Aquamarine}\Verb$_$}{\color{Aquamarine}\Verb$r$}{\color{Aquamarine}\Verb$e$}{\color{Aquamarine}\Verb$s$}{\color{Aquamarine}\Verb$u$}{\color{Aquamarine}\Verb$l$}{\color{Aquamarine}\Verb$t$}\\
{\color{red}\Verb$r$}{\color{red}\Verb$u$}{\color{red}\Verb$l$}{\color{red}\Verb$e$}{\color{Aquamarine}\Verb$:$}{\color{Aquamarine}\Verb$v$}& {\color{red}\Verb$r$}{\color{red}\Verb$u$}{\color{red}\Verb$l$}{\color{red}\Verb$e$}\Verb$ ${\color{Tan}\Verb${$}{\color{Aquamarine}\Verb$@$}{\color{Aquamarine}\Verb$@$}{\color{Aquamarine}\Verb$_$}{\color{Aquamarine}\Verb$r$}{\color{Aquamarine}\Verb$e$}{\color{Aquamarine}\Verb$s$}{\color{Aquamarine}\Verb$u$}{\color{Aquamarine}\Verb$l$}{\color{Aquamarine}\Verb$t$}{\color{Tan}\Verb$}$}{\color{Aquamarine}\Verb$:$}{\color{Aquamarine}\Verb$v$}\verb$                                  $\\
\hline
\end{tabular}\vskip 0.2em

\newpage

\section{Parametrization} \label{sec:parameter}

As Ruby functions can take {\color{green} parameters} so can amethyst rules.
 The syntax for simple form of parametrization is the following:

\noindent\begin{tabular}{| l  || p{9cm} |}
\hline
Pattern & Description \\
\hline
{\color{red}\Verb$n$}{\color{red}\Verb$a$}{\color{red}\Verb$m$}{\color{red}\Verb$e$}{\color{green}\Verb$($}{\color{green}\Verb$p$}{\color{green}\Verb$1$}{\color{green}\Verb$,$}{\color{green}\Verb$p$}{\color{green}\Verb$2$}{\color{green}\Verb$,$}{\color{green}\Verb$.$}{\color{green}\Verb$.$}{\color{green}\Verb$.$}{\color{green}\Verb$)$}\Verb$ $\Verb$=$\Verb$ ${\color{red}\Verb$e$}{\color{red}\Verb$x$}{\color{red}\Verb$p$}  & Define rule with parameters p1,p2...\\
{\color{red}\Verb$n$}{\color{red}\Verb$a$}{\color{red}\Verb$m$}{\color{red}\Verb$e$}{\color{green}\Verb$($}{\color{green}\Verb$p$}{\color{green}\Verb$1$}{\color{green}\Verb$,$}{\color{green}\Verb$p$}{\color{green}\Verb$2$}{\color{green}\Verb$,$}{\color{green}\Verb$.$}{\color{green}\Verb$.$}{\color{green}\Verb$.$}{\color{green}\Verb$)$}  & Call rule with parameters p1,p2...\\
\hline
\end{tabular}\vskip 0.2em

This construction is space sensitive. Placing a space after name is always interpreted  as sequencing of rule and expression.

Parametrization in its full generality is more complicated as will be explained in Section \ref{paramcont}. Here we will give several examples of using parametrization.

Simplest example of parametrization is the following:

\vskip -1.8em\begin{spacing}{0.8}
{\small
\begin{exambox}
\Verb$a$\Verb$m$\Verb$e$\Verb$t$\Verb$h$\Verb$y$\Verb$s$\Verb$t$\Verb$ $\Verb$A$\Verb$d$\Verb$d$\Verb$e$\Verb$r$\Verb$ $\Verb${$\\\Verb$ $\Verb$ ${\color{red}\Verb$a$}{\color{red}\Verb$d$}{\color{red}\Verb$d$}{\color{green}\Verb$($}{\color{green}\Verb$x$}{\color{green}\Verb$,$}{\color{green}\Verb$y$}{\color{green}\Verb$)$}\Verb$ $\Verb$=$\Verb$ ${\color{Tan}\Verb$-$}{\color{Tan}\Verb$>$}{\color{Tan}\Verb$ $}{\color{Tan}\Verb$x$}{\color{Tan}\Verb$+$}{\color{Tan}\Verb$y$}{\color{Tan}\\}{\color{Tan}\Verb$ $}\Verb$ ${\color{red}\Verb$f$}{\color{red}\Verb$o$}{\color{red}\Verb$u$}{\color{red}\Verb$r$}\Verb$ $\Verb$=$\Verb$ ${\color{red}\Verb$a$}{\color{red}\Verb$d$}{\color{red}\Verb$d$}{\color{green}\Verb$($}{\color{green}\Verb$2$}{\color{green}\Verb$,$}{\color{green}\Verb$2$}{\color{green}\Verb$)$}\\\Verb$}$
\end{exambox}
}
\end{spacing}\vskip -0.4em

Parametrized rule can be called from Ruby with an input string followed by rule parameters:

\begin{verbatim}
Adder.add("",2,2) #-> 4
\end{verbatim}

Several builtin parametrized rules are included in amethyst. We have already seen a parametrized rules {\color{red}\Verb$s$}{\color{red}\Verb$e$}{\color{red}\Verb$q$} and {\color{red}\Verb$t$}{\color{red}\Verb$o$}{\color{red}\Verb$k$}{\color{red}\Verb$e$}{\color{red}\Verb$n$}. 

Replacing is text is common task. Say we want replace \extext{foo} with \extext{FOO} and \extext{bar} with \extext{BAR}. We can use builtin rule {\color{red}\Verb$r$}{\color{red}\Verb$e$}{\color{red}\Verb$p$}{\color{red}\Verb$l$}{\color{red}\Verb$a$}{\color{red}\Verb$c$}{\color{red}\Verb$e$}:

\vskip -1.8em\begin{spacing}{0.8}
{\small
\begin{exambox}
\Verb$A$\Verb$m$\Verb$e$\Verb$t$\Verb$h$\Verb$y$\Verb$s$\Verb$t$\Verb$.$\Verb$r$\Verb$e$\Verb$p$\Verb$l$\Verb$a$\Verb$c$\Verb$e$\Verb$($\Verb$"$\Verb$f$\Verb$o$\Verb$o$\Verb$o$\Verb$b$\Verb$a$\Verb$r$\Verb$s$\Verb$"$\Verb$,${\color{Violet}\Verb$($}{\color{Violet}\Verb$|$}\Verb$ $\Verb$(${\color{black}\Verb$"$}{\color{black}\Verb$f$}{\color{black}\Verb$o$}{\color{black}\Verb$o$}{\color{black}\Verb$"$}\Verb$ $\Verb$|$\Verb$ ${\color{black}\Verb$"$}{\color{black}\Verb$b$}{\color{black}\Verb$a$}{\color{black}\Verb$r$}{\color{black}\Verb$"$}\Verb$)${\color{Aquamarine}\Verb$:$}{\color{Tan}\Verb${$}{\color{Aquamarine}\Verb$i$}{\color{Aquamarine}\Verb$t$}{\color{Tan}\Verb$.$}{\color{Tan}\Verb$u$}{\color{Tan}\Verb$p$}{\color{Tan}\Verb$c$}{\color{Tan}\Verb$a$}{\color{Tan}\Verb$s$}{\color{Tan}\Verb$e$}{\color{Tan}\Verb$}$}\Verb$ ${\color{Violet}\Verb$|$}{\color{Violet}\Verb$)$}\Verb$)$\\\Verb$#$\Verb$-$\Verb$>$\Verb$ $\Verb$"$\Verb$F$\Verb$O$\Verb$O$\Verb$o$\Verb$B$\Verb$A$\Verb$R$\Verb$s$\Verb$"$
\end{exambox}
}
\end{spacing}\vskip -0.4em

Note that amethyst expressions can be passed also inside grammars. \\
Example above can be also written as:

\vskip -1.8em\begin{spacing}{0.8}
{\small
\begin{exambox}
\Verb$a$\Verb$m$\Verb$e$\Verb$t$\Verb$h$\Verb$y$\Verb$s$\Verb$t$\Verb$ $\Verb$P$\Verb$a$\Verb$r$\Verb$a$\Verb$m$\Verb$ $\Verb${$\\\Verb$ $\Verb$ ${\color{red}\Verb$r$}{\color{red}\Verb$e$}{\color{red}\Verb$p$}{\color{red}\Verb$l$}{\color{red}\Verb$a$}{\color{red}\Verb$c$}{\color{red}\Verb$e$}{\color{red}\Verb$_$}{\color{red}\Verb$f$}{\color{red}\Verb$o$}{\color{red}\Verb$o$}{\color{red}\Verb$b$}{\color{red}\Verb$a$}{\color{red}\Verb$r$}\Verb$ $\Verb$=$\Verb$ ${\color{red}\Verb$r$}{\color{red}\Verb$e$}{\color{red}\Verb$p$}{\color{red}\Verb$l$}{\color{red}\Verb$a$}{\color{red}\Verb$c$}{\color{red}\Verb$e$}{\color{green}\Verb$($}{\color{green}\Verb$ $}{\color{Violet}\Verb$($}{\color{Violet}\Verb$|$}{\color{green}\Verb$ $}{\color{green}\Verb$($}{\color{black}\Verb$"$}{\color{black}\Verb$f$}{\color{black}\Verb$o$}{\color{black}\Verb$o$}{\color{black}\Verb$"$}{\color{green}\Verb$ $}{\color{green}\Verb$|$}{\color{green}\Verb$ $}{\color{black}\Verb$"$}{\color{black}\Verb$b$}{\color{black}\Verb$a$}{\color{black}\Verb$r$}{\color{black}\Verb$"$}{\color{green}\Verb$)$}{\color{Aquamarine}\Verb$:$}{\color{Tan}\Verb${$}{\color{Aquamarine}\Verb$i$}{\color{Aquamarine}\Verb$t$}{\color{Tan}\Verb$.$}{\color{Tan}\Verb$u$}{\color{Tan}\Verb$p$}{\color{Tan}\Verb$c$}{\color{Tan}\Verb$a$}{\color{Tan}\Verb$s$}{\color{Tan}\Verb$e$}{\color{Tan}\Verb$}$}{\color{green}\Verb$ $}{\color{Violet}\Verb$|$}{\color{Violet}\Verb$)$}{\color{green}\Verb$ $}{\color{green}\Verb$)$}\\\Verb$}$
\end{exambox}
}
\end{spacing}\vskip -0.4em

Construction above is called {\it lambda} \cite{lambdakalkul}. Rule calls inside lambda are resolved lexically  \cite{scope}. We form a closure for enclosing amethyst rule as is expected from lambda.

Only two parametrized rules are core atomic expressions:

\noindent\begin{tabular}{| l | p{12cm} |}
\hline
Rule & Description \\
\hline
{\color{red}\Verb$a$}{\color{red}\Verb$p$}{\color{red}\Verb$p$}{\color{red}\Verb$l$}{\color{red}\Verb$y$}{\color{green}\Verb$($}{\color{green}\Verb$x$}{\color{green}\Verb$)$}  & Apply lambda in parameter\\
{\color{red}\Verb$s$}{\color{red}\Verb$e$}{\color{red}\Verb$q$}{\color{green}\Verb$($}{\color{green}\Verb$x$}{\color{green}\Verb$)$}  & Match string or apply lambda in parameter\\
\hline
\end{tabular}\vskip 0.2em

Reason why {\color{red}\Verb$a$}{\color{red}\Verb$p$}{\color{red}\Verb$p$}{\color{red}\Verb$l$}{\color{red}\Verb$y$} does not accept string as a parameter is that we want to do resolving in the callers scope.

Other parametrized rules just use {\color{red}\Verb$s$}{\color{red}\Verb$e$}{\color{red}\Verb$q$} and {\color{red}\Verb$a$}{\color{red}\Verb$p$}{\color{red}\Verb$p$}{\color{red}\Verb$l$}{\color{red}\Verb$y$}.
In standard prologue we follow good practice that rule that accepts string as a parameter accepts lambda too. 

As an example consider how rules {\color{red}\Verb$f$}{\color{red}\Verb$i$}{\color{red}\Verb$n$}{\color{red}\Verb$d$} and {\color{red}\Verb$r$}{\color{red}\Verb$e$}{\color{red}\Verb$p$}{\color{red}\Verb$l$}{\color{red}\Verb$a$}{\color{red}\Verb$c$}{\color{red}\Verb$e$} are implemented in the standard prologue:

{\small
\begin{spacing}{0.8}
\begin{prolbox}
{\color{red}\Verb$f$}{\color{red}\Verb$i$}{\color{red}\Verb$n$}{\color{red}\Verb$d$}{\color{green}\Verb$($}{\color{green}\Verb$e$}{\color{green}\Verb$x$}{\color{green}\Verb$p$}{\color{green}\Verb$)$}\Verb$ $\Verb$ $\Verb$ $\Verb$ $\Verb$=$\Verb$ $\Verb$($\Verb$ $\Verb$ ${\color{red}\Verb$s$}{\color{red}\Verb$e$}{\color{red}\Verb$q$}{\color{green}\Verb$($}{\color{green}\Verb$e$}{\color{green}\Verb$x$}{\color{green}\Verb$p$}{\color{green}\Verb$)$}{\color{Aquamarine}\Verb$:$}{\color{Aquamarine}\Verb$x$}\Verb$ $\Verb$b$\Verb$r$\Verb$e$\Verb$a$\Verb$k$\Verb$ $\Verb$|$\Verb$ $\Verb$.$\Verb$)${\color{black}\Verb$*$}\Verb$ $\Verb$.${\color{black}\Verb$*$}\Verb$ ${\color{Tan}\Verb$-$}{\color{Tan}\Verb$>$}{\color{Tan}\Verb$ $}{\color{Aquamarine}\Verb$x$}{\color{Tan}\\}{\color{Tan}\Verb$r$}{\color{red}\Verb$e$}{\color{red}\Verb$p$}{\color{red}\Verb$l$}{\color{red}\Verb$a$}{\color{red}\Verb$c$}{\color{red}\Verb$e$}{\color{green}\Verb$($}{\color{green}\Verb$e$}{\color{green}\Verb$x$}{\color{green}\Verb$p$}{\color{green}\Verb$)$}\Verb$ $\Verb$=$\Verb$ $\Verb$(${\color{red}\Verb$a$}{\color{red}\Verb$p$}{\color{red}\Verb$p$}{\color{red}\Verb$l$}{\color{red}\Verb$y$}{\color{green}\Verb$($}{\color{green}\Verb$e$}{\color{green}\Verb$x$}{\color{green}\Verb$p$}{\color{green}\Verb$)$}\Verb$ $\Verb$|$\Verb$ $\Verb$.$\Verb$)${\color{black}\Verb$*$}{\color{Aquamarine}\Verb$:$}{\color{Tan}\Verb${$}{\color{Aquamarine}\Verb$i$}{\color{Aquamarine}\Verb$t$}{\color{Tan}\Verb$.$}{\color{Tan}\Verb$j$}{\color{Tan}\Verb$o$}{\color{Tan}\Verb$i$}{\color{Tan}\Verb$n$}{\color{Tan}\Verb$}$}
\end{prolbox}\end{spacing}
}

\subsection*{Closer look at lambda}
Amethyst lambdas form a closure as is illustrated in the following example:

\vskip -1.8em\begin{spacing}{0.8}
{\small
\begin{exambox}
\Verb$a$\Verb$m$\Verb$e$\Verb$t$\Verb$h$\Verb$y$\Verb$s$\Verb$t$\Verb$ $\Verb$C$\Verb$l$\Verb$o$\Verb$s$\Verb$u$\Verb$r$\Verb$e$\Verb${$\\\Verb$ $\Verb$ ${\color{red}\Verb$l$}{\color{red}\Verb$a$}{\color{red}\Verb$m$}{\color{red}\Verb$b$}{\color{red}\Verb$d$}{\color{red}\Verb$a$}{\color{green}\Verb$($}{\color{green}\Verb$z$}{\color{green}\Verb$)$}\Verb$ $\Verb$=$\Verb$ ${\color{Tan}\Verb${$}{\color{Tan}\Verb$ $}{\color{Violet}\Verb$($}{\color{Violet}\Verb$|$}{\color{Tan}\Verb$ $}{\color{Tan}\Verb$ $}{\color{Tan}\Verb${$}{\color{Tan}\Verb$p$}{\color{Tan}\Verb$u$}{\color{Tan}\Verb$t$}{\color{Tan}\Verb$s$}{\color{Tan}\Verb$ $}{\color{Tan}\Verb$z$}{\color{Tan}\Verb$+$}{\color{Tan}\Verb$=$}{\color{Tan}\Verb$1$}{\color{Tan}\Verb$}$}{\color{Tan}\Verb$ $}{\color{Violet}\Verb$|$}{\color{Violet}\Verb$)$}{\color{Tan}\Verb$ $}{\color{Tan}\Verb$}$}\\\\\Verb$ $\Verb$ ${\color{red}\Verb$t$}{\color{red}\Verb$e$}{\color{red}\Verb$s$}{\color{red}\Verb$t$}\Verb$ $\Verb$=$\Verb$ ${\color{red}\Verb$l$}{\color{red}\Verb$a$}{\color{red}\Verb$m$}{\color{red}\Verb$b$}{\color{red}\Verb$d$}{\color{red}\Verb$a$}{\color{green}\Verb$($}{\color{green}\Verb$3$}{\color{green}\Verb$)$}{\color{Aquamarine}\Verb$:$}{\color{Aquamarine}\Verb$x$}\Verb$ $\Verb$ ${\color{red}\Verb$a$}{\color{red}\Verb$p$}{\color{red}\Verb$p$}{\color{red}\Verb$l$}{\color{red}\Verb$y$}{\color{green}\Verb$($}{\color{Aquamarine}\Verb$x$}{\color{green}\Verb$)$}\Verb$ ${\color{red}\Verb$a$}{\color{red}\Verb$p$}{\color{red}\Verb$p$}{\color{red}\Verb$l$}{\color{red}\Verb$y$}{\color{green}\Verb$($}{\color{Aquamarine}\Verb$x$}{\color{green}\Verb$)$}\Verb$ ${\color{red}\Verb$a$}{\color{red}\Verb$p$}{\color{red}\Verb$p$}{\color{red}\Verb$l$}{\color{red}\Verb$y$}{\color{green}\Verb$($}{\color{Aquamarine}\Verb$x$}{\color{green}\Verb$)$}\\\Verb$}$\\\Verb$C$\Verb$l$\Verb$o$\Verb$s$\Verb$u$\Verb$r$\Verb$e$\Verb$.$\Verb$t$\Verb$e$\Verb$s$\Verb$t$\Verb$($\Verb$"$\Verb$"$\Verb$)$\Verb$ $\Verb$#$\Verb$-$\Verb$>$\Verb$ $\Verb$4$\Verb$ $\Verb$5$\Verb$ $\Verb$6$
\end{exambox}
}
\end{spacing}\vskip -0.4em

Lambda can receive arguments. We can read arguments by calling {\color{red}\Verb$_$}{\color{red}\Verb$_$} method. This allows implicit syntax for partial application.

\vskip -1.8em\begin{spacing}{0.8}
{\small
\begin{exambox}
\Verb$p$\Verb$a$\Verb$r$\Verb$($\Verb$x$\Verb$,$\Verb$y$\Verb$,$\Verb$z$\Verb$)$\Verb$ $\Verb$=$\Verb$ $\Verb$-$\Verb$>$\Verb$ $\Verb$p$\Verb$u$\Verb$t$\Verb$s$\Verb$($\Verb$x$\Verb$ $\Verb$+$\Verb$y$\Verb$*$\Verb$z$\Verb$ $\Verb$)$\Verb$ $\Verb$ $\\\Verb$f$\Verb$o$\Verb$o$\Verb$($\Verb$x$\Verb$)$\Verb$ $\Verb$ $\Verb$ $\Verb$ $\Verb$ $\Verb$=$\Verb$ $\Verb$-$\Verb$>$\Verb$ ${\color{Violet}\Verb$($}{\color{Violet}\Verb$|$}\Verb$ ${\color{red}\Verb$p$}{\color{red}\Verb$a$}{\color{red}\Verb$r$}{\color{green}\Verb$($}{\color{green}\Verb$_$}{\color{green}\Verb$_$}{\color{green}\Verb$,$}{\color{green}\Verb$x$}{\color{green}\Verb$,$}{\color{green}\Verb$_$}{\color{green}\Verb$_$}{\color{green}\Verb$)$}\Verb$ ${\color{Violet}\Verb$|$}{\color{Violet}\Verb$)$}
\end{exambox}
}
\end{spacing}\vskip -0.4em

\subsection*{Example: Parser combinators}
Parser combinators \cite{combinators} are a popular way to implement parsers by people with a functional programming background. They allow the construction of parser expressions by using the host language operators. A combinator support is easy to add by defining operators for amethyst lambda. An implementation follows:

\vskip -1.8em\begin{spacing}{0.8}
{\small
\begin{exambox}
\Verb$a$\Verb$m$\Verb$e$\Verb$t$\Verb$h$\Verb$y$\Verb$s$\Verb$t$\Verb$ $\Verb$C$\Verb$o$\Verb$m$\Verb$b$\Verb$i$\Verb$n$\Verb$a$\Verb$t$\Verb$o$\Verb$r$\Verb$s$\Verb$ $\Verb${$\\\Verb$ $\Verb$ ${\color{red}\Verb$p$}{\color{red}\Verb$l$}{\color{red}\Verb$u$}{\color{red}\Verb$s$}{\color{green}\Verb$($}{\color{green}\Verb$x$}{\color{green}\Verb$,$}{\color{green}\Verb$y$}{\color{green}\Verb$)$}\Verb$ $\Verb$=$\Verb$ ${\color{Tan}\Verb$-$}{\color{Tan}\Verb$>$}{\color{Tan}\Verb$ $}{\color{Violet}\Verb$($}{\color{Violet}\Verb$|$}{\color{Tan}\Verb$ $}{\color{red}\Verb$s$}{\color{red}\Verb$e$}{\color{red}\Verb$q$}{\color{green}\Verb$($}{\color{green}\Verb$x$}{\color{green}\Verb$)$}{\color{Tan}\Verb$ $}{\color{Tan}\Verb$ $}{\color{Tan}\Verb$ $}{\color{red}\Verb$s$}{\color{red}\Verb$e$}{\color{red}\Verb$q$}{\color{green}\Verb$($}{\color{green}\Verb$y$}{\color{green}\Verb$)$}{\color{Tan}\Verb$ $}{\color{Violet}\Verb$|$}{\color{Violet}\Verb$)$}{\color{Tan}\\}{\color{Tan}\Verb$ $}\Verb$ ${\color{red}\Verb$o$}{\color{red}\Verb$r$}{\color{green}\Verb$($}{\color{green}\Verb$x$}{\color{green}\Verb$,$}{\color{green}\Verb$y$}{\color{green}\Verb$)$}\Verb$ $\Verb$ $\Verb$ $\Verb$=$\Verb$ ${\color{Tan}\Verb$-$}{\color{Tan}\Verb$>$}{\color{Tan}\Verb$ $}{\color{Violet}\Verb$($}{\color{Violet}\Verb$|$}{\color{Tan}\Verb$ $}{\color{red}\Verb$s$}{\color{red}\Verb$e$}{\color{red}\Verb$q$}{\color{green}\Verb$($}{\color{green}\Verb$x$}{\color{green}\Verb$)$}{\color{Tan}\Verb$ $}{\color{Tan}\Verb$|$}{\color{Tan}\Verb$ $}{\color{red}\Verb$s$}{\color{red}\Verb$e$}{\color{red}\Verb$q$}{\color{green}\Verb$($}{\color{green}\Verb$y$}{\color{green}\Verb$)$}{\color{Tan}\Verb$ $}{\color{Violet}\Verb$|$}{\color{Violet}\Verb$)$}{\color{Tan}\\}{\color{Tan}\Verb$ $}\Verb$ ${\color{red}\Verb$a$}{\color{red}\Verb$n$}{\color{red}\Verb$d$}{\color{green}\Verb$($}{\color{green}\Verb$x$}{\color{green}\Verb$,$}{\color{green}\Verb$y$}{\color{green}\Verb$)$}\Verb$ $\Verb$ $\Verb$=$\Verb$ ${\color{Tan}\Verb$-$}{\color{Tan}\Verb$>$}{\color{Tan}\Verb$ $}{\color{Violet}\Verb$($}{\color{Violet}\Verb$|$}{\color{Tan}\Verb$ $}{\color{red}\Verb$s$}{\color{red}\Verb$e$}{\color{red}\Verb$q$}{\color{green}\Verb$($}{\color{green}\Verb$x$}{\color{green}\Verb$)$}{\color{Tan}\Verb$ $}{\color{Violet}\Verb$&$}{\color{Violet}\Verb$ $}{\color{Tan}\Verb$s$}{\color{Tan}\Verb$e$}{\color{Tan}\Verb$q$}{\color{Tan}\Verb$($}{\color{Tan}\Verb$y$}{\color{Tan}\Verb$)$}{\color{Tan}\Verb$ $}{\color{Tan}\Verb$|$}{\color{Tan}\Verb$)$}{\color{Tan}\\}{\color{Tan}\Verb$ $}\Verb$ ${\color{red}\Verb$n$}{\color{red}\Verb$o$}{\color{red}\Verb$t$}{\color{green}\Verb$($}{\color{green}\Verb$x$}{\color{green}\Verb$)$}\Verb$ $\Verb$ $\Verb$ $\Verb$ $\Verb$=$\Verb$ ${\color{Tan}\Verb$-$}{\color{Tan}\Verb$>$}{\color{Tan}\Verb$ $}{\color{Violet}\Verb$($}{\color{Violet}\Verb$|$}{\color{Tan}\Verb$ $}{\color{Violet}\Verb$~$}{\color{Violet}\Verb$s$}{\color{red}\Verb$e$}{\color{red}\Verb$q$}{\color{green}\Verb$($}{\color{green}\Verb$x$}{\color{Violet}\Verb$)$}{\color{Tan}\Verb$ $}{\color{Tan}\Verb$ $}{\color{Tan}\Verb$ $}{\color{Tan}\Verb$ $}{\color{Tan}\Verb$ $}{\color{Tan}\Verb$ $}{\color{Tan}\Verb$ $}{\color{Tan}\Verb$ $}{\color{Tan}\Verb$ $}{\color{Violet}\Verb$|$}{\color{Violet}\Verb$)$}{\color{Tan}\\}{\color{Tan}\Verb$ $}\Verb$ ${\color{red}\Verb$s$}{\color{red}\Verb$t$}{\color{red}\Verb$a$}{\color{red}\Verb$r$}{\color{green}\Verb$($}{\color{green}\Verb$x$}{\color{green}\Verb$)$}\Verb$ $\Verb$ $\Verb$ $\Verb$=$\Verb$ ${\color{Tan}\Verb$-$}{\color{Tan}\Verb$>$}{\color{Tan}\Verb$ $}{\color{Violet}\Verb$($}{\color{Violet}\Verb$|$}{\color{Tan}\Verb$ $}{\color{red}\Verb$s$}{\color{red}\Verb$e$}{\color{red}\Verb$q$}{\color{green}\Verb$($}{\color{green}\Verb$x$}{\color{green}\Verb$)$}{\color{black}\Verb$*$}{\color{Tan}\Verb$ $}{\color{Tan}\Verb$ $}{\color{Tan}\Verb$ $}{\color{Tan}\Verb$ $}{\color{Tan}\Verb$ $}{\color{Tan}\Verb$ $}{\color{Tan}\Verb$ $}{\color{Tan}\Verb$ $}{\color{Tan}\Verb$ $}{\color{Violet}\Verb$|$}{\color{Violet}\Verb$)$}{\color{Tan}\\}{\color{Tan}\Verb$}$}\\\\\Verb$c$\Verb$l$\Verb$a$\Verb$s$\Verb$s$\Verb$ $\Verb$A$\Verb$m$\Verb$e$\Verb$t$\Verb$h$\Verb$y$\Verb$s$\Verb$t$\Verb$L$\Verb$a$\Verb$m$\Verb$b$\Verb$d$\Verb$a$\\\Verb$ $\Verb$ $\Verb$b$\Verb$i$\Verb$n$\Verb$_$\Verb$o$\Verb$p$\Verb$=$\Verb$[$\Verb$[$\Verb$'$\Verb$+$\Verb$'$\Verb$,$\Verb$:$\Verb$p$\Verb$l$\Verb$u$\Verb$s$\Verb$]$\Verb$,$\Verb$[$\Verb$'$\Verb$|$\Verb$'$\Verb$,$\Verb$:$\Verb$o$\Verb$r$\Verb$]$\Verb$,$\Verb$[$\Verb$'$\Verb$&$\Verb$'$\Verb$,$\Verb$:$\Verb$a$\Verb$n$\Verb$d$\Verb$]$\Verb$]$\\\Verb$ $\Verb$ $\Verb$u$\Verb$n$\Verb$_$\Verb$o$\Verb$p$\Verb$ $\Verb$=$\Verb$[$\Verb$[$\Verb$'$\Verb$s$\Verb$t$\Verb$a$\Verb$r$\Verb$'$\Verb$,$\Verb$:$\Verb$s$\Verb$t$\Verb$a$\Verb$r$\Verb$]$\Verb$,$\Verb$[$\Verb$'$\Verb$~$\Verb$'$\Verb$,$\Verb$:$\Verb$n$\Verb$o$\Verb$t$\Verb$]$\Verb$]$\\\Verb$ $\Verb$ $\Verb$b$\Verb$i$\Verb$n$\Verb$_$\Verb$o$\Verb$p$\Verb$.$\Verb$e$\Verb$a$\Verb$c$\Verb$h$\Verb${$\Verb$|$\Verb$s$\Verb$y$\Verb$m$\Verb$,$\Verb$n$\Verb$a$\Verb$m$\Verb$e$\Verb$|$\\\Verb$ $\Verb$ $\Verb$ $\Verb$ $\Verb$d$\Verb$e$\Verb$f$\Verb$i$\Verb$n$\Verb$e$\Verb$_$\Verb$m$\Verb$e$\Verb$t$\Verb$h$\Verb$o$\Verb$d$\Verb$($\Verb$s$\Verb$y$\Verb$m$\Verb$)$\Verb${$\Verb$|$\Verb$x$\Verb$|$\Verb$ $\Verb$C$\Verb$o$\Verb$m$\Verb$b$\Verb$i$\Verb$n$\Verb$a$\Verb$t$\Verb$o$\Verb$r$\Verb$s$\Verb$.$\Verb$s$\Verb$e$\Verb$n$\Verb$d$\Verb$($\Verb$n$\Verb$a$\Verb$m$\Verb$e$\Verb$,$\Verb$n$\Verb$i$\Verb$l$\Verb$,$\Verb$s$\Verb$e$\Verb$l$\Verb$f$\Verb$,$\Verb$x$\Verb$)$\Verb$}$\\\Verb$ $\Verb$ $\Verb$}$\\\Verb$ $\Verb$ $\Verb$u$\Verb$n$\Verb$_$\Verb$o$\Verb$p$\Verb$.$\Verb$e$\Verb$a$\Verb$c$\Verb$h$\Verb${$\Verb$|$\Verb$s$\Verb$y$\Verb$m$\Verb$,$\Verb$n$\Verb$a$\Verb$m$\Verb$e$\Verb$|$\\\Verb$ $\Verb$ $\Verb$ $\Verb$ $\Verb$d$\Verb$e$\Verb$f$\Verb$i$\Verb$n$\Verb$e$\Verb$_$\Verb$m$\Verb$e$\Verb$t$\Verb$h$\Verb$o$\Verb$d$\Verb$($\Verb$s$\Verb$y$\Verb$m$\Verb$)$\Verb${$\Verb$ $\Verb$ $\Verb$ $\Verb$ $\Verb$C$\Verb$o$\Verb$m$\Verb$b$\Verb$i$\Verb$n$\Verb$a$\Verb$t$\Verb$o$\Verb$r$\Verb$s$\Verb$.$\Verb$s$\Verb$e$\Verb$n$\Verb$d$\Verb$($\Verb$n$\Verb$a$\Verb$m$\Verb$e$\Verb$,$\Verb$n$\Verb$i$\Verb$l$\Verb$,$\Verb$s$\Verb$e$\Verb$l$\Verb$f$\Verb$)$\Verb$}$\\\Verb$ $\Verb$ $\Verb$}$\\\Verb$e$\Verb$n$\Verb$d$
\end{exambox}
}
\end{spacing}\vskip -0.4em

A ``hello world'' example when we use parser combinators becomes:

\vskip -1.8em\begin{spacing}{0.8}
{\small
\begin{exambox}
{\color{Violet}\Verb$($}{\color{Violet}\Verb$|$}{\color{black}\Verb$'$}{\color{black}\Verb$h$}{\color{black}\Verb$e$}{\color{black}\Verb$l$}{\color{black}\Verb$l$}{\color{black}\Verb$o$}{\color{black}\Verb$'$}{\color{Violet}\Verb$|$}{\color{Violet}\Verb$)$}\Verb$ $\Verb$+$\Verb$ ${\color{Violet}\Verb$($}{\color{Violet}\Verb$|$}{\color{black}\Verb$'$}{\color{black}\Verb$ $}{\color{black}\Verb$'$}{\color{Violet}\Verb$|$}{\color{Violet}\Verb$)$}\Verb$ $\Verb$+$\Verb$ ${\color{Violet}\Verb$($}{\color{Violet}\Verb$|$}{\color{black}\Verb$'$}{\color{black}\Verb$w$}{\color{black}\Verb$o$}{\color{black}\Verb$r$}{\color{black}\Verb$l$}{\color{black}\Verb$d$}{\color{black}\Verb$'$}{\color{Violet}\Verb$|$}{\color{Violet}\Verb$)$}
\end{exambox}
}
\end{spacing}\vskip -0.4em

Moreover, a user can write:

\vskip -1.8em\begin{spacing}{0.8}
{\small
\begin{exambox}
{\color{Violet}\Verb$($}{\color{Violet}\Verb$|$}{\color{black}\Verb$'$}{\color{black}\Verb$|$}{\color{black}\Verb$'$}{\color{Violet}\Verb$|$}{\color{Violet}\Verb$)$}\Verb$|${\color{Violet}\Verb$($}{\color{Violet}\Verb$|$}{\color{black}\Verb$'$}{\color{black}\Verb$|$}{\color{black}\Verb$'$}{\color{Violet}\Verb$|$}{\color{Violet}\Verb$)$}
\end{exambox}
}
\end{spacing}\vskip -0.4em

\noindent instead of
\vskip -1.8em\begin{spacing}{0.8}
{\small
\begin{exambox}
\Verb$'$\Verb$|$\Verb$'$\Verb$ $\Verb$|$\Verb$ $\Verb$'$\Verb$|$\Verb$'$
\end{exambox}
}
\end{spacing}\vskip -0.4em

\newpage
\section{Amethyst extends \peg} \label{extpeg}

Parsing expression grammars (\peg) were introduced by Ford \cite{ford}. Our parser started as a \peg{} parser but evolved into a more general language. In this section we explain the similarities and differences between \peg{} and our parser.

\subsection*{\peg{} operators}

A typical \peg{} parser defines expressions formed by the following operators:

\vskip 0.2em\noindent\begin{tabular}{| l || p{11.8cm} |}
\hline
Operation & Description \\
\hline
{\color{blue}\Verb$e$}{\color{blue}\Verb$1$}\Verb$ ${\color{blue}\Verb$e$}{\color{blue}\Verb$2$} & Sequencing\\
{\color{blue}\Verb$e$}{\color{blue}\Verb$1$}\Verb$|${\color{blue}\Verb$e$}{\color{blue}\Verb$2$} & Ordered choice\\
{\color{blue}\Verb$e$}{\color{black}\Verb$*$},{\color{blue}\Verb$e$}{\color{black}\Verb$+$},{\color{blue}\Verb$e$}{\color{black}\Verb$?$} &Iteration\\
{\color{Violet}\Verb$&$}{\color{blue}\Verb$e$},{\color{Violet}\Verb$~$}{\color{blue}\Verb$e$}& Lookaheads\\
\hline
\end{tabular}\vskip 0.2em

\subsection*{Sequencing and choice}\label{ss:choice}
Parsing expression grammars achieve linear time by making choice deterministic and by memoization. In \peg{} {\it ordered choice} tries alternatives at left to right order and when an alternative succeeds it does not try further alternatives. 

Amethyst extends this choice to {\it priorized choice} that does backtracking. Linear time is obtained by adding several natural conditions to recursion as is described in Chapter \ref{regreg}.

To describe semantic of our parser we chosen to define auxiliary constructs \verb$Cut$ and \verb$Stop$ that simplify description \footnote{similar situation is extending reals to complex numbers.}. A compiler may use different representation for example one defined in \ref{regreg}.

Our choice operator tries alternatives in left to right order. When an alternative succeeds it does not try further alternatives. We extend choice with \verb$Cut$ operator that when encountered it prevents parser form trying other choices. This allows more trackable description of the lookaheads.

Then behavior of operators from the Section \ref{opers} can be described by the following table:

\vskip 0.2em\noindent\begin{tabular}{| l | l || p{7.2cm} |}
\hline
Operation & Expansion & Description \\
\hline
{\color{blue}\Verb$e$}{\color{black}\Verb$?$}  & {\color{blue}\Verb$e$}\Verb$|${\color{Tan}\Verb${$}{\color{Tan}\Verb$n$}{\color{Tan}\Verb$i$}{\color{Tan}\Verb$l$}{\color{Tan}\Verb$}$} & Make {\color{blue}\Verb$e$} optional.  \\
{\color{red}\Verb$C$}{\color{red}\Verb$u$}{\color{red}\Verb$t$}  & auxiliary & Like ! in prolog\\
{\color{Violet}\Verb$~$}{\color{blue}\Verb$e$}  &{\color{blue}\Verb$e$}\Verb$ ${\color{red}\Verb$C$}{\color{red}\Verb$u$}{\color{red}\Verb$t$}\Verb$ ${\color{red}\Verb$f$}{\color{red}\Verb$a$}{\color{red}\Verb$i$}{\color{red}\Verb$l$}{\color{red}\Verb$s$}\Verb$ $\Verb$|$\Verb$ ${\color{Tan}\Verb${$}{\color{Tan}\Verb$n$}{\color{Tan}\Verb$i$}{\color{Tan}\Verb$l$}{\color{Tan}\Verb$}$} & Negative lookahead. \\
{\color{blue}\Verb$e$}{\color{blue}\Verb$1$}\Verb$ ${\color{Violet}\Verb$&$}{\color{Violet}\Verb$ $}{\color{blue}\Verb$e$}{\color{blue}\Verb$2$}  &{\color{Violet}\Verb$~$}{\color{Violet}\Verb$~$}{\color{blue}\Verb$e$}{\color{blue}\Verb$1$}\Verb$ ${\color{blue}\Verb$e$}{\color{blue}\Verb$2$} & Positive lookahead. \\
\hline
\end{tabular}\vskip 0.2em

\newpage
\subsection*{Iteration}\label{ss:iter}

To describe iteration terminating conditions we define an additional atomic expression:

\vskip 0.2em\noindent\begin{tabular}{| l | l || p{11.3cm} |}
\hline
\Verb$b$\Verb$r$\Verb$e$\Verb$a$\Verb$k$         &{\color{red}\Verb$c$}{\color{red}\Verb$o$}{\color{red}\Verb$r$}{\color{red}\Verb$e$}          & To terminate iteration.\\
\hline
\end{tabular}\vskip 0.2em

 We explain iteration by auxiliary  {\it repeat-until} operator. Repeat-until repeatedly tries to match its body and ends only after \verb$Stop$ is encountered. If that iteration fails then repeat-until fails.

\vskip 0.2em\noindent\begin{tabular}{| l | l || p{10cm} |}
\hline
{\color{blue}\Verb$e$}{\color{black}\Verb$*$}{\color{black}\Verb$*$}         & auxiliary & repeat-until\\
{\color{red}\Verb$S$}{\color{red}\Verb$t$}{\color{red}\Verb$o$}{\color{red}\Verb$p$}         & auxiliary & Stop iteration\\
{\color{blue}\Verb$e$}{\color{black}\Verb$*$}         &{\color{blue}\Verb$e$}{\color{black}\Verb$*$}{\color{black}\Verb$*$}          & When {\color{blue}\Verb$e$} contains \verb$Stop$,\\
{\color{blue}\Verb$e$}{\color{black}\Verb$*$}         &\Verb$(${\color{blue}\Verb$e$}\Verb$|${\color{red}\Verb$S$}{\color{red}\Verb$t$}{\color{red}\Verb$o$}{\color{red}\Verb$p$}\Verb$)${\color{black}\Verb$*$}{\color{black}\Verb$*$}          & otherwise. \\
\Verb$b$\Verb$r$\Verb$e$\Verb$a$\Verb$k$         &{\color{red}\Verb$C$}{\color{red}\Verb$u$}{\color{red}\Verb$t$}\Verb$ ${\color{red}\Verb$S$}{\color{red}\Verb$t$}{\color{red}\Verb$o$}{\color{red}\Verb$p$}          & Possible expansion.\\
\hline
\end{tabular}\vskip 0.2em

\subsection*{Examples}
Operators  \verb$Cut$ and \verb$Stop$ were introduced to describe a semantic of \verb$break$. A common task is to collect characters until certain character occurs. An {\color{red}\Verb$u$}{\color{red}\Verb$n$}{\color{red}\Verb$t$}{\color{red}\Verb$i$}{\color{red}\Verb$l$} rule defined in standard prologue has following implementation:

{\small
\begin{spacing}{0.8}
\begin{prolbox}
{\color{red}\Verb$u$}{\color{red}\Verb$n$}{\color{red}\Verb$t$}{\color{red}\Verb$i$}{\color{red}\Verb$l$}{\color{green}\Verb$($}{\color{green}\Verb$c$}{\color{green}\Verb$h$}{\color{green}\Verb$r$}{\color{green}\Verb$)$}\Verb$ $\Verb$=$\Verb$ $\Verb$($\Verb$ ${\color{red}\Verb$s$}{\color{red}\Verb$e$}{\color{red}\Verb$q$}{\color{green}\Verb$($}{\color{green}\Verb$c$}{\color{green}\Verb$h$}{\color{green}\Verb$r$}{\color{green}\Verb$)$}\Verb$ $\Verb$b$\Verb$r$\Verb$e$\Verb$a$\Verb$k$\\\Verb$ $\Verb$ $\Verb$ $\Verb$ $\Verb$ $\Verb$ $\Verb$ $\Verb$ $\Verb$ $\Verb$ $\Verb$ $\Verb$ $\Verb$ $\Verb$|$\Verb$ ${\color{black}\Verb$'$}{\color{black}\Verb$\$}{\color{black}\Verb$\$}{\color{black}\Verb$'$}{\color{Aquamarine}\Verb$:$}{\color{Aquamarine}\Verb$[$}{\color{Aquamarine}\Verb$x$}{\color{Aquamarine}\Verb$]$}\Verb$ $\Verb$.${\color{Aquamarine}\Verb$:$}{\color{Aquamarine}\Verb$[$}{\color{Aquamarine}\Verb$x$}{\color{Aquamarine}\Verb$]$}\\\Verb$ $\Verb$ $\Verb$ $\Verb$ $\Verb$ $\Verb$ $\Verb$ $\Verb$ $\Verb$ $\Verb$ $\Verb$ $\Verb$ $\Verb$ $\Verb$|$\Verb$ $\Verb$.${\color{Aquamarine}\Verb$:$}{\color{Aquamarine}\Verb$[$}{\color{Aquamarine}\Verb$x$}{\color{Aquamarine}\Verb$]$}\\\Verb$ $\Verb$ $\Verb$ $\Verb$ $\Verb$ $\Verb$ $\Verb$ $\Verb$ $\Verb$ $\Verb$ $\Verb$ $\Verb$ $\Verb$ $\Verb$)${\color{black}\Verb$*$}\Verb$ ${\color{Tan}\Verb$-$}{\color{Tan}\Verb$>$}{\color{Tan}\Verb$ $}{\color{Aquamarine}\Verb$x$}{\color{Tan}\Verb$.$}{\color{Tan}\Verb$j$}{\color{Tan}\Verb$o$}{\color{Tan}\Verb$i$}{\color{Tan}\Verb$n$}
\end{prolbox}\end{spacing}
}

For example, rule {\color{red}\Verb$l$}{\color{red}\Verb$i$}{\color{red}\Verb$n$}{\color{red}\Verb$e$} can be implemented as:
 
{\small
\begin{spacing}{0.8}
\begin{prolbox}
{\color{red}\Verb$n$}{\color{red}\Verb$e$}{\color{red}\Verb$w$}{\color{red}\Verb$l$}{\color{red}\Verb$i$}{\color{red}\Verb$n$}{\color{red}\Verb$e$}\Verb$ $\Verb$=$\Verb$ ${\color{black}\Verb$'$}{\color{black}\Verb$\$}{\color{black}\Verb$r$}{\color{black}\Verb$\$}{\color{black}\Verb$n$}{\color{black}\Verb$'$}\Verb$ $\Verb$|$\Verb$ ${\color{black}\Verb$'$}{\color{black}\Verb$\$}{\color{black}\Verb$r$}{\color{black}\Verb$'$}\Verb$ $\Verb$|$\Verb$ ${\color{black}\Verb$'$}{\color{black}\Verb$\$}{\color{black}\Verb$n$}{\color{black}\Verb$'$}\Verb$ $\\{\color{red}\Verb$l$}{\color{red}\Verb$i$}{\color{red}\Verb$n$}{\color{red}\Verb$e$}\Verb$ $\Verb$ $\Verb$ $\Verb$ $\Verb$=$\Verb$ ${\color{red}\Verb$u$}{\color{red}\Verb$n$}{\color{red}\Verb$t$}{\color{red}\Verb$i$}{\color{red}\Verb$l$}{\color{Violet}\Verb$($}{\color{Violet}\Verb$|$}{\color{green}\Verb$ $}{\color{red}\Verb$n$}{\color{red}\Verb$e$}{\color{red}\Verb$w$}{\color{red}\Verb$l$}{\color{red}\Verb$i$}{\color{red}\Verb$n$}{\color{red}\Verb$e$}{\color{green}\Verb$ $}{\color{Violet}\Verb$|$}{\color{Violet}\Verb$)$}
\end{prolbox}\end{spacing}
}

Some functionality of \verb$C$ standard library can be translated into amethyst as:

\vskip 0.2em\noindent\begin{tabular}{| l | l |}
\hline
C variant & Amethyst variant \\
\hline
\verb$scanf("%i")$  & {\color{red}\Verb$i$}{\color{red}\Verb$n$}{\color{red}\Verb$t$} \\
\verb$scanf("%f")$  & {\color{red}\Verb$f$}{\color{red}\Verb$l$}{\color{red}\Verb$o$}{\color{red}\Verb$a$}{\color{red}\Verb$t$}  \\
\verb$scanf("%[xyz]")$& {\color{red}\Verb$u$}{\color{red}\Verb$n$}{\color{red}\Verb$t$}{\color{red}\Verb$i$}{\color{red}\Verb$l$}{\color{Violet}\Verb$($}{\color{Violet}\Verb$|$}{\color{green}\Verb$ $}{\color{green}\Verb$<$}{\color{green}\Verb$x$}{\color{green}\Verb$y$}{\color{green}\Verb$z$}{\color{green}\Verb$>$}{\color{green}\Verb$ $}{\color{Violet}\Verb$|$}{\color{Violet}\Verb$)$} \\
\verb$scanf("%s")$  & {\color{red}\Verb$u$}{\color{red}\Verb$n$}{\color{red}\Verb$t$}{\color{red}\Verb$i$}{\color{red}\Verb$l$}{\color{Violet}\Verb$($}{\color{Violet}\Verb$|$}{\color{green}\Verb$ $}{\color{red}\Verb$_$}{\color{green}\Verb$ $}{\color{Violet}\Verb$|$}{\color{Violet}\Verb$)$} \\
\verb$gets$         & {\color{red}\Verb$l$}{\color{red}\Verb$i$}{\color{red}\Verb$n$}{\color{red}\Verb$e$}\Verb$ $\Verb$=$\Verb$ ${\color{red}\Verb$u$}{\color{red}\Verb$n$}{\color{red}\Verb$t$}{\color{red}\Verb$i$}{\color{red}\Verb$l$}{\color{Violet}\Verb$($}{\color{Violet}\Verb$|$}{\color{green}\Verb$ $}{\color{red}\Verb$n$}{\color{red}\Verb$e$}{\color{red}\Verb$w$}{\color{red}\Verb$l$}{\color{red}\Verb$i$}{\color{red}\Verb$n$}{\color{red}\Verb$e$}{\color{green}\Verb$ $}{\color{Violet}\Verb$|$}{\color{Violet}\Verb$)$} \\
\hline
\end{tabular}\vskip 0.2em

\newpage
\section{Inheritance}
In object oriented languages  inheritance is a form of reusing code by subtyping existing objects.
In Ruby class names must be capitalized.
Ruby has simple inheritance with mixins as is shown it the following example:

\vskip -1.8em\begin{spacing}{0.8}
{\small
\begin{exambox}
\Verb$c$\Verb$l$\Verb$a$\Verb$s$\Verb$s$\Verb$ $\Verb$F$\Verb$o$\Verb$o$\\\Verb$ $\Verb$ $\Verb$d$\Verb$e$\Verb$f$\Verb$ $\Verb$f$\Verb$o$\Verb$o$\Verb$;$\Verb$ $\Verb$4$\Verb$2$\Verb$ $\Verb$ $\Verb$ $\Verb$ $\Verb$ $\Verb$ $\Verb$;$\Verb$e$\Verb$n$\Verb$d$\\\Verb$e$\Verb$n$\Verb$d$\\\Verb$p$\Verb$u$\Verb$t$\Verb$s$\Verb$ $\Verb$F$\Verb$o$\Verb$o$\Verb$.$\Verb$n$\Verb$e$\Verb$w$\Verb$.$\Verb$f$\Verb$o$\Verb$o$\Verb$ $\Verb$#$\Verb$-$\Verb$>$\Verb$ $\Verb$4$\Verb$2$\\\Verb$c$\Verb$l$\Verb$a$\Verb$s$\Verb$s$\Verb$ $\Verb$B$\Verb$a$\Verb$r$\Verb$ $\Verb$<$\Verb$ $\Verb$F$\Verb$o$\Verb$o$\\\Verb$ $\Verb$ $\Verb$d$\Verb$e$\Verb$f$\Verb$ $\Verb$f$\Verb$o$\Verb$o$\Verb$;$\Verb$ $\Verb$s$\Verb$u$\Verb$p$\Verb$e$\Verb$r$\Verb$+$\Verb$1$\Verb$ $\Verb$;$\Verb$e$\Verb$n$\Verb$d$\\\Verb$e$\Verb$n$\Verb$d$\\\Verb$p$\Verb$u$\Verb$t$\Verb$s$\Verb$ $\Verb$B$\Verb$a$\Verb$r$\Verb$.$\Verb$n$\Verb$e$\Verb$w$\Verb$.$\Verb$f$\Verb$o$\Verb$o$\Verb$ $\Verb$#$\Verb$-$\Verb$>$\Verb$ $\Verb$4$\Verb$3$\\\Verb$m$\Verb$o$\Verb$d$\Verb$u$\Verb$l$\Verb$e$\Verb$ $\Verb$B$\Verb$a$\Verb$z$\\\Verb$ $\Verb$ $\Verb$d$\Verb$e$\Verb$f$\Verb$ $\Verb$f$\Verb$o$\Verb$o$\Verb$;$\Verb$ $\Verb$s$\Verb$u$\Verb$p$\Verb$e$\Verb$r$\Verb$*$\Verb$2$\Verb$ $\Verb$;$\Verb$e$\Verb$n$\Verb$d$\\\Verb$e$\Verb$n$\Verb$d$\\\Verb$c$\Verb$l$\Verb$a$\Verb$s$\Verb$s$\Verb$ $\Verb$B$\Verb$a$\Verb$r$\Verb$ $\Verb$<$\Verb$ $\Verb$F$\Verb$o$\Verb$o$\Verb$ $\Verb$#$\Verb$c$\Verb$l$\Verb$a$\Verb$s$\Verb$s$\Verb$ $\Verb$c$\Verb$a$\Verb$n$\Verb$ $\Verb$b$\Verb$e$\Verb$ $\Verb$d$\Verb$e$\Verb$f$\Verb$i$\Verb$n$\Verb$e$\Verb$d$\Verb$ $\Verb$p$\Verb$i$\Verb$e$\Verb$c$\Verb$e$\Verb$w$\Verb$i$\Verb$s$\Verb$e$\Verb$.$\\\Verb$ $\Verb$ $\Verb$i$\Verb$n$\Verb$c$\Verb$l$\Verb$u$\Verb$d$\Verb$e$\Verb$ $\Verb$B$\Verb$a$\Verb$z$\Verb$ $\Verb$ $\Verb$ $\Verb$#$\Verb$i$\Verb$n$\Verb$c$\Verb$l$\Verb$u$\Verb$d$\Verb$e$\Verb$ $\Verb$m$\Verb$o$\Verb$d$\Verb$u$\Verb$l$\Verb$e$\\\Verb$e$\Verb$n$\Verb$d$\\\Verb$p$\Verb$u$\Verb$t$\Verb$s$\Verb$ $\Verb$B$\Verb$a$\Verb$r$\Verb$.$\Verb$n$\Verb$e$\Verb$w$\Verb$.$\Verb$f$\Verb$o$\Verb$o$\Verb$ $\Verb$#$\Verb$-$\Verb$>$\Verb$ $\Verb$8$\Verb$5$\\\Verb$#$\Verb$R$\Verb$u$\Verb$b$\Verb$y$\Verb$ $\Verb$i$\Verb$m$\Verb$p$\Verb$l$\Verb$e$\Verb$m$\Verb$e$\Verb$n$\Verb$t$\Verb$s$\Verb$ $\Verb$m$\Verb$i$\Verb$x$\Verb$i$\Verb$n$\Verb$ $\Verb$b$\Verb$y$\Verb$ $\Verb$i$\Verb$n$\Verb$s$\Verb$e$\Verb$r$\Verb$t$\Verb$i$\Verb$n$\Verb$g$\Verb$ $\Verb$c$\Verb$l$\Verb$a$\Verb$s$\Verb$s$\Verb$ $\Verb$b$\Verb$e$\Verb$t$\Verb$w$\Verb$e$\Verb$e$\Verb$n$\Verb$ $\\\Verb$#$\Verb$c$\Verb$u$\Verb$r$\Verb$r$\Verb$e$\Verb$n$\Verb$t$\Verb$ $\Verb$a$\Verb$n$\Verb$d$\Verb$ $\Verb$p$\Verb$a$\Verb$r$\Verb$e$\Verb$n$\Verb$t$\Verb$ $\Verb$c$\Verb$l$\Verb$a$\Verb$s$\Verb$s$
\end{exambox}
}
\end{spacing}\vskip -0.4em

\subsection*{Inheritance in amethyst}
We reuse Ruby class system for inheritance. Ruby class names must be capitalized.

\vskip -1.8em\begin{spacing}{0.8}
{\small
\begin{exambox}
\Verb$a$\Verb$m$\Verb$e$\Verb$t$\Verb$h$\Verb$y$\Verb$s$\Verb$t$\Verb$ $\Verb$F$\Verb$o$\Verb$o$\Verb$ $\Verb${$\\\Verb$ $\Verb$ ${\color{red}\Verb$f$}{\color{red}\Verb$o$}{\color{red}\Verb$o$}\Verb$ $\Verb$=$\Verb$ ${\color{Tan}\Verb${$}{\color{Tan}\Verb$4$}{\color{Tan}\Verb$2$}{\color{Tan}\Verb$ $}{\color{Tan}\Verb$}$}\\\Verb$}$\\\Verb$a$\Verb$m$\Verb$e$\Verb$t$\Verb$h$\Verb$y$\Verb$s$\Verb$t$\Verb$_$\Verb$m$\Verb$o$\Verb$d$\Verb$u$\Verb$l$\Verb$e$\Verb$ $\Verb$M$\Verb$o$\Verb$d$\Verb$ $\Verb${$\\\Verb$ $\Verb$ $\Verb$f$\Verb$o$\Verb$o$\Verb$ $\Verb$=$\Verb$ $\Verb$s$\Verb$u$\Verb$p$\Verb$e$\Verb$r$\Verb$:$\Verb$x$\Verb$ $\Verb${$\Verb$ $\Verb$2$\Verb$*$\Verb$x$\Verb$ $\Verb$}$\\\Verb$}$\\\Verb$a$\Verb$m$\Verb$e$\Verb$t$\Verb$h$\Verb$y$\Verb$s$\Verb$t$\Verb$_$\Verb$m$\Verb$o$\Verb$d$\Verb$u$\Verb$l$\Verb$e$\Verb$ $\Verb$B$\Verb$a$\Verb$z$\Verb$ $\Verb${$\\\Verb$ $\Verb$ $\Verb$b$\Verb$a$\Verb$z$\Verb$ $\Verb$=$\Verb$ $\Verb${$\Verb$"$\Verb$b$\Verb$a$\Verb$z$\Verb$"$\Verb$}$\\\Verb$}$\\\Verb$c$\Verb$l$\Verb$a$\Verb$s$\Verb$s$\Verb$ $\Verb$B$\Verb$a$\Verb$r$\Verb$ $\Verb$<$\Verb$ $\Verb$F$\Verb$o$\Verb$o$\\\Verb$ $\Verb$ $\Verb$i$\Verb$n$\Verb$c$\Verb$l$\Verb$u$\Verb$d$\Verb$e$\Verb$ $\Verb$M$\Verb$o$\Verb$d$\\\Verb$e$\Verb$n$\Verb$d$\\\Verb$a$\Verb$m$\Verb$e$\Verb$t$\Verb$h$\Verb$y$\Verb$s$\Verb$t$\Verb$ $\Verb$B$\Verb$a$\Verb$r$\Verb$ $\Verb$<$\Verb$ $\Verb$F$\Verb$o$\Verb$o$\Verb$ $\Verb${$\\\Verb$ $\Verb$ ${\color{red}\Verb$f$}{\color{red}\Verb$o$}{\color{red}\Verb$o$}\Verb$ $\Verb$ $\Verb$ $\Verb$ $\Verb$ $\Verb$ $\Verb$=$\Verb$ ${\color{red}\Verb$s$}{\color{red}\Verb$u$}{\color{red}\Verb$p$}{\color{red}\Verb$e$}{\color{red}\Verb$r$}{\color{Aquamarine}\Verb$:$}{\color{Aquamarine}\Verb$x$}\Verb$ ${\color{Tan}\Verb${$}{\color{Aquamarine}\Verb$x$}{\color{Tan}\Verb$+$}{\color{Tan}\Verb$1$}{\color{Tan}\Verb$}$}\\\Verb$ $\Verb$ ${\color{red}\Verb$f$}{\color{red}\Verb$o$}{\color{red}\Verb$o$}{\color{red}\Verb$_$}{\color{red}\Verb$o$}{\color{red}\Verb$r$}{\color{red}\Verb$i$}{\color{red}\Verb$g$}\Verb$ $\Verb$=$\Verb$ ${\color{red}\Verb$F$}{\color{red}\Verb$o$}{\color{red}\Verb$o$}{\color{red}\Verb$:$}{\color{red}\Verb$:$}{\color{red}\Verb$f$}{\color{red}\Verb$o$}{\color{red}\Verb$o$}\\\Verb$ $\Verb$ ${\color{red}\Verb$b$}{\color{red}\Verb$a$}{\color{red}\Verb$z$}\Verb$ $\Verb$ $\Verb$ $\Verb$ $\Verb$ $\Verb$ $\Verb$=$\Verb$ ${\color{red}\Verb$B$}{\color{red}\Verb$a$}{\color{red}\Verb$z$}{\color{red}\Verb$:$}{\color{red}\Verb$:$}{\color{red}\Verb$b$}{\color{red}\Verb$a$}{\color{red}\Verb$z$}\\\Verb$}$
\end{exambox}
}
\end{spacing}\vskip -0.4em

One can use \verb$Grammar::rule$ syntax to call rule from ancestor or rule from module. Calling rule of arbitrary grammar is not allowed.

\newpage
\section{Pattern matching of tree-like structures} \label{matchtree}
Amethyst takes inspiration from an OMeta (2007) \cite{ometa} which extended parsing expression grammars (2002).
One of extensions made in OMeta is pattern matching of tree-like data structures. We further extend this work in several respects. One described in the next section is extending pattern matching to arbitrary data structures with possible cyclic references. 

All operators defined so far carry over into this setting. An \verb$Enter$ operator and parametrized rules are essential for this transition.

\subsection{Pattern matching in functional languages}

Most functional languages offer limited form of pattern matching. While syntax is different it usually boils down to the following constructs:

\vskip 0.2em\noindent\begin{tabular}{| l | p{11.2cm} |}
\hline
Expression & Description \\
\hline
\verb$Struct$ & Match when it is described structure with given name\\
\verb$:x   $ & Bind head to variable x\\
\verb$exp1   exp2$ & Sequencing\\
\verb$exp1 | exp2$ & Choice\\
\verb$[ exp ]    $ & Enter - take head and match it recursively with exp\\
\hline
\end{tabular}\vskip 0.2em

Our framework extends these operations with iteration constructs and other features.

\subsection{Matching nested arrays in amethyst}
Recall following amethyst operators. For matching arrays \verb$Enter$ can omit leading \extext{.} as syntax sugar.

\vskip 0.2em\noindent\begin{tabular}{| l | l || p{5cm} |}
\hline
Expression & Expansion & Description \\ 
\hline   
\Verb$.$ & core & Match single element\\
{\color{blue}\Verb$e$}{\color{blue}\Verb$1$}{\color{blue}\Verb$[$}\Verb$ ${\color{blue}\Verb$e$}{\color{blue}\Verb$2$}\Verb$ ${\color{blue}\Verb$]$} & core & Enter operator.\\
{\color{blue}\Verb$[$}\Verb$ ${\color{blue}\Verb$e$}\Verb$ ${\color{blue}\Verb$]$} & \Verb$.${\color{blue}\Verb$[$}\Verb$ ${\color{blue}\Verb$e$}\Verb$ ${\color{blue}\Verb$]$}\verb$                       $& For nested arrays.\\
\hline
\end{tabular}\vskip 0.2em

Then matching of arrays is quite natural:

\vskip -1.8em\begin{spacing}{0.8}
{\small
\begin{exambox}
{\color{Violet}\Verb$($}{\color{Violet}\Verb$|$}\Verb$ $\Verb$.${\color{Aquamarine}\Verb$:$}{\color{Aquamarine}\Verb$x$}\Verb$ ${\color{blue}\Verb$[$}\Verb$.${\color{Aquamarine}\Verb$:$}{\color{Aquamarine}\Verb$y$}\Verb$ $\Verb$.${\color{Aquamarine}\Verb$:$}{\color{Aquamarine}\Verb$z$}\Verb$ ${\color{blue}\Verb$]$}\Verb$ ${\color{Violet}\Verb$|$}{\color{Violet}\Verb$)$}\Verb$.$\Verb$m$\Verb$a$\Verb$t$\Verb$c$\Verb$h$\Verb$($\Verb$[$\Verb$1$\Verb$,$\Verb$[$\Verb$2$\Verb$,$\Verb$3$\Verb$]$\Verb$]$\Verb$)$\\\Verb$p$\Verb$u$\Verb$t$\Verb$s$\Verb$ $\Verb$x$\Verb$,$\Verb$y$\Verb$,$\Verb$z$\Verb$ $\Verb$#$\Verb$-$\Verb$>$\Verb$ $\Verb$1$\Verb$2$\Verb$3$
\end{exambox}
}
\end{spacing}\vskip -0.4em

\subsection{Classes and pattern matching in Ruby}

A common convention to construct tree like structures in Ruby is to define \verb$[]$ class method. One way how to construct syntax trees in Ruby source code is the following:

\vskip -1.8em\begin{spacing}{0.8}
{\small
\begin{exambox}
\Verb$P$\Verb$l$\Verb$u$\Verb$s$\Verb$[$\Verb$1$\Verb$,$\Verb$T$\Verb$i$\Verb$m$\Verb$e$\Verb$s$\Verb$[$\Verb$2$\Verb$,$\Verb$3$\Verb$,$\Verb$P$\Verb$l$\Verb$u$\Verb$s$\Verb$[$\Verb$4$\Verb$,$\Verb$P$\Verb$l$\Verb$u$\Verb$s$\Verb$[$\Verb$1$\Verb$,$\Verb$5$\Verb$]$\Verb$]$\Verb$]$\Verb$,$\Verb$3$\Verb$]$
\end{exambox}
}
\end{spacing}\vskip -0.4em

In Ruby membership is tested by case construct. We demonstrate it on contrived implementation of logarithm:

\vskip -1.8em\begin{spacing}{0.8}
{\small
\begin{exambox}
\Verb$d$\Verb$e$\Verb$f$\Verb$ $\Verb$l$\Verb$o$\Verb$g$\Verb$1$\Verb$0$\Verb$($\Verb$x$\Verb$)$\\\Verb$ $\Verb$ $\Verb$ $\Verb$ $\Verb$c$\Verb$a$\Verb$s$\Verb$e$\Verb$ $\Verb$x$\\\Verb$ $\Verb$ $\Verb$ $\Verb$ $\Verb$w$\Verb$h$\Verb$e$\Verb$n$\Verb$ $\Verb$0$\Verb$ $\Verb$ $\Verb$ $\Verb$ $\Verb$ $\Verb$ $\Verb$ $\Verb$ $\Verb$ $\Verb$;$\Verb$ $\Verb$r$\Verb$a$\Verb$i$\Verb$s$\Verb$e$\Verb$ $\Verb$"$\Verb$n$\Verb$o$\Verb$t$\Verb$ $\Verb$d$\Verb$e$\Verb$f$\Verb$i$\Verb$n$\Verb$e$\Verb$d$\Verb$"$\\\Verb$ $\Verb$ $\Verb$ $\Verb$ $\Verb$w$\Verb$h$\Verb$e$\Verb$n$\Verb$ $\Verb$1$\Verb$.$\Verb$.$\Verb$9$\Verb$ $\Verb$ $\Verb$ $\Verb$ $\Verb$ $\Verb$ $\Verb$;$\Verb$ $\Verb$1$\\\Verb$ $\Verb$ $\Verb$ $\Verb$ $\Verb$w$\Verb$h$\Verb$e$\Verb$n$\Verb$ $\Verb$1$\Verb$0$\Verb$.$\Verb$.$\Verb$9$\Verb$9$\Verb$ $\Verb$ $\Verb$ $\Verb$ $\Verb$;$\Verb$ $\Verb$2$\\\Verb$ $\Verb$ $\Verb$ $\Verb$ $\Verb$w$\Verb$h$\Verb$e$\Verb$n$\Verb$ $\Verb$F$\Verb$l$\Verb$o$\Verb$a$\Verb$t$\Verb$ $\Verb$ $\Verb$ $\Verb$ $\Verb$ $\Verb$;$\Verb$ $\Verb$l$\Verb$o$\Verb$g$\Verb$1$\Verb$0$\Verb$($\Verb$x$\Verb$.$\Verb$t$\Verb$o$\Verb$_$\Verb$i$\Verb$)$\\\Verb$ $\Verb$ $\Verb$ $\Verb$ $\Verb$e$\Verb$l$\Verb$s$\Verb$e$\Verb$ $\Verb$ $\Verb$ $\Verb$ $\Verb$ $\Verb$ $\Verb$ $\Verb$ $\Verb$ $\Verb$ $\Verb$ $\Verb$;$\Verb$ $\Verb$l$\Verb$o$\Verb$g$\Verb$1$\Verb$0$\Verb$($\Verb$x$\Verb$/$\Verb$1$\Verb$0$\Verb$)$\Verb$+$\Verb$1$\\\Verb$ $\Verb$ $\Verb$ $\Verb$ $\Verb$e$\Verb$n$\Verb$d$\\\Verb$ $\Verb$ $\Verb$e$\Verb$n$\Verb$d$
\end{exambox}
}
\end{spacing}\vskip -0.4em

A case match is done by invoking \Verb$=$\Verb$=$\Verb$=$ method. Use cases of this method are diverse as demonstrated by the following examples:

\vskip 0.2em\noindent\begin{tabular}{| l | l |}
\hline
Left argument & Test performed \\
\hline
\verb$true,false,nil$     & Equality.\\
\verb$42, 3.14$           & Equality.\\
\verb$-42..42$ & Range membership.\\
\verb$Class$   & Class membership.\\
\verb$/exp/$   & Regular expression match.\\

\hline
\end{tabular}\vskip 0.2em

\subsection{Class membership}

Implementation of matching of basic types depends on the host language. We inform amethyst about this by defining a parametrized rule {\color{red}\Verb$m$}{\color{red}\Verb$e$}{\color{red}\Verb$m$}{\color{red}\Verb$b$}{\color{red}\Verb$e$}{\color{red}\Verb$r$}.
 To implement {\color{red}\Verb$m$}{\color{red}\Verb$e$}{\color{red}\Verb$m$}{\color{red}\Verb$b$}{\color{red}\Verb$e$}{\color{red}\Verb$r$} rule in Ruby we use \verb$===$ operator from previous section:

{\small
\begin{spacing}{0.8}
\begin{prolbox}
{\color{red}\Verb$m$}{\color{red}\Verb$e$}{\color{red}\Verb$m$}{\color{red}\Verb$b$}{\color{red}\Verb$e$}{\color{red}\Verb$r$}{\color{green}\Verb$($}{\color{green}\Verb$x$}{\color{green}\Verb$)$}\Verb$ $\Verb$=$\Verb$ $\Verb$.${\color{Aquamarine}\Verb$:$}{\color{Aquamarine}\Verb$a$}\Verb$ ${\color{Violet}\Verb$&$}{\color{Violet}\Verb${$}{\color{Tan}\Verb$x$}{\color{Tan}\Verb$ $}{\color{Tan}\Verb$=$}{\color{Tan}\Verb$=$}{\color{Tan}\Verb$=$}{\color{Tan}\Verb$ $}{\color{Aquamarine}\Verb$a$}{\color{Violet}\Verb$}$}\Verb$ ${\color{Tan}\Verb${$}{\color{Aquamarine}\Verb$a$}{\color{Tan}\Verb$}$}
\end{prolbox}\end{spacing}
}

We define tests for following basic types:

\vskip 0.2em\noindent\begin{tabular}{| l | l |}
\hline
Expression & Expansion \\
\hline
\verb$true,false,nil$     & {\color{red}\Verb$m$}{\color{red}\Verb$e$}{\color{red}\Verb$m$}{\color{red}\Verb$b$}{\color{red}\Verb$e$}{\color{red}\Verb$r$}{\color{green}\Verb$($}{\color{green}\Verb$t$}{\color{green}\Verb$r$}{\color{green}\Verb$u$}{\color{green}\Verb$e$}{\color{green}\Verb$)$},...\\
\verb$42$                 & {\color{red}\Verb$m$}{\color{red}\Verb$e$}{\color{red}\Verb$m$}{\color{red}\Verb$b$}{\color{red}\Verb$e$}{\color{red}\Verb$r$}{\color{green}\Verb$($}{\color{green}\Verb$4$}{\color{green}\Verb$2$}{\color{green}\Verb$)$}\verb$                                         $\\
\verb$-42..42$            & {\color{red}\Verb$m$}{\color{red}\Verb$e$}{\color{red}\Verb$m$}{\color{red}\Verb$b$}{\color{red}\Verb$e$}{\color{red}\Verb$r$}{\color{green}\Verb$($}{\color{green}\Verb$-$}{\color{green}\Verb$4$}{\color{green}\Verb$2$}{\color{green}\Verb$.$}{\color{green}\Verb$.$}{\color{green}\Verb$4$}{\color{green}\Verb$2$}{\color{green}\Verb$)$}\\
\verb$Class$              & {\color{red}\Verb$m$}{\color{red}\Verb$e$}{\color{red}\Verb$m$}{\color{red}\Verb$b$}{\color{red}\Verb$e$}{\color{red}\Verb$r$}{\color{green}\Verb$($}{\color{green}\Verb$C$}{\color{green}\Verb$l$}{\color{green}\Verb$a$}{\color{green}\Verb$s$}{\color{green}\Verb$s$}{\color{green}\Verb$)$}\\
\hline
\end{tabular}\vskip 0.2em

\newpage
\subsection{Building abstract syntax trees} \label{calcast}
When we parse we typically build some abstract syntax tree. The following atomic expression makes creation of AST more convenient.

\vskip 0.2em\noindent\begin{tabular}{| l | l || p{4.5cm} |}
\hline
Expression & Expansion & Description \\
\hline
{\color{Aquamarine}\Verb$@$}{\color{Aquamarine}\Verb$C$}{\color{Aquamarine}\Verb$l$}{\color{Aquamarine}\Verb$a$}{\color{Aquamarine}\Verb$s$}{\color{Aquamarine}\Verb$s$}& {\color{Tan}\Verb${$}{\color{Tan}\Verb$C$}{\color{Tan}\Verb$l$}{\color{Tan}\Verb$a$}{\color{Tan}\Verb$s$}{\color{Tan}\Verb$s$}{\color{Tan}\Verb$.$}{\color{Tan}\Verb$c$}{\color{Tan}\Verb$r$}{\color{Tan}\Verb$e$}{\color{Tan}\Verb$a$}{\color{Tan}\Verb$t$}{\color{Tan}\Verb$e$}{\color{Tan}\Verb$($}{\color{Tan}\Verb$l$}{\color{Tan}\Verb$o$}{\color{Tan}\Verb$c$}{\color{Tan}\Verb$a$}{\color{Tan}\Verb$l$}{\color{Tan}\Verb$_$}{\color{Tan}\Verb$v$}{\color{Tan}\Verb$a$}{\color{Tan}\Verb$r$}{\color{Tan}\Verb$i$}{\color{Tan}\Verb$a$}{\color{Tan}\Verb$b$}{\color{Tan}\Verb$l$}{\color{Tan}\Verb$e$}{\color{Tan}\Verb$s$}{\color{Tan}\Verb$)$}{\color{Tan}\Verb$}$} & Create object. \\
\hline
\end{tabular}\vskip 0.2em

This syntax also encourages proper naming of variables. Assume we want to change calculator from first section to produce a syntax tree. Possible implementation is:

\vskip -1.8em\begin{spacing}{0.8}
{\small
\begin{exambox}
\Verb$c$\Verb$l$\Verb$a$\Verb$s$\Verb$s$\Verb$ $\Verb$A$\Verb$d$\Verb$d$\Verb$ $\\\Verb$ $\Verb$ $\Verb$d$\Verb$e$\Verb$f$\Verb$ $\Verb$s$\Verb$e$\Verb$l$\Verb$f$\Verb$.$\Verb$c$\Verb$r$\Verb$e$\Verb$a$\Verb$t$\Verb$e$\Verb$($\Verb$h$\Verb$a$\Verb$s$\Verb$h$\Verb$)$\\\Verb$ $\Verb$ $\Verb$ $\Verb$ $\Verb$a$\Verb$=$\Verb$A$\Verb$d$\Verb$d$\Verb$.$\Verb$n$\Verb$e$\Verb$w$\\\Verb$ $\Verb$ $\Verb$ $\Verb$ $\Verb$a$\Verb$.$\Verb$x$\Verb$=$\Verb$h$\Verb$a$\Verb$s$\Verb$h$\Verb$[$\Verb$:$\Verb$x$\Verb$]$\\\Verb$ $\Verb$ $\Verb$ $\Verb$ $\Verb$a$\Verb$.$\Verb$y$\Verb$=$\Verb$h$\Verb$a$\Verb$s$\Verb$h$\Verb$[$\Verb$:$\Verb$y$\Verb$]$\\\Verb$ $\Verb$ $\Verb$ $\Verb$ $\Verb$r$\Verb$e$\Verb$t$\Verb$u$\Verb$r$\Verb$n$\Verb$ $\Verb$a$\\\Verb$ $\Verb$ $\Verb$e$\Verb$n$\Verb$d$\\\Verb$ $\Verb$ $\Verb$d$\Verb$e$\Verb$f$\Verb$ $\Verb$a$\Verb$m$\Verb$e$\Verb$t$\Verb$h$\Verb$y$\Verb$s$\Verb$t$\Verb$_$\Verb$a$\Verb$r$\Verb$r$\Verb$a$\Verb$y$\Verb$ $\\\Verb$ $\Verb$ $\Verb$ $\Verb$ $\Verb$[$\Verb$@$\Verb$x$\Verb$,$\Verb$@$\Verb$y$\Verb$]$\\\Verb$ $\Verb$ $\Verb$e$\Verb$n$\Verb$d$\\\Verb$e$\Verb$n$\Verb$d$\\\Verb$#$\Verb$ $\Verb$.$\Verb$.$\Verb$.$\\\Verb$a$\Verb$m$\Verb$e$\Verb$t$\Verb$h$\Verb$y$\Verb$s$\Verb$t$\Verb$ $\Verb$C$\Verb$a$\Verb$l$\Verb$c$\Verb$u$\Verb$l$\Verb$a$\Verb$t$\Verb$o$\Verb$r$\Verb$_$\Verb$A$\Verb$S$\Verb$T$\Verb$ $\Verb${$\\\Verb$ $\Verb$ ${\color{red}\Verb$c$}{\color{red}\Verb$a$}{\color{red}\Verb$l$}{\color{red}\Verb$c$}{\color{red}\Verb$u$}{\color{red}\Verb$l$}{\color{red}\Verb$a$}{\color{red}\Verb$t$}{\color{red}\Verb$e$}{\color{red}\Verb$_$}{\color{red}\Verb$a$}{\color{red}\Verb$s$}{\color{red}\Verb$t$}\Verb$ $\Verb$=$\Verb$ ${\color{red}\Verb$a$}{\color{red}\Verb$d$}{\color{red}\Verb$d$}{\color{red}\Verb$_$}{\color{red}\Verb$e$}{\color{red}\Verb$x$}{\color{red}\Verb$p$}{\color{red}\Verb$r$}\Verb$ $\Verb$ $\\\\\Verb$ $\Verb$ ${\color{red}\Verb$a$}{\color{red}\Verb$d$}{\color{red}\Verb$d$}{\color{red}\Verb$_$}{\color{red}\Verb$e$}{\color{red}\Verb$x$}{\color{red}\Verb$p$}{\color{red}\Verb$r$}\Verb$ $\Verb$ $\Verb$=$\Verb$ ${\color{red}\Verb$a$}{\color{red}\Verb$d$}{\color{red}\Verb$d$}{\color{red}\Verb$_$}{\color{red}\Verb$e$}{\color{red}\Verb$x$}{\color{red}\Verb$p$}{\color{red}\Verb$r$}{\color{Aquamarine}\Verb$:$}{\color{Aquamarine}\Verb$x$}\Verb$ ${\color{black}\Verb$"$}{\color{black}\Verb$+$}{\color{black}\Verb$"$}\Verb$ ${\color{red}\Verb$m$}{\color{red}\Verb$u$}{\color{red}\Verb$l$}{\color{red}\Verb$_$}{\color{red}\Verb$e$}{\color{red}\Verb$x$}{\color{red}\Verb$p$}{\color{red}\Verb$r$}{\color{Aquamarine}\Verb$:$}{\color{Aquamarine}\Verb$y$}\Verb$ $\Verb$ $\Verb$ ${\color{Aquamarine}\Verb$@$}{\color{Aquamarine}\Verb$A$}{\color{Aquamarine}\Verb$d$}{\color{Aquamarine}\Verb$d$}\\\Verb$ $\Verb$ $\Verb$ $\Verb$ $\Verb$ $\Verb$ $\Verb$ $\Verb$ $\Verb$ $\Verb$ $\Verb$ $\Verb$ $\Verb$|$\Verb$ ${\color{red}\Verb$a$}{\color{red}\Verb$d$}{\color{red}\Verb$d$}{\color{red}\Verb$_$}{\color{red}\Verb$e$}{\color{red}\Verb$x$}{\color{red}\Verb$p$}{\color{red}\Verb$r$}{\color{Aquamarine}\Verb$:$}{\color{Aquamarine}\Verb$x$}\Verb$ ${\color{black}\Verb$"$}{\color{black}\Verb$-$}{\color{black}\Verb$"$}\Verb$ ${\color{red}\Verb$m$}{\color{red}\Verb$u$}{\color{red}\Verb$l$}{\color{red}\Verb$_$}{\color{red}\Verb$e$}{\color{red}\Verb$x$}{\color{red}\Verb$p$}{\color{red}\Verb$r$}{\color{Aquamarine}\Verb$:$}{\color{Aquamarine}\Verb$y$}\Verb$ $\Verb$ $\Verb$ ${\color{Aquamarine}\Verb$@$}{\color{Aquamarine}\Verb$S$}{\color{Aquamarine}\Verb$u$}{\color{Aquamarine}\Verb$b$}{\color{Aquamarine}\Verb$s$}{\color{Aquamarine}\Verb$t$}{\color{Aquamarine}\Verb$r$}{\color{Aquamarine}\Verb$a$}{\color{Aquamarine}\Verb$c$}{\color{Aquamarine}\Verb$t$}\\\Verb$ $\Verb$ $\Verb$ $\Verb$ $\Verb$ $\Verb$ $\Verb$ $\Verb$ $\Verb$ $\Verb$ $\Verb$ $\Verb$ $\Verb$|$\Verb$ ${\color{red}\Verb$m$}{\color{red}\Verb$u$}{\color{red}\Verb$l$}{\color{red}\Verb$_$}{\color{red}\Verb$e$}{\color{red}\Verb$x$}{\color{red}\Verb$p$}{\color{red}\Verb$r$}\\\\\Verb$ $\Verb$ ${\color{red}\Verb$m$}{\color{red}\Verb$u$}{\color{red}\Verb$l$}{\color{red}\Verb$_$}{\color{red}\Verb$e$}{\color{red}\Verb$x$}{\color{red}\Verb$p$}{\color{red}\Verb$r$}\Verb$ $\Verb$ $\Verb$=$\Verb$ ${\color{red}\Verb$m$}{\color{red}\Verb$u$}{\color{red}\Verb$l$}{\color{red}\Verb$_$}{\color{red}\Verb$e$}{\color{red}\Verb$x$}{\color{red}\Verb$p$}{\color{red}\Verb$r$}{\color{Aquamarine}\Verb$:$}{\color{Aquamarine}\Verb$x$}\Verb$ ${\color{black}\Verb$"$}{\color{black}\Verb$*$}{\color{black}\Verb$"$}\Verb$ ${\color{red}\Verb$a$}{\color{red}\Verb$t$}{\color{red}\Verb$o$}{\color{red}\Verb$m$}{\color{red}\Verb$_$}{\color{red}\Verb$e$}{\color{red}\Verb$x$}{\color{red}\Verb$p$}{\color{red}\Verb$r$}{\color{Aquamarine}\Verb$:$}{\color{Aquamarine}\Verb$y$}\Verb$ $\Verb$ ${\color{Aquamarine}\Verb$@$}{\color{Aquamarine}\Verb$M$}{\color{Aquamarine}\Verb$u$}{\color{Aquamarine}\Verb$l$}{\color{Aquamarine}\Verb$t$}{\color{Aquamarine}\Verb$i$}{\color{Aquamarine}\Verb$p$}{\color{Aquamarine}\Verb$l$}{\color{Aquamarine}\Verb$y$}\\\Verb$ $\Verb$ $\Verb$ $\Verb$ $\Verb$ $\Verb$ $\Verb$ $\Verb$ $\Verb$ $\Verb$ $\Verb$ $\Verb$ $\Verb$|$\Verb$ ${\color{red}\Verb$m$}{\color{red}\Verb$u$}{\color{red}\Verb$l$}{\color{red}\Verb$_$}{\color{red}\Verb$e$}{\color{red}\Verb$x$}{\color{red}\Verb$p$}{\color{red}\Verb$r$}{\color{Aquamarine}\Verb$:$}{\color{Aquamarine}\Verb$x$}\Verb$ ${\color{black}\Verb$"$}{\color{black}\Verb$/$}{\color{black}\Verb$"$}\Verb$ ${\color{red}\Verb$a$}{\color{red}\Verb$t$}{\color{red}\Verb$o$}{\color{red}\Verb$m$}{\color{red}\Verb$_$}{\color{red}\Verb$e$}{\color{red}\Verb$x$}{\color{red}\Verb$p$}{\color{red}\Verb$r$}{\color{Aquamarine}\Verb$:$}{\color{Aquamarine}\Verb$y$}\Verb$ $\Verb$ ${\color{Aquamarine}\Verb$@$}{\color{Aquamarine}\Verb$D$}{\color{Aquamarine}\Verb$i$}{\color{Aquamarine}\Verb$v$}{\color{Aquamarine}\Verb$i$}{\color{Aquamarine}\Verb$d$}{\color{Aquamarine}\Verb$e$}\\\Verb$ $\Verb$ $\Verb$ $\Verb$ $\Verb$ $\Verb$ $\Verb$ $\Verb$ $\Verb$ $\Verb$ $\Verb$ $\Verb$ $\Verb$|$\Verb$ ${\color{red}\Verb$a$}{\color{red}\Verb$t$}{\color{red}\Verb$o$}{\color{red}\Verb$m$}{\color{red}\Verb$_$}{\color{red}\Verb$e$}{\color{red}\Verb$x$}{\color{red}\Verb$p$}{\color{red}\Verb$r$}\\\Verb$ $\\\Verb$ $\Verb$ ${\color{red}\Verb$a$}{\color{red}\Verb$t$}{\color{red}\Verb$o$}{\color{red}\Verb$m$}{\color{red}\Verb$_$}{\color{red}\Verb$e$}{\color{red}\Verb$x$}{\color{red}\Verb$p$}{\color{red}\Verb$r$}\Verb$ $\Verb$=$\Verb$ ${\color{black}\Verb$"$}{\color{black}\Verb$($}{\color{black}\Verb$"$}\Verb$ ${\color{red}\Verb$a$}{\color{red}\Verb$d$}{\color{red}\Verb$d$}{\color{red}\Verb$_$}{\color{red}\Verb$e$}{\color{red}\Verb$x$}{\color{red}\Verb$p$}{\color{red}\Verb$r$}{\color{Aquamarine}\Verb$:$}{\color{Aquamarine}\Verb$x$}\Verb$ ${\color{black}\Verb$"$}{\color{black}\Verb$)$}{\color{black}\Verb$"$}\Verb$ $\Verb$ $\Verb$ $\Verb$ $\Verb$ $\Verb$ $\Verb$ ${\color{Tan}\Verb$-$}{\color{Tan}\Verb$>$}{\color{Tan}\Verb$ $}{\color{Aquamarine}\Verb$x$}{\color{Tan}\\}{\color{Tan}\Verb$ $}\Verb$ $\Verb$ $\Verb$ $\Verb$ $\Verb$ $\Verb$ $\Verb$ $\Verb$ $\Verb$ $\Verb$ $\Verb$ $\Verb$|$\Verb$ ${\color{red}\Verb$f$}{\color{red}\Verb$l$}{\color{red}\Verb$o$}{\color{red}\Verb$a$}{\color{red}\Verb$t$}\\\Verb$}$
\end{exambox}
}
\end{spacing}\vskip -0.4em

\newpage
\subsection{Pattern matching of tree like structures}
Amethyst can match any object as an array. First\footnote{Unless object is a String or Array} amethyst tries to call method \verb$amethyst_array$ and to do matching on returned array. If \verb$amethyst_array$ method is not defined it pretends that empty array was returned. This is useful when we match arbitrary objects.

 \verb$Enter$ operator combined with property testing allows us write evaluator to calculator in the following way:

\vskip -1.8em\begin{spacing}{0.8}
{\small
\begin{exambox}
\Verb$a$\Verb$m$\Verb$e$\Verb$t$\Verb$h$\Verb$y$\Verb$s$\Verb$t$\Verb$ $\Verb$E$\Verb$v$\Verb$a$\Verb$l$\Verb$u$\Verb$a$\Verb$t$\Verb$o$\Verb$r$\Verb$ $\Verb$<$\Verb$ $\Verb$C$\Verb$a$\Verb$l$\Verb$c$\Verb$u$\Verb$l$\Verb$a$\Verb$t$\Verb$o$\Verb$r$\Verb$_$\Verb$A$\Verb$S$\Verb$T$\Verb$ $\Verb${$\\\Verb$ $\Verb$ ${\color{red}\Verb$e$}{\color{red}\Verb$v$}{\color{red}\Verb$a$}{\color{red}\Verb$l$}\Verb$ $\Verb$=$\Verb$ ${\color{red}\Verb$A$}{\color{red}\Verb$d$}{\color{red}\Verb$d$}{\color{blue}\Verb$[$}\Verb$ $\Verb$ $\Verb$ $\Verb$ $\Verb$ $\Verb$ ${\color{red}\Verb$e$}{\color{red}\Verb$v$}{\color{red}\Verb$a$}{\color{red}\Verb$l$}{\color{Aquamarine}\Verb$:$}{\color{Aquamarine}\Verb$x$}\Verb$ ${\color{red}\Verb$e$}{\color{red}\Verb$v$}{\color{red}\Verb$a$}{\color{red}\Verb$l$}{\color{Aquamarine}\Verb$:$}{\color{Aquamarine}\Verb$y$}\Verb$ ${\color{blue}\Verb$]$}\Verb$ ${\color{Tan}\Verb$-$}{\color{Tan}\Verb$>$}{\color{Tan}\Verb$ $}{\color{Aquamarine}\Verb$x$}{\color{Tan}\Verb$+$}{\color{Aquamarine}\Verb$y$}{\color{Tan}\\}{\color{Tan}\Verb$ $}\Verb$ $\Verb$ $\Verb$ $\Verb$ $\Verb$ $\Verb$ $\Verb$|$\Verb$ ${\color{red}\Verb$T$}{\color{red}\Verb$i$}{\color{red}\Verb$m$}{\color{red}\Verb$e$}{\color{red}\Verb$s$}{\color{blue}\Verb$[$}\Verb$ $\Verb$ $\Verb$ $\Verb$ ${\color{red}\Verb$e$}{\color{red}\Verb$v$}{\color{red}\Verb$a$}{\color{red}\Verb$l$}{\color{Aquamarine}\Verb$:$}{\color{Aquamarine}\Verb$x$}\Verb$ ${\color{red}\Verb$e$}{\color{red}\Verb$v$}{\color{red}\Verb$a$}{\color{red}\Verb$l$}{\color{Aquamarine}\Verb$:$}{\color{Aquamarine}\Verb$y$}\Verb$ ${\color{blue}\Verb$]$}\Verb$ ${\color{Tan}\Verb$-$}{\color{Tan}\Verb$>$}{\color{Tan}\Verb$ $}{\color{Aquamarine}\Verb$x$}{\color{Tan}\Verb$*$}{\color{Aquamarine}\Verb$y$}{\color{Tan}\\}{\color{Tan}\Verb$ $}\Verb$ $\Verb$ $\Verb$ $\Verb$ $\Verb$ $\Verb$ $\Verb$|$\Verb$ ${\color{red}\Verb$M$}{\color{red}\Verb$u$}{\color{red}\Verb$l$}{\color{red}\Verb$t$}{\color{red}\Verb$i$}{\color{red}\Verb$p$}{\color{red}\Verb$l$}{\color{red}\Verb$y$}{\color{blue}\Verb$[$}\Verb$ ${\color{red}\Verb$e$}{\color{red}\Verb$v$}{\color{red}\Verb$a$}{\color{red}\Verb$l$}{\color{Aquamarine}\Verb$:$}{\color{Aquamarine}\Verb$x$}\Verb$ ${\color{red}\Verb$e$}{\color{red}\Verb$v$}{\color{red}\Verb$a$}{\color{red}\Verb$l$}{\color{Aquamarine}\Verb$:$}{\color{Aquamarine}\Verb$y$}\Verb$ ${\color{blue}\Verb$]$}\Verb$ ${\color{Tan}\Verb$-$}{\color{Tan}\Verb$>$}{\color{Tan}\Verb$ $}{\color{Aquamarine}\Verb$x$}{\color{Tan}\Verb$*$}{\color{Aquamarine}\Verb$y$}{\color{Tan}\\}{\color{Tan}\Verb$ $}\Verb$ $\Verb$ $\Verb$ $\Verb$ $\Verb$ $\Verb$ $\Verb$|$\Verb$ ${\color{red}\Verb$N$}{\color{red}\Verb$u$}{\color{red}\Verb$m$}{\color{red}\Verb$e$}{\color{red}\Verb$r$}{\color{red}\Verb$i$}{\color{red}\Verb$c$}\\\\\Verb$ $\Verb$ ${\color{red}\Verb$c$}{\color{red}\Verb$a$}{\color{red}\Verb$l$}{\color{red}\Verb$c$}{\color{red}\Verb$u$}{\color{red}\Verb$l$}{\color{red}\Verb$a$}{\color{red}\Verb$t$}{\color{red}\Verb$e$}\Verb$ $\Verb$=$\Verb$ ${\color{red}\Verb$c$}{\color{red}\Verb$a$}{\color{red}\Verb$l$}{\color{red}\Verb$c$}{\color{red}\Verb$u$}{\color{red}\Verb$l$}{\color{red}\Verb$a$}{\color{red}\Verb$t$}{\color{red}\Verb$e$}{\color{red}\Verb$_$}{\color{red}\Verb$a$}{\color{red}\Verb$s$}{\color{red}\Verb$t$}\Verb$=$\Verb$>$\Verb$e$\Verb$v$\Verb$a$\Verb$l$\\\Verb$}$\\\\\Verb$E$\Verb$v$\Verb$a$\Verb$l$\Verb$u$\Verb$a$\Verb$t$\Verb$o$\Verb$r$\Verb$.$\Verb$c$\Verb$a$\Verb$l$\Verb$c$\Verb$u$\Verb$l$\Verb$a$\Verb$t$\Verb$e$\Verb$($\Verb$"$\Verb$2$\Verb$+$\Verb$2$\Verb$"$\Verb$)$\Verb$ $\Verb$#$\Verb$-$\Verb$>$\Verb$ $\Verb$4$
\end{exambox}
}
\end{spacing}\vskip -0.4em

Example above was created to show possibility of the following simplification:

\vskip -1.8em\begin{spacing}{0.8}
{\small
\begin{exambox}
\Verb$a$\Verb$m$\Verb$e$\Verb$t$\Verb$h$\Verb$y$\Verb$s$\Verb$t$\Verb$ $\Verb$E$\Verb$v$\Verb$a$\Verb$l$\Verb$u$\Verb$a$\Verb$t$\Verb$o$\Verb$r$\Verb$ $\Verb$<$\Verb$ $\Verb$C$\Verb$a$\Verb$l$\Verb$c$\Verb$u$\Verb$l$\Verb$a$\Verb$t$\Verb$o$\Verb$r$\Verb$_$\Verb$A$\Verb$S$\Verb$T$\Verb$ $\Verb${$\\\Verb$ $\Verb$ ${\color{red}\Verb$e$}{\color{red}\Verb$v$}{\color{red}\Verb$a$}{\color{red}\Verb$l$}\Verb$ $\Verb$=$\Verb$ ${\color{red}\Verb$A$}{\color{red}\Verb$d$}{\color{red}\Verb$d$}{\color{blue}\Verb$[$}\Verb$ $\Verb$ $\Verb$ $\Verb$ $\Verb$ $\Verb$ $\Verb$ $\Verb$ $\Verb$ $\Verb$ $\Verb$ $\Verb$ $\Verb$ $\Verb$ $\Verb$ $\Verb$ ${\color{red}\Verb$e$}{\color{red}\Verb$v$}{\color{red}\Verb$a$}{\color{red}\Verb$l$}{\color{Aquamarine}\Verb$:$}{\color{Aquamarine}\Verb$x$}\Verb$ ${\color{red}\Verb$e$}{\color{red}\Verb$v$}{\color{red}\Verb$a$}{\color{red}\Verb$l$}{\color{Aquamarine}\Verb$:$}{\color{Aquamarine}\Verb$y$}\Verb$ ${\color{blue}\Verb$]$}\Verb$ ${\color{Tan}\Verb$-$}{\color{Tan}\Verb$>$}{\color{Tan}\Verb$ $}{\color{Aquamarine}\Verb$x$}{\color{Tan}\Verb$+$}{\color{Aquamarine}\Verb$y$}{\color{Tan}\\}{\color{Tan}\Verb$ $}\Verb$ $\Verb$ $\Verb$ $\Verb$ $\Verb$ $\Verb$ $\Verb$|$\Verb$ $\Verb$(${\color{red}\Verb$T$}{\color{red}\Verb$i$}{\color{red}\Verb$m$}{\color{red}\Verb$e$}{\color{red}\Verb$s$}\Verb$ $\Verb$|$\Verb$ ${\color{red}\Verb$M$}{\color{red}\Verb$u$}{\color{red}\Verb$l$}{\color{red}\Verb$t$}{\color{red}\Verb$i$}{\color{red}\Verb$p$}{\color{red}\Verb$l$}{\color{red}\Verb$y$}\Verb$)${\color{blue}\Verb$[$}\Verb$ ${\color{red}\Verb$e$}{\color{red}\Verb$v$}{\color{red}\Verb$a$}{\color{red}\Verb$l$}{\color{Aquamarine}\Verb$:$}{\color{Aquamarine}\Verb$x$}\Verb$ ${\color{red}\Verb$e$}{\color{red}\Verb$v$}{\color{red}\Verb$a$}{\color{red}\Verb$l$}{\color{Aquamarine}\Verb$:$}{\color{Aquamarine}\Verb$y$}\Verb$ ${\color{blue}\Verb$]$}\Verb$ ${\color{Tan}\Verb$-$}{\color{Tan}\Verb$>$}{\color{Tan}\Verb$ $}{\color{Aquamarine}\Verb$x$}{\color{Tan}\Verb$*$}{\color{Aquamarine}\Verb$y$}{\color{Tan}\\}{\color{Tan}\Verb$ $}\Verb$ $\Verb$ $\Verb$ $\Verb$ $\Verb$ $\Verb$ $\Verb$|$\Verb$ ${\color{red}\Verb$N$}{\color{red}\Verb$u$}{\color{red}\Verb$m$}{\color{red}\Verb$e$}{\color{red}\Verb$r$}{\color{red}\Verb$i$}{\color{red}\Verb$c$}\\\\\Verb$ $\Verb$ ${\color{red}\Verb$c$}{\color{red}\Verb$a$}{\color{red}\Verb$l$}{\color{red}\Verb$c$}{\color{red}\Verb$u$}{\color{red}\Verb$l$}{\color{red}\Verb$a$}{\color{red}\Verb$t$}{\color{red}\Verb$e$}\Verb$ $\Verb$=$\Verb$ ${\color{red}\Verb$c$}{\color{red}\Verb$a$}{\color{red}\Verb$l$}{\color{red}\Verb$c$}{\color{red}\Verb$u$}{\color{red}\Verb$l$}{\color{red}\Verb$a$}{\color{red}\Verb$t$}{\color{red}\Verb$e$}{\color{red}\Verb$_$}{\color{red}\Verb$a$}{\color{red}\Verb$s$}{\color{red}\Verb$t$}\Verb$=$\Verb$>$\Verb$e$\Verb$v$\Verb$a$\Verb$l$\\\Verb$}$\\\\\Verb$E$\Verb$v$\Verb$a$\Verb$l$\Verb$u$\Verb$a$\Verb$t$\Verb$o$\Verb$r$\Verb$.$\Verb$c$\Verb$a$\Verb$l$\Verb$c$\Verb$u$\Verb$l$\Verb$a$\Verb$t$\Verb$e$\Verb$($\Verb$"$\Verb$2$\Verb$+$\Verb$2$\Verb$"$\Verb$)$\Verb$ $\Verb$#$\Verb$-$\Verb$>$\Verb$ $\Verb$4$
\end{exambox}
}
\end{spacing}\vskip -0.4em

If we wanted to also represent addition as a $n$-ary operation we could extend previous example in the following way:

\vskip -1.8em\begin{spacing}{0.8}
{\small
\begin{exambox}
\Verb$a$\Verb$m$\Verb$e$\Verb$t$\Verb$h$\Verb$y$\Verb$s$\Verb$t$\Verb$ $\Verb$E$\Verb$v$\Verb$a$\Verb$l$\Verb$u$\Verb$a$\Verb$t$\Verb$o$\Verb$r$\Verb$ $\Verb$<$\Verb$ $\Verb$C$\Verb$a$\Verb$l$\Verb$c$\Verb$u$\Verb$l$\Verb$a$\Verb$t$\Verb$o$\Verb$r$\Verb$_$\Verb$A$\Verb$S$\Verb$T$\Verb${$\\\Verb$ $\Verb$ ${\color{red}\Verb$e$}{\color{red}\Verb$v$}{\color{red}\Verb$a$}{\color{red}\Verb$l$}\Verb$ $\Verb$=$\Verb$ ${\color{red}\Verb$p$}{\color{red}\Verb$l$}{\color{red}\Verb$u$}{\color{red}\Verb$s$}{\color{blue}\Verb$[$}\Verb$ $\Verb$ $\Verb$ $\Verb$ $\Verb$ $\Verb$ $\Verb$ $\Verb$ $\Verb$ $\Verb$ $\Verb$ $\Verb$ $\Verb$ $\Verb$ $\Verb$ $\Verb$ $\Verb$ $\Verb$ ${\color{red}\Verb$e$}{\color{red}\Verb$v$}{\color{red}\Verb$a$}{\color{red}\Verb$l$}{\color{Aquamarine}\Verb$:$}{\color{Aquamarine}\Verb$x$}\Verb$ ${\color{red}\Verb$e$}{\color{red}\Verb$v$}{\color{red}\Verb$a$}{\color{red}\Verb$l$}{\color{Aquamarine}\Verb$:$}{\color{Aquamarine}\Verb$y$}{\color{blue}\Verb$]$}\Verb$ ${\color{Tan}\Verb$-$}{\color{Tan}\Verb$>$}{\color{Tan}\Verb$ $}{\color{Aquamarine}\Verb$x$}{\color{Tan}\Verb$+$}{\color{Aquamarine}\Verb$y$}{\color{Tan}\\}{\color{Tan}\Verb$ $}\Verb$ $\Verb$ $\Verb$ $\Verb$ $\Verb$ $\Verb$ $\Verb$|$\Verb$ $\Verb$(${\color{red}\Verb$T$}{\color{red}\Verb$i$}{\color{red}\Verb$m$}{\color{red}\Verb$e$}{\color{red}\Verb$s$}\Verb$ $\Verb$|$\Verb$ ${\color{red}\Verb$M$}{\color{red}\Verb$u$}{\color{red}\Verb$l$}{\color{red}\Verb$t$}{\color{red}\Verb$i$}{\color{red}\Verb$p$}{\color{red}\Verb$l$}{\color{red}\Verb$y$}\Verb$)${\color{blue}\Verb$[$}\Verb$ $\Verb$ $\Verb$ $\Verb$ ${\color{red}\Verb$e$}{\color{red}\Verb$v$}{\color{red}\Verb$a$}{\color{red}\Verb$l$}{\color{Aquamarine}\Verb$:$}{\color{Aquamarine}\Verb$x$}\Verb$ ${\color{red}\Verb$e$}{\color{red}\Verb$v$}{\color{red}\Verb$a$}{\color{red}\Verb$l$}{\color{Aquamarine}\Verb$:$}{\color{Aquamarine}\Verb$y$}{\color{blue}\Verb$]$}\Verb$ ${\color{Tan}\Verb$-$}{\color{Tan}\Verb$>$}{\color{Tan}\Verb$ $}{\color{Aquamarine}\Verb$x$}{\color{Tan}\Verb$*$}{\color{Aquamarine}\Verb$y$}{\color{Tan}\\}{\color{Tan}\Verb$ $}\Verb$ $\Verb$ $\Verb$ $\Verb$ $\Verb$ $\Verb$ $\Verb$|$\Verb$ ${\color{red}\Verb$N$}{\color{red}\Verb$u$}{\color{red}\Verb$m$}{\color{red}\Verb$e$}{\color{red}\Verb$r$}{\color{red}\Verb$i$}{\color{red}\Verb$c$}\\\\\Verb$ $\Verb$ ${\color{red}\Verb$p$}{\color{red}\Verb$l$}{\color{red}\Verb$u$}{\color{red}\Verb$s$}\Verb$ $\Verb$=$\Verb$ ${\color{red}\Verb$A$}{\color{red}\Verb$d$}{\color{red}\Verb$d$}\\\Verb$ $\Verb$ $\Verb$ $\Verb$ $\Verb$ $\Verb$ $\Verb$ $\Verb$|$\Verb$ ${\color{red}\Verb$P$}{\color{red}\Verb$l$}{\color{red}\Verb$u$}{\color{red}\Verb$s$}{\color{blue}\Verb$[$}\Verb$ $\Verb$.${\color{Aquamarine}\Verb$:$}{\color{Aquamarine}\Verb$f$}{\color{Aquamarine}\Verb$i$}{\color{Aquamarine}\Verb$r$}{\color{Aquamarine}\Verb$s$}{\color{Aquamarine}\Verb$t$}\Verb$ $\Verb$.${\color{black}\Verb$+$}{\color{Aquamarine}\Verb$:$}{\color{Aquamarine}\Verb$r$}{\color{Aquamarine}\Verb$e$}{\color{Aquamarine}\Verb$s$}{\color{Aquamarine}\Verb$t$}\Verb$ ${\color{blue}\Verb$]$}\Verb$ ${\color{Tan}\Verb$-$}{\color{Tan}\Verb$>$}{\color{Tan}\Verb$ $}{\color{Tan}\Verb$A$}{\color{Tan}\Verb$d$}{\color{Tan}\Verb$d$}{\color{Tan}\Verb$[$}{\color{Aquamarine}\Verb$f$}{\color{Aquamarine}\Verb$i$}{\color{Aquamarine}\Verb$r$}{\color{Aquamarine}\Verb$s$}{\color{Aquamarine}\Verb$t$}{\color{Tan}\Verb$,$}{\color{Tan}\Verb$P$}{\color{Tan}\Verb$l$}{\color{Tan}\Verb$u$}{\color{Tan}\Verb$s$}{\color{Tan}\Verb$[$}{\color{Tan}\Verb$*$}{\color{Aquamarine}\Verb$r$}{\color{Aquamarine}\Verb$e$}{\color{Aquamarine}\Verb$s$}{\color{Aquamarine}\Verb$t$}{\color{Tan}\Verb$]$}{\color{Tan}\Verb$]$}{\color{Tan}\\}{\color{Tan}\Verb$ $}\Verb$ $\Verb$ $\Verb$ $\Verb$ $\Verb$ $\Verb$ $\Verb$|$\Verb$ ${\color{red}\Verb$P$}{\color{red}\Verb$l$}{\color{red}\Verb$u$}{\color{red}\Verb$s$}{\color{blue}\Verb$[$}\Verb$ $\Verb$.${\color{Aquamarine}\Verb$:$}{\color{Aquamarine}\Verb$f$}{\color{Aquamarine}\Verb$i$}{\color{Aquamarine}\Verb$r$}{\color{Aquamarine}\Verb$s$}{\color{Aquamarine}\Verb$t$}\Verb$ ${\color{blue}\Verb$]$}\Verb$ $\Verb$ $\Verb$ $\Verb$ $\Verb$ $\Verb$ $\Verb$ $\Verb$ $\Verb$ ${\color{Tan}\Verb$-$}{\color{Tan}\Verb$>$}{\color{Tan}\Verb$ $}{\color{Aquamarine}\Verb$f$}{\color{Aquamarine}\Verb$i$}{\color{Aquamarine}\Verb$r$}{\color{Aquamarine}\Verb$s$}{\color{Aquamarine}\Verb$t$}{\color{Tan}\\}{\color{Tan}\\}\Verb$ $\Verb$ ${\color{red}\Verb$c$}{\color{red}\Verb$a$}{\color{red}\Verb$l$}{\color{red}\Verb$c$}{\color{red}\Verb$u$}{\color{red}\Verb$l$}{\color{red}\Verb$a$}{\color{red}\Verb$t$}{\color{red}\Verb$e$}\Verb$ $\Verb$=$\Verb$ ${\color{red}\Verb$c$}{\color{red}\Verb$a$}{\color{red}\Verb$l$}{\color{red}\Verb$c$}{\color{red}\Verb$u$}{\color{red}\Verb$l$}{\color{red}\Verb$a$}{\color{red}\Verb$t$}{\color{red}\Verb$e$}{\color{red}\Verb$_$}{\color{red}\Verb$a$}{\color{red}\Verb$s$}{\color{red}\Verb$t$}\Verb$=$\Verb$>$\Verb$e$\Verb$v$\Verb$a$\Verb$l$\\\Verb$}$\\\Verb$ $\Verb$ $\Verb$ $\Verb$ $\Verb$ $\Verb$ $\Verb$ $\Verb$ $\Verb$ $\Verb$ $\Verb$ $\Verb$ $\Verb$ $\Verb$ $\Verb$ $\Verb$#$\Verb$ $\Verb$ $\Verb$ $\Verb$ $\Verb$2$\Verb$ $\Verb$+$\Verb$ $\Verb$3$\Verb$ $\Verb$+$\Verb$ $\Verb$ $\Verb$ $\Verb$2$\Verb$*$\Verb$2$\Verb$ $\Verb$ $\Verb$+$\Verb$ $\Verb$ $\Verb$1$\Verb$ $\Verb$+$\Verb$ $\Verb$2$\Verb$*$\Verb$2$\\\Verb$E$\Verb$v$\Verb$a$\Verb$l$\Verb$u$\Verb$a$\Verb$t$\Verb$o$\Verb$r$\Verb$.$\Verb$e$\Verb$v$\Verb$a$\Verb$l$\Verb$($\Verb$P$\Verb$l$\Verb$u$\Verb$s$\Verb$[$\Verb$2$\Verb$,$\Verb$3$\Verb$,$\Verb$M$\Verb$u$\Verb$l$\Verb$t$\Verb$i$\Verb$p$\Verb$l$\Verb$y$\Verb$[$\Verb$2$\Verb$,$\Verb$2$\Verb$]$\Verb$,$\Verb$1$\Verb$,$\Verb$T$\Verb$i$\Verb$m$\Verb$e$\Verb$s$\Verb$[$\Verb$2$\Verb$,$\Verb$2$\Verb$]$\Verb$]$\Verb$)$\Verb$ $\Verb$#$\Verb$-$\Verb$>$\Verb$ $\Verb$1$\Verb$4$\\\\\Verb$E$\Verb$v$\Verb$a$\Verb$l$\Verb$u$\Verb$a$\Verb$t$\Verb$o$\Verb$r$\Verb$.$\Verb$c$\Verb$a$\Verb$l$\Verb$c$\Verb$u$\Verb$l$\Verb$a$\Verb$t$\Verb$e$\Verb$($\Verb$ $\Verb$"$\Verb$2$\Verb$+$\Verb$2$\Verb$"$\Verb$ $\Verb$)$\Verb$#$\Verb$-$\Verb$>$\Verb$ $\Verb$4$
\end{exambox}
}
\end{spacing}\vskip -0.4em

Rule {\color{red}\Verb$p$}{\color{red}\Verb$l$}{\color{red}\Verb$u$}{\color{red}\Verb$s$} shows a way to archive an independence of representation. It allows us to freely switch between a representation of addition as an array of summands and a recursive representation.

\newpage 
\section{Matching arbitrary objects} \label{objmatch}

Most of the rules are written to match element of an array. The {\color{blue} \verb$Pass$} operator that behaves like the \verb$Enter$ operator except it wraps the first result into one element array.

\vskip 0.2em\noindent\begin{tabular}{| l | l | p{9cm} |}
\hline
Name & Expansion & Description\\
\hline
{\color{blue}\Verb$e$}{\color{blue}\Verb$1$}\Verb$=$\Verb$>${\color{blue}\Verb$e$}{\color{blue}\Verb$2$} & {\color{blue}\Verb$e$}{\color{blue}\Verb$1$}{\color{Aquamarine}\Verb$:$}{\color{Aquamarine}\Verb$a$}\Verb$ ${\color{Tan}\Verb${$}{\color{Tan}\Verb$[$}{\color{Aquamarine}\Verb$a$}{\color{Tan}\Verb$]$}{\color{Tan}\Verb$}$}{\color{blue}\Verb$[$}{\color{blue}\Verb$e$}{\color{blue}\Verb$2$}{\color{blue}\Verb$]$} & Pass operator.\\
\hline
\end{tabular}\vskip 0.2em

As Ruby is object oriented language you can discover state of object only by method calls. Inside \verb$Enter$ operator you can call matched object methods. The syntax is the following:

\vskip 0.2em\noindent\begin{tabular}{| l  | p{10.5cm} |}
\hline
Atomic expression & Description \\
\hline
{\color{Aquamarine}\Verb$@$}{\color{Aquamarine}\Verb$m$}{\color{Aquamarine}\Verb$e$}{\color{Aquamarine}\Verb$t$}{\color{Aquamarine}\Verb$h$}{\color{Aquamarine}\Verb$o$}{\color{Aquamarine}\Verb$d$} & Call method of matched object.\\
{\color{Aquamarine}\Verb$@$}{\color{Aquamarine}\Verb$m$}{\color{Aquamarine}\Verb$e$}{\color{Aquamarine}\Verb$t$}{\color{Aquamarine}\Verb$h$}{\color{Aquamarine}\Verb$o$}{\color{Aquamarine}\Verb$d$}{\color{Aquamarine}\Verb$($}{\color{Aquamarine}\Verb$a$}{\color{Aquamarine}\Verb$1$}{\color{Aquamarine}\Verb$,$}{\color{Aquamarine}\Verb$a$}{\color{Aquamarine}\Verb$2$}{\color{Aquamarine}\Verb$)$} & Call method with arguments.\\
\hline
\end{tabular}\vskip 0.2em

You can call matched object methods inside semantic acts with same syntax.

Note that disambiguation between method call or object creation is based on the fact that in ruby all class names are capitalized.

\subsection*{Example:}

We can also write evaluator by accessing object methods:

\vskip -1.8em\begin{spacing}{0.8}
{\small
\begin{exambox}
\Verb$a$\Verb$m$\Verb$e$\Verb$t$\Verb$h$\Verb$y$\Verb$s$\Verb$t$\Verb$ $\Verb$E$\Verb$v$\Verb$a$\Verb$l$\Verb$u$\Verb$a$\Verb$t$\Verb$o$\Verb$r$\Verb$ $\Verb${$\\\Verb$ $\Verb$ ${\color{red}\Verb$e$}{\color{red}\Verb$v$}{\color{red}\Verb$a$}{\color{red}\Verb$l$}\Verb$ $\Verb$=$\Verb$ $\Verb$(${\color{red}\Verb$A$}{\color{red}\Verb$d$}{\color{red}\Verb$d$}\Verb$ $\Verb$|$\Verb$ ${\color{red}\Verb$T$}{\color{red}\Verb$i$}{\color{red}\Verb$m$}{\color{red}\Verb$e$}{\color{red}\Verb$s$}\Verb$ $\Verb$|$\Verb$ ${\color{red}\Verb$M$}{\color{red}\Verb$u$}{\color{red}\Verb$l$}{\color{red}\Verb$t$}{\color{red}\Verb$i$}{\color{red}\Verb$p$}{\color{red}\Verb$l$}{\color{red}\Verb$y$}\Verb$)${\color{blue}\Verb$[$}\Verb$ ${\color{Aquamarine}\Verb$@$}{\color{Aquamarine}\Verb$x$}{\color{blue}\Verb$=$}{\color{blue}\Verb$>$}{\color{red}\Verb$e$}{\color{red}\Verb$v$}{\color{red}\Verb$a$}{\color{red}\Verb$l$}{\color{Aquamarine}\Verb$:$}{\color{Aquamarine}\Verb$x$}\Verb$ ${\color{Aquamarine}\Verb$@$}{\color{Aquamarine}\Verb$y$}{\color{blue}\Verb$=$}{\color{blue}\Verb$>$}{\color{red}\Verb$e$}{\color{red}\Verb$v$}{\color{red}\Verb$a$}{\color{red}\Verb$l$}{\color{Aquamarine}\Verb$:$}{\color{Aquamarine}\Verb$y$}\Verb$ $\\\Verb$ $\Verb$ $\Verb$ $\Verb$ $\Verb$ $\Verb$ $\Verb$ $\Verb$ $\Verb$ $\Verb$ $\Verb$ $\Verb$ $\Verb$ $\Verb$ $\Verb$ $\Verb$ $\Verb$ $\Verb$ $\Verb$ $\Verb$ $\Verb$ $\Verb$ $\Verb$ $\Verb$ $\Verb$ $\Verb$ $\Verb$ $\Verb$ $\Verb$ $\Verb$ $\Verb$ $\Verb$ $\Verb$ $\Verb$ $\Verb$ ${\color{Tan}\Verb$-$}{\color{Tan}\Verb$>$}{\color{Tan}\Verb$ $}{\color{Aquamarine}\Verb$@$}{\color{Aquamarine}\Verb$i$}{\color{Aquamarine}\Verb$s$}{\color{Aquamarine}\Verb$_$}{\color{Aquamarine}\Verb$a$}{\color{Tan}\Verb$?$}{\color{Tan}\Verb$($}{\color{Tan}\Verb$A$}{\color{Tan}\Verb$d$}{\color{Tan}\Verb$d$}{\color{Tan}\Verb$)$}{\color{Tan}\Verb$ $}{\color{Tan}\Verb$?$}{\color{Tan}\Verb$ $}{\color{Aquamarine}\Verb$x$}{\color{Tan}\Verb$+$}{\color{Aquamarine}\Verb$y$}{\color{Tan}\Verb$ $}{\color{Tan}\Verb$:$}{\color{Tan}\Verb$ $}{\color{Aquamarine}\Verb$x$}{\color{Tan}\Verb$*$}{\color{Aquamarine}\Verb$y$}{\color{Tan}\\}{\color{Tan}\Verb$ $}\Verb$ $\Verb$ $\Verb$ $\Verb$ $\Verb$ $\Verb$ $\Verb$ $\Verb$ $\Verb$ $\Verb$ $\Verb$ $\Verb$ $\Verb$ $\Verb$ $\Verb$ $\Verb$ $\Verb$ $\Verb$ $\Verb$ $\Verb$ $\Verb$ $\Verb$ $\Verb$ $\Verb$ $\Verb$ $\Verb$ $\Verb$ $\Verb$ $\Verb$ $\Verb$ $\Verb$ $\Verb$ ${\color{blue}\Verb$]$}\\\Verb$ $\Verb$ $\Verb$ $\Verb$ $\Verb$ $\Verb$ $\Verb$ $\Verb$|$\Verb$ ${\color{red}\Verb$N$}{\color{red}\Verb$u$}{\color{red}\Verb$m$}{\color{red}\Verb$e$}{\color{red}\Verb$r$}{\color{red}\Verb$i$}{\color{red}\Verb$c$}\\\Verb$}$\\\\\Verb$E$\Verb$v$\Verb$a$\Verb$l$\Verb$u$\Verb$a$\Verb$t$\Verb$o$\Verb$r$\Verb$.$\Verb$e$\Verb$v$\Verb$a$\Verb$l$\Verb$($\Verb$C$\Verb$a$\Verb$l$\Verb$c$\Verb$u$\Verb$l$\Verb$a$\Verb$t$\Verb$o$\Verb$r$\Verb$_$\Verb$A$\Verb$S$\Verb$T$\Verb$.$\Verb$c$\Verb$a$\Verb$l$\Verb$c$\Verb$u$\Verb$l$\Verb$a$\Verb$t$\Verb$e$\Verb$_$\Verb$a$\Verb$s$\Verb$t$\Verb$($\Verb$ $\Verb$"$\Verb$2$\Verb$+$\Verb$2$\Verb$*$\Verb$2$\Verb$"$\Verb$ $\Verb$)$\Verb$)$\Verb$#$\Verb$-$\Verb$>$\Verb$ $\Verb$6$
\end{exambox}
}
\end{spacing}\vskip -0.4em

We can match arbitrary objects, for example hashes.
\vskip -1.8em\begin{spacing}{0.8}
{\small
\begin{exambox}
\Verb$a$\Verb$m$\Verb$e$\Verb$t$\Verb$h$\Verb$y$\Verb$s$\Verb$t$\Verb$ $\Verb$M$\Verb$a$\Verb$t$\Verb$c$\Verb$h$\Verb$_$\Verb$H$\Verb$a$\Verb$s$\Verb$h$\Verb$ $\Verb${$\\\Verb$ $\Verb$ ${\color{red}\Verb$m$}{\color{red}\Verb$a$}{\color{red}\Verb$t$}{\color{red}\Verb$c$}{\color{red}\Verb$h$}\Verb$ $\Verb$=$\Verb$ ${\color{Aquamarine}\Verb$@$}{\color{Aquamarine}\Verb$f$}{\color{Aquamarine}\Verb$e$}{\color{Aquamarine}\Verb$t$}{\color{Aquamarine}\Verb$c$}{\color{Aquamarine}\Verb$h$}{\color{Aquamarine}\Verb$($}{\color{Aquamarine}\Verb$:$}{\color{Aquamarine}\Verb$b$}{\color{Aquamarine}\Verb$)$}{\color{blue}\Verb$=$}{\color{blue}\Verb$>$}{\color{blue}\Verb$[$}\Verb$ $\Verb$.${\color{Aquamarine}\Verb$:$}{\color{Aquamarine}\Verb$x$}\Verb$ $\Verb$.${\color{Aquamarine}\Verb$:$}{\color{Aquamarine}\Verb$y$}\Verb$ ${\color{blue}\Verb$]$}\\\Verb$ $\Verb$ $\Verb$ $\Verb$ $\Verb$ $\Verb$ $\Verb$ $\Verb$ $\Verb$ ${\color{Tan}\Verb$-$}{\color{Tan}\Verb$>$}{\color{Tan}\Verb$ $}{\color{Aquamarine}\Verb$x$}{\color{Tan}\Verb$*$}{\color{Aquamarine}\Verb$y$}{\color{Tan}\Verb$+$}{\color{Aquamarine}\Verb$@$}{\color{Aquamarine}\Verb$f$}{\color{Aquamarine}\Verb$e$}{\color{Aquamarine}\Verb$t$}{\color{Aquamarine}\Verb$c$}{\color{Aquamarine}\Verb$h$}{\color{Aquamarine}\Verb$($}{\color{Aquamarine}\Verb$:$}{\color{Aquamarine}\Verb$a$}{\color{Aquamarine}\Verb$)$}{\color{Tan}\Verb$+$}{\color{Aquamarine}\Verb$@$}{\color{Aquamarine}\Verb$f$}{\color{Aquamarine}\Verb$e$}{\color{Aquamarine}\Verb$t$}{\color{Aquamarine}\Verb$c$}{\color{Aquamarine}\Verb$h$}{\color{Aquamarine}\Verb$($}{\color{Aquamarine}\Verb$:$}{\color{Aquamarine}\Verb$c$}{\color{Aquamarine}\Verb$)$}{\color{Tan}\\}{\color{Tan}\Verb$}$}\\\Verb$h$\Verb$ $\Verb$=$\Verb$ $\Verb${$\Verb$:$\Verb$a$\Verb$=$\Verb$>$\Verb$1$\Verb$,$\Verb$:$\Verb$b$\Verb$=$\Verb$>$\Verb$[$\Verb$2$\Verb$,$\Verb$2$\Verb$]$\Verb$,$\Verb$:$\Verb$c$\Verb$=$\Verb$>$\Verb$4$\Verb$}$\\\Verb$M$\Verb$a$\Verb$t$\Verb$c$\Verb$h$\Verb$_$\Verb$H$\Verb$a$\Verb$s$\Verb$h$\Verb$.$\Verb$m$\Verb$a$\Verb$t$\Verb$c$\Verb$h$\Verb$($\Verb$h$\Verb$)$\Verb$ $\Verb$#$\Verb$-$\Verb$>$\Verb$ $\Verb$9$
\end{exambox}
}
\end{spacing}\vskip -0.4em

\newpage
\section{Dataflow analysis generalizes tree traversal} \label{dflow}

Dataflow analysis \cite{dataflow} is important technique in compiler optimization. It generalizes tree traversal to handle cyclic dependencies.

We illustrate dataflow analysis on real world example that amethyst needs to solve. We first start with two simpler problems where simpler approach is sufficient until we get into a situation where dataflow analysis is necessary. 

\subsection*{First example: In regular expressions}
We are given a regular expression and want to know a minimal size of string that matches this expression. For simplicity we are given expression as syntax tree consisting only of immutable \verb$Or$, \verb$Seq$, \verb$Char$ nodes for binary choice, sequencing and to match character.

\vskip -1.8em\begin{spacing}{0.8}
{\small
\begin{exambox}
\Verb$a$\Verb$m$\Verb$e$\Verb$t$\Verb$h$\Verb$y$\Verb$s$\Verb$t$\Verb$ $\Verb$R$\Verb$e$\Verb$g$\Verb$e$\Verb$x$\Verb$p$\Verb$_$\Verb$m$\Verb$i$\Verb$n$\Verb$i$\Verb$m$\Verb$a$\Verb$l$\Verb$_$\Verb$s$\Verb$i$\Verb$z$\Verb$e$\Verb$ $\Verb${$\\\Verb$ $\Verb$ ${\color{red}\Verb$v$}{\color{red}\Verb$a$}{\color{red}\Verb$l$}{\color{red}\Verb$u$}{\color{red}\Verb$e$}\Verb$ $\Verb$=$\Verb$ ${\color{red}\Verb$S$}{\color{red}\Verb$e$}{\color{red}\Verb$q$}{\color{blue}\Verb$[$}\Verb$ ${\color{red}\Verb$v$}{\color{red}\Verb$a$}{\color{red}\Verb$l$}{\color{red}\Verb$u$}{\color{red}\Verb$e$}{\color{Aquamarine}\Verb$:$}{\color{Aquamarine}\Verb$v$}{\color{Aquamarine}\Verb$1$}\Verb$ ${\color{red}\Verb$v$}{\color{red}\Verb$a$}{\color{red}\Verb$l$}{\color{red}\Verb$u$}{\color{red}\Verb$e$}{\color{Aquamarine}\Verb$:$}{\color{Aquamarine}\Verb$v$}{\color{Aquamarine}\Verb$2$}\Verb$ ${\color{blue}\Verb$]$}\Verb$ ${\color{Tan}\Verb$-$}{\color{Tan}\Verb$>$}{\color{Tan}\Verb$ $}{\color{Aquamarine}\Verb$v$}{\color{Aquamarine}\Verb$1$}{\color{Tan}\Verb$+$}{\color{Aquamarine}\Verb$v$}{\color{Aquamarine}\Verb$2$}{\color{Tan}\\}{\color{Tan}\Verb$ $}\Verb$ $\Verb$ $\Verb$ $\Verb$ $\Verb$ $\Verb$ $\Verb$ $\Verb$|$\Verb$ ${\color{red}\Verb$O$}{\color{red}\Verb$r$}{\color{blue}\Verb$[$}\Verb$ $\Verb$ ${\color{red}\Verb$v$}{\color{red}\Verb$a$}{\color{red}\Verb$l$}{\color{red}\Verb$u$}{\color{red}\Verb$e$}{\color{Aquamarine}\Verb$:$}{\color{Aquamarine}\Verb$v$}{\color{Aquamarine}\Verb$1$}\Verb$ ${\color{red}\Verb$v$}{\color{red}\Verb$a$}{\color{red}\Verb$l$}{\color{red}\Verb$u$}{\color{red}\Verb$e$}{\color{Aquamarine}\Verb$:$}{\color{Aquamarine}\Verb$v$}{\color{Aquamarine}\Verb$2$}\Verb$ ${\color{blue}\Verb$]$}\Verb$ ${\color{Tan}\Verb$-$}{\color{Tan}\Verb$>$}{\color{Tan}\Verb$ $}{\color{Tan}\Verb$m$}{\color{Tan}\Verb$i$}{\color{Tan}\Verb$n$}{\color{Tan}\Verb$($}{\color{Aquamarine}\Verb$v$}{\color{Aquamarine}\Verb$1$}{\color{Tan}\Verb$,$}{\color{Aquamarine}\Verb$v$}{\color{Aquamarine}\Verb$2$}{\color{Tan}\Verb$)$}{\color{Tan}\\}{\color{Tan}\Verb$ $}\Verb$ $\Verb$ $\Verb$ $\Verb$ $\Verb$ $\Verb$ $\Verb$ $\Verb$|$\Verb$ ${\color{red}\Verb$C$}{\color{red}\Verb$h$}{\color{red}\Verb$a$}{\color{red}\Verb$r$}\Verb$ $\Verb$ $\Verb$ $\Verb$ $\Verb$ $\Verb$ $\Verb$ $\Verb$ $\Verb$ $\Verb$ $\Verb$ $\Verb$ $\Verb$ $\Verb$ $\Verb$ $\Verb$ $\Verb$ $\Verb$ $\Verb$ $\Verb$ $\Verb$ ${\color{Tan}\Verb$-$}{\color{Tan}\Verb$>$}{\color{Tan}\Verb$ $}{\color{Tan}\Verb$1$}{\color{Tan}\\}{\color{Tan}\Verb$}$}\\\\\Verb$p$\Verb$u$\Verb$t$\Verb$s$\Verb$ $\Verb$R$\Verb$e$\Verb$g$\Verb$e$\Verb$x$\Verb$p$\Verb$_$\Verb$m$\Verb$i$\Verb$n$\Verb$i$\Verb$m$\Verb$a$\Verb$l$\Verb$_$\Verb$s$\Verb$i$\Verb$z$\Verb$e$\Verb$.$\Verb$v$\Verb$a$\Verb$l$\Verb$u$\Verb$e$\Verb$($\Verb$ $\Verb$#$\Verb$a$\Verb$b$\Verb$c$\Verb$|$\Verb$d$\Verb$e$\\\Verb$ $\Verb$ $\Verb$O$\Verb$r$\Verb$[$\Verb$S$\Verb$e$\Verb$q$\Verb$[$\Verb$C$\Verb$h$\Verb$a$\Verb$r$\Verb$[$\Verb$'$\Verb$a$\Verb$'$\Verb$]$\Verb$,$\Verb$C$\Verb$h$\Verb$a$\Verb$r$\Verb$[$\Verb$'$\Verb$b$\Verb$'$\Verb$]$\Verb$,$\Verb$C$\Verb$h$\Verb$a$\Verb$r$\Verb$[$\Verb$'$\Verb$c$\Verb$'$\Verb$]$\Verb$]$\Verb$,$\\\Verb$ $\Verb$ $\Verb$ $\Verb$ $\Verb$ $\Verb$S$\Verb$e$\Verb$q$\Verb$[$\Verb$C$\Verb$h$\Verb$a$\Verb$r$\Verb$[$\Verb$'$\Verb$d$\Verb$'$\Verb$]$\Verb$,$\Verb$C$\Verb$h$\Verb$a$\Verb$r$\Verb$[$\Verb$'${\color{blue}\Verb$e$}\Verb$'$\Verb$]$\Verb$]$\Verb$]$\Verb$)$\\\Verb$#$\Verb$-$\Verb$>$\Verb$ $\Verb$2$
\end{exambox}
}
\end{spacing}\vskip -0.4em

\subsection*{Second example: Adding rules}
Now we add rule calls represented as an immutable node \verb$Rule$ containing link to body to execute. An example follows:

\vskip -1.8em\begin{spacing}{0.8}
{\small
\begin{exambox}
\Verb$#$\Verb$ $\Verb$f$\Verb$o$\Verb$o$\Verb$ $\Verb$ $\Verb$ $\Verb$ $\Verb$=$\Verb$ $\Verb$'$\Verb$f$\Verb$o$\Verb$o$\Verb$'$\\\Verb$#$\Verb$ $\Verb$b$\Verb$a$\Verb$r$\Verb$ $\Verb$ $\Verb$ $\Verb$ $\Verb$=$\Verb$ $\Verb$'$\Verb$b$\Verb$a$\Verb$r$\Verb$'$\\\Verb$#$\Verb$ $\Verb$f$\Verb$o$\Verb$o$\Verb$b$\Verb$a$\Verb$r$\Verb$ $\Verb$=$\Verb$ $\Verb$f$\Verb$o$\Verb$o$\Verb$ $\Verb$b$\Verb$a$\Verb$r$\\\Verb$f$\Verb$o$\Verb$o$\Verb$ $\Verb$ $\Verb$ $\Verb$ $\Verb$=$\Verb$ $\Verb$S$\Verb$e$\Verb$q$\Verb$[$\Verb$C$\Verb$h$\Verb$a$\Verb$r$\Verb$[$\Verb$'$\Verb$f$\Verb$'$\Verb$]$\Verb$,$\Verb$C$\Verb$h$\Verb$a$\Verb$r$\Verb$[$\Verb$'$\Verb$o$\Verb$'$\Verb$]$\Verb$,$\Verb$C$\Verb$h$\Verb$a$\Verb$r$\Verb$[$\Verb$'$\Verb$o$\Verb$'$\Verb$]$\Verb$]$\\\Verb$b$\Verb$a$\Verb$r$\Verb$ $\Verb$ $\Verb$ $\Verb$ $\Verb$=$\Verb$ $\Verb$S$\Verb$e$\Verb$q$\Verb$[$\Verb$C$\Verb$h$\Verb$a$\Verb$r$\Verb$[$\Verb$'$\Verb$b$\Verb$'$\Verb$]$\Verb$,$\Verb$C$\Verb$h$\Verb$a$\Verb$r$\Verb$[$\Verb$'$\Verb$a$\Verb$'$\Verb$]$\Verb$,$\Verb$C$\Verb$h$\Verb$a$\Verb$r$\Verb$[$\Verb$'$\Verb$r$\Verb$'$\Verb$]$\Verb$]$\\\Verb$f$\Verb$o$\Verb$o$\Verb$b$\Verb$a$\Verb$r$\Verb$ $\Verb$=$\Verb$ $\Verb$S$\Verb$e$\Verb$q$\Verb$[$\Verb$R$\Verb$u$\Verb$l$\Verb$e$\Verb$[$\Verb$f$\Verb$o$\Verb$o$\Verb$]$\Verb$,$\Verb$ $\Verb$R$\Verb$u$\Verb$l$\Verb$e$\Verb$[$\Verb$b$\Verb$a$\Verb$r$\Verb$]$\Verb$]$
\end{exambox}
}
\end{spacing}\vskip -0.4em

As far as no recursion is present we can modify our traverser into:

\vskip -1.8em\begin{spacing}{0.8}
{\small
\begin{exambox}
\Verb$a$\Verb$m$\Verb$e$\Verb$t$\Verb$h$\Verb$y$\Verb$s$\Verb$t$\Verb$ $\Verb$R$\Verb$u$\Verb$l$\Verb$e$\Verb$s$\Verb$_$\Verb$m$\Verb$i$\Verb$n$\Verb$i$\Verb$m$\Verb$a$\Verb$l$\Verb$_$\Verb$s$\Verb$i$\Verb$z$\Verb$e$\Verb$ $\Verb${$\\\Verb$ $\Verb$ ${\color{red}\Verb$v$}{\color{red}\Verb$a$}{\color{red}\Verb$l$}{\color{red}\Verb$u$}{\color{red}\Verb$e$}\Verb$ $\Verb$=$\Verb$ ${\color{red}\Verb$R$}{\color{red}\Verb$u$}{\color{red}\Verb$l$}{\color{red}\Verb$e$}{\color{blue}\Verb$[$}\Verb$ ${\color{red}\Verb$v$}{\color{red}\Verb$a$}{\color{red}\Verb$l$}{\color{red}\Verb$u$}{\color{red}\Verb$e$}{\color{Aquamarine}\Verb$:$}{\color{Aquamarine}\Verb$v$}\Verb$ $\Verb$ $\Verb$ $\Verb$ $\Verb$ $\Verb$ $\Verb$ $\Verb$ $\Verb$ $\Verb$ ${\color{blue}\Verb$]$}\Verb$ ${\color{Tan}\Verb$-$}{\color{Tan}\Verb$>$}{\color{Tan}\Verb$ $}{\color{Aquamarine}\Verb$v$}{\color{Tan}\\}{\color{Tan}\Verb$ $}\Verb$ $\Verb$ $\Verb$ $\Verb$ $\Verb$ $\Verb$ $\Verb$ $\Verb$|$\Verb$ ${\color{red}\Verb$S$}{\color{red}\Verb$e$}{\color{red}\Verb$q$}{\color{blue}\Verb$[$}\Verb$ ${\color{red}\Verb$v$}{\color{red}\Verb$a$}{\color{red}\Verb$l$}{\color{red}\Verb$u$}{\color{red}\Verb$e$}{\color{Aquamarine}\Verb$:$}{\color{Aquamarine}\Verb$v$}{\color{Aquamarine}\Verb$1$}\Verb$ ${\color{red}\Verb$v$}{\color{red}\Verb$a$}{\color{red}\Verb$l$}{\color{red}\Verb$u$}{\color{red}\Verb$e$}{\color{Aquamarine}\Verb$:$}{\color{Aquamarine}\Verb$v$}{\color{Aquamarine}\Verb$2$}\Verb$ ${\color{blue}\Verb$]$}\Verb$ ${\color{Tan}\Verb$-$}{\color{Tan}\Verb$>$}{\color{Tan}\Verb$ $}{\color{Aquamarine}\Verb$v$}{\color{Aquamarine}\Verb$1$}{\color{Tan}\Verb$+$}{\color{Aquamarine}\Verb$v$}{\color{Aquamarine}\Verb$2$}{\color{Tan}\\}{\color{Tan}\Verb$ $}\Verb$ $\Verb$ $\Verb$ $\Verb$ $\Verb$ $\Verb$ $\Verb$ $\Verb$|$\Verb$ ${\color{red}\Verb$O$}{\color{red}\Verb$r$}{\color{blue}\Verb$[$}\Verb$ $\Verb$ ${\color{red}\Verb$v$}{\color{red}\Verb$a$}{\color{red}\Verb$l$}{\color{red}\Verb$u$}{\color{red}\Verb$e$}{\color{Aquamarine}\Verb$:$}{\color{Aquamarine}\Verb$v$}{\color{Aquamarine}\Verb$1$}\Verb$ ${\color{red}\Verb$v$}{\color{red}\Verb$a$}{\color{red}\Verb$l$}{\color{red}\Verb$u$}{\color{red}\Verb$e$}{\color{Aquamarine}\Verb$:$}{\color{Aquamarine}\Verb$v$}{\color{Aquamarine}\Verb$2$}\Verb$ ${\color{blue}\Verb$]$}\Verb$ ${\color{Tan}\Verb$-$}{\color{Tan}\Verb$>$}{\color{Tan}\Verb$ $}{\color{Tan}\Verb$m$}{\color{Tan}\Verb$i$}{\color{Tan}\Verb$n$}{\color{Tan}\Verb$($}{\color{Aquamarine}\Verb$v$}{\color{Aquamarine}\Verb$1$}{\color{Tan}\Verb$,$}{\color{Aquamarine}\Verb$v$}{\color{Aquamarine}\Verb$2$}{\color{Tan}\Verb$)$}{\color{Tan}\\}{\color{Tan}\Verb$ $}\Verb$ $\Verb$ $\Verb$ $\Verb$ $\Verb$ $\Verb$ $\Verb$ $\Verb$|$\Verb$ ${\color{red}\Verb$C$}{\color{red}\Verb$h$}{\color{red}\Verb$a$}{\color{red}\Verb$r$}\Verb$ $\Verb$ $\Verb$ $\Verb$ $\Verb$ $\Verb$ $\Verb$ $\Verb$ $\Verb$ $\Verb$ $\Verb$ $\Verb$ $\Verb$ $\Verb$ $\Verb$ $\Verb$ $\Verb$ $\Verb$ $\Verb$ $\Verb$ $\Verb$ ${\color{Tan}\Verb$-$}{\color{Tan}\Verb$>$}{\color{Tan}\Verb$ $}{\color{Tan}\Verb$1$}{\color{Tan}\\}{\color{Tan}\Verb$}$}\\\\\Verb$R$\Verb$u$\Verb$l$\Verb$e$\Verb$s$\Verb$_$\Verb$m$\Verb$i$\Verb$n$\Verb$i$\Verb$m$\Verb$a$\Verb$l$\Verb$_$\Verb$s$\Verb$i$\Verb$z$\Verb$e$\Verb$.$\Verb$v$\Verb$a$\Verb$l$\Verb$u$\Verb$e$\Verb$($\Verb$f$\Verb$o$\Verb$o$\Verb$b$\Verb$a$\Verb$r$\Verb$)$\Verb$ $\Verb$#$\Verb$-$\Verb$>$\Verb$ $\Verb$6$
\end{exambox}
}
\end{spacing}\vskip -0.4em

And it will always terminate and produce the correct result. 
\newpage
\subsection*{Dealing with recursion}
When recursion is present then dataflow analysis becomes necessary.

Dataflow analysis is a method of solving sets of monotonic equations over arbitrary lattice. 
In our case we use the lattice associated to ordering of natural numbers.
We interpret \verb$value$ rule in previous example as an inequality that bound a size of expression based on sizes of its subexpressions. 
We implement the well known worklist algorithm \cite{dataflow} using it to find a minimal solutions of the dataflow equations that correspond to our inequalities.

The algorithm starts with a setting everywhere a value zero. This violates some inequalities. When an inequality is violated we increase left size to value of right side. We repeat this until all inequalities are satisfied. We use algorithm by inheriting from \verb$Dataflow$ grammar that is implemented in the next section. 
Algorithm terminates when all inequalities are satisfied and each \verb$value$ attains minimum among all solutions.

We do not have to change our code much to use this analysis:

\vskip -1.8em\begin{spacing}{0.8}
{\small
\begin{exambox}
\Verb$a$\Verb$m$\Verb$e$\Verb$t$\Verb$h$\Verb$y$\Verb$s$\Verb$t$\Verb$ $\Verb$R$\Verb$u$\Verb$l$\Verb$e$\Verb$s$\Verb$_$\Verb$m$\Verb$i$\Verb$n$\Verb$i$\Verb$m$\Verb$a$\Verb$l$\Verb$_$\Verb$s$\Verb$i$\Verb$z$\Verb$e$\Verb$ $\Verb$<$\Verb$ $\Verb$D$\Verb$a$\Verb$t$\Verb$a$\Verb$f$\Verb$l$\Verb$o$\Verb$w$\Verb$ $\Verb${$\\\Verb$ ${\color{red}\Verb$f$}{\color{red}\Verb$l$}{\color{red}\Verb$o$}{\color{red}\Verb$w$}\Verb$ $\Verb$ $\Verb$=$\Verb$ ${\color{red}\Verb$R$}{\color{red}\Verb$u$}{\color{red}\Verb$l$}{\color{red}\Verb$e$}{\color{blue}\Verb$[$}\Verb$ ${\color{red}\Verb$v$}{\color{red}\Verb$i$}{\color{red}\Verb$s$}{\color{red}\Verb$i$}{\color{red}\Verb$t$}{\color{Aquamarine}\Verb$:$}{\color{Aquamarine}\Verb$v$}\Verb$ $\Verb$ $\Verb$ $\Verb$ $\Verb$ $\Verb$ $\Verb$ $\Verb$ $\Verb$ $\Verb$ ${\color{blue}\Verb$]$}\Verb$ $\Verb$ ${\color{Tan}\Verb$-$}{\color{Tan}\Verb$>$}{\color{Tan}\Verb$ $}{\color{Aquamarine}\Verb$v$}{\color{Tan}\\}{\color{Tan}\Verb$ $}\Verb$ $\Verb$ $\Verb$ $\Verb$ $\Verb$ $\Verb$ $\Verb$|$\Verb$ ${\color{red}\Verb$S$}{\color{red}\Verb$e$}{\color{red}\Verb$q$}{\color{blue}\Verb$[$}\Verb$ $\Verb$ ${\color{red}\Verb$v$}{\color{red}\Verb$i$}{\color{red}\Verb$s$}{\color{red}\Verb$i$}{\color{red}\Verb$t$}{\color{Aquamarine}\Verb$:$}{\color{Aquamarine}\Verb$v$}{\color{Aquamarine}\Verb$1$}\Verb$ ${\color{red}\Verb$v$}{\color{red}\Verb$i$}{\color{red}\Verb$s$}{\color{red}\Verb$i$}{\color{red}\Verb$t$}{\color{Aquamarine}\Verb$:$}{\color{Aquamarine}\Verb$v$}{\color{Aquamarine}\Verb$2$}\Verb$ ${\color{blue}\Verb$]$}\Verb$ ${\color{Tan}\Verb$-$}{\color{Tan}\Verb$>$}{\color{Tan}\Verb$ $}{\color{Aquamarine}\Verb$v$}{\color{Aquamarine}\Verb$1$}{\color{Tan}\Verb$+$}{\color{Aquamarine}\Verb$v$}{\color{Aquamarine}\Verb$2$}{\color{Tan}\\}{\color{Tan}\Verb$ $}\Verb$ $\Verb$ $\Verb$ $\Verb$ $\Verb$ $\Verb$ $\Verb$|$\Verb$ ${\color{red}\Verb$O$}{\color{red}\Verb$r$}{\color{blue}\Verb$[$}\Verb$ $\Verb$ $\Verb$ ${\color{red}\Verb$v$}{\color{red}\Verb$i$}{\color{red}\Verb$s$}{\color{red}\Verb$i$}{\color{red}\Verb$t$}{\color{Aquamarine}\Verb$:$}{\color{Aquamarine}\Verb$v$}{\color{Aquamarine}\Verb$1$}\Verb$ ${\color{red}\Verb$v$}{\color{red}\Verb$i$}{\color{red}\Verb$s$}{\color{red}\Verb$i$}{\color{red}\Verb$t$}{\color{Aquamarine}\Verb$:$}{\color{Aquamarine}\Verb$v$}{\color{Aquamarine}\Verb$2$}\Verb$ ${\color{blue}\Verb$]$}\Verb$ ${\color{Tan}\Verb$-$}{\color{Tan}\Verb$>$}{\color{Tan}\Verb$ $}{\color{Tan}\Verb$m$}{\color{Tan}\Verb$i$}{\color{Tan}\Verb$n$}{\color{Tan}\Verb$($}{\color{Aquamarine}\Verb$v$}{\color{Aquamarine}\Verb$1$}{\color{Tan}\Verb$,$}{\color{Aquamarine}\Verb$v$}{\color{Aquamarine}\Verb$2$}{\color{Tan}\Verb$)$}{\color{Tan}\\}{\color{Tan}\Verb$ $}\Verb$ $\Verb$ $\Verb$ $\Verb$ $\Verb$ $\Verb$ $\Verb$|$\Verb$ ${\color{red}\Verb$C$}{\color{red}\Verb$h$}{\color{red}\Verb$a$}{\color{red}\Verb$r$}\Verb$ $\Verb$ $\Verb$ $\Verb$ $\Verb$ $\Verb$ $\Verb$ $\Verb$ $\Verb$ $\Verb$ $\Verb$ $\Verb$ $\Verb$ $\Verb$ $\Verb$ $\Verb$ $\Verb$ $\Verb$ $\Verb$ $\Verb$ $\Verb$ $\Verb$ ${\color{Tan}\Verb$-$}{\color{Tan}\Verb$>$}{\color{Tan}\Verb$ $}{\color{Tan}\Verb$1$}{\color{Tan}\\}{\color{Tan}\Verb$}$}\\\\\Verb$c$\Verb$l$\Verb$a$\Verb$s$\Verb$s$\Verb$ $\Verb$R$\Verb$u$\Verb$l$\Verb$e$\Verb$s$\Verb$_$\Verb$m$\Verb$i$\Verb$n$\Verb$i$\Verb$m$\Verb$a$\Verb$l$\Verb$_$\Verb$s$\Verb$i$\Verb$z$\Verb$e$\Verb$ $\Verb$<$\Verb$ $\Verb$D$\Verb$a$\Verb$t$\Verb$a$\Verb$f$\Verb$l$\Verb$o$\Verb$w$\\\Verb$ $\Verb$ $\Verb$d$\Verb$e$\Verb$f$\Verb$ $\Verb$l$\Verb$a$\Verb$t$\Verb$t$\Verb$i$\Verb$c$\Verb$e$\Verb$_$\Verb$b$\Verb$o$\Verb$t$\Verb$t$\Verb$o$\Verb$m$\Verb$ $\Verb$ $\Verb$ $\Verb$;$\Verb$ $\Verb$0$\Verb$ $\Verb$ $\Verb$ $\Verb$ $\Verb$ $\Verb$ $\Verb$ $\Verb$;$\Verb$ $\Verb$e$\Verb$n$\Verb$d$\Verb$ $\Verb$#$\Verb$ $\Verb$S$\Verb$t$\Verb$a$\Verb$r$\Verb$t$\Verb$i$\Verb$n$\Verb$g$\Verb$ $\Verb$s$\Verb$o$\Verb$l$\Verb$u$\Verb$t$\Verb$i$\Verb$o$\Verb$n$\Verb$.$\\\Verb$ $\Verb$ $\Verb$d$\Verb$e$\Verb$f$\Verb$ $\Verb$l$\Verb$a$\Verb$t$\Verb$t$\Verb$i$\Verb$c$\Verb$e$\Verb$_$\Verb$j$\Verb$o$\Verb$i$\Verb$n$\Verb$($\Verb$x$\Verb$,$\Verb$y$\Verb$)$\Verb$;$\Verb$ $\Verb$m$\Verb$a$\Verb$x$\Verb$($\Verb$x$\Verb$,$\Verb$y$\Verb$)$\Verb$;$\Verb$ $\Verb$e$\Verb$n$\Verb$d$\\\Verb$e$\Verb$n$\Verb$d$
\end{exambox}
}
\end{spacing}\vskip -0.4em

A monotonicity in our case means that if we increase \verb$value$ in right side then corresponding value at left side can not decrease. This is in our case true.

Conditions in which this algorithm terminates is finite height of a lattice. This means that for every value we can bound number of increases until we reach this value by same constant. If no recursive rule without terminating condition is present then we know that minimal solution satisfies this condition and our algorithm terminates.

\subsection{Implementing dataflow analysis} \label{implement_df}
This analysis deals only with immutable objects. It is possible to support mutable objects if you add timestamps to inform if object was changed or not.

In this section we describe a variant of incremental dataflow analysis \cite{incrementaldf}. We added two new properties.

First property is that analysis is dynamic. User does not have to construct data flow graph. A graph is learned automatically and dependencies vary based for different value assignments. This property can naturally describe concepts like shortcircuit evaluation or flow-sensitive analysis.

Second is that analysis is lazy in sense that it does not compute values until they are necessary to compute.

A simple implementation of our analysis follows. 

An amethyst interface is the following:

\vskip -1.8em\begin{spacing}{0.8}
{\small
\begin{exambox}
\Verb$a$\Verb$m$\Verb$e$\Verb$t$\Verb$h$\Verb$y$\Verb$s$\Verb$t$\Verb$ $\Verb$D$\Verb$a$\Verb$t$\Verb$a$\Verb$f$\Verb$l$\Verb$o$\Verb$w$\Verb$ $\Verb${$\\\Verb$ $\Verb$ ${\color{red}\Verb$v$}{\color{red}\Verb$i$}{\color{red}\Verb$s$}{\color{red}\Verb$i$}{\color{red}\Verb$t$}\Verb$ $\Verb$=$\Verb$ $\Verb$ $\Verb$.${\color{Aquamarine}\Verb$:$}{\color{Aquamarine}\Verb$x$}\Verb$ $\Verb$ ${\color{Tan}\Verb${$}{\color{Tan}\Verb$d$}{\color{Tan}\Verb$e$}{\color{Tan}\Verb$p$}{\color{Tan}\Verb$e$}{\color{Tan}\Verb$n$}{\color{Tan}\Verb$d$}{\color{Tan}\Verb$s$}{\color{Tan}\Verb$($}{\color{Aquamarine}\Verb$x$}{\color{Tan}\Verb$)$}{\color{Tan}\Verb$;$}{\color{Aquamarine}\Verb$@$}{\color{Aquamarine}\Verb$@$}{\color{Aquamarine}\Verb$v$}{\color{Aquamarine}\Verb$a$}{\color{Aquamarine}\Verb$l$}{\color{Aquamarine}\Verb$s$}{\color{Tan}\Verb$[$}{\color{Aquamarine}\Verb$x$}{\color{Tan}\Verb$]$}{\color{Tan}\Verb$}$}\\\\\Verb$ $\Verb$ ${\color{red}\Verb$r$}{\color{red}\Verb$o$}{\color{red}\Verb$o$}{\color{red}\Verb$t$}\Verb$ $\Verb$=$\Verb$ $\Verb$.${\color{Aquamarine}\Verb$:$}{\color{Aquamarine}\Verb$x$}\Verb$ ${\color{red}\Verb$a$}{\color{red}\Verb$n$}{\color{red}\Verb$a$}{\color{red}\Verb$l$}{\color{red}\Verb$y$}{\color{red}\Verb$z$}{\color{red}\Verb$e$}{\color{green}\Verb$($}{\color{Aquamarine}\Verb$x$}{\color{green}\Verb$)$}\\\Verb$ $\Verb$ $\Verb$ $\\\Verb$ $\Verb$ ${\color{red}\Verb$g$}{\color{red}\Verb$e$}{\color{red}\Verb$t$}{\color{red}\Verb$v$}{\color{red}\Verb$a$}{\color{red}\Verb$l$}{\color{red}\Verb$u$}{\color{red}\Verb$e$}{\color{green}\Verb$($}{\color{green}\Verb$v$}{\color{green}\Verb$)$}\Verb$ $\Verb$=$\Verb$ ${\color{Tan}\Verb${$}{\color{Aquamarine}\Verb$@$}{\color{Aquamarine}\Verb$@$}{\color{Aquamarine}\Verb$v$}{\color{Aquamarine}\Verb$i$}{\color{Aquamarine}\Verb$s$}{\color{Tan}\Verb$=$}{\color{Tan}\Verb$v$}{\color{Tan}\Verb$;$}{\color{Tan}\Verb$ $}{\color{Tan}\Verb$v$}{\color{Tan}\Verb$}$}{\color{blue}\Verb$=$}{\color{blue}\Verb$>$}{\color{red}\Verb$v$}{\color{red}\Verb$i$}{\color{red}\Verb$s$}{\color{red}\Verb$i$}{\color{red}\Verb$t$}\\\Verb$}$
\end{exambox}
}
\end{spacing}\vskip -0.4em

And a simple analysis based on worklist algorithm follows.

\vskip -1.8em\begin{spacing}{0.8}
{\small
\begin{exambox}
\Verb$c$\Verb$l$\Verb$a$\Verb$s$\Verb$s$\Verb$ $\Verb$D$\Verb$a$\Verb$t$\Verb$a$\Verb$f$\Verb$l$\Verb$o$\Verb$w$\Verb$ $\Verb$<$\Verb$ $\Verb$A$\Verb$m$\Verb$e$\Verb$t$\Verb$h$\Verb$y$\Verb$s$\Verb$t$\\\Verb$ $\Verb$ $\Verb$d$\Verb$e$\Verb$f$\Verb$ $\Verb$v$\Verb$a$\Verb$l$\Verb$u$\Verb$e$\Verb$(${\color{blue}\Verb$e$}\Verb$)$\\\Verb$ $\Verb$ $\Verb$ $\Verb$ $\Verb$@$\Verb$a$\Verb$c$\Verb$t$\Verb$i$\Verb$v$\Verb$e$\Verb$=$\Verb${$\Verb$}$\\\Verb$ $\Verb$ $\Verb$ $\Verb$ $\Verb$@$\Verb$a$\Verb$c$\Verb$t$\Verb$i$\Verb$v$\Verb$e$\Verb$a$\Verb$=$\Verb$[${\color{blue}\Verb$e$}\Verb$]$\\\Verb$ $\Verb$ $\Verb$ $\Verb$ $\Verb$w$\Verb$h$\Verb$i$\Verb$l$\Verb$e$\Verb$ $\Verb$e$\Verb$l$\Verb$=$\Verb$@$\Verb$a$\Verb$c$\Verb$t$\Verb$i$\Verb$v$\Verb$e$\Verb$a$\Verb$.$\Verb$p$\Verb$o$\Verb$p$\\\Verb$ $\Verb$ $\Verb$ $\Verb$ $\Verb$ $\Verb$ $\Verb$@$\Verb$a$\Verb$c$\Verb$t$\Verb$i$\Verb$v$\Verb$e$\Verb$.$\Verb$d$\Verb$e$\Verb$l$\Verb$e$\Verb$t$\Verb$e$\Verb$($\Verb$e$\Verb$l$\Verb$)$\\\Verb$ $\Verb$ $\Verb$ $\Verb$ $\Verb$ $\Verb$ $\Verb$@$\Verb$d$\Verb$e$\Verb$p$\Verb$e$\Verb$n$\Verb$d$\Verb$.$\Verb$d$\Verb$e$\Verb$l$\Verb$e$\Verb$t$\Verb$e$\Verb$_$\Verb$a$\Verb$l$\Verb$l$\Verb$_$\Verb$e$\Verb$d$\Verb$g$\Verb$e$\Verb$s$\Verb$_$\Verb$t$\Verb$o$\Verb$($\Verb$e$\Verb$l$\Verb$)$\\\Verb$ $\Verb$ $\Verb$ $\Verb$ $\Verb$ $\Verb$ $\\\Verb$ $\Verb$ $\Verb$ $\Verb$ $\Verb$ $\Verb$ $\Verb$v$\Verb$a$\Verb$l$\Verb$=$\Verb$g$\Verb$e$\Verb$t$\Verb$v$\Verb$a$\Verb$l$\Verb$u$\Verb$e$\Verb$($\Verb$e$\Verb$l$\Verb$)$\\\Verb$ $\Verb$ $\Verb$ $\Verb$ $\Verb$ $\Verb$ $\Verb$v$\Verb$a$\Verb$l$\Verb$=$\Verb$l$\Verb$a$\Verb$t$\Verb$t$\Verb$i$\Verb$c$\Verb$e$\Verb$_$\Verb$j$\Verb$o$\Verb$i$\Verb$n$\Verb$($\Verb$v$\Verb$a$\Verb$l$\Verb$,$\Verb$@$\Verb$v$\Verb$a$\Verb$l$\Verb$s$\Verb$[$\Verb$e$\Verb$l$\Verb$]$\Verb$)$\\\Verb$ $\Verb$ $\Verb$ $\Verb$ $\Verb$ $\Verb$ $\Verb$i$\Verb$f$\Verb$ $\Verb$v$\Verb$a$\Verb$l$\Verb$ $\Verb$>$\Verb$ $\Verb$@$\Verb$v$\Verb$a$\Verb$l$\Verb$s$\Verb$[$\Verb$e$\Verb$l$\Verb$]$\\\Verb$ $\Verb$ $\Verb$ $\Verb$ $\Verb$ $\Verb$ $\Verb$ $\Verb$ $\Verb$@$\Verb$v$\Verb$a$\Verb$l$\Verb$s$\Verb$[$\Verb$e$\Verb$l$\Verb$]$\Verb$=$\Verb$v$\Verb$a$\Verb$l$\\\Verb$ $\Verb$ $\Verb$ $\Verb$ $\Verb$ $\Verb$ $\Verb$ $\Verb$ $\Verb$@$\Verb$d$\Verb$e$\Verb$p$\Verb$e$\Verb$n$\Verb$d$\Verb$.$\Verb$e$\Verb$d$\Verb$g$\Verb$e$\Verb$s$\Verb$[$\Verb$e$\Verb$l$\Verb$]$\Verb$.$\Verb$e$\Verb$a$\Verb$c$\Verb$h$\Verb${$\Verb$|$\Verb$d$\Verb$|$\Verb$ $\Verb$a$\Verb$d$\Verb$d$\Verb$a$\Verb$c$\Verb$t$\Verb$i$\Verb$v$\Verb$e$\Verb$($\Verb$d$\Verb$)$\Verb$}$\\\Verb$ $\Verb$ $\Verb$ $\Verb$ $\Verb$ $\Verb$ $\Verb$e$\Verb$n$\Verb$d$\\\Verb$ $\Verb$ $\Verb$ $\Verb$ $\Verb$e$\Verb$n$\Verb$d$\\\Verb$ $\Verb$ $\Verb$ $\Verb$ $\Verb$@$\Verb$v$\Verb$a$\Verb$l$\Verb$s$\Verb$[${\color{blue}\Verb$e$}\Verb$]$\\\Verb$ $\Verb$ $\Verb$e$\Verb$n$\Verb$d$\Verb$ $\\\\\Verb$ $\Verb$ $\Verb$d$\Verb$e$\Verb$f$\Verb$ $\Verb$d$\Verb$e$\Verb$p$\Verb$e$\Verb$n$\Verb$d$\Verb$s$\Verb$(${\color{blue}\Verb$e$}\Verb$)$\\\Verb$ $\Verb$ $\Verb$ $\Verb$ $\Verb$@$\Verb$d$\Verb$e$\Verb$p$\Verb$e$\Verb$n$\Verb$d$\Verb$.$\Verb$a$\Verb$d$\Verb$d$\Verb$(${\color{blue}\Verb$e$}\Verb$,$\Verb$@$\Verb$v$\Verb$i$\Verb$s$\Verb$)$\Verb$ $\Verb$i$\Verb$f$\Verb$ $\Verb$!$\Verb$@$\Verb$d$\Verb$e$\Verb$p$\Verb$e$\Verb$n$\Verb$d$\Verb$.$\Verb$e$\Verb$d$\Verb$g$\Verb$e$\Verb$s$\Verb$[${\color{blue}\Verb$e$}\Verb$]$\Verb$.$\Verb$i$\Verb$n$\Verb$c$\Verb$l$\Verb$u$\Verb$d$\Verb$e$\Verb$?$\Verb$($\Verb$@$\Verb$v$\Verb$i$\Verb$s$\Verb$)$\\\Verb$ $\Verb$ $\Verb$ $\Verb$ $\Verb$i$\Verb$f$\Verb$ $\Verb$!$\Verb$@$\Verb$v$\Verb$i$\Verb$s$\Verb$i$\Verb$t$\Verb$e$\Verb$d$\Verb$[${\color{blue}\Verb$e$}\Verb$]$\\\Verb$ $\Verb$ $\Verb$ $\Verb$ $\Verb$ $\Verb$ $\Verb$@$\Verb$v$\Verb$i$\Verb$s$\Verb$i$\Verb$t$\Verb$e$\Verb$d$\Verb$[${\color{blue}\Verb$e$}\Verb$]$\Verb$=$\Verb$t$\Verb$r$\Verb$u$\Verb$e$\\\Verb$ $\Verb$ $\Verb$ $\Verb$ $\Verb$ $\Verb$ $\Verb$a$\Verb$d$\Verb$d$\Verb$a$\Verb$c$\Verb$t$\Verb$i$\Verb$v$\Verb$e$\Verb$(${\color{blue}\Verb$e$}\Verb$)$\\\Verb$ $\Verb$ $\Verb$ $\Verb$ $\Verb$e$\Verb$n$\Verb$d$\\\Verb$ $\Verb$ $\Verb$e$\Verb$n$\Verb$d$\\\\\Verb$ $\Verb$ $\Verb$d$\Verb$e$\Verb$f$\Verb$ $\Verb$a$\Verb$d$\Verb$d$\Verb$a$\Verb$c$\Verb$t$\Verb$i$\Verb$v$\Verb$e$\Verb$(${\color{blue}\Verb$e$}\Verb$)$\\\Verb$ $\Verb$ $\Verb$ $\Verb$ $\Verb$i$\Verb$f$\Verb$ $\Verb$!$\Verb$@$\Verb$a$\Verb$c$\Verb$t$\Verb$i$\Verb$v$\Verb$e$\Verb$[${\color{blue}\Verb$e$}\Verb$]$\\\Verb$ $\Verb$ $\Verb$ $\Verb$ $\Verb$ $\Verb$ $\Verb$@$\Verb$a$\Verb$c$\Verb$t$\Verb$i$\Verb$v$\Verb$e$\Verb$[${\color{blue}\Verb$e$}\Verb$]$\Verb$=$\Verb$t$\Verb$r$\Verb$u$\Verb$e$\\\Verb$ $\Verb$ $\Verb$ $\Verb$ $\Verb$ $\Verb$ $\Verb$@$\Verb$a$\Verb$c$\Verb$t$\Verb$i$\Verb$v$\Verb$e$\Verb$a$\Verb$<$\Verb$<${\color{blue}\Verb$e$}\\\Verb$ $\Verb$ $\Verb$ $\Verb$ $\Verb$e$\Verb$n$\Verb$d$\\\Verb$ $\Verb$ $\Verb$e$\Verb$n$\Verb$d$\\\\\Verb$ $\Verb$ $\Verb$d$\Verb$e$\Verb$f$\Verb$ $\Verb$i$\Verb$n$\Verb$i$\Verb$t$\Verb$i$\Verb$a$\Verb$l$\Verb$i$\Verb$z$\Verb$e$\\\Verb$ $\Verb$ $\Verb$ $\Verb$ $\Verb$@$\Verb$d$\Verb$e$\Verb$p$\Verb$e$\Verb$n$\Verb$d$\Verb$=$\Verb$O$\Verb$r$\Verb$i$\Verb$e$\Verb$n$\Verb$t$\Verb$e$\Verb$d$\Verb$_$\Verb$G$\Verb$r$\Verb$a$\Verb$p$\Verb$h$\Verb$.$\Verb$n$\Verb$e$\Verb$w$\\\Verb$ $\Verb$ $\Verb$ $\Verb$ $\Verb$@$\Verb$v$\Verb$a$\Verb$l$\Verb$s$\Verb$=$\Verb$H$\Verb$a$\Verb$s$\Verb$h$\Verb$.$\Verb$n$\Verb$e$\Verb$w$\Verb$($\Verb$l$\Verb$a$\Verb$t$\Verb$t$\Verb$i$\Verb$c$\Verb$e$\Verb$_$\Verb$b$\Verb$o$\Verb$t$\Verb$t$\Verb$o$\Verb$m$\Verb$)$\\\Verb$ $\Verb$ $\Verb$ $\Verb$ $\Verb$@$\Verb$v$\Verb$i$\Verb$s$\Verb$i$\Verb$t$\Verb$e$\Verb$d$\Verb$=$\Verb${$\Verb$}$\\\Verb$ $\Verb$ $\Verb$e$\Verb$n$\Verb$d$\\\Verb$e$\Verb$n$\Verb$d$
\end{exambox}
}
\end{spacing}\vskip -0.4em

\newpage
\section{Parameters as an object}\label{paramcont}
Now we covered enough background to decribe the full form of amethyst\\
 parametrization.

In Ruby we can pass parameters in several ways:

\vskip -1.8em\begin{spacing}{0.8}
{\small
\begin{exambox}
\Verb$d$\Verb$e$\Verb$f$\Verb$ $\Verb$a$\Verb$($\Verb$x$\Verb$,$\Verb$y$\Verb$)$\\\Verb$ $\Verb$ $\Verb$x$\Verb$+$\Verb$y$\\\Verb$e$\Verb$n$\Verb$d$\\\Verb$p$\Verb$u$\Verb$t$\Verb$s$\Verb$ $\Verb$a$\Verb$($\Verb$2$\Verb$+$\Verb$2$\Verb$)$\Verb$ $\Verb$#$\Verb$-$\Verb$>$\Verb$4$\\\Verb$d$\Verb$e$\Verb$f$\Verb$ $\Verb$b$\Verb$($\Verb$x$\Verb$=$\Verb$1$\Verb$,$\Verb$y$\Verb$=$\Verb$2$\Verb$)$\\\Verb$ $\Verb$ $\Verb$x$\Verb$,$\Verb$y$\\\Verb$e$\Verb$n$\Verb$d$\\\Verb$p$\Verb$u$\Verb$t$\Verb$s$\Verb$ $\Verb$b$\Verb$($\Verb$2$\Verb$)$\Verb$ $\Verb$ $\Verb$ $\Verb$#$\Verb$-$\Verb$>$\Verb$4$\\\Verb$d$\Verb$e$\Verb$f$\Verb$ $\Verb$c$\Verb$($\Verb$x$\Verb$,$\Verb$y$\Verb$,$\Verb$*$\Verb$a$\Verb$r$\Verb$y$\Verb$)$\\\Verb$ $\Verb$ $\Verb$a$\Verb$r$\Verb$y$\Verb$.$\Verb$i$\Verb$n$\Verb$s$\Verb$p$\Verb$e$\Verb$c$\Verb$t$\\\Verb$e$\Verb$n$\Verb$d$\\\Verb$p$\Verb$u$\Verb$t$\Verb$s$\Verb$ $\Verb$c$\Verb$($\Verb$1$\Verb$,$\Verb$2$\Verb$,$\Verb$3$\Verb$,$\Verb$4$\Verb$,$\Verb$5$\Verb$)$\Verb$ $\Verb$#$\Verb$-$\Verb$>$\Verb$ $\Verb$[$\Verb$3$\Verb$,$\Verb$4$\Verb$,$\Verb$5$\Verb$]$\\\Verb$d$\Verb$e$\Verb$f$\Verb$ $\Verb$w$\Verb$i$\Verb$t$\Verb$h$\Verb$_$\Verb$b$\Verb$l$\Verb$o$\Verb$c$\Verb$k$\\\Verb$ $\Verb$ $\Verb$y$\Verb$i$\Verb$e$\Verb$l$\Verb$d$\Verb$($\Verb$1$\Verb$)$\Verb$ $\Verb$+$\Verb$ $\Verb$y$\Verb$i$\Verb$e$\Verb$l$\Verb$d$\Verb$($\Verb$3$\Verb$)$\\\Verb$e$\Verb$n$\Verb$d$\\\Verb$w$\Verb$i$\Verb$t$\Verb$h$\Verb$_$\Verb$b$\Verb$l$\Verb$o$\Verb$c$\Verb$k$\Verb${$\Verb$|$\Verb$x$\Verb$|$\Verb$ $\Verb$x$\Verb$+$\Verb$3$\Verb$}$\Verb$ $\Verb$#$\Verb$ $\Verb$-$\Verb$>$\Verb$ $\Verb$1$\Verb$0$\\\Verb$#$\Verb$ $\Verb$r$\Verb$u$\Verb$b$\Verb$y$\Verb$1$\Verb$.$\Verb$9$\Verb$ $\Verb$e$\Verb$m$\Verb$u$\Verb$l$\Verb$a$\Verb$t$\Verb$e$\Verb$s$\Verb$ $\Verb$n$\Verb$a$\Verb$m$\Verb$e$\Verb$d$\Verb$ $\Verb$p$\Verb$a$\Verb$r$\Verb$a$\Verb$m$\Verb$e$\Verb$t$\Verb$e$\Verb$r$\Verb$ $\Verb$b$\Verb$y$\Verb$ $\Verb$p$\Verb$a$\Verb$s$\Verb$s$\Verb$i$\Verb$n$\Verb$g$\Verb$ $\Verb$l$\Verb$a$\Verb$s$\Verb$t$\Verb$ $\Verb$p$\Verb$a$\Verb$r$\Verb$a$\Verb$m$\Verb$e$\Verb$t$\Verb$e$\Verb$r$\Verb$ $\\\Verb$#$\Verb$ $\Verb$a$\Verb$s$\Verb$ $\Verb$h$\Verb$a$\Verb$s$\Verb$h$\Verb$ $\Verb$a$\Verb$n$\Verb$d$\Verb$ $\Verb$a$\Verb$l$\Verb$l$\Verb$o$\Verb$w$\Verb$i$\Verb$n$\Verb$g$\Verb$ $\Verb$t$\Verb$o$\Verb$ $\Verb$o$\Verb$m$\Verb$i$\Verb$t$\Verb$ $\Verb${$\Verb$}$\Verb$.$\\\Verb$d$\Verb$e$\Verb$f$\Verb$ $\Verb$n$\Verb$a$\Verb$m$\Verb$e$\Verb$d$\Verb$($\Verb$x$\Verb$)$\\\Verb$ $\Verb$ $\Verb$p$\Verb$u$\Verb$t$\Verb$s$\Verb$ $\Verb$x$\Verb$.$\Verb$i$\Verb$n$\Verb$s$\Verb$p$\Verb$e$\Verb$c$\Verb$t$\\\Verb$e$\Verb$n$\Verb$d$\\\Verb$n$\Verb$a$\Verb$m$\Verb$e$\Verb$d$\Verb$($\Verb$:$\Verb$f$\Verb$o$\Verb$o$\Verb$=$\Verb$>$\Verb$1$\Verb$,$\Verb$:$\Verb$b$\Verb$a$\Verb$r$\Verb$=$\Verb$>$\Verb$2$\Verb$)$\Verb$ $\Verb$#$\Verb$-$\Verb$>$\Verb$ $\Verb${$\Verb$:$\Verb$f$\Verb$o$\Verb$o$\Verb$=$\Verb$>$\Verb$1$\Verb$,$\Verb$:$\Verb$b$\Verb$a$\Verb$r$\Verb$=$\Verb$>$\Verb$2$\Verb$}$
\end{exambox}
}
\end{spacing}\vskip -0.4em

Amethyst parametrized calls are done by creating special object and pattern matching it againist definition.

We can describe them by the following shortcuts:

\vskip 0.2em\noindent\begin{tabular}{|l | l |}
\hline
Pattern & Description \\
\hline
{\color{red}\Verb$n$}{\color{red}\Verb$a$}{\color{red}\Verb$m$}{\color{red}\Verb$e$}{\color{green}\Verb$($}{\color{green}\Verb$p$}{\color{green}\Verb$a$}{\color{green}\Verb$t$}{\color{green}\Verb$t$}{\color{green}\Verb$e$}{\color{green}\Verb$r$}{\color{green}\Verb$n$}{\color{green}\Verb$)$}\Verb$=${\color{blue}\Verb$e$} &  {\color{red}\Verb$_$}{\color{red}\Verb$n$}{\color{red}\Verb$a$}{\color{red}\Verb$m$}{\color{red}\Verb$e$}{\color{green}\Verb$($}{\color{green}\Verb$a$}{\color{green}\Verb$r$}{\color{green}\Verb$g$}{\color{green}\Verb$s$}{\color{green}\Verb$)$}\Verb$ $\Verb$=$\Verb$ ${\color{Tan}\Verb${$}{\color{Tan}\Verb$a$}{\color{Tan}\Verb$r$}{\color{Tan}\Verb$g$}{\color{Tan}\Verb$s$}{\color{Tan}\Verb$}$}{\color{blue}\Verb$=$}{\color{blue}\Verb$>$}{\color{red}\Verb$p$}{\color{red}\Verb$a$}{\color{red}\Verb$t$}{\color{red}\Verb$t$}{\color{red}\Verb$e$}{\color{red}\Verb$r$}{\color{red}\Verb$n$}\Verb$ ${\color{blue}\Verb$e$} \\
{\color{red}\Verb$n$}{\color{red}\Verb$a$}{\color{red}\Verb$m$}{\color{red}\Verb$e$}{\color{green}\Verb$($}{\color{green}\Verb$a$}{\color{green}\Verb$1$}{\color{green}\Verb$,$}{\color{green}\Verb$a$}{\color{green}\Verb$2$}{\color{green}\Verb$,$}{\color{green}\Verb$k$}{\color{green}\Verb$e$}{\color{green}\Verb$y$}{\color{green}\Verb$1$}{\color{green}\Verb$:$}{\color{green}\Verb$v$}{\color{green}\Verb$a$}{\color{green}\Verb$l$}{\color{green}\Verb$1$}{\color{green}\Verb$)$}\verb$$ & {\color{red}\Verb$_$}{\color{red}\Verb$n$}{\color{red}\Verb$a$}{\color{red}\Verb$m$}{\color{red}\Verb$e$}{\color{green}\Verb$($}{\color{green}\Verb$A$}{\color{green}\Verb$r$}{\color{green}\Verb$g$}{\color{green}\Verb$u$}{\color{green}\Verb$m$}{\color{green}\Verb$e$}{\color{green}\Verb$n$}{\color{green}\Verb$t$}{\color{green}\Verb$s$}{\color{green}\Verb$[$}{\color{green}\Verb$[$}{\color{green}\Verb$a$}{\color{green}\Verb$1$}{\color{green}\Verb$,$}{\color{green}\Verb$a$}{\color{green}\Verb$2$}{\color{green}\Verb$]$}{\color{green}\Verb$,$}{\color{green}\Verb${$}{\color{green}\Verb$k$}{\color{green}\Verb$e$}{\color{green}\Verb$y$}{\color{green}\Verb$1$}{\color{green}\Verb$=$}{\color{green}\Verb$>$}{\color{green}\Verb$v$}{\color{green}\Verb$a$}{\color{green}\Verb$l$}{\color{green}\Verb$1$}{\color{green}\Verb$}$}{\color{green}\Verb$]$}{\color{green}\Verb$)$}\verb$     $\\
\hline
\end{tabular}\vskip 0.2em

We use a convention similar to block passing in Ruby.

\vskip 0.2em\noindent\begin{tabular}{ | l | l | }
\hline
{\color{blue}\Verb$e$}{\color{green}\Verb$($}{\color{green}\Verb$c$}{\color{green}\Verb$1$}{\color{green}\Verb$,$}{\color{green}\Verb$c$}{\color{green}\Verb$2$}{\color{green}\Verb$)$}{\color{Violet}\Verb$($}{\color{Violet}\Verb$|$}\Verb$ ${\color{blue}\Verb$e$}{\color{blue}\Verb$2$}\Verb$ ${\color{Violet}\Verb$|$}{\color{Violet}\Verb$)$}\verb$     $ & {\color{blue}\Verb$e$}{\color{green}\Verb$($}{\color{green}\Verb$c$}{\color{green}\Verb$1$}{\color{green}\Verb$,$}{\color{green}\Verb$c$}{\color{green}\Verb$2$}{\color{green}\Verb$,$}{\color{Violet}\Verb$($}{\color{Violet}\Verb$|$}{\color{green}\Verb$ $}{\color{blue}\Verb$e$}{\color{blue}\Verb$2$}{\color{green}\Verb$ $}{\color{Violet}\Verb$|$}{\color{Violet}\Verb$)$}{\color{green}\Verb$)$} \verb$                         $\\
\hline
\end{tabular}\vskip 0.2em

In matching arguments we use different syntax. Following syntax express idioms common in argument passing more directly and extends syntax of Ruby argument passing. 

\vskip 0.2em\noindent\begin{tabular}{|l | l | p{8cm} |}
\hline
{\color{green}\Verb$n$}{\color{green}\Verb$a$}{\color{green}\Verb$m$}{\color{green}\Verb$e$}
&\Verb$.${\color{Aquamarine}\Verb$:$}{\color{Aquamarine}\Verb$n$}{\color{Aquamarine}\Verb$a$}{\color{Aquamarine}\Verb$m$}{\color{Aquamarine}\Verb$e$}& Positional argument\\
{\color{green}\Verb$*$}{\color{green}\Verb$n$}{\color{green}\Verb$a$}{\color{green}\Verb$m$}{\color{green}\Verb$e$}
&\Verb$.${\color{black}\Verb$*$}{\color{Aquamarine}\Verb$:$}{\color{Aquamarine}\Verb$n$}{\color{Aquamarine}\Verb$a$}{\color{Aquamarine}\Verb$m$}{\color{Aquamarine}\Verb$e$}& Splat operator\\
{\color{green}\Verb$n$}{\color{green}\Verb$a$}{\color{green}\Verb$m$}{\color{green}\Verb$e$}{\color{green}\Verb$:$}{\color{blue}\Verb$e$}
&{\color{blue}\Verb$e$}{\color{Aquamarine}\Verb$:$}{\color{Aquamarine}\Verb$n$}{\color{Aquamarine}\Verb$a$}{\color{Aquamarine}\Verb$m$}{\color{Aquamarine}\Verb$e$}& Match amethyst expression\\
{\color{Aquamarine}\Verb$@$}{\color{Aquamarine}\Verb$n$}{\color{Aquamarine}\Verb$a$}{\color{Aquamarine}\Verb$m$}{\color{Aquamarine}\Verb$e$}
&{\color{Aquamarine}\Verb$@$}{\color{Aquamarine}\Verb$n$}{\color{Aquamarine}\Verb$a$}{\color{Aquamarine}\Verb$m$}{\color{Aquamarine}\Verb$e$}{\color{Aquamarine}\Verb$:$}{\color{Aquamarine}\Verb$n$}{\color{Aquamarine}\Verb$a$}{\color{Aquamarine}\Verb$m$}{\color{Aquamarine}\Verb$e$}& Named argument\\
{\color{Aquamarine}\Verb$@$}{\color{Aquamarine}\Verb$n$}{\color{Aquamarine}\Verb$a$}{\color{Aquamarine}\Verb$m$}{\color{Aquamarine}\Verb$e$}{\color{green}\Verb$:$}{\color{blue}\Verb$e$}
&{\color{Aquamarine}\Verb$@$}{\color{Aquamarine}\Verb$n$}{\color{Aquamarine}\Verb$a$}{\color{Aquamarine}\Verb$m$}{\color{Aquamarine}\Verb$e$}{\color{blue}\Verb$=$}{\color{blue}\Verb$>$}{\color{blue}\Verb$e$}{\color{Aquamarine}\Verb$:$}{\color{Aquamarine}\Verb$n$}{\color{Aquamarine}\Verb$a$}{\color{Aquamarine}\Verb$m$}{\color{Aquamarine}\Verb$e$}& Match named argument with expression\\
{\color{green}\Verb$n$}{\color{green}\Verb$a$}{\color{green}\Verb$m$}{\color{green}\Verb$e$}{\color{green}\Verb$=$}{\color{green}\Verb$v$}{\color{green}\Verb$a$}{\color{green}\Verb$l$}
&\Verb$($\Verb$.$\Verb$|${\color{Tan}\Verb${$}{\color{Tan}\Verb$v$}{\color{Tan}\Verb$a$}{\color{Tan}\Verb$l$}{\color{Tan}\Verb$}$}\Verb$)${\color{Aquamarine}\Verb$:$}{\color{Aquamarine}\Verb$n$}{\color{Aquamarine}\Verb$a$}{\color{Aquamarine}\Verb$m$}{\color{Aquamarine}\Verb$e$}& Optional argument\\
{\color{Aquamarine}\Verb$@$}{\color{Aquamarine}\Verb$n$}{\color{Aquamarine}\Verb$a$}{\color{Aquamarine}\Verb$m$}{\color{Aquamarine}\Verb$e$}{\color{green}\Verb$=$}{\color{green}\Verb$v$}{\color{green}\Verb$a$}{\color{green}\Verb$l$}
&& Optional named argument \\
\hline
\end{tabular}\vskip 0.2em
\newpage
Example with use cases follows:

\vskip -1.8em\begin{spacing}{0.8}
{\small
\begin{exambox}
\Verb$a$\Verb$m$\Verb$e$\Verb$t$\Verb$h$\Verb$y$\Verb$s$\Verb$t$\Verb$ $\Verb$P$\Verb$a$\Verb$r$\Verb$a$\Verb$m$\Verb$e$\Verb$t$\Verb$r$\Verb$i$\Verb$z$\Verb$a$\Verb$t$\Verb$i$\Verb$o$\Verb$n$\Verb${$\\\Verb$ $\Verb$ ${\color{red}\Verb$o$}{\color{red}\Verb$p$}{\color{red}\Verb$t$}{\color{green}\Verb$($}{\color{green}\Verb$x$}{\color{green}\Verb$,$}{\color{green}\Verb$y$}{\color{green}\Verb$=$}{\color{green}\Verb$1$}{\color{green}\Verb$)$}\Verb$ $\Verb$=$\Verb$ ${\color{red}\Verb$x$}{\color{black}\Verb$+$}{\color{red}\Verb$y$}\\\Verb$ $\Verb$ ${\color{red}\Verb$u$}{\color{red}\Verb$s$}{\color{red}\Verb$e$}{\color{red}\Verb$_$}{\color{red}\Verb$o$}{\color{red}\Verb$p$}{\color{red}\Verb$t$}\Verb$ $\Verb$ $\Verb$ $\Verb$ $\Verb$=$\Verb$ ${\color{red}\Verb$o$}{\color{red}\Verb$p$}{\color{red}\Verb$t$}{\color{green}\Verb$($}{\color{green}\Verb$1$}{\color{green}\Verb$,$}{\color{green}\Verb$2$}{\color{green}\Verb$)$}\Verb$ ${\color{gray}\Verb$#$}{\color{gray}\Verb$-$}{\color{gray}\Verb$>$}{\color{gray}\Verb$ $}{\color{gray}\Verb$3$}{\color{gray}\\}\Verb$ $\Verb$ ${\color{red}\Verb$u$}{\color{red}\Verb$s$}{\color{red}\Verb$e$}{\color{red}\Verb$_$}{\color{red}\Verb$o$}{\color{red}\Verb$p$}{\color{red}\Verb$t$}{\color{red}\Verb$2$}\Verb$ $\Verb$ $\Verb$ $\Verb$=$\Verb$ ${\color{red}\Verb$o$}{\color{red}\Verb$p$}{\color{red}\Verb$t$}{\color{green}\Verb$($}{\color{green}\Verb$1$}{\color{green}\Verb$)$}\Verb$ $\Verb$ $\Verb$ ${\color{gray}\Verb$#$}{\color{gray}\Verb$-$}{\color{gray}\Verb$>$}{\color{gray}\Verb$ $}{\color{gray}\Verb$2$}{\color{gray}\\}\Verb$ $\Verb$ ${\color{red}\Verb$m$}{\color{red}\Verb$u$}{\color{red}\Verb$l$}{\color{red}\Verb$t$}{\color{red}\Verb$i$}{\color{green}\Verb$($}{\color{green}\Verb$x$}{\color{green}\Verb$,$}{\color{green}\Verb$*$}{\color{green}\Verb$y$}{\color{green}\Verb$)$}\Verb$ $\Verb$=$\Verb$ ${\color{Tan}\Verb$-$}{\color{Tan}\Verb$>$}{\color{Tan}\Verb$ $}{\color{Tan}\Verb$y$}{\color{Tan}\\}{\color{Tan}\Verb$ $}\Verb$ ${\color{red}\Verb$u$}{\color{red}\Verb$s$}{\color{red}\Verb$e$}{\color{red}\Verb$_$}{\color{red}\Verb$m$}{\color{red}\Verb$u$}{\color{red}\Verb$l$}{\color{red}\Verb$t$}{\color{red}\Verb$i$}\Verb$ $\Verb$=$\Verb$ ${\color{red}\Verb$m$}{\color{red}\Verb$u$}{\color{red}\Verb$l$}{\color{red}\Verb$t$}{\color{red}\Verb$i$}{\color{green}\Verb$($}{\color{green}\Verb$1$}{\color{green}\Verb$,$}{\color{green}\Verb$2$}{\color{green}\Verb$,$}{\color{green}\Verb$3$}{\color{green}\Verb$)$}\Verb$ ${\color{gray}\Verb$#$}{\color{gray}\Verb$-$}{\color{gray}\Verb$>$}{\color{gray}\Verb$[$}{\color{gray}\Verb$2$}{\color{gray}\Verb$,$}{\color{gray}\Verb$3$}{\color{gray}\Verb$]$}{\color{gray}\\}\\\Verb$ $\Verb$ ${\color{red}\Verb$c$}{\color{red}\Verb$h$}{\color{red}\Verb$e$}{\color{red}\Verb$c$}{\color{red}\Verb$k$}{\color{green}\Verb$($}{\color{green}\Verb$x$}{\color{green}\Verb$:$}{\color{green}\Verb$S$}{\color{green}\Verb$t$}{\color{green}\Verb$r$}{\color{green}\Verb$i$}{\color{green}\Verb$n$}{\color{green}\Verb$g$}{\color{green}\Verb$,$}{\color{green}\Verb$y$}{\color{green}\Verb$:$}{\color{green}\Verb$S$}{\color{green}\Verb$t$}{\color{green}\Verb$r$}{\color{green}\Verb$i$}{\color{green}\Verb$n$}{\color{green}\Verb$g$}{\color{green}\Verb$)$}\Verb$ $\Verb$=$\Verb$ ${\color{Tan}\Verb$-$}{\color{Tan}\Verb$>$}{\color{Tan}\Verb$ $}{\color{Tan}\Verb$x$}{\color{Tan}\Verb$+$}{\color{Tan}\Verb$y$}{\color{Tan}\\}{\color{Tan}\Verb$ $}\Verb$ ${\color{red}\Verb$u$}{\color{red}\Verb$s$}{\color{red}\Verb$e$}{\color{red}\Verb$_$}{\color{red}\Verb$c$}{\color{red}\Verb$h$}{\color{red}\Verb$e$}{\color{red}\Verb$c$}{\color{red}\Verb$k$}\Verb$ $\Verb$ $\Verb$=$\Verb$ ${\color{red}\Verb$c$}{\color{red}\Verb$h$}{\color{red}\Verb$e$}{\color{red}\Verb$c$}{\color{red}\Verb$k$}{\color{green}\Verb$($}{\color{green}\Verb$"$}{\color{green}\Verb$a$}{\color{green}\Verb$"$}{\color{green}\Verb$,$}{\color{green}\Verb$"$}{\color{green}\Verb$b$}{\color{green}\Verb$"$}{\color{green}\Verb$)$}\Verb$ ${\color{gray}\Verb$#$}{\color{gray}\Verb$-$}{\color{gray}\Verb$>$}{\color{gray}\Verb$"$}{\color{gray}\Verb$a$}{\color{gray}\Verb$b$}{\color{gray}\Verb$"$}{\color{gray}\\}\Verb$ $\Verb$ ${\color{red}\Verb$u$}{\color{red}\Verb$s$}{\color{red}\Verb$e$}{\color{red}\Verb$_$}{\color{red}\Verb$c$}{\color{red}\Verb$h$}{\color{red}\Verb$e$}{\color{red}\Verb$c$}{\color{red}\Verb$k$}{\color{red}\Verb$2$}\Verb$ $\Verb$=$\Verb$ ${\color{red}\Verb$c$}{\color{red}\Verb$h$}{\color{red}\Verb$e$}{\color{red}\Verb$c$}{\color{red}\Verb$k$}{\color{green}\Verb$($}{\color{green}\Verb$1$}{\color{green}\Verb$,$}{\color{green}\Verb$2$}{\color{green}\Verb$)$}\Verb$ $\\\Verb$ $\Verb$ $\Verb$ $\Verb$ $\Verb$ $\Verb$ $\Verb$ $\Verb$ $\Verb$ $\Verb$ $\Verb$ $\Verb$ $\Verb$ $\Verb$|$\Verb$ ${\color{Tan}\Verb$-$}{\color{Tan}\Verb$>$}{\color{Tan}\Verb$ $}{\color{Tan}\Verb$"$}{\color{Tan}\Verb$f$}{\color{Tan}\Verb$a$}{\color{Tan}\Verb$i$}{\color{Tan}\Verb$l$}{\color{Tan}\Verb$e$}{\color{Tan}\Verb$d$}{\color{Tan}\Verb$"$}{\color{Tan}\Verb$ $}{\color{Tan}\Verb$ $}{\color{Tan}\Verb$#$}{\color{Tan}\Verb$-$}{\color{Tan}\Verb$>$}{\color{Tan}\Verb$ $}{\color{Tan}\Verb$"$}{\color{Tan}\Verb$f$}{\color{Tan}\Verb$a$}{\color{Tan}\Verb$i$}{\color{Tan}\Verb$l$}{\color{Tan}\Verb$e$}{\color{Tan}\Verb$d$}{\color{Tan}\Verb$"$}{\color{Tan}\\}{\color{Tan}\Verb$ $}\Verb$ $\Verb$ $\Verb$ $\Verb$ $\Verb$ $\Verb$ $\Verb$ $\Verb$ $\Verb$ $\Verb$ $\Verb$ $\Verb$ $\\\Verb$ $\Verb$ ${\color{red}\Verb$n$}{\color{red}\Verb$a$}{\color{red}\Verb$m$}{\color{red}\Verb$e$}{\color{red}\Verb$d$}{\color{green}\Verb$($}{\color{Aquamarine}\Verb$@$}{\color{Aquamarine}\Verb$x$}{\color{green}\Verb$=$}{\color{green}\Verb$1$}{\color{green}\Verb$,$}{\color{Aquamarine}\Verb$@$}{\color{Aquamarine}\Verb$y$}{\color{green}\Verb$=$}{\color{green}\Verb$2$}{\color{green}\Verb$)$}\Verb$ $\Verb$=$\Verb$ ${\color{Tan}\Verb$-$}{\color{Tan}\Verb$>$}{\color{Tan}\Verb$ $}{\color{Tan}\Verb$x$}{\color{Tan}\Verb$+$}{\color{Tan}\Verb$y$}{\color{Tan}\\}{\color{Tan}\Verb$ $}\Verb$ ${\color{red}\Verb$u$}{\color{red}\Verb$s$}{\color{red}\Verb$e$}{\color{red}\Verb$_$}{\color{red}\Verb$n$}{\color{red}\Verb$a$}{\color{red}\Verb$m$}{\color{red}\Verb$e$}{\color{red}\Verb$d$}\Verb$ $\Verb$ $\Verb$=$\Verb$ ${\color{red}\Verb$n$}{\color{red}\Verb$a$}{\color{red}\Verb$m$}{\color{red}\Verb$e$}{\color{red}\Verb$d$}{\color{green}\Verb$($}{\color{green}\Verb$x$}{\color{green}\Verb$:$}{\color{green}\Verb$3$}{\color{green}\Verb$,$}{\color{green}\Verb$y$}{\color{green}\Verb$:$}{\color{green}\Verb$3$}{\color{green}\Verb$)$}\Verb$ ${\color{gray}\Verb$#$}{\color{gray}\Verb$-$}{\color{gray}\Verb$>$}{\color{gray}\Verb$ $}{\color{gray}\Verb$6$}{\color{gray}\\}\Verb$ $\Verb$ ${\color{red}\Verb$u$}{\color{red}\Verb$s$}{\color{red}\Verb$e$}{\color{red}\Verb$_$}{\color{red}\Verb$n$}{\color{red}\Verb$a$}{\color{red}\Verb$m$}{\color{red}\Verb$e$}{\color{red}\Verb$d$}{\color{red}\Verb$2$}\Verb$ $\Verb$=$\Verb$ ${\color{red}\Verb$n$}{\color{red}\Verb$a$}{\color{red}\Verb$m$}{\color{red}\Verb$e$}{\color{red}\Verb$d$}{\color{green}\Verb$($}{\color{green}\Verb$x$}{\color{green}\Verb$:$}{\color{green}\Verb$2$}{\color{green}\Verb$)$}\Verb$ $\Verb$ $\Verb$ $\Verb$ $\Verb$ ${\color{gray}\Verb$#$}{\color{gray}\Verb$-$}{\color{gray}\Verb$>$}{\color{gray}\Verb$ $}{\color{gray}\Verb$4$}{\color{gray}\Verb$ $}{\color{gray}\Verb$ $}{\color{gray}\Verb$ $}{\color{gray}\Verb$ $}{\color{gray}\Verb$ $}{\color{gray}\Verb$ $}{\color{gray}\Verb$ $}{\color{gray}\Verb$ $}{\color{gray}\Verb$ $}{\color{gray}\Verb$ $}{\color{gray}\Verb$ $}{\color{gray}\Verb$ $}{\color{gray}\Verb$ $}{\color{gray}\\}\Verb$ $\Verb$ ${\color{red}\Verb$u$}{\color{red}\Verb$s$}{\color{red}\Verb$e$}{\color{red}\Verb$_$}{\color{red}\Verb$n$}{\color{red}\Verb$a$}{\color{red}\Verb$m$}{\color{red}\Verb$e$}{\color{red}\Verb$d$}{\color{red}\Verb$3$}\Verb$ $\Verb$=$\Verb$ ${\color{red}\Verb$n$}{\color{red}\Verb$a$}{\color{red}\Verb$m$}{\color{red}\Verb$e$}{\color{red}\Verb$d$}{\color{green}\Verb$($}{\color{green}\Verb$y$}{\color{green}\Verb$:$}{\color{green}\Verb$1$}{\color{green}\Verb$)$}\Verb$ $\Verb$ $\Verb$ $\Verb$ $\Verb$ ${\color{gray}\Verb$#$}{\color{gray}\Verb$-$}{\color{gray}\Verb$>$}{\color{gray}\Verb$ $}{\color{gray}\Verb$2$}{\color{gray}\\}\Verb$}$
\end{exambox}
}
\end{spacing}\vskip -0.4em

\subsection*{Example: Syntax highlighting} \label{ex:highlight}
A syntax highlighting in this thesis was relatively simple to implement by amethyst parser. This example relies on parametrized rules.

Consider the following simplified part of amethyst grammar:

\begin{grambox}
\begin{spacing}{0.8}
{\small
{\color{red}\Verb$p$}{\color{red}\Verb$o$}{\color{red}\Verb$s$}{\color{red}\Verb$t$}{\color{red}\Verb$f$}{\color{red}\Verb$i$}{\color{red}\Verb$x$}{\color{red}\Verb$e$}{\color{red}\Verb$d$}\Verb$ $\Verb$=$\Verb$ $\Verb$ ${\color{red}\Verb$t$}{\color{red}\Verb$e$}{\color{red}\Verb$r$}{\color{red}\Verb$m$}\\\Verb$ $\Verb$ $\Verb$ $\Verb$ $\Verb$ $\Verb$ $\Verb$ $\Verb$ $\Verb$ $\Verb$ $\Verb$ $\Verb$ $\Verb$ $\Verb$ $\Verb$($\Verb$ ${\color{black}\Verb$'$}{\color{black}\Verb$=$}{\color{black}\Verb$>$}{\color{black}\Verb$'$}\Verb$ ${\color{red}\Verb$t$}{\color{red}\Verb$e$}{\color{red}\Verb$r$}{\color{red}\Verb$m$}\\\Verb$ $\Verb$ $\Verb$ $\Verb$ $\Verb$ $\Verb$ $\Verb$ $\Verb$ $\Verb$ $\Verb$ $\Verb$ $\Verb$ $\Verb$ $\Verb$ $\Verb$|$\Verb$ ${\color{black}\Verb$'$}{\color{black}\Verb$[$}{\color{black}\Verb$'$}\Verb$ ${\color{red}\Verb$e$}{\color{red}\Verb$x$}{\color{red}\Verb$p$}{\color{red}\Verb$r$}{\color{red}\Verb$e$}{\color{red}\Verb$s$}{\color{red}\Verb$s$}{\color{red}\Verb$i$}{\color{red}\Verb$o$}{\color{red}\Verb$n$}\Verb$ ${\color{black}\Verb$"$}{\color{black}\Verb$"$}\Verb$ ${\color{black}\Verb$'$}{\color{black}\Verb$]$}{\color{black}\Verb$'$}\Verb$ $\\\Verb$ $\Verb$ $\Verb$ $\Verb$ $\Verb$ $\Verb$ $\Verb$ $\Verb$ $\Verb$ $\Verb$ $\Verb$ $\Verb$ $\Verb$ $\Verb$ $\Verb$|$\Verb$ $\Verb$<$\Verb$+$\Verb$*$\Verb$?$\Verb$>$\Verb$ $\Verb$ $\Verb$ $\Verb$ $\Verb$ $\Verb$ $\Verb$ $\\\Verb$ $\Verb$ $\Verb$ $\Verb$ $\Verb$ $\Verb$ $\Verb$ $\Verb$ $\Verb$ $\Verb$ $\Verb$ $\Verb$ $\Verb$ $\Verb$ $\Verb$|$\Verb$ ${\color{black}\Verb$'$}{\color{black}\Verb$:$}{\color{black}\Verb$'$}\Verb$ ${\color{black}\Verb$'$}{\color{black}\Verb$[$}{\color{black}\Verb$'$}\Verb$ $\Verb$(${\color{red}\Verb$k$}{\color{red}\Verb$e$}{\color{red}\Verb$y$}\Verb$ $\Verb$|$\Verb$ ${\color{red}\Verb$n$}{\color{red}\Verb$a$}{\color{red}\Verb$m$}{\color{red}\Verb$e$}\Verb$)$\Verb$ ${\color{black}\Verb$'$}{\color{black}\Verb$]$}{\color{black}\Verb$'$}\Verb$ $\\\Verb$ $\Verb$ $\Verb$ $\Verb$ $\Verb$ $\Verb$ $\Verb$ $\Verb$ $\Verb$ $\Verb$ $\Verb$ $\Verb$ $\Verb$ $\Verb$ $\Verb$|$\Verb$ ${\color{black}\Verb$'$}{\color{black}\Verb$:$}{\color{black}\Verb$'$}\Verb$ $\Verb$ $\Verb$ $\Verb$ $\Verb$ $\Verb$(${\color{red}\Verb$k$}{\color{red}\Verb$e$}{\color{red}\Verb$y$}\Verb$ $\Verb$|$\Verb$ ${\color{red}\Verb$n$}{\color{red}\Verb$a$}{\color{red}\Verb$m$}{\color{red}\Verb$e$}\Verb$)$\\\Verb$ $\Verb$ $\Verb$ $\Verb$ $\Verb$ $\Verb$ $\Verb$ $\Verb$ $\Verb$ $\Verb$ $\Verb$ $\Verb$ $\Verb$ $\Verb$ $\Verb$|$\Verb$ ${\color{red}\Verb$i$}{\color{red}\Verb$n$}{\color{red}\Verb$l$}{\color{red}\Verb$i$}{\color{red}\Verb$n$}{\color{red}\Verb$e$}{\color{red}\Verb$_$}{\color{red}\Verb$h$}{\color{red}\Verb$o$}{\color{red}\Verb$s$}{\color{red}\Verb$t$}{\color{red}\Verb$_$}{\color{red}\Verb$e$}{\color{red}\Verb$x$}{\color{red}\Verb$p$}{\color{red}\Verb$r$}\\\Verb$ $\Verb$ $\Verb$ $\Verb$ $\Verb$ $\Verb$ $\Verb$ $\Verb$ $\Verb$ $\Verb$ $\Verb$ $\Verb$ $\Verb$ $\Verb$ $\Verb$)${\color{black}\Verb$*$}
}\end{spacing}

\end{grambox}

This grammar can be annotated by colors in the following way:

\begin{exambox}
\begin{spacing}{0.8}
{\small
{\color{red}\Verb$p$}{\color{red}\Verb$o$}{\color{red}\Verb$s$}{\color{red}\Verb$t$}{\color{red}\Verb$f$}{\color{red}\Verb$i$}{\color{red}\Verb$x$}{\color{red}\Verb$e$}{\color{red}\Verb$d$}\Verb$ $\Verb$=$\Verb$ $\Verb$ ${\color{red}\Verb$t$}{\color{red}\Verb$e$}{\color{red}\Verb$r$}{\color{red}\Verb$m$}\\\Verb$ $\Verb$ $\Verb$ $\Verb$ $\Verb$ $\Verb$ $\Verb$ $\Verb$ $\Verb$ $\Verb$ $\Verb$ $\Verb$ $\Verb$ $\Verb$ $\Verb$ $\Verb$($\Verb$ ${\color{red}\Verb$c$}{\color{red}\Verb$o$}{\color{red}\Verb$l$}{\color{red}\Verb$o$}{\color{red}\Verb$r$}{\color{green}\Verb$($}{\color{green}\Verb$"$}{\color{green}\Verb$b$}{\color{green}\Verb$l$}{\color{green}\Verb$u$}{\color{green}\Verb$e$}{\color{green}\Verb$"$}{\color{green}\Verb$ $}{\color{green}\Verb$)$}{\color{Violet}\Verb$($}{\color{Violet}\Verb$|$}{\color{black}\Verb$'$}{\color{black}\Verb$=$}{\color{black}\Verb$>$}{\color{black}\Verb$'$}{\color{Violet}\Verb$|$}{\color{Violet}\Verb$)$}\Verb$ $\Verb$ $\Verb$ $\Verb$ $\Verb$ ${\color{red}\Verb$t$}{\color{red}\Verb$e$}{\color{red}\Verb$r$}{\color{red}\Verb$m$}{\color{Aquamarine}\Verb$:$}{\color{blue}\Verb$e$}\Verb$ $\Verb$ $\Verb$ $\Verb$ $\Verb$ $\Verb$ $\Verb$ $\Verb$ $\Verb$ $\Verb$ $\Verb$ $\\\Verb$ $\Verb$ $\Verb$ $\Verb$ $\Verb$ $\Verb$ $\Verb$ $\Verb$ $\Verb$ $\Verb$ $\Verb$ $\Verb$ $\Verb$ $\Verb$ $\Verb$ $\Verb$|$\Verb$ ${\color{red}\Verb$c$}{\color{red}\Verb$o$}{\color{red}\Verb$l$}{\color{red}\Verb$o$}{\color{red}\Verb$r$}{\color{green}\Verb$($}{\color{green}\Verb$"$}{\color{green}\Verb$b$}{\color{green}\Verb$l$}{\color{green}\Verb$u$}{\color{green}\Verb$e$}{\color{green}\Verb$"$}{\color{green}\Verb$ $}{\color{green}\Verb$)$}{\color{Violet}\Verb$($}{\color{Violet}\Verb$|$}{\color{black}\Verb$'$}{\color{black}\Verb$[$}{\color{black}\Verb$'$}\Verb$ ${\color{red}\Verb$e$}{\color{red}\Verb$x$}{\color{red}\Verb$p$}{\color{red}\Verb$r$}{\color{red}\Verb$e$}{\color{red}\Verb$s$}{\color{red}\Verb$s$}{\color{red}\Verb$i$}{\color{red}\Verb$o$}{\color{red}\Verb$n$}{\color{Aquamarine}\Verb$:$}{\color{blue}\Verb$e$}\Verb$ ${\color{black}\Verb$"$}{\color{black}\Verb$"$}\Verb$ ${\color{black}\Verb$'$}{\color{black}\Verb$]$}{\color{black}\Verb$'$}\Verb$ ${\color{Violet}\Verb$|$}{\color{Violet}\Verb$)$}\\\Verb$ $\Verb$ $\Verb$ $\Verb$ $\Verb$ $\Verb$ $\Verb$ $\Verb$ $\Verb$ $\Verb$ $\Verb$ $\Verb$ $\Verb$ $\Verb$ $\Verb$ $\Verb$|$\Verb$ ${\color{red}\Verb$c$}{\color{red}\Verb$o$}{\color{red}\Verb$l$}{\color{red}\Verb$o$}{\color{red}\Verb$r$}{\color{green}\Verb$($}{\color{green}\Verb$"$}{\color{green}\Verb$b$}{\color{green}\Verb$l$}{\color{green}\Verb$a$}{\color{green}\Verb$c$}{\color{green}\Verb$k$}{\color{green}\Verb$"$}{\color{green}\Verb$)$}{\color{Violet}\Verb$($}{\color{Violet}\Verb$|$}\Verb$ $\Verb$<$\Verb$+$\Verb$*$\Verb$?$\Verb$>$\Verb$ ${\color{Violet}\Verb$|$}{\color{Violet}\Verb$)$}\Verb$ $\\\Verb$ $\Verb$ $\Verb$ $\Verb$ $\Verb$ $\Verb$ $\Verb$ $\Verb$ $\Verb$ $\Verb$ $\Verb$ $\Verb$ $\Verb$ $\Verb$ $\Verb$ $\Verb$|$\Verb$ ${\color{red}\Verb$c$}{\color{red}\Verb$o$}{\color{red}\Verb$l$}{\color{red}\Verb$o$}{\color{red}\Verb$r$}{\color{green}\Verb$($}{\color{green}\Verb$"$}{\color{green}\Verb$g$}{\color{green}\Verb$r$}{\color{green}\Verb$e$}{\color{green}\Verb$e$}{\color{green}\Verb$n$}{\color{green}\Verb$"$}{\color{green}\Verb$)$}{\color{Violet}\Verb$($}{\color{Violet}\Verb$|$}\Verb$ ${\color{black}\Verb$'$}{\color{black}\Verb$:$}{\color{black}\Verb$'$}\Verb$ ${\color{black}\Verb$'$}{\color{black}\Verb$[$}{\color{black}\Verb$'$}\Verb$ $\Verb$(${\color{red}\Verb$k$}{\color{red}\Verb$e$}{\color{red}\Verb$y$}\Verb$ $\Verb$|$\Verb$ ${\color{red}\Verb$n$}{\color{red}\Verb$a$}{\color{red}\Verb$m$}{\color{red}\Verb$e$}\Verb$)$\Verb$ ${\color{black}\Verb$'$}{\color{black}\Verb$]$}{\color{black}\Verb$'$}\Verb$ ${\color{Violet}\Verb$|$}{\color{Violet}\Verb$)$}\\\Verb$ $\Verb$ $\Verb$ $\Verb$ $\Verb$ $\Verb$ $\Verb$ $\Verb$ $\Verb$ $\Verb$ $\Verb$ $\Verb$ $\Verb$ $\Verb$ $\Verb$ $\Verb$|$\Verb$ ${\color{red}\Verb$c$}{\color{red}\Verb$o$}{\color{red}\Verb$l$}{\color{red}\Verb$o$}{\color{red}\Verb$r$}{\color{green}\Verb$($}{\color{green}\Verb$"$}{\color{green}\Verb$g$}{\color{green}\Verb$r$}{\color{green}\Verb$e$}{\color{green}\Verb$e$}{\color{green}\Verb$n$}{\color{green}\Verb$"$}{\color{green}\Verb$)$}{\color{Violet}\Verb$($}{\color{Violet}\Verb$|$}\Verb$ ${\color{black}\Verb$'$}{\color{black}\Verb$:$}{\color{black}\Verb$'$}\Verb$ $\Verb$ $\Verb$ $\Verb$ $\Verb$ $\Verb$(${\color{red}\Verb$k$}{\color{red}\Verb$e$}{\color{red}\Verb$y$}\Verb$ $\Verb$|$\Verb$ ${\color{red}\Verb$n$}{\color{red}\Verb$a$}{\color{red}\Verb$m$}{\color{red}\Verb$e$}\Verb$)$\Verb$ $\Verb$ $\Verb$ $\Verb$ $\Verb$ ${\color{Violet}\Verb$|$}{\color{Violet}\Verb$)$}\\\Verb$ $\Verb$ $\Verb$ $\Verb$ $\Verb$ $\Verb$ $\Verb$ $\Verb$ $\Verb$ $\Verb$ $\Verb$ $\Verb$ $\Verb$ $\Verb$ $\Verb$ $\Verb$|$\Verb$ ${\color{red}\Verb$i$}{\color{red}\Verb$n$}{\color{red}\Verb$l$}{\color{red}\Verb$i$}{\color{red}\Verb$n$}{\color{red}\Verb$e$}{\color{red}\Verb$_$}{\color{red}\Verb$h$}{\color{red}\Verb$o$}{\color{red}\Verb$s$}{\color{red}\Verb$t$}{\color{red}\Verb$_$}{\color{red}\Verb$e$}{\color{red}\Verb$x$}{\color{red}\Verb$p$}{\color{red}\Verb$r$}\\\Verb$ $\Verb$ $\Verb$ $\Verb$ $\Verb$ $\Verb$ $\Verb$ $\Verb$ $\Verb$ $\Verb$ $\Verb$ $\Verb$ $\Verb$ $\Verb$ $\Verb$ $\Verb$)${\color{black}\Verb$*$}\\\\{\color{gray}\Verb$#$}{\color{gray}\Verb$a$}{\color{gray}\Verb$ $}{\color{gray}\Verb$s$}{\color{gray}\Verb$a$}{\color{gray}\Verb$m$}{\color{gray}\Verb$p$}{\color{gray}\Verb$l$}{\color{gray}\Verb$e$}{\color{gray}\Verb$ $}{\color{gray}\Verb$i$}{\color{gray}\Verb$m$}{\color{gray}\Verb$p$}{\color{gray}\Verb$l$}{\color{gray}\Verb$e$}{\color{gray}\Verb$m$}{\color{gray}\Verb$e$}{\color{gray}\Verb$n$}{\color{gray}\Verb$t$}{\color{gray}\Verb$a$}{\color{gray}\Verb$t$}{\color{gray}\Verb$i$}{\color{gray}\Verb$o$}{\color{gray}\Verb$n$}{\color{gray}\Verb$ $}{\color{gray}\Verb$o$}{\color{gray}\Verb$f$}{\color{gray}\Verb$ $}{\color{gray}\Verb$c$}{\color{gray}\Verb$o$}{\color{gray}\Verb$l$}{\color{gray}\Verb$o$}{\color{gray}\Verb$r$}{\color{gray}\Verb$ $}{\color{gray}\Verb$c$}{\color{gray}\Verb$a$}{\color{gray}\Verb$n$}{\color{gray}\Verb$ $}{\color{gray}\Verb$b$}{\color{gray}\Verb$e$}{\color{gray}\\}{\color{red}\Verb$c$}{\color{red}\Verb$o$}{\color{red}\Verb$l$}{\color{red}\Verb$o$}{\color{red}\Verb$r$}{\color{green}\Verb$($}{\color{green}\Verb$c$}{\color{green}\Verb$o$}{\color{green}\Verb$l$}{\color{green}\Verb$,$}{\color{green}\Verb$l$}{\color{green}\Verb$a$}{\color{green}\Verb$m$}{\color{green}\Verb$)$}\Verb$ $\Verb$=$\Verb$ ${\color{Tan}\Verb${$}{\color{Tan}\Verb$p$}{\color{Tan}\Verb$o$}{\color{Tan}\Verb$s$}{\color{Tan}\Verb$}$}{\color{Aquamarine}\Verb$:$}{\color{Aquamarine}\Verb$o$}{\color{Aquamarine}\Verb$l$}{\color{Aquamarine}\Verb$d$}{\color{Aquamarine}\Verb$p$}{\color{Aquamarine}\Verb$o$}{\color{Aquamarine}\Verb$s$}\Verb$ ${\color{red}\Verb$a$}{\color{red}\Verb$p$}{\color{red}\Verb$p$}{\color{red}\Verb$l$}{\color{red}\Verb$y$}{\color{green}\Verb$($}{\color{green}\Verb$l$}{\color{green}\Verb$a$}{\color{green}\Verb$m$}{\color{green}\Verb$)$}{\color{Aquamarine}\Verb$:$}{\color{Aquamarine}\Verb$r$}\Verb$ $\\\Verb$ $\Verb$ $\Verb$ $\Verb$ $\Verb$ $\Verb$ $\Verb$ $\Verb$ $\Verb$ $\Verb$ $\Verb$ $\Verb$ $\Verb$ $\Verb$ $\Verb$ $\Verb$ $\Verb$ ${\color{Tan}\Verb${$}{\color{Tan}\Verb$c$}{\color{Tan}\Verb$o$}{\color{Tan}\Verb$l$}{\color{Tan}\Verb$o$}{\color{Tan}\Verb$r$}{\color{Tan}\Verb$_$}{\color{Tan}\Verb$b$}{\color{Tan}\Verb$y$}{\color{Tan}\Verb$($}{\color{Tan}\Verb$c$}{\color{Tan}\Verb$o$}{\color{Tan}\Verb$l$}{\color{Tan}\Verb$,$}{\color{Aquamarine}\Verb$o$}{\color{Aquamarine}\Verb$l$}{\color{Aquamarine}\Verb$d$}{\color{Aquamarine}\Verb$p$}{\color{Aquamarine}\Verb$o$}{\color{Aquamarine}\Verb$s$}{\color{Tan}\Verb$,$}{\color{Tan}\Verb$p$}{\color{Tan}\Verb$o$}{\color{Tan}\Verb$s$}{\color{Tan}\Verb$)$}{\color{Tan}\Verb$}$}\Verb$ ${\color{Tan}\Verb$-$}{\color{Tan}\Verb$>$}{\color{Tan}\Verb$ $}{\color{Aquamarine}\Verb$r$}
}\end{spacing}

\end{exambox}

This approach prevents any changes to the actual text representation of the input (as opposed to translating abstract syntax trees back to text form). The annotation is straightforward. It is realistic to expect advanced users of IDE to write new grammars if amethyst was used as a syntax highlighting engine. This approach also benefits from a dynamic parsing (Chapter \ref{dynparsing}).

\newpage
\section{Taming state} \label{tamingstate}
Purity is important concept in programming languages. We say that function is pure when it can not produce any side effect. Advantage of pure functions is that they are easy to reason about. 
When function is not pure then its behaviour depends on operations made before that function, also known as {\it  state}. Often we have to add  state to function as a necessary evil. We will present several constructions that make state behavior more predictable. 

We take inspiration from several earlier attempts.  We could view Warth's worlds \cite{ometa}  as first attempt. However as worlds are applied only for position tracking so all work is left to programmer. The rats parser \cite{rats} recognizes problem and proposes transaction. Again bookkeeping is left to programmer.
In general setting Tanter's contextual values \cite{tanter} are more general than Warth's worlds \cite{ometa}. 

Amethyst uses similar idea. For modality reasons we must split contextual values to two cases: Contextual argument and return.

\subsection*{Local state}
{\it Local state} refers to how can values of local variables inside function change.
Functional languages use notion of referential transparency \cite{reftrans}.
We use weaker notion. When an alternative fails the we revert all local variables to a state just before alternative was tried.

Lookaheads are especially dangerous because they break assumptions programmer makes about state we always revert to state before lookahead.

Reverting of local state may come as a surprise in a context of initialing variables:

\begin{exambox}
\begin{spacing}{0.8}
{\small
{\color{red}\Verb$f$}{\color{red}\Verb$o$}{\color{red}\Verb$o$}\Verb$ $\Verb$=$\Verb$ ${\color{Tan}\Verb${$}{\color{Tan}\Verb$x$}{\color{Tan}\Verb$=$}{\color{Tan}\Verb$4$}{\color{Tan}\Verb$}$}\Verb$ $\Verb$ ${\color{red}\Verb$f$}{\color{red}\Verb$a$}{\color{red}\Verb$i$}{\color{red}\Verb$l$}\Verb$ $\Verb$|$\Verb$ ${\color{red}\Verb$s$}{\color{red}\Verb$u$}{\color{red}\Verb$c$}{\color{red}\Verb$c$}{\color{red}\Verb$e$}{\color{red}\Verb$s$}{\color{red}\Verb$s$}\Verb$ ${\color{Tan}\Verb${$}{\color{Tan}\Verb$p$}{\color{Tan}\Verb$u$}{\color{Tan}\Verb$t$}{\color{Tan}\Verb$s$}{\color{Tan}\Verb$ $}{\color{Tan}\Verb$x$}{\color{Tan}\Verb$}$}\Verb$ $\Verb$ ${\color{gray}\Verb$#$}{\color{gray}\Verb$-$}{\color{gray}\Verb$>$}{\color{gray}\Verb$ $}{\color{gray}\Verb$n$}{\color{gray}\Verb$i$}{\color{gray}\Verb$l$}{\color{gray}\Verb$ $}{\color{gray}\Verb$b$}{\color{gray}\Verb$e$}{\color{gray}\Verb$c$}{\color{gray}\Verb$a$}{\color{gray}\Verb$u$}{\color{gray}\Verb$s$}{\color{gray}\Verb$e$}{\color{gray}\Verb$ $}{\color{gray}\Verb$x$}{\color{gray}\Verb$=$}{\color{gray}\Verb$4$}{\color{gray}\Verb$ $}{\color{gray}\Verb$w$}{\color{gray}\Verb$a$}{\color{gray}\Verb$s$}{\color{gray}\Verb$ $}{\color{gray}\Verb$r$}{\color{gray}\Verb$e$}{\color{gray}\Verb$v$}{\color{gray}\Verb$e$}{\color{gray}\Verb$r$}{\color{gray}\Verb$t$}{\color{gray}\Verb$e$}{\color{gray}\Verb$d$}{\color{gray}\Verb$.$}{\color{gray}\\}{\color{red}\Verb$f$}{\color{red}\Verb$o$}{\color{red}\Verb$o$}\Verb$ $\Verb$=$\Verb$ ${\color{Tan}\Verb${$}{\color{Tan}\Verb$x$}{\color{Tan}\Verb$=$}{\color{Tan}\Verb$4$}{\color{Tan}\Verb$}$}\Verb$ $\Verb$(${\color{red}\Verb$f$}{\color{red}\Verb$a$}{\color{red}\Verb$i$}{\color{red}\Verb$l$}\Verb$ $\Verb$|$\Verb$ ${\color{red}\Verb$s$}{\color{red}\Verb$u$}{\color{red}\Verb$c$}{\color{red}\Verb$c$}{\color{red}\Verb$e$}{\color{red}\Verb$s$}{\color{red}\Verb$s$}\Verb$ ${\color{Tan}\Verb${$}{\color{Tan}\Verb$p$}{\color{Tan}\Verb$u$}{\color{Tan}\Verb$t$}{\color{Tan}\Verb$s$}{\color{Tan}\Verb$ $}{\color{Tan}\Verb$x$}{\color{Tan}\Verb$}$}\Verb$)$\Verb$ ${\color{gray}\Verb$#$}{\color{gray}\Verb$-$}{\color{gray}\Verb$>$}{\color{gray}\Verb$ $}{\color{gray}\Verb$4$}
}\end{spacing}

\end{exambox}

This can be done effectively by data structures that do a backtracking persistence.
\subsection*{Global state}
Handling global state is more tricky. Memoizing parsers cause objects to be shared unexpectedly. Following example returns a modified object instead of correct unmodified one.

\begin{exambox}
\begin{spacing}{0.8}
{\small
{\color{red}\Verb$f$}{\color{red}\Verb$o$}{\color{red}\Verb$o$}\Verb$:$\Verb$x$\Verb$ $\Verb${$\Verb$x$\Verb$.$\Verb$a$\Verb$=$\Verb$4$\Verb$}$\Verb$ $\Verb$b$\Verb$a$\Verb$r$\Verb$ $\Verb$|$\Verb$ $\Verb$f$\Verb$o$\Verb$o$
}\end{spacing}

\end{exambox}

Here backtracking persistence can not help as memoized value would be reverted back to nil. 

\newpage

There are ways how mitigate this problem.
\noindent\begin{itemize}
\item Blame the programmer.

\item Recursively clone everything. When naively done we are about as slow as if we would recalculate everything. Using full persistence can typically reduce overhead to constant factor \cite{persistent}. Disadvantage is that all user structures must support persistence.

\item Recursively make every result immutable. This preserves time complexity as we make every object immutable at most once. 

We chosen the last alternative as it is conceptually simplest alternative. Dynamic parsing benefits from immutability as we will see in Chapter \ref{dynparsing}.
\end{itemize}

\subsection{Contextual arguments and return}
Are a more transparent way to model a global state than by global variable. 

For supplying context we use contextual argument \verb$@>name$. A contextual argument is accessible to all rules that current rule calls. However a change of contextual argument in son does not change parent's contextual argument. We illustrate this on example:

\begin{exambox}
\begin{spacing}{0.8}
{\small
{\color{red}\Verb$f$}{\color{red}\Verb$o$}{\color{red}\Verb$o$}{\color{red}\Verb$1$}\Verb$ $\Verb$=$\Verb$ ${\color{Tan}\Verb${$}{\color{Aquamarine}\Verb$@$}{\color{Aquamarine}\Verb$>$}{\color{Aquamarine}\Verb$n$}{\color{Aquamarine}\Verb$a$}{\color{Aquamarine}\Verb$m$}{\color{Aquamarine}\Verb$e$}{\color{Tan}\Verb$=$}{\color{Tan}\Verb$"$}{\color{Tan}\Verb$f$}{\color{Tan}\Verb$o$}{\color{Tan}\Verb$o$}{\color{Tan}\Verb$"$}{\color{Tan}\Verb$}$}\Verb$ ${\color{red}\Verb$f$}{\color{red}\Verb$o$}{\color{red}\Verb$o$}{\color{red}\Verb$2$}\Verb$ $\Verb$ $\Verb$ $\Verb$ $\Verb$ $\Verb$ $\Verb$ $\Verb$ ${\color{Tan}\Verb${$}{\color{Tan}\Verb$p$}{\color{Tan}\Verb$u$}{\color{Tan}\Verb$t$}{\color{Tan}\Verb$s$}{\color{Tan}\Verb$ $}{\color{Aquamarine}\Verb$@$}{\color{Aquamarine}\Verb$>$}{\color{Aquamarine}\Verb$n$}{\color{Aquamarine}\Verb$a$}{\color{Aquamarine}\Verb$m$}{\color{Aquamarine}\Verb$e$}{\color{Tan}\Verb$}$}\Verb$ ${\color{gray}\Verb$#$}{\color{gray}\Verb$-$}{\color{gray}\Verb$>$}{\color{gray}\Verb$ $}{\color{gray}\Verb$f$}{\color{gray}\Verb$o$}{\color{gray}\Verb$o$}{\color{gray}\\}{\color{red}\Verb$f$}{\color{red}\Verb$o$}{\color{red}\Verb$o$}{\color{red}\Verb$2$}\Verb$ $\Verb$=$\Verb$ ${\color{red}\Verb$f$}{\color{red}\Verb$o$}{\color{red}\Verb$o$}{\color{red}\Verb$3$}\\{\color{red}\Verb$f$}{\color{red}\Verb$o$}{\color{red}\Verb$o$}{\color{red}\Verb$3$}\Verb$ $\Verb$=$\Verb$ ${\color{Tan}\Verb${$}{\color{Aquamarine}\Verb$@$}{\color{Aquamarine}\Verb$>$}{\color{Aquamarine}\Verb$n$}{\color{Aquamarine}\Verb$a$}{\color{Aquamarine}\Verb$m$}{\color{Aquamarine}\Verb$e$}{\color{Tan}\Verb$=$}{\color{Aquamarine}\Verb$@$}{\color{Aquamarine}\Verb$>$}{\color{Aquamarine}\Verb$n$}{\color{Aquamarine}\Verb$a$}{\color{Aquamarine}\Verb$m$}{\color{Aquamarine}\Verb$e$}{\color{Tan}\Verb$+$}{\color{Tan}\Verb$"$}{\color{Tan}\Verb$b$}{\color{Tan}\Verb$a$}{\color{Tan}\Verb$r$}{\color{Tan}\Verb$"$}{\color{Tan}\Verb$}$}\Verb$ ${\color{red}\Verb$f$}{\color{red}\Verb$o$}{\color{red}\Verb$o$}{\color{red}\Verb$4$}\Verb$ ${\color{Tan}\Verb${$}{\color{Tan}\Verb$p$}{\color{Tan}\Verb$u$}{\color{Tan}\Verb$t$}{\color{Tan}\Verb$s$}{\color{Tan}\Verb$ $}{\color{Aquamarine}\Verb$@$}{\color{Aquamarine}\Verb$>$}{\color{Aquamarine}\Verb$n$}{\color{Aquamarine}\Verb$a$}{\color{Aquamarine}\Verb$m$}{\color{Aquamarine}\Verb$e$}{\color{Tan}\Verb$}$}\Verb$ ${\color{gray}\Verb$#$}{\color{gray}\Verb$-$}{\color{gray}\Verb$>$}{\color{gray}\Verb$ $}{\color{gray}\Verb$f$}{\color{gray}\Verb$o$}{\color{gray}\Verb$o$}{\color{gray}\Verb$b$}{\color{gray}\Verb$a$}{\color{gray}\Verb$r$}{\color{gray}\\}{\color{red}\Verb$f$}{\color{red}\Verb$o$}{\color{red}\Verb$o$}{\color{red}\Verb$4$}\Verb$ $\Verb$=$\Verb$ $\Verb$ $\Verb$ $\Verb$ $\Verb$ $\Verb$ $\Verb$ $\Verb$ $\Verb$ $\Verb$ $\Verb$ $\Verb$ $\Verb$ $\Verb$ $\Verb$ $\Verb$ $\Verb$ $\Verb$ $\Verb$ $\Verb$ $\Verb$ $\Verb$ $\Verb$ $\Verb$ $\Verb$ $\Verb$ $\Verb$ $\Verb$ ${\color{Tan}\Verb${$}{\color{Tan}\Verb$p$}{\color{Tan}\Verb$u$}{\color{Tan}\Verb$t$}{\color{Tan}\Verb$s$}{\color{Tan}\Verb$ $}{\color{Aquamarine}\Verb$@$}{\color{Aquamarine}\Verb$>$}{\color{Aquamarine}\Verb$n$}{\color{Aquamarine}\Verb$a$}{\color{Aquamarine}\Verb$m$}{\color{Aquamarine}\Verb$e$}{\color{Tan}\Verb$}$}\Verb$ ${\color{gray}\Verb$#$}{\color{gray}\Verb$-$}{\color{gray}\Verb$>$}{\color{gray}\Verb$ $}{\color{gray}\Verb$f$}{\color{gray}\Verb$o$}{\color{gray}\Verb$o$}{\color{gray}\Verb$b$}{\color{gray}\Verb$a$}{\color{gray}\Verb$r$}
}\end{spacing}

\end{exambox}

Second most frequent use of global state is to collect some values that are inconvenient to collect directly.

A contextual return \verb$@<name$ is concept dual to contextual arguments and can be viewed as a set such that every parent gets union of contextual returns of his sons. This also elegantly handles case when contextual return does not return anything. Again we illustrate contextual return on example:

\begin{exambox}
\begin{spacing}{0.8}
{\small
{\color{red}\Verb$f$}{\color{red}\Verb$o$}{\color{red}\Verb$o$}{\color{red}\Verb$1$}\Verb$ $\Verb$=$\Verb$ ${\color{red}\Verb$f$}{\color{red}\Verb$o$}{\color{red}\Verb$o$}{\color{red}\Verb$2$}\Verb$ ${\color{Tan}\Verb${$}{\color{Tan}\Verb$p$}{\color{Tan}\Verb$u$}{\color{Tan}\Verb$t$}{\color{Tan}\Verb$s$}{\color{Tan}\Verb$ $}{\color{Aquamarine}\Verb$@$}{\color{Aquamarine}\Verb$<$}{\color{Aquamarine}\Verb$n$}{\color{Aquamarine}\Verb$a$}{\color{Aquamarine}\Verb$m$}{\color{Aquamarine}\Verb$e$}{\color{Aquamarine}\Verb$s$}{\color{Tan}\Verb$}$}\Verb$ ${\color{gray}\Verb$#$}{\color{gray}\Verb$-$}{\color{gray}\Verb$>$}{\color{gray}\Verb$ $}{\color{gray}\Verb$[$}{\color{gray}\Verb$"$}{\color{gray}\Verb$f$}{\color{gray}\Verb$o$}{\color{gray}\Verb$o$}{\color{gray}\Verb$"$}{\color{gray}\Verb$,$}{\color{gray}\Verb$"$}{\color{gray}\Verb$b$}{\color{gray}\Verb$a$}{\color{gray}\Verb$r$}{\color{gray}\Verb$"$}{\color{gray}\Verb$,$}{\color{gray}\Verb$"$}{\color{gray}\Verb$b$}{\color{gray}\Verb$a$}{\color{gray}\Verb$z$}{\color{gray}\Verb$"$}{\color{gray}\Verb$]$}{\color{gray}\\}{\color{red}\Verb$f$}{\color{red}\Verb$o$}{\color{red}\Verb$o$}{\color{red}\Verb$2$}\Verb$ $\Verb$=$\Verb$ ${\color{red}\Verb$f$}{\color{red}\Verb$o$}{\color{red}\Verb$o$}{\color{red}\Verb$3$}\\{\color{red}\Verb$f$}{\color{red}\Verb$o$}{\color{red}\Verb$o$}{\color{red}\Verb$3$}\Verb$ $\Verb$=$\Verb$ ${\color{red}\Verb$f$}{\color{red}\Verb$o$}{\color{red}\Verb$o$}{\color{red}\Verb$4$}\Verb$ ${\color{red}\Verb$s$}{\color{red}\Verb$u$}{\color{red}\Verb$p$}{\color{red}\Verb$p$}{\color{red}\Verb$r$}{\color{red}\Verb$e$}{\color{red}\Verb$s$}{\color{red}\Verb$s$}\Verb$ ${\color{red}\Verb$b$}{\color{red}\Verb$a$}{\color{red}\Verb$r$}\\{\color{red}\Verb$f$}{\color{red}\Verb$o$}{\color{red}\Verb$o$}{\color{red}\Verb$4$}\Verb$ $\Verb$=$\Verb$ ${\color{Tan}\Verb${$}{\color{Tan}\Verb$ $}{\color{Aquamarine}\Verb$@$}{\color{Aquamarine}\Verb$<$}{\color{Aquamarine}\Verb$n$}{\color{Aquamarine}\Verb$a$}{\color{Aquamarine}\Verb$m$}{\color{Aquamarine}\Verb$e$}{\color{Aquamarine}\Verb$s$}{\color{Tan}\Verb$ $}{\color{Tan}\Verb$<$}{\color{Tan}\Verb$<$}{\color{Tan}\Verb$ $}{\color{Tan}\Verb$"$}{\color{Tan}\Verb$f$}{\color{Tan}\Verb$o$}{\color{Tan}\Verb$o$}{\color{Tan}\Verb$"$}{\color{Tan}\Verb$ $}{\color{Tan}\Verb$}$}\\{\color{red}\Verb$s$}{\color{red}\Verb$u$}{\color{red}\Verb$p$}{\color{red}\Verb$p$}{\color{red}\Verb$r$}{\color{red}\Verb$e$}{\color{red}\Verb$s$}{\color{red}\Verb$s$}\Verb$ $\Verb$=$\Verb$ ${\color{red}\Verb$s$}{\color{red}\Verb$u$}{\color{red}\Verb$p$}\Verb$ ${\color{Tan}\Verb${$}{\color{Aquamarine}\Verb$@$}{\color{Aquamarine}\Verb$<$}{\color{Aquamarine}\Verb$n$}{\color{Aquamarine}\Verb$a$}{\color{Aquamarine}\Verb$m$}{\color{Aquamarine}\Verb$e$}{\color{Aquamarine}\Verb$s$}{\color{Tan}\Verb$=$}{\color{Tan}\Verb$[$}{\color{Tan}\Verb$]$}{\color{Tan}\Verb$}$}\\{\color{red}\Verb$s$}{\color{red}\Verb$u$}{\color{red}\Verb$p$}\Verb$ $\Verb$ $\Verb$=$\Verb$ ${\color{Tan}\Verb${$}{\color{Tan}\Verb$ $}{\color{Aquamarine}\Verb$@$}{\color{Aquamarine}\Verb$<$}{\color{Aquamarine}\Verb$n$}{\color{Aquamarine}\Verb$a$}{\color{Aquamarine}\Verb$m$}{\color{Aquamarine}\Verb$e$}{\color{Aquamarine}\Verb$s$}{\color{Tan}\Verb$ $}{\color{Tan}\Verb$<$}{\color{Tan}\Verb$<$}{\color{Tan}\Verb$ $}{\color{Tan}\Verb$"$}{\color{Tan}\Verb$s$}{\color{Tan}\Verb$u$}{\color{Tan}\Verb$p$}{\color{Tan}\Verb$p$}{\color{Tan}\Verb$r$}{\color{Tan}\Verb$e$}{\color{Tan}\Verb$s$}{\color{Tan}\Verb$s$}{\color{Tan}\Verb$e$}{\color{Tan}\Verb$d$}{\color{Tan}\Verb$"$}{\color{Tan}\Verb$ $}{\color{Tan}\Verb$}$}\\{\color{red}\Verb$b$}{\color{red}\Verb$a$}{\color{red}\Verb$r$}\Verb$ $\Verb$ $\Verb$=$\Verb$ ${\color{Tan}\Verb${$}{\color{Tan}\Verb$ $}{\color{Aquamarine}\Verb$@$}{\color{Aquamarine}\Verb$<$}{\color{Aquamarine}\Verb$n$}{\color{Aquamarine}\Verb$a$}{\color{Aquamarine}\Verb$m$}{\color{Aquamarine}\Verb$e$}{\color{Aquamarine}\Verb$s$}{\color{Tan}\Verb$ $}{\color{Tan}\Verb$<$}{\color{Tan}\Verb$<$}{\color{Tan}\Verb$ $}{\color{Tan}\Verb$"$}{\color{Tan}\Verb$b$}{\color{Tan}\Verb$a$}{\color{Tan}\Verb$r$}{\color{Tan}\Verb$"$}{\color{Tan}\Verb$ $}{\color{Tan}\Verb$}$}\Verb$ ${\color{red}\Verb$b$}{\color{red}\Verb$a$}{\color{red}\Verb$z$}\\{\color{red}\Verb$b$}{\color{red}\Verb$a$}{\color{red}\Verb$z$}\Verb$ $\Verb$ $\Verb$=$\Verb$ ${\color{Tan}\Verb${$}{\color{Tan}\Verb$ $}{\color{Aquamarine}\Verb$@$}{\color{Aquamarine}\Verb$<$}{\color{Aquamarine}\Verb$n$}{\color{Aquamarine}\Verb$a$}{\color{Aquamarine}\Verb$m$}{\color{Aquamarine}\Verb$e$}{\color{Aquamarine}\Verb$s$}{\color{Tan}\Verb$ $}{\color{Tan}\Verb$<$}{\color{Tan}\Verb$<$}{\color{Tan}\Verb$ $}{\color{Tan}\Verb$"$}{\color{Tan}\Verb$b$}{\color{Tan}\Verb$a$}{\color{Tan}\Verb$z$}{\color{Tan}\Verb$"$}{\color{Tan}\Verb$ $}{\color{Tan}\Verb$}$}
}\end{spacing}

\end{exambox}

By defining contextual arguments and returns in this way a memoization respecting global state becomes trackable. 

We defer describing how to implement these concepts into  Section \ref{tamingstate}

\newpage

\section{Maintainance}
One of the design goals of amethyst is to allow users write general purpose grammars that can be extended as the described language or protocol evolves.

To get a specification of a language or protocol write:

\vskip -1.8em\begin{spacing}{0.8}
{\small
\begin{exambox}
\Verb$A$\Verb$m$\Verb$e$\Verb$t$\Verb$h$\Verb$y$\Verb$s$\Verb$t$\Verb$:$\Verb$:$\Verb$p$\Verb$u$\Verb$l$\Verb$l$\Verb$ $\Verb$'$\Verb$g$\Verb$r$\Verb$a$\Verb$m$\Verb$m$\Verb$a$\Verb$r$\Verb$:$\Verb$v$\Verb$e$\Verb$r$\Verb$s$\Verb$i$\Verb$o$\Verb$n$\Verb$'$
\end{exambox}
}
\end{spacing}\vskip -0.4em

Which loads given version of grammar, downloading it from central repository if necessary. Grammar obtained in this way is immutable and will be always same on all machines. We expect from grammars in repository to be stable and do not change often.

However we expect that protocols will evolve. We want to make updating easier by migrations. A proposed command is:

\vskip -1.8em\begin{spacing}{0.8}
{\small
\begin{exambox}
\Verb$a$\Verb$m$\Verb$e$\Verb$t$\Verb$h$\Verb$y$\Verb$s$\Verb$t$\Verb$_$\Verb$m$\Verb$i$\Verb$g$\Verb$r$\Verb$a$\Verb$t$\Verb$e$\Verb$ $\Verb$f$\Verb$i$\Verb$l$\Verb$e$\Verb$ $\Verb$g$\Verb$r$\Verb$a$\Verb$m$\Verb$m$\Verb$a$\Verb$r$\Verb$:$\Verb$n$\Verb$e$\Verb$w$\Verb$v$\Verb$e$\Verb$r$\Verb$s$\Verb$i$\Verb$o$\Verb$n$
\end{exambox}
}
\end{spacing}\vskip -0.4em

Which will replay refactorings (for example renaming a rule) described in migration files to new version. This could not be always possible\footnote{Fox example code that evaluates strings from standard input.} in this case we ask programmer to do migration manually.

\subsection*{Migration from regular expressions}
We also want to make transition from other framework easier. As a simple example we implemented an functor that convert subset of regular expressions into amethyst expressions.
Usage is the following:

\vskip -1.8em\begin{spacing}{0.8}
{\small
\begin{exambox}
\Verb$r$\Verb$e$\Verb$g$\Verb$e$\Verb$x$\Verb$p$\Verb$=$\Verb$ $\Verb$/$\Verb$[$\Verb$H$\Verb$h$\Verb$]$\Verb$e$\Verb$l$\Verb$l$\Verb$o$\Verb$ $\Verb$($\Verb$w$\Verb$o$\Verb$r$\Verb$l$\Verb$d$\Verb$|$\Verb$w$\Verb$o$\Verb$r$\Verb$l$\Verb$d$\Verb$s$\Verb$)$\Verb$/$\\\Verb$r$\Verb$e$\Verb$g$\Verb$2$\Verb$a$\Verb$m$\Verb$e$\Verb$($\Verb$r$\Verb$e$\Verb$g$\Verb$e$\Verb$x$\Verb$p$\Verb$)$\Verb$.$\Verb$i$\Verb$n$\Verb$s$\Verb$p$\Verb$e$\Verb$c$\Verb$t$\Verb$ $\Verb$#$\Verb$-$\Verb$>$\Verb$ ${\color{Violet}\Verb$($}{\color{Violet}\Verb$|$}\Verb$ $\Verb$<$\Verb$H$\Verb$h$\Verb$>$\Verb$ ${\color{black}\Verb$'$}{\color{black}\Verb$e$}{\color{black}\Verb$l$}{\color{black}\Verb$l$}{\color{black}\Verb$o$}{\color{black}\Verb$ $}{\color{black}\Verb$'$}\Verb$ $\Verb$(${\color{black}\Verb$'$}{\color{black}\Verb$w$}{\color{black}\Verb$o$}{\color{black}\Verb$r$}{\color{black}\Verb$l$}{\color{black}\Verb$d$}{\color{black}\Verb$'$}\Verb$|${\color{black}\Verb$'$}{\color{black}\Verb$w$}{\color{black}\Verb$o$}{\color{black}\Verb$r$}{\color{black}\Verb$l$}{\color{black}\Verb$d$}{\color{black}\Verb$s$}{\color{black}\Verb$'$}\Verb$)$\Verb$ ${\color{Violet}\Verb$|$}{\color{Violet}\Verb$)$}\Verb$ $\\\Verb$r$\Verb$e$\Verb$g$\Verb$2$\Verb$a$\Verb$m$\Verb$e$\Verb$($\Verb$r$\Verb$e$\Verb$g$\Verb$e$\Verb$x$\Verb$p$\Verb$)$\Verb$ $\Verb$=$\Verb$=$\Verb$=$\Verb$ $\Verb$"$\Verb$h$\Verb$e$\Verb$l$\Verb$l$\Verb$o$\Verb$ $\Verb$w$\Verb$o$\Verb$r$\Verb$l$\Verb$d$\Verb$"$\Verb$ $\Verb$#$\Verb$-$\Verb$>$\Verb$ $\Verb$t$\Verb$r$\Verb$u$\Verb$e$
\end{exambox}
}
\end{spacing}\vskip -0.4em

\section{Error handling}
An error detection is important topic on its own. We implemented only a simple strategy that detects misplaced parethness and suggest probable causes. This is a type of problem that that needs global error recovery. It can be formulated as problem that a given sequence of parentheses what is minimal number of parentheses we have to change to get properly parenthised expression. A simple strategy to guess most probable places can be found in files \verb$amethyst/error_recovery.ame$ and \verb$lib/repair_errors.rb$.

\subsection*{Position tracking}
For position tracking our approach is simple. We subclass string to the class \verb$Origin_Tracking_String$. Information about position automatically propagates through parser and subsequent pipeline. This also allows to wrap and recursively parse substrings while preserving position information.

\newpage
\section{Example: Parser of amethyst}
We conclude this chapter by explaining amethyst in terms of itself by providing amethyst parser in amethyst. Summary of constructions used is in Appendix \ref{syntsummary}. We omit several parts that are too technical. 

Our first task is parse rule and variable names. 

\begin{grambox}
\noindent \begin{spacing}{0.8}
{\small
{\color{red}\Verb$n$}{\color{red}\Verb$a$}{\color{red}\Verb$m$}{\color{red}\Verb$e$}\Verb$ $\Verb$ $\Verb$ $\Verb$ $\Verb$ $\Verb$ $\Verb$ $\Verb$ $\Verb$ $\Verb$ $\Verb$ $\Verb$=$\Verb$ $\Verb$($\Verb$<$\Verb$_$\Verb$a$\Verb$-$\Verb$z$\Verb$A$\Verb$-$\Verb$Z$\Verb$>$\Verb$ $\Verb$<$\Verb$_$\Verb$a$\Verb$-$\Verb$z$\Verb$A$\Verb$-$\Verb$Z$\Verb$0$\Verb$-$\Verb$9$\Verb$>${\color{black}\Verb$*$}\Verb$)$\Verb$[$\Verb$]${\color{Aquamarine}\Verb$:$}{\color{Tan}\Verb${$}{\color{Aquamarine}\Verb$i$}{\color{Aquamarine}\Verb$t$}{\color{Tan}\Verb$.$}{\color{Tan}\Verb$j$}{\color{Tan}\Verb$o$}{\color{Tan}\Verb$i$}{\color{Tan}\Verb$n$}{\color{Tan}\Verb$}$}\Verb$ $\\{\color{red}\Verb$c$}{\color{red}\Verb$l$}{\color{red}\Verb$a$}{\color{red}\Verb$s$}{\color{red}\Verb$s$}{\color{red}\Verb$N$}{\color{red}\Verb$a$}{\color{red}\Verb$m$}{\color{red}\Verb$e$}\Verb$ $\Verb$ $\Verb$ $\Verb$ $\Verb$ $\Verb$ $\Verb$=$\Verb$ $\Verb$($\Verb$ $\Verb$ $\Verb$ $\Verb$ $\Verb$<$\Verb$A$\Verb$-$\Verb$Z$\Verb$>$\Verb$ $\Verb$<$\Verb$_$\Verb$a$\Verb$-$\Verb$z$\Verb$A$\Verb$-$\Verb$Z$\Verb$0$\Verb$-$\Verb$9$\Verb$>${\color{black}\Verb$*$}\Verb$)$\Verb$[$\Verb$]${\color{Aquamarine}\Verb$:$}{\color{Tan}\Verb${$}{\color{Aquamarine}\Verb$i$}{\color{Aquamarine}\Verb$t$}{\color{Tan}\Verb$.$}{\color{Tan}\Verb$j$}{\color{Tan}\Verb$o$}{\color{Tan}\Verb$i$}{\color{Tan}\Verb$n$}{\color{Tan}\Verb$}$}
}\end{spacing}

\end{grambox}

\subsection*{File structure}
Amethyst file consist of grammars and host language code. We make ``\verb$amethyst$'' a keyword otherwise grammar with error would be interpreted verbatim.

\begin{grambox}
\noindent \begin{spacing}{0.8}
{\small
{\color{red}\Verb$f$}{\color{red}\Verb$i$}{\color{red}\Verb$l$}{\color{red}\Verb$e$}\Verb$ $\Verb$ $\Verb$ $\Verb$ $\Verb$ $\Verb$=$\Verb$($\Verb$ ${\color{red}\Verb$g$}{\color{red}\Verb$r$}{\color{red}\Verb$a$}{\color{red}\Verb$m$}{\color{red}\Verb$m$}{\color{red}\Verb$a$}{\color{red}\Verb$r$}\\\Verb$ $\Verb$ $\Verb$ $\Verb$ $\Verb$ $\Verb$ $\Verb$ $\Verb$ $\Verb$ $\Verb$ $\Verb$|$\Verb$ ${\color{red}\Verb$l$}{\color{red}\Verb$a$}{\color{red}\Verb$m$}{\color{red}\Verb$b$}{\color{red}\Verb$d$}{\color{red}\Verb$a$}\\\Verb$ $\Verb$ $\Verb$ $\Verb$ $\Verb$ $\Verb$ $\Verb$ $\Verb$ $\Verb$ $\Verb$ $\Verb$|$\Verb$ ${\color{Violet}\Verb$~$}{\color{Violet}\Verb$($}{\color{black}\Verb$'$}{\color{black}\Verb$a$}{\color{black}\Verb$m$}{\color{black}\Verb$e$}{\color{black}\Verb$t$}{\color{black}\Verb$h$}{\color{black}\Verb$y$}{\color{black}\Verb$s$}{\color{black}\Verb$t$}{\color{black}\Verb$'$}\Verb$ ${\color{red}\Verb$_$}{\color{Violet}\Verb$)$}\Verb$ $\Verb$.$\Verb$ $\Verb$)${\color{black}\Verb$*$}
}\end{spacing}

\end{grambox}

\subsection*{Grammars}

Amethyst grammar consist from rules. 

\begin{grambox}
\noindent \begin{spacing}{0.8}
{\small
{\color{red}\Verb$g$}{\color{red}\Verb$r$}{\color{red}\Verb$a$}{\color{red}\Verb$m$}{\color{red}\Verb$m$}{\color{red}\Verb$a$}{\color{red}\Verb$r$}\Verb$ $\Verb$=$\Verb$ ${\color{black}\Verb$'$}{\color{black}\Verb$a$}{\color{black}\Verb$m$}{\color{black}\Verb$e$}{\color{black}\Verb$t$}{\color{black}\Verb$h$}{\color{black}\Verb$y$}{\color{black}\Verb$s$}{\color{black}\Verb$t$}{\color{black}\Verb$'$}\Verb$ ${\color{black}\Verb$"$}{\color{black}\Verb$"$}\Verb$ ${\color{red}\Verb$n$}{\color{red}\Verb$a$}{\color{red}\Verb$m$}{\color{red}\Verb$e$}\Verb$ $\Verb$(${\color{black}\Verb$"$}{\color{black}\Verb$<$}{\color{black}\Verb$"$}\Verb$ ${\color{black}\Verb$"$}{\color{black}\Verb$"$}\Verb$ ${\color{red}\Verb$n$}{\color{red}\Verb$a$}{\color{red}\Verb$m$}{\color{red}\Verb$e$}\Verb$ $\Verb$|$\Verb$ ${\color{Tan}\Verb${$}{\color{Tan}\Verb$"$}{\color{Tan}\Verb$A$}{\color{Tan}\Verb$m$}{\color{Tan}\Verb$e$}{\color{Tan}\Verb$t$}{\color{Tan}\Verb$h$}{\color{Tan}\Verb$y$}{\color{Tan}\Verb$s$}{\color{Tan}\Verb$t$}{\color{Tan}\Verb$"$}{\color{Tan}\Verb$}$}\Verb$ $\Verb$)${\color{Aquamarine}\Verb$:$}{\color{Aquamarine}\Verb$p$}{\color{Aquamarine}\Verb$a$}{\color{Aquamarine}\Verb$r$}{\color{Aquamarine}\Verb$e$}{\color{Aquamarine}\Verb$n$}{\color{Aquamarine}\Verb$t$}\\\Verb$ $\Verb$ $\Verb$ $\Verb$ $\Verb$ $\Verb$ $\Verb$ $\Verb$ $\Verb$ $\Verb$ ${\color{black}\Verb$"$}{\color{black}\Verb${$}{\color{black}\Verb$"$}\Verb$ ${\color{red}\Verb$r$}{\color{red}\Verb$u$}{\color{red}\Verb$l$}{\color{red}\Verb$e$}{\color{black}\Verb$*$}{\color{Aquamarine}\Verb$:$}{\color{Aquamarine}\Verb$r$}{\color{Aquamarine}\Verb$u$}{\color{Aquamarine}\Verb$l$}{\color{Aquamarine}\Verb$e$}{\color{Aquamarine}\Verb$s$}\Verb$ ${\color{black}\Verb$"$}{\color{black}\Verb$}$}{\color{black}\Verb$"$}\Verb$ $\Verb$ ${\color{Aquamarine}\Verb$@$}{\color{Aquamarine}\Verb$G$}{\color{Aquamarine}\Verb$r$}{\color{Aquamarine}\Verb$a$}{\color{Aquamarine}\Verb$m$}{\color{Aquamarine}\Verb$m$}{\color{Aquamarine}\Verb$a$}{\color{Aquamarine}\Verb$r$}
}\end{spacing}

\end{grambox}

We specify optional parts of grammar by a ``\verb$?$'' operator. When we also want to supply default value we use an idiom \Verb$(${\color{black}\Verb$"$}{\color{black}\Verb$<$}{\color{black}\Verb$"$}\Verb$ ${\color{black}\Verb$"$}{\color{black}\Verb$"$}\Verb$ ${\color{red}\Verb$n$}{\color{red}\Verb$a$}{\color{red}\Verb$m$}{\color{red}\Verb$e$}\Verb$ $\Verb$|$\Verb$ ${\color{Tan}\Verb${$}{\color{Tan}\Verb$"$}{\color{Tan}\Verb$A$}{\color{Tan}\Verb$m$}{\color{Tan}\Verb$e$}{\color{Tan}\Verb$t$}{\color{Tan}\Verb$h$}{\color{Tan}\Verb$y$}{\color{Tan}\Verb$s$}{\color{Tan}\Verb$t$}{\color{Tan}\Verb$"$}{\color{Tan}\Verb$}$}\Verb$ $\Verb$)$.

\subsection*{Rules}

\begin{grambox}
\noindent \begin{spacing}{0.8}
{\small
{\color{red}\Verb$a$}{\color{red}\Verb$r$}{\color{red}\Verb$g$}{\color{red}\Verb$s$}{\color{red}\Verb$O$}{\color{red}\Verb$p$}{\color{red}\Verb$t$}\Verb$ $\Verb$=$\Verb$ ${\color{red}\Verb$a$}{\color{red}\Verb$r$}{\color{red}\Verb$g$}{\color{red}\Verb$s$}{\color{green}\Verb$($}{\color{green}\Verb$'$}{\color{green}\Verb$($}{\color{green}\Verb$'$}{\color{green}\Verb$,$}{\color{green}\Verb$'$}{\color{green}\Verb$)$}{\color{green}\Verb$'$}{\color{green}\Verb$)$}\Verb$ $\Verb$|$\Verb$ ${\color{Tan}\Verb${$}{\color{Tan}\Verb$[$}{\color{Tan}\Verb$]$}{\color{Tan}\Verb$}$}\\{\color{gray}\Verb$#$}{\color{gray}\Verb$F$}{\color{gray}\Verb$o$}{\color{gray}\Verb$r$}{\color{gray}\Verb$ $}{\color{gray}\Verb$n$}{\color{gray}\Verb$o$}{\color{gray}\Verb$w$}{\color{gray}\Verb$ $}{\color{gray}\Verb$y$}{\color{gray}\Verb$o$}{\color{gray}\Verb$u$}{\color{gray}\Verb$ $}{\color{gray}\Verb$c$}{\color{gray}\Verb$a$}{\color{gray}\Verb$n$}{\color{gray}\Verb$ $}{\color{gray}\Verb$i$}{\color{gray}\Verb$m$}{\color{gray}\Verb$a$}{\color{gray}\Verb$g$}{\color{gray}\Verb$i$}{\color{gray}\Verb$n$}{\color{gray}\Verb$e$}{\color{gray}\Verb$ $}{\color{gray}\Verb$t$}{\color{gray}\Verb$h$}{\color{gray}\Verb$a$}{\color{gray}\Verb$t$}{\color{gray}\Verb$ $}{\color{gray}\Verb$a$}{\color{gray}\Verb$r$}{\color{gray}\Verb$g$}{\color{gray}\Verb$s$}{\color{gray}\Verb$($}{\color{gray}\Verb$'$}{\color{gray}\Verb$($}{\color{gray}\Verb$'$}{\color{gray}\Verb$,$}{\color{gray}\Verb$'$}{\color{gray}\Verb$)$}{\color{gray}\Verb$'$}{\color{gray}\Verb$)$}{\color{gray}\Verb$ $}{\color{gray}\Verb$m$}{\color{gray}\Verb$a$}{\color{gray}\Verb$t$}{\color{gray}\Verb$c$}{\color{gray}\Verb$h$}{\color{gray}\Verb$e$}{\color{gray}\Verb$s$}{\color{gray}\Verb$ $}{\color{gray}\Verb$p$}{\color{gray}\Verb$r$}{\color{gray}\Verb$o$}{\color{gray}\Verb$p$}{\color{gray}\Verb$e$}{\color{gray}\Verb$r$}{\color{gray}\Verb$l$}{\color{gray}\Verb$y$}{\color{gray}\Verb$ $}{\color{gray}\\}{\color{gray}\Verb$#$}{\color{gray}\Verb$n$}{\color{gray}\Verb$e$}{\color{gray}\Verb$s$}{\color{gray}\Verb$t$}{\color{gray}\Verb$e$}{\color{gray}\Verb$d$}{\color{gray}\Verb$ $}{\color{gray}\Verb$p$}{\color{gray}\Verb$a$}{\color{gray}\Verb$r$}{\color{gray}\Verb$e$}{\color{gray}\Verb$n$}{\color{gray}\Verb$t$}{\color{gray}\Verb$h$}{\color{gray}\Verb$e$}{\color{gray}\Verb$s$}{\color{gray}\Verb$e$}{\color{gray}\Verb$s$}{\color{gray}\Verb$.$}{\color{gray}\\}\\{\color{red}\Verb$r$}{\color{red}\Verb$u$}{\color{red}\Verb$l$}{\color{red}\Verb$e$}\Verb$ $\Verb$=$\Verb$ ${\color{black}\Verb$"$}{\color{black}\Verb$"$}\Verb$ ${\color{red}\Verb$n$}{\color{red}\Verb$a$}{\color{red}\Verb$m$}{\color{red}\Verb$e$}{\color{Aquamarine}\Verb$:$}{\color{Aquamarine}\Verb$n$}{\color{Aquamarine}\Verb$a$}{\color{Aquamarine}\Verb$m$}{\color{Aquamarine}\Verb$e$}\Verb$ ${\color{Violet}\Verb$~$}{\color{Violet}\Verb$_$}\Verb$ ${\color{red}\Verb$a$}{\color{red}\Verb$r$}{\color{red}\Verb$g$}{\color{red}\Verb$s$}{\color{red}\Verb$O$}{\color{red}\Verb$p$}{\color{red}\Verb$t$}{\color{Aquamarine}\Verb$:$}{\color{Aquamarine}\Verb$a$}{\color{Aquamarine}\Verb$r$}{\color{Aquamarine}\Verb$g$}{\color{Aquamarine}\Verb$s$}\Verb$ ${\color{black}\Verb$"$}{\color{black}\Verb$=$}{\color{black}\Verb$"$}\Verb$ ${\color{red}\Verb$e$}{\color{red}\Verb$x$}{\color{red}\Verb$p$}{\color{red}\Verb$r$}{\color{red}\Verb$e$}{\color{red}\Verb$s$}{\color{red}\Verb$s$}{\color{red}\Verb$i$}{\color{red}\Verb$o$}{\color{red}\Verb$n$}{\color{Aquamarine}\Verb$:$}{\color{Aquamarine}\Verb$b$}{\color{Aquamarine}\Verb$o$}{\color{Aquamarine}\Verb$d$}{\color{Aquamarine}\Verb$y$}\Verb$ ${\color{Aquamarine}\Verb$@$}{\color{Aquamarine}\Verb$R$}{\color{Aquamarine}\Verb$u$}{\color{Aquamarine}\Verb$l$}{\color{Aquamarine}\Verb$e$}\\\\{\color{red}\Verb$c$}{\color{red}\Verb$a$}{\color{red}\Verb$l$}{\color{red}\Verb$l$}\Verb$ $\Verb$=$\Verb$ ${\color{red}\Verb$c$}{\color{red}\Verb$l$}{\color{red}\Verb$a$}{\color{red}\Verb$s$}{\color{red}\Verb$s$}{\color{red}\Verb$N$}{\color{red}\Verb$a$}{\color{red}\Verb$m$}{\color{red}\Verb$e$}{\color{Aquamarine}\Verb$:$}{\color{Aquamarine}\Verb$k$}{\color{Aquamarine}\Verb$l$}{\color{Aquamarine}\Verb$a$}{\color{Aquamarine}\Verb$s$}\Verb$ ${\color{black}\Verb$'$}{\color{black}\Verb$:$}{\color{black}\Verb$:$}{\color{black}\Verb$'$}\Verb$ ${\color{red}\Verb$n$}{\color{red}\Verb$a$}{\color{red}\Verb$m$}{\color{red}\Verb$e$}{\color{Aquamarine}\Verb$:$}{\color{Aquamarine}\Verb$n$}{\color{Aquamarine}\Verb$a$}{\color{Aquamarine}\Verb$m$}{\color{Aquamarine}\Verb$e$}\Verb$ ${\color{Violet}\Verb$~$}{\color{Violet}\Verb$_$}\Verb$ ${\color{red}\Verb$a$}{\color{red}\Verb$r$}{\color{red}\Verb$g$}{\color{red}\Verb$s$}{\color{red}\Verb$O$}{\color{red}\Verb$p$}{\color{red}\Verb$t$}{\color{Aquamarine}\Verb$:$}{\color{Aquamarine}\Verb$a$}{\color{Aquamarine}\Verb$r$}{\color{Aquamarine}\Verb$g$}\Verb$ ${\color{gray}\Verb$#$}{\color{gray}\Verb$F$}{\color{gray}\Verb$o$}{\color{gray}\Verb$r$}{\color{gray}\Verb$e$}{\color{gray}\Verb$i$}{\color{gray}\Verb$g$}{\color{gray}\Verb$n$}{\color{gray}\Verb$ $}{\color{gray}\Verb$r$}{\color{gray}\Verb$u$}{\color{gray}\Verb$l$}{\color{gray}\Verb$e$}{\color{gray}\\}\Verb$ $\Verb$ $\Verb$ $\Verb$ $\Verb$ $\Verb$|$\Verb$ $\Verb$ $\Verb$ $\Verb$ $\Verb$ $\Verb$ $\Verb$ $\Verb$ $\Verb$ $\Verb$ $\Verb$ $\Verb$ $\Verb$ $\Verb$ $\Verb$ $\Verb$ $\Verb$ $\Verb$ $\Verb$ $\Verb$ $\Verb$ ${\color{red}\Verb$n$}{\color{red}\Verb$a$}{\color{red}\Verb$m$}{\color{red}\Verb$e$}{\color{Aquamarine}\Verb$:$}{\color{Aquamarine}\Verb$n$}{\color{Aquamarine}\Verb$a$}{\color{Aquamarine}\Verb$m$}{\color{Aquamarine}\Verb$e$}\Verb$ ${\color{Violet}\Verb$~$}{\color{Violet}\Verb$_$}\Verb$ ${\color{red}\Verb$a$}{\color{red}\Verb$r$}{\color{red}\Verb$g$}{\color{red}\Verb$s$}{\color{red}\Verb$O$}{\color{red}\Verb$p$}{\color{red}\Verb$t$}{\color{Aquamarine}\Verb$:$}{\color{Aquamarine}\Verb$a$}{\color{Aquamarine}\Verb$r$}{\color{Aquamarine}\Verb$g$}\Verb$ ${\color{Tan}\Verb$-$}{\color{Tan}\Verb$>$}{\color{Tan}\Verb$ $}{\color{Tan}\Verb$A$}{\color{Tan}\Verb$p$}{\color{Tan}\Verb$p$}{\color{Tan}\Verb$l$}{\color{Tan}\Verb$y$}{\color{Tan}\Verb$[$}{\color{Aquamarine}\Verb$n$}{\color{Aquamarine}\Verb$a$}{\color{Aquamarine}\Verb$m$}{\color{Aquamarine}\Verb$e$}{\color{Tan}\Verb$,$}{\color{Aquamarine}\Verb$a$}{\color{Aquamarine}\Verb$r$}{\color{Aquamarine}\Verb$g$}{\color{Tan}\Verb$]$}
}\end{spacing}

\end{grambox}

When we want to tell an user reading grammars that whitespaces are forbidden at certain place we use an {\color{Violet}\Verb$~$}{\color{Violet}\Verb$_$} idiom even if it is not neccessary for parser.
\newline
\section*{Sequencing and choice}

Sequencing and choice have the usual precedence. At choice we need to forbid interpreting end of lambda as choice.    Whitespaces separate sequence elements. We use negative lookahead  {\color{Violet}\Verb$~$}{\color{Violet}\Verb$r$}{\color{red}\Verb$u$}{\color{red}\Verb$l$}{\color{red}\Verb$e$}{\color{red}\Verb$_$}{\color{red}\Verb$h$}{\color{red}\Verb$e$}{\color{red}\Verb$a$}{\color{Violet}\Verb$d$} to separate rules.

\begin{grambox}
\noindent \begin{spacing}{0.8}
{\small
{\color{red}\Verb$e$}{\color{red}\Verb$x$}{\color{red}\Verb$p$}{\color{red}\Verb$r$}{\color{red}\Verb$e$}{\color{red}\Verb$s$}{\color{red}\Verb$s$}{\color{red}\Verb$i$}{\color{red}\Verb$o$}{\color{red}\Verb$n$}\Verb$ $\Verb$=$\Verb$ ${\color{red}\Verb$l$}{\color{red}\Verb$i$}{\color{red}\Verb$s$}{\color{red}\Verb$t$}{\color{red}\Verb$O$}{\color{red}\Verb$f$}{\color{green}\Verb$($}{\color{Violet}\Verb$($}{\color{Violet}\Verb$|$}{\color{red}\Verb$s$}{\color{red}\Verb$e$}{\color{red}\Verb$q$}{\color{red}\Verb$u$}{\color{red}\Verb$e$}{\color{red}\Verb$n$}{\color{red}\Verb$c$}{\color{red}\Verb$e$}{\color{Violet}\Verb$|$}{\color{Violet}\Verb$)$}{\color{green}\Verb$,$}{\color{Violet}\Verb$($}{\color{Violet}\Verb$|$}{\color{black}\Verb$"$}{\color{black}\Verb$|$}{\color{black}\Verb$"$}{\color{green}\Verb$ $}{\color{Violet}\Verb$~$}{\color{Violet}\Verb$'$}{\color{black}\Verb$)$}{\color{Violet}\Verb$'$}{\color{Violet}\Verb$|$}{\color{Violet}\Verb$)$}{\color{green}\Verb$)$}{\color{Aquamarine}\Verb$:$}{\color{Aquamarine}\Verb$a$}{\color{Aquamarine}\Verb$r$}{\color{Aquamarine}\Verb$y$}\Verb$ ${\color{Aquamarine}\Verb$@$}{\color{Aquamarine}\Verb$O$}{\color{Aquamarine}\Verb$r$}{\color{Aquamarine}\Verb$_$}{\color{Aquamarine}\Verb$A$}{\color{Aquamarine}\Verb$S$}{\color{Aquamarine}\Verb$T$}\\\\{\color{red}\Verb$s$}{\color{red}\Verb$e$}{\color{red}\Verb$q$}{\color{red}\Verb$u$}{\color{red}\Verb$e$}{\color{red}\Verb$n$}{\color{red}\Verb$c$}{\color{red}\Verb$e$}\Verb$ $\Verb$ $\Verb$ $\Verb$=$\Verb$ $\Verb$ $\Verb$(${\color{Violet}\Verb$~$}{\color{Violet}\Verb$r$}{\color{red}\Verb$u$}{\color{red}\Verb$l$}{\color{red}\Verb$e$}{\color{red}\Verb$_$}{\color{red}\Verb$h$}{\color{red}\Verb$e$}{\color{red}\Verb$a$}{\color{Violet}\Verb$d$}\Verb$ ${\color{red}\Verb$l$}{\color{red}\Verb$o$}{\color{red}\Verb$o$}{\color{red}\Verb$k$}{\color{red}\Verb$a$}{\color{red}\Verb$h$}{\color{red}\Verb$e$}{\color{red}\Verb$a$}{\color{red}\Verb$d$}{\color{red}\Verb$s$}\Verb$)${\color{black}\Verb$*$}{\color{Aquamarine}\Verb$:$}{\color{Aquamarine}\Verb$a$}{\color{Aquamarine}\Verb$r$}{\color{Aquamarine}\Verb$y$}\Verb$ $\Verb$ ${\color{Aquamarine}\Verb$@$}{\color{Aquamarine}\Verb$S$}{\color{Aquamarine}\Verb$e$}{\color{Aquamarine}\Verb$q$}{\color{Aquamarine}\Verb$_$}{\color{Aquamarine}\Verb$A$}{\color{Aquamarine}\Verb$S$}{\color{Aquamarine}\Verb$T$}\\\\{\color{red}\Verb$r$}{\color{red}\Verb$u$}{\color{red}\Verb$l$}{\color{red}\Verb$e$}{\color{red}\Verb$_$}{\color{red}\Verb$h$}{\color{red}\Verb$e$}{\color{red}\Verb$a$}{\color{red}\Verb$d$}\Verb$ $\Verb$ $\Verb$=$\Verb$ ${\color{black}\Verb$"$}{\color{black}\Verb$"$}\Verb$ ${\color{red}\Verb$n$}{\color{red}\Verb$a$}{\color{red}\Verb$m$}{\color{red}\Verb$e$}{\color{Aquamarine}\Verb$:$}{\color{Aquamarine}\Verb$n$}{\color{Aquamarine}\Verb$a$}{\color{Aquamarine}\Verb$m$}{\color{Aquamarine}\Verb$e$}\Verb$ ${\color{Violet}\Verb$~$}{\color{Violet}\Verb$_$}\Verb$ ${\color{red}\Verb$a$}{\color{red}\Verb$r$}{\color{red}\Verb$g$}{\color{red}\Verb$s$}{\color{red}\Verb$O$}{\color{red}\Verb$p$}{\color{red}\Verb$t$}{\color{Aquamarine}\Verb$:$}{\color{Aquamarine}\Verb$a$}{\color{Aquamarine}\Verb$r$}{\color{Aquamarine}\Verb$g$}{\color{Aquamarine}\Verb$s$}\Verb$ ${\color{black}\Verb$"$}{\color{black}\Verb$=$}{\color{black}\Verb$"$}
}\end{spacing}

\end{grambox}

\subsection*{Lookaheads}
Negative and positive lookaheads are recognized by the following expressions:

\begin{grambox}
\noindent \begin{spacing}{0.8}
{\small
{\color{red}\Verb$l$}{\color{red}\Verb$o$}{\color{red}\Verb$o$}{\color{red}\Verb$k$}{\color{red}\Verb$a$}{\color{red}\Verb$h$}{\color{red}\Verb$e$}{\color{red}\Verb$a$}{\color{red}\Verb$d$}{\color{red}\Verb$s$}\Verb$ $\Verb$=$\Verb$ ${\color{black}\Verb$"$}{\color{black}\Verb$"$}\Verb$ ${\color{red}\Verb$n$}{\color{red}\Verb$e$}{\color{red}\Verb$g$}{\color{red}\Verb$_$}{\color{red}\Verb$l$}{\color{red}\Verb$a$}{\color{red}\Verb$h$}{\color{red}\Verb$e$}{\color{red}\Verb$a$}{\color{red}\Verb$d$}{\color{Aquamarine}\Verb$:$}{\color{Aquamarine}\Verb$[$}{\color{Aquamarine}\Verb$s$}{\color{Aquamarine}\Verb$]$}\Verb$ $\Verb$(${\color{black}\Verb$"$}{\color{black}\Verb$&$}{\color{black}\Verb$"$}\Verb$ ${\color{Violet}\Verb$~$}{\color{Violet}\Verb$"$}{\color{black}\Verb${$}{\color{Violet}\Verb$"$}\Verb$ ${\color{black}\Verb$"$}{\color{black}\Verb$"$}\Verb$ ${\color{red}\Verb$n$}{\color{red}\Verb$e$}{\color{red}\Verb$g$}{\color{red}\Verb$_$}{\color{red}\Verb$l$}{\color{red}\Verb$a$}{\color{red}\Verb$h$}{\color{red}\Verb$e$}{\color{red}\Verb$a$}{\color{red}\Verb$d$}{\color{Aquamarine}\Verb$:$}{\color{Aquamarine}\Verb$[$}{\color{Aquamarine}\Verb$s$}{\color{Aquamarine}\Verb$]$}\Verb$)${\color{black}\Verb$*$}\Verb$ $\\\\{\color{red}\Verb$n$}{\color{red}\Verb$e$}{\color{red}\Verb$g$}{\color{red}\Verb$_$}{\color{red}\Verb$l$}{\color{red}\Verb$a$}{\color{red}\Verb$h$}{\color{red}\Verb$e$}{\color{red}\Verb$a$}{\color{red}\Verb$d$}\Verb$ $\Verb$=$\Verb$ ${\color{black}\Verb$'$}{\color{black}\Verb$~$}{\color{black}\Verb$'$}\Verb$ ${\color{Violet}\Verb$~$}{\color{Violet}\Verb$"$}{\color{black}\Verb${$}{\color{Violet}\Verb$"$}\Verb$ ${\color{red}\Verb$n$}{\color{red}\Verb$e$}{\color{red}\Verb$g$}{\color{red}\Verb$_$}{\color{red}\Verb$l$}{\color{red}\Verb$a$}{\color{red}\Verb$h$}{\color{red}\Verb$e$}{\color{red}\Verb$a$}{\color{red}\Verb$d$}{\color{Aquamarine}\Verb$:$}{\color{Aquamarine}\Verb$m$}\Verb$ $\Verb$ $\Verb$ $\Verb$ ${\color{Tan}\Verb$-$}{\color{Tan}\Verb$>$}{\color{Tan}\Verb$ $}{\color{Tan}\Verb$L$}{\color{Tan}\Verb$o$}{\color{Tan}\Verb$o$}{\color{Tan}\Verb$k$}{\color{Tan}\Verb$a$}{\color{Tan}\Verb$h$}{\color{Tan}\Verb$e$}{\color{Tan}\Verb$a$}{\color{Tan}\Verb$d$}{\color{Tan}\Verb$[$}{\color{Aquamarine}\Verb$m$}{\color{Tan}\Verb$,$}{\color{Tan}\Verb$t$}{\color{Tan}\Verb$r$}{\color{Tan}\Verb$u$}{\color{Tan}\Verb$e$}{\color{Tan}\Verb$]$}{\color{Tan}\\}{\color{Tan}\Verb$ $}\Verb$ $\Verb$ $\Verb$ $\Verb$ $\Verb$ $\Verb$ $\Verb$ $\Verb$ $\Verb$ $\Verb$ $\Verb$|$\Verb$ $\Verb$ $\Verb$<$\Verb$&$\Verb$~$\Verb$>${\color{Aquamarine}\Verb$:$}{\color{Aquamarine}\Verb$n$}\Verb$ ${\color{Violet}\Verb$~$}{\color{Violet}\Verb$_$}\Verb$ ${\color{red}\Verb$i$}{\color{red}\Verb$n$}{\color{red}\Verb$l$}{\color{red}\Verb$i$}{\color{red}\Verb$n$}{\color{red}\Verb$e$}{\color{red}\Verb$_$}{\color{red}\Verb$h$}{\color{red}\Verb$o$}{\color{red}\Verb$s$}{\color{red}\Verb$t$}{\color{red}\Verb$_$}{\color{red}\Verb$e$}{\color{red}\Verb$x$}{\color{red}\Verb$p$}{\color{red}\Verb$r$}{\color{Aquamarine}\Verb$:$}{\color{blue}\Verb$e$}\Verb$ ${\color{Tan}\Verb$-$}{\color{Tan}\Verb$>$}{\color{Tan}\Verb$ $}{\color{Tan}\Verb$P$}{\color{Tan}\Verb$r$}{\color{Tan}\Verb$e$}{\color{Tan}\Verb$d$}{\color{Tan}\Verb$[$}{\color{blue}\Verb$e$}{\color{Tan}\Verb$,$}{\color{Aquamarine}\Verb$n$}{\color{Tan}\Verb$=$}{\color{Tan}\Verb$=$}{\color{Tan}\Verb$'$}{\color{Tan}\Verb$&$}{\color{Tan}\Verb$'$}{\color{Tan}\Verb$]$}{\color{Tan}\\}{\color{Tan}\Verb$ $}\Verb$ $\Verb$ $\Verb$ $\Verb$ $\Verb$ $\Verb$ $\Verb$ $\Verb$ $\Verb$ $\Verb$ $\Verb$|$\Verb$ ${\color{red}\Verb$p$}{\color{red}\Verb$o$}{\color{red}\Verb$s$}{\color{red}\Verb$t$}{\color{red}\Verb$f$}{\color{red}\Verb$i$}{\color{red}\Verb$x$}{\color{red}\Verb$e$}{\color{red}\Verb$d$}
}\end{spacing}

\end{grambox}

\subsection*{Postfixes}

Note that postfixes are left-associative. In particular {\color{red}\Verb$a$}\Verb$=$\Verb$>$\Verb$b$\Verb$?$  is equivalent to \Verb$($\Verb$a$\Verb$=$\Verb$>$\Verb$b$\Verb$)$\Verb$?$.

\begin{grambox}
\noindent \begin{spacing}{0.8}
{\small
{\color{red}\Verb$p$}{\color{red}\Verb$o$}{\color{red}\Verb$s$}{\color{red}\Verb$t$}{\color{red}\Verb$f$}{\color{red}\Verb$i$}{\color{red}\Verb$x$}{\color{red}\Verb$e$}{\color{red}\Verb$d$}\Verb$ $\Verb$ $\Verb$ $\Verb$ $\Verb$ $\Verb$ $\Verb$=$\Verb$ ${\color{red}\Verb$t$}{\color{red}\Verb$e$}{\color{red}\Verb$r$}{\color{red}\Verb$m$}{\color{Aquamarine}\Verb$:$}{\color{Aquamarine}\Verb$f$}{\color{Aquamarine}\Verb$r$}{\color{Aquamarine}\Verb$o$}{\color{Aquamarine}\Verb$m$}\\\Verb$ $\Verb$ $\Verb$ $\Verb$ $\Verb$ $\Verb$ $\Verb$ $\Verb$ $\Verb$ $\Verb$ $\Verb$ $\Verb$ $\Verb$ $\Verb$ $\Verb$ $\Verb$($\Verb$ $\Verb$ $\Verb$<$\Verb$*$\Verb$+$\Verb$?$\Verb$>$\\\Verb$ $\Verb$ $\Verb$ $\Verb$ $\Verb$ $\Verb$ $\Verb$ $\Verb$ $\Verb$ $\Verb$ $\Verb$ $\Verb$ $\Verb$ $\Verb$ $\Verb$ $\Verb$|$\Verb$ ${\color{black}\Verb$'$}{\color{black}\Verb$[$}{\color{black}\Verb$'$}\Verb$ $\Verb$ ${\color{red}\Verb$e$}{\color{red}\Verb$x$}{\color{red}\Verb$p$}{\color{red}\Verb$r$}{\color{red}\Verb$e$}{\color{red}\Verb$s$}{\color{red}\Verb$s$}{\color{red}\Verb$i$}{\color{red}\Verb$o$}{\color{red}\Verb$n$}{\color{Aquamarine}\Verb$:$}{\color{blue}\Verb$e$}\Verb$ ${\color{black}\Verb$"$}{\color{black}\Verb$]$}{\color{black}\Verb$"$}\Verb$ ${\color{Tan}\Verb$-$}{\color{Tan}\Verb$>$}{\color{Tan}\Verb$ $}{\color{Tan}\Verb$E$}{\color{Tan}\Verb$n$}{\color{Tan}\Verb$t$}{\color{Tan}\Verb$e$}{\color{Tan}\Verb$r$}{\color{Tan}\Verb$[$}{\color{Aquamarine}\Verb$f$}{\color{Aquamarine}\Verb$r$}{\color{Aquamarine}\Verb$o$}{\color{Aquamarine}\Verb$m$}{\color{Tan}\Verb$,$}{\color{blue}\Verb$e$}{\color{Tan}\Verb$]$}{\color{Tan}\\}{\color{Tan}\Verb$ $}\Verb$ $\Verb$ $\Verb$ $\Verb$ $\Verb$ $\Verb$ $\Verb$ $\Verb$ $\Verb$ $\Verb$ $\Verb$ $\Verb$ $\Verb$ $\Verb$ $\Verb$|$\Verb$ ${\color{black}\Verb$'$}{\color{black}\Verb$=$}{\color{black}\Verb$>$}{\color{black}\Verb$'$}\Verb$ ${\color{red}\Verb$e$}{\color{red}\Verb$x$}{\color{red}\Verb$p$}{\color{red}\Verb$r$}{\color{red}\Verb$e$}{\color{red}\Verb$s$}{\color{red}\Verb$s$}{\color{red}\Verb$i$}{\color{red}\Verb$o$}{\color{red}\Verb$n$}{\color{Aquamarine}\Verb$:$}{\color{blue}\Verb$e$}\Verb$ $\Verb$ $\Verb$ $\Verb$ $\Verb$ ${\color{Tan}\Verb$-$}{\color{Tan}\Verb$>$}{\color{Tan}\Verb$ $}{\color{Tan}\Verb$ $}{\color{Tan}\Verb$P$}{\color{Tan}\Verb$a$}{\color{Tan}\Verb$s$}{\color{Tan}\Verb$s$}{\color{Tan}\Verb$[$}{\color{Aquamarine}\Verb$f$}{\color{Aquamarine}\Verb$r$}{\color{Aquamarine}\Verb$o$}{\color{Aquamarine}\Verb$m$}{\color{Tan}\Verb$,$}{\color{blue}\Verb$e$}{\color{Tan}\Verb$]$}{\color{Tan}\\}{\color{Tan}\Verb$ $}\Verb$ $\Verb$ $\Verb$ $\Verb$ $\Verb$ $\Verb$ $\Verb$ $\Verb$ $\Verb$ $\Verb$ $\Verb$ $\Verb$ $\Verb$ $\Verb$ $\Verb$|$\Verb$ ${\color{black}\Verb$'$}{\color{black}\Verb$:$}{\color{black}\Verb$'$}\Verb$ ${\color{black}\Verb$'$}{\color{black}\Verb$[$}{\color{black}\Verb$'$}\Verb$ ${\color{red}\Verb$n$}{\color{red}\Verb$a$}{\color{red}\Verb$m$}{\color{red}\Verb$e$}\Verb$ ${\color{black}\Verb$'$}{\color{black}\Verb$]$}{\color{black}\Verb$'$}\Verb$ $\Verb$|$\Verb$ ${\color{black}\Verb$'$}{\color{black}\Verb$:$}{\color{black}\Verb$'$}\Verb$ ${\color{black}\Verb$'$}{\color{black}\Verb$"$}{\color{black}\Verb$'$}\Verb$ ${\color{red}\Verb$n$}{\color{red}\Verb$a$}{\color{red}\Verb$m$}{\color{red}\Verb$e$}\Verb$ ${\color{black}\Verb$'$}{\color{black}\Verb$"$}{\color{black}\Verb$'$}\Verb$ $\Verb$|$\Verb$ ${\color{black}\Verb$'$}{\color{black}\Verb$:$}{\color{black}\Verb$'$}\Verb$ ${\color{red}\Verb$n$}{\color{red}\Verb$a$}{\color{red}\Verb$m$}{\color{red}\Verb$e$}\\\Verb$ $\Verb$ $\Verb$ $\Verb$ $\Verb$ $\Verb$ $\Verb$ $\Verb$ $\Verb$ $\Verb$ $\Verb$ $\Verb$ $\Verb$ $\Verb$ $\Verb$ $\Verb$|$\Verb$ ${\color{black}\Verb$'$}{\color{black}\Verb$:$}{\color{black}\Verb$'$}\Verb$ ${\color{red}\Verb$i$}{\color{red}\Verb$n$}{\color{red}\Verb$l$}{\color{red}\Verb$i$}{\color{red}\Verb$n$}{\color{red}\Verb$e$}{\color{red}\Verb$_$}{\color{red}\Verb$h$}{\color{red}\Verb$o$}{\color{red}\Verb$s$}{\color{red}\Verb$t$}{\color{red}\Verb$_$}{\color{red}\Verb$e$}{\color{red}\Verb$x$}{\color{red}\Verb$p$}{\color{red}\Verb$r$}{\color{Aquamarine}\Verb$:$}{\color{Tan}\Verb${$}{\color{Tan}\Verb$S$}{\color{Tan}\Verb$e$}{\color{Tan}\Verb$q$}{\color{Tan}\Verb$[$}{\color{Tan}\Verb$B$}{\color{Tan}\Verb$i$}{\color{Tan}\Verb$n$}{\color{Tan}\Verb$d$}{\color{Tan}\Verb$[$}{\color{Tan}\Verb$"$}{\color{Tan}\Verb$i$}{\color{Tan}\Verb$t$}{\color{Tan}\Verb$"$}{\color{Tan}\Verb$,$}{\color{Aquamarine}\Verb$f$}{\color{Aquamarine}\Verb$r$}{\color{Aquamarine}\Verb$o$}{\color{Aquamarine}\Verb$m$}{\color{Tan}\Verb$]$}{\color{Tan}\Verb$ $}{\color{Tan}\Verb$,$}{\color{Tan}\Verb$ $}{\color{Tan}\Verb$A$}{\color{Tan}\Verb$c$}{\color{Tan}\Verb$t$}{\color{Tan}\Verb$[$}{\color{Aquamarine}\Verb$i$}{\color{Aquamarine}\Verb$t$}{\color{Tan}\Verb$]$}{\color{Tan}\Verb$]$}{\color{Tan}\Verb$}$}\\\Verb$ $\Verb$ $\Verb$ $\Verb$ $\Verb$ $\Verb$ $\Verb$ $\Verb$ $\Verb$ $\Verb$ $\Verb$ $\Verb$ $\Verb$ $\Verb$ $\Verb$ $\Verb$)${\color{Aquamarine}\Verb$:$}{\color{Aquamarine}\Verb$f$}{\color{Aquamarine}\Verb$r$}{\color{Aquamarine}\Verb$o$}{\color{Aquamarine}\Verb$m$}\Verb$ $\Verb$ $\Verb$)$\Verb$*$\Verb$ $\Verb$-$\Verb$>$\Verb$ $\Verb$f$\Verb$r$\Verb$o$\Verb$m$
}\end{spacing}

\end{grambox}

\subsection*{Atomic expressions}

Various atomic expressions are handled by the following rules:

\begin{grambox}
\noindent \begin{spacing}{0.8}
{\small
{\color{red}\Verb$c$}{\color{red}\Verb$a$}{\color{red}\Verb$s$}{\color{red}\Verb$e$}{\color{red}\Verb$s$}\Verb$ $\Verb$=$\Verb$ $\Verb$ ${\color{red}\Verb$c$}{\color{red}\Verb$l$}{\color{red}\Verb$a$}{\color{red}\Verb$s$}{\color{red}\Verb$s$}{\color{red}\Verb$N$}{\color{red}\Verb$a$}{\color{red}\Verb$m$}{\color{red}\Verb$e$}{\color{Aquamarine}\Verb$:$}{\color{Aquamarine}\Verb$k$}{\color{Aquamarine}\Verb$l$}{\color{Aquamarine}\Verb$a$}{\color{Aquamarine}\Verb$s$}\Verb$ $\Verb$ ${\color{Violet}\Verb$~$}{\color{Violet}\Verb$'$}{\color{black}\Verb$:$}{\color{black}\Verb$:$}{\color{Violet}\Verb$'$}\Verb$ $\Verb$ $\Verb$ $\Verb$ $\Verb$ $\Verb$ ${\color{Tan}\Verb$-$}{\color{Tan}\Verb$>$}{\color{Tan}\Verb$ $}{\color{Tan}\Verb$A$}{\color{Tan}\Verb$p$}{\color{Tan}\Verb$p$}{\color{Tan}\Verb$l$}{\color{Tan}\Verb$y$}{\color{Tan}\Verb$[$}{\color{Tan}\Verb$"$}{\color{Tan}\Verb$c$}{\color{Tan}\Verb$l$}{\color{Tan}\Verb$a$}{\color{Tan}\Verb$s$}{\color{Tan}\Verb$"$}{\color{Tan}\Verb$,$}{\color{Aquamarine}\Verb$k$}{\color{Aquamarine}\Verb$l$}{\color{Aquamarine}\Verb$a$}{\color{Aquamarine}\Verb$s$}{\color{Tan}\Verb$]$}{\color{Tan}\\}{\color{Tan}\Verb$ $}\Verb$ $\Verb$ $\Verb$ $\Verb$ $\Verb$ $\Verb$|$\Verb$ $\Verb$(${\color{red}\Verb$n$}{\color{red}\Verb$u$}{\color{red}\Verb$m$}{\color{red}\Verb$b$}{\color{red}\Verb$e$}{\color{red}\Verb$r$}\Verb$ $\Verb$(${\color{black}\Verb$'$}{\color{black}\Verb$.$}{\color{black}\Verb$.$}{\color{black}\Verb$.$}{\color{black}\Verb$'$}\Verb$|${\color{black}\Verb$'$}{\color{black}\Verb$.$}{\color{black}\Verb$.$}{\color{black}\Verb$'$}\Verb$)$\Verb$ ${\color{red}\Verb$n$}{\color{red}\Verb$u$}{\color{red}\Verb$m$}{\color{red}\Verb$b$}{\color{red}\Verb$e$}{\color{red}\Verb$r$}\Verb$ $\\\Verb$ $\Verb$ $\Verb$ $\Verb$ $\Verb$ $\Verb$ $\Verb$ $\Verb$ $\Verb$|$\Verb$ ${\color{red}\Verb$n$}{\color{red}\Verb$u$}{\color{red}\Verb$m$}{\color{red}\Verb$b$}{\color{red}\Verb$e$}{\color{red}\Verb$r$}\Verb$)$\Verb$[$\Verb$]${\color{Aquamarine}\Verb$:$}{\color{Aquamarine}\Verb$n$}{\color{Aquamarine}\Verb$u$}{\color{Aquamarine}\Verb$m$}\Verb$ $\Verb$ $\Verb$ $\Verb$ $\Verb$ $\Verb$ $\Verb$ $\Verb$ $\Verb$ $\Verb$ $\Verb$ $\Verb$ $\Verb$ ${\color{Tan}\Verb$-$}{\color{Tan}\Verb$>$}{\color{Tan}\Verb$ $}{\color{Tan}\Verb$A$}{\color{Tan}\Verb$p$}{\color{Tan}\Verb$p$}{\color{Tan}\Verb$l$}{\color{Tan}\Verb$y$}{\color{Tan}\Verb$[$}{\color{Tan}\Verb$"$}{\color{Tan}\Verb$m$}{\color{Tan}\Verb$e$}{\color{Tan}\Verb$m$}{\color{Tan}\Verb$b$}{\color{Tan}\Verb$e$}{\color{Tan}\Verb$r$}{\color{Tan}\Verb$"$}{\color{Tan}\Verb$,$}{\color{Aquamarine}\Verb$n$}{\color{Aquamarine}\Verb$u$}{\color{Aquamarine}\Verb$m$}{\color{Tan}\Verb$.$}{\color{Tan}\Verb$j$}{\color{Tan}\Verb$o$}{\color{Tan}\Verb$i$}{\color{Tan}\Verb$n$}{\color{Tan}\Verb$]$}{\color{Tan}\\}{\color{Tan}\Verb$ $}\Verb$ $\Verb$ $\Verb$ $\Verb$ $\Verb$ $\Verb$|$\Verb$ ${\color{black}\Verb$'$}{\color{black}\Verb$<$}{\color{black}\Verb$'$}\Verb$ $\Verb$ ${\color{red}\Verb$u$}{\color{red}\Verb$n$}{\color{red}\Verb$t$}{\color{red}\Verb$i$}{\color{red}\Verb$l$}{\color{green}\Verb$($}{\color{green}\Verb$'$}{\color{green}\Verb$>$}{\color{green}\Verb$'$}{\color{green}\Verb$ $}{\color{green}\Verb$)$}{\color{Aquamarine}\Verb$:$}{\color{Aquamarine}\Verb$s$}\Verb$ $\Verb$ $\Verb$ $\Verb$ $\Verb$ $\Verb$ $\Verb$ $\Verb$ $\Verb$ $\Verb$ ${\color{Tan}\Verb$-$}{\color{Tan}\Verb$>$}{\color{Tan}\Verb$ $}{\color{Tan}\Verb$A$}{\color{Tan}\Verb$p$}{\color{Tan}\Verb$p$}{\color{Tan}\Verb$l$}{\color{Tan}\Verb$y$}{\color{Tan}\Verb$[$}{\color{Tan}\Verb$"$}{\color{Tan}\Verb$r$}{\color{Tan}\Verb$e$}{\color{Tan}\Verb$g$}{\color{Tan}\Verb$c$}{\color{Tan}\Verb$h$}{\color{Tan}\Verb$"$}{\color{Tan}\Verb$,$}{\color{Tan}\Verb$"$}{\color{Tan}\Verb$/$}{\color{Tan}\Verb$[$}{\color{Tan}\Verb$"$}{\color{Tan}\Verb$+$}{\color{Aquamarine}\Verb$s$}{\color{Tan}\Verb$+$}{\color{Tan}\Verb$"$}{\color{Tan}\Verb$]$}{\color{Tan}\Verb$/$}{\color{Tan}\Verb$"$}{\color{Tan}\Verb$]$}{\color{Tan}\\}{\color{Tan}\\}{\color{red}\Verb$k$}{\color{red}\Verb$e$}{\color{red}\Verb$y$}\Verb$ $\Verb$ $\Verb$ $\Verb$=$\Verb$ ${\color{black}\Verb$'$}{\color{black}\Verb$($}{\color{black}\Verb$|$}{\color{black}\Verb$'$}\Verb$ ${\color{red}\Verb$e$}{\color{red}\Verb$x$}{\color{red}\Verb$p$}{\color{red}\Verb$r$}{\color{red}\Verb$e$}{\color{red}\Verb$s$}{\color{red}\Verb$s$}{\color{red}\Verb$i$}{\color{red}\Verb$o$}{\color{red}\Verb$n$}{\color{Aquamarine}\Verb$:$}{\color{Aquamarine}\Verb$e$}{\color{Aquamarine}\Verb$x$}{\color{Aquamarine}\Verb$p$}\Verb$ ${\color{black}\Verb$'$}{\color{black}\Verb$|$}{\color{black}\Verb$)$}{\color{black}\Verb$'$}\Verb$ $\Verb$ $\Verb$ $\Verb$ ${\color{Tan}\Verb$-$}{\color{Tan}\Verb$>$}{\color{Tan}\Verb$ $}{\color{Tan}\Verb$L$}{\color{Tan}\Verb$a$}{\color{Tan}\Verb$m$}{\color{Tan}\Verb$b$}{\color{Tan}\Verb$d$}{\color{Tan}\Verb$a$}{\color{Tan}\Verb$[$}{\color{Aquamarine}\Verb$e$}{\color{Aquamarine}\Verb$x$}{\color{Aquamarine}\Verb$p$}{\color{Tan}\Verb$]$}{\color{Tan}\\}{\color{Tan}\Verb$ $}\Verb$ $\Verb$ $\Verb$ $\Verb$ $\Verb$ $\Verb$|$\Verb$ ${\color{black}\Verb$'$}{\color{black}\Verb$@$}{\color{black}\Verb$'$}\Verb$ $\Verb$ ${\color{red}\Verb$c$}{\color{red}\Verb$l$}{\color{red}\Verb$a$}{\color{red}\Verb$s$}{\color{red}\Verb$s$}{\color{red}\Verb$N$}{\color{red}\Verb$a$}{\color{red}\Verb$m$}{\color{red}\Verb$e$}\Verb$ $\Verb$ $\Verb$ $\Verb$ $\Verb$ $\Verb$ $\Verb$ $\Verb$ $\Verb$ $\Verb$ $\Verb$ $\Verb$ $\Verb$ $\Verb$ $\Verb$ ${\color{gray}\Verb$#$}{\color{gray}\Verb$ $}{\color{gray}\Verb$T$}{\color{gray}\Verb$e$}{\color{gray}\Verb$c$}{\color{gray}\Verb$h$}{\color{gray}\Verb$n$}{\color{gray}\Verb$i$}{\color{gray}\Verb$c$}{\color{gray}\Verb$a$}{\color{gray}\Verb$l$}{\color{gray}\\}\Verb$ $\Verb$ $\Verb$ $\Verb$ $\Verb$ $\Verb$ $\Verb$|$\Verb$ ${\color{black}\Verb$'$}{\color{black}\Verb$@$}{\color{black}\Verb$'$}\Verb$ $\Verb$ ${\color{red}\Verb$n$}{\color{red}\Verb$a$}{\color{red}\Verb$m$}{\color{red}\Verb$e$}{\color{Aquamarine}\Verb$:$}{\color{Aquamarine}\Verb$n$}{\color{Aquamarine}\Verb$a$}{\color{Aquamarine}\Verb$m$}{\color{Aquamarine}\Verb$e$}\Verb$ ${\color{red}\Verb$a$}{\color{red}\Verb$r$}{\color{red}\Verb$g$}{\color{red}\Verb$s$}{\color{red}\Verb$O$}{\color{red}\Verb$p$}{\color{red}\Verb$t$}{\color{Aquamarine}\Verb$:$}{\color{Aquamarine}\Verb$a$}{\color{Aquamarine}\Verb$r$}{\color{Aquamarine}\Verb$g$}\Verb$ $\Verb$ ${\color{Tan}\Verb$-$}{\color{Tan}\Verb$>$}{\color{Tan}\Verb$ $}{\color{Tan}\Verb$K$}{\color{Tan}\Verb$e$}{\color{Tan}\Verb$y$}{\color{Tan}\Verb$[$}{\color{Aquamarine}\Verb$n$}{\color{Aquamarine}\Verb$a$}{\color{Aquamarine}\Verb$m$}{\color{Aquamarine}\Verb$e$}{\color{Tan}\Verb$,$}{\color{Aquamarine}\Verb$a$}{\color{Aquamarine}\Verb$r$}{\color{Aquamarine}\Verb$g$}{\color{Tan}\Verb$]$}{\color{Tan}\\}{\color{Tan}\Verb$ $}\Verb$ $\Verb$ $\Verb$ $\Verb$ $\Verb$ $\Verb$|$\Verb$ ${\color{black}\Verb$'$}{\color{black}\Verb$@$}{\color{black}\Verb$@$}{\color{black}\Verb$'$}\Verb$ ${\color{red}\Verb$n$}{\color{red}\Verb$a$}{\color{red}\Verb$m$}{\color{red}\Verb$e$}{\color{Aquamarine}\Verb$:$}{\color{Aquamarine}\Verb$n$}{\color{Aquamarine}\Verb$a$}{\color{Aquamarine}\Verb$m$}{\color{Aquamarine}\Verb$e$}\Verb$ $\Verb$ $\Verb$ $\Verb$ $\Verb$ $\Verb$ $\Verb$ $\Verb$ $\Verb$ $\Verb$ $\Verb$ $\Verb$ $\Verb$ $\Verb$ ${\color{Tan}\Verb$-$}{\color{Tan}\Verb$>$}{\color{Tan}\Verb$ $}{\color{Tan}\Verb$G$}{\color{Tan}\Verb$l$}{\color{Tan}\Verb$o$}{\color{Tan}\Verb$b$}{\color{Tan}\Verb$a$}{\color{Tan}\Verb$l$}{\color{Tan}\Verb$[$}{\color{Aquamarine}\Verb$n$}{\color{Aquamarine}\Verb$a$}{\color{Aquamarine}\Verb$m$}{\color{Aquamarine}\Verb$e$}{\color{Tan}\Verb$]$}
}\end{spacing}

\end{grambox}

\newpage
\subsection*{Semantic actions}
Recognizing semantic actions is dependent on the host language. We do not have to understand whole grammar, recognizing pairing tags suffices\footnote{We use more complicated rules to recognize local variables}. Implementation specific to Ruby follows:

\begin{grambox}
\noindent \begin{spacing}{0.8}
{\small
{\color{red}\Verb$a$}{\color{red}\Verb$r$}{\color{red}\Verb$g$}{\color{red}\Verb$s$}{\color{green}\Verb$($}{\color{green}\Verb$o$}{\color{green}\Verb$,$}{\color{green}\Verb$c$}{\color{green}\Verb$)$}\Verb$ $\Verb$=$\Verb$ ${\color{red}\Verb$s$}{\color{red}\Verb$e$}{\color{red}\Verb$q$}{\color{green}\Verb$($}{\color{green}\Verb$o$}{\color{green}\Verb$)$}\Verb$ ${\color{red}\Verb$h$}{\color{red}\Verb$o$}{\color{red}\Verb$s$}{\color{red}\Verb$t$}{\color{red}\Verb$a$}{\color{red}\Verb$r$}{\color{red}\Verb$g$}{\color{black}\Verb$*$}{\color{Aquamarine}\Verb$:$}{\color{Aquamarine}\Verb$r$}\Verb$ ${\color{red}\Verb$s$}{\color{red}\Verb$e$}{\color{red}\Verb$q$}{\color{green}\Verb$($}{\color{green}\Verb$c$}{\color{green}\Verb$)$}\Verb$ ${\color{Tan}\Verb$-$}{\color{Tan}\Verb$>$}{\color{Tan}\Verb$ $}{\color{Aquamarine}\Verb$r$}{\color{Tan}\\}{\color{Tan}\\}{\color{red}\Verb$h$}{\color{red}\Verb$o$}{\color{red}\Verb$s$}{\color{red}\Verb$t$}{\color{red}\Verb$a$}{\color{red}\Verb$r$}{\color{red}\Verb$g$}\Verb$ $\Verb$ $\Verb$=$\Verb$ ${\color{red}\Verb$k$}{\color{red}\Verb$e$}{\color{red}\Verb$y$}\Verb$ $\\\Verb$ $\Verb$ $\Verb$ $\Verb$ $\Verb$ $\Verb$ $\Verb$ $\Verb$ $\Verb$ $\Verb$|$\Verb$ ${\color{red}\Verb$a$}{\color{red}\Verb$r$}{\color{red}\Verb$g$}{\color{red}\Verb$s$}{\color{green}\Verb$($}{\color{green}\Verb$'$}{\color{green}\Verb$($}{\color{green}\Verb$'$}{\color{green}\Verb$,$}{\color{green}\Verb$'$}{\color{green}\Verb$)$}{\color{green}\Verb$'$}{\color{green}\Verb$)$}\Verb$ $\Verb$ $\Verb$ $\Verb$|$\Verb$ ${\color{red}\Verb$a$}{\color{red}\Verb$r$}{\color{red}\Verb$g$}{\color{red}\Verb$s$}{\color{green}\Verb$($}{\color{green}\Verb$'$}{\color{green}\Verb$[$}{\color{green}\Verb$'$}{\color{green}\Verb$,$}{\color{green}\Verb$'$}{\color{green}\Verb$]$}{\color{green}\Verb$'$}{\color{green}\Verb$)$}\Verb$ $\Verb$ $\Verb$ $\Verb$ $\Verb$|$\Verb$ ${\color{red}\Verb$a$}{\color{red}\Verb$r$}{\color{red}\Verb$g$}{\color{red}\Verb$s$}{\color{green}\Verb$($}{\color{green}\Verb$'$}{\color{green}\Verb${$}{\color{green}\Verb$'$}{\color{green}\Verb$,$}{\color{green}\Verb$'$}{\color{green}\Verb$}$}{\color{green}\Verb$'$}{\color{green}\Verb$)$}\\\Verb$ $\Verb$ $\Verb$ $\Verb$ $\Verb$ $\Verb$ $\Verb$ $\Verb$ $\Verb$ $\Verb$|$\Verb$ $\Verb$ ${\color{black}\Verb$'$}{\color{black}\Verb$\$}{\color{black}\Verb$'$}{\color{black}\Verb$'$}\Verb$ ${\color{red}\Verb$u$}{\color{red}\Verb$n$}{\color{red}\Verb$t$}{\color{red}\Verb$i$}{\color{red}\Verb$l$}{\color{green}\Verb$($}{\color{green}\Verb$'$}{\color{green}\Verb$\$}{\color{green}\Verb$'$}{\color{green}\Verb$'$}{\color{green}\Verb$)$}\\\Verb$ $\Verb$ $\Verb$ $\Verb$ $\Verb$ $\Verb$ $\Verb$ $\Verb$ $\Verb$ $\Verb$|$\Verb$ ${\color{black}\Verb$'$}{\color{black}\Verb$"$}{\color{black}\Verb$'$}\Verb$ ${\color{red}\Verb$i$}{\color{red}\Verb$n$}{\color{red}\Verb$t$}{\color{red}\Verb$e$}{\color{red}\Verb$r$}{\color{red}\Verb$p$}{\color{red}\Verb$o$}{\color{red}\Verb$l$}{\color{red}\Verb$a$}{\color{red}\Verb$t$}{\color{red}\Verb$e$}{\color{red}\Verb$d$}\\\Verb$ $\Verb$ $\Verb$ $\Verb$ $\Verb$ $\Verb$ $\Verb$ $\Verb$ $\Verb$ $\Verb$|$\Verb$ ${\color{black}\Verb$'$}{\color{black}\Verb$#$}{\color{black}\Verb$'$}\Verb$ ${\color{red}\Verb$l$}{\color{red}\Verb$i$}{\color{red}\Verb$n$}{\color{red}\Verb$e$}\\\Verb$ $\Verb$ $\Verb$ $\Verb$ $\Verb$ $\Verb$ $\Verb$ $\Verb$ $\Verb$ $\Verb$|$\Verb$ $\Verb$<$\Verb$^$\Verb$`$\Verb$'$\Verb$"$\Verb$($\Verb$)$\Verb$[$\Verb$]$\Verb${$\Verb$}$\Verb$>$\\\\\\{\color{red}\Verb$i$}{\color{red}\Verb$n$}{\color{red}\Verb$t$}{\color{red}\Verb$e$}{\color{red}\Verb$r$}{\color{red}\Verb$p$}{\color{red}\Verb$o$}{\color{red}\Verb$l$}{\color{red}\Verb$a$}{\color{red}\Verb$t$}{\color{red}\Verb$e$}{\color{red}\Verb$d$}\Verb$ $\Verb$ $\Verb$=$\Verb$ $\Verb$($\Verb$ ${\color{black}\Verb$'$}{\color{black}\Verb$"$}{\color{black}\Verb$'$}\Verb$ $\Verb$b$\Verb$r$\Verb$e$\Verb$a$\Verb$k$\\\Verb$ $\Verb$ $\Verb$ $\Verb$ $\Verb$ $\Verb$ $\Verb$ $\Verb$ $\Verb$ $\Verb$ $\Verb$ $\Verb$ $\Verb$ $\Verb$ $\Verb$ $\Verb$ $\Verb$|$\Verb$ ${\color{red}\Verb$a$}{\color{red}\Verb$r$}{\color{red}\Verb$g$}{\color{red}\Verb$s$}{\color{green}\Verb$($}{\color{green}\Verb$'$}{\color{green}\Verb$#$}{\color{green}\Verb${$}{\color{green}\Verb$'$}{\color{green}\Verb$,$}{\color{green}\Verb$'$}{\color{green}\Verb$}$}{\color{green}\Verb$'$}{\color{green}\Verb$)$}\\\Verb$ $\Verb$ $\Verb$ $\Verb$ $\Verb$ $\Verb$ $\Verb$ $\Verb$ $\Verb$ $\Verb$ $\Verb$ $\Verb$ $\Verb$ $\Verb$ $\Verb$ $\Verb$ $\Verb$|$\Verb$ ${\color{black}\Verb$'$}{\color{black}\Verb$\$}{\color{black}\Verb$\$}{\color{black}\Verb$'$}{\color{black}\Verb$?$}\Verb$ $\Verb$.$\Verb$)${\color{black}\Verb$*$}\\\\{\color{red}\Verb$i$}{\color{red}\Verb$n$}{\color{red}\Verb$l$}{\color{red}\Verb$i$}{\color{red}\Verb$n$}{\color{red}\Verb$e$}{\color{red}\Verb$_$}{\color{red}\Verb$h$}{\color{red}\Verb$o$}{\color{red}\Verb$s$}{\color{red}\Verb$t$}{\color{red}\Verb$_$}{\color{red}\Verb$e$}{\color{red}\Verb$x$}{\color{red}\Verb$p$}{\color{red}\Verb$r$}\Verb$ $\Verb$=$\Verb$ ${\color{red}\Verb$a$}{\color{red}\Verb$r$}{\color{red}\Verb$g$}{\color{red}\Verb$s$}{\color{green}\Verb$($}{\color{green}\Verb$'$}{\color{green}\Verb${$}{\color{green}\Verb$'$}{\color{green}\Verb$,$}{\color{green}\Verb$'$}{\color{green}\Verb$}$}{\color{green}\Verb$'$}{\color{green}\Verb$)$}\\{\color{red}\Verb$h$}{\color{red}\Verb$o$}{\color{red}\Verb$s$}{\color{red}\Verb$t$}{\color{red}\Verb$_$}{\color{red}\Verb$e$}{\color{red}\Verb$x$}{\color{red}\Verb$p$}{\color{red}\Verb$r$}\Verb$ $\Verb$ $\Verb$ $\Verb$ $\Verb$ $\Verb$ $\Verb$ $\Verb$ $\Verb$=$\Verb$ ${\color{red}\Verb$i$}{\color{red}\Verb$n$}{\color{red}\Verb$l$}{\color{red}\Verb$i$}{\color{red}\Verb$n$}{\color{red}\Verb$e$}{\color{red}\Verb$_$}{\color{red}\Verb$h$}{\color{red}\Verb$o$}{\color{red}\Verb$s$}{\color{red}\Verb$t$}{\color{red}\Verb$_$}{\color{red}\Verb$e$}{\color{red}\Verb$x$}{\color{red}\Verb$p$}{\color{red}\Verb$r$}\\\Verb$ $\Verb$ $\Verb$ $\Verb$ $\Verb$ $\Verb$ $\Verb$ $\Verb$ $\Verb$ $\Verb$ $\Verb$ $\Verb$ $\Verb$ $\Verb$ $\Verb$ $\Verb$ $\Verb$ $\Verb$|$\Verb$ ${\color{black}\Verb$'$}{\color{black}\Verb$-$}{\color{black}\Verb$>$}{\color{black}\Verb$'$}\Verb$ ${\color{red}\Verb$l$}{\color{red}\Verb$i$}{\color{red}\Verb$n$}{\color{red}\Verb$e$}{\color{Aquamarine}\Verb$:$}{\color{Aquamarine}\Verb$s$}\Verb$ ${\color{Tan}\Verb${$}{\color{Tan}\Verb$"$}{\color{Tan}\Verb${$}{\color{Tan}\Verb$"$}{\color{Tan}\Verb$+$}{\color{Aquamarine}\Verb$s$}{\color{Tan}\Verb$+$}{\color{Tan}\Verb$"$}{\color{Tan}\Verb$}$}{\color{Tan}\Verb$"$}{\color{Tan}\Verb$}$}{\color{blue}\Verb$=$}{\color{blue}\Verb$>$}{\color{blue}\Verb$[$}\Verb$ ${\color{red}\Verb$i$}{\color{red}\Verb$n$}{\color{red}\Verb$l$}{\color{red}\Verb$i$}{\color{red}\Verb$n$}{\color{red}\Verb$e$}{\color{red}\Verb$_$}{\color{red}\Verb$h$}{\color{red}\Verb$o$}{\color{red}\Verb$s$}{\color{red}\Verb$t$}{\color{red}\Verb$_$}{\color{red}\Verb$e$}{\color{red}\Verb$x$}{\color{red}\Verb$p$}{\color{red}\Verb$r$}\Verb$ ${\color{blue}\Verb$]$}
}\end{spacing}

\end{grambox}
Note that rule args as example of parametrized application. Also note how we in {\color{red}\Verb$h$}{\color{red}\Verb$o$}{\color{red}\Verb$s$}{\color{red}\Verb$t$}{\color{red}\Verb$_$}{\color{red}\Verb$e$}{\color{red}\Verb$x$}{\color{red}\Verb$p$}{\color{red}\Verb$r$} we parse recursively.

Now we can put everything together to form term:

\begin{grambox}
\noindent \begin{spacing}{0.8}
{\small
{\color{red}\Verb$t$}{\color{red}\Verb$e$}{\color{red}\Verb$r$}{\color{red}\Verb$m$}\Verb$ $\Verb$ $\Verb$=$\Verb$ ${\color{red}\Verb$c$}{\color{red}\Verb$a$}{\color{red}\Verb$s$}{\color{red}\Verb$e$}{\color{red}\Verb$s$}\\\Verb$ $\Verb$ $\Verb$ $\Verb$ $\Verb$ $\Verb$ $\Verb$|$\Verb$ ${\color{red}\Verb$c$}{\color{red}\Verb$a$}{\color{red}\Verb$l$}{\color{red}\Verb$l$}\\\Verb$ $\Verb$ $\Verb$ $\Verb$ $\Verb$ $\Verb$ $\Verb$|$\Verb$ ${\color{red}\Verb$k$}{\color{red}\Verb$e$}{\color{red}\Verb$y$}{\color{Aquamarine}\Verb$:$}{\color{Tan}\Verb${$}{\color{Tan}\Verb$A$}{\color{Tan}\Verb$c$}{\color{Tan}\Verb$t$}{\color{Tan}\Verb$[$}{\color{Aquamarine}\Verb$i$}{\color{Aquamarine}\Verb$t$}{\color{Tan}\Verb$]$}{\color{Tan}\Verb$}$}\\\Verb$ $\Verb$ $\Verb$ $\Verb$ $\Verb$ $\Verb$ $\Verb$|$\Verb$ ${\color{red}\Verb$h$}{\color{red}\Verb$o$}{\color{red}\Verb$s$}{\color{red}\Verb$t$}{\color{red}\Verb$_$}{\color{red}\Verb$e$}{\color{red}\Verb$x$}{\color{red}\Verb$p$}{\color{red}\Verb$r$}\\\Verb$ $\Verb$ $\Verb$ $\Verb$ $\Verb$ $\Verb$ $\Verb$|$\Verb$ ${\color{black}\Verb$'$}{\color{black}\Verb$.$}{\color{black}\Verb$'$}\Verb$ $\Verb$ $\Verb$ $\Verb$ $\Verb$ $\Verb$ $\Verb$ $\Verb$ $\Verb$ $\Verb$ $\Verb$ $\Verb$ $\Verb$ $\Verb$ $\Verb$ $\Verb$ $\Verb$ $\Verb$ $\Verb$ $\Verb$ ${\color{Tan}\Verb$-$}{\color{Tan}\Verb$>$}{\color{Tan}\Verb$ $}{\color{Tan}\Verb$A$}{\color{Tan}\Verb$p$}{\color{Tan}\Verb$p$}{\color{Tan}\Verb$l$}{\color{Tan}\Verb$y$}{\color{Tan}\Verb$[$}{\color{Tan}\Verb$"$}{\color{Tan}\Verb$a$}{\color{Tan}\Verb$n$}{\color{Tan}\Verb$y$}{\color{Tan}\Verb$t$}{\color{Tan}\Verb$h$}{\color{Tan}\Verb$i$}{\color{Tan}\Verb$n$}{\color{Tan}\Verb$g$}{\color{Tan}\Verb$"$}{\color{Tan}\Verb$]$}{\color{Tan}\\}{\color{Tan}\Verb$ $}\Verb$ $\Verb$ $\Verb$ $\Verb$ $\Verb$ $\Verb$|$\Verb$ ${\color{black}\Verb$'$}{\color{black}\Verb$[$}{\color{black}\Verb$'$}\Verb$ ${\color{red}\Verb$e$}{\color{red}\Verb$x$}{\color{red}\Verb$p$}{\color{red}\Verb$r$}{\color{red}\Verb$e$}{\color{red}\Verb$s$}{\color{red}\Verb$s$}{\color{red}\Verb$i$}{\color{red}\Verb$o$}{\color{red}\Verb$n$}{\color{Aquamarine}\Verb$:$}{\color{blue}\Verb$e$}\Verb$ ${\color{black}\Verb$"$}{\color{black}\Verb$]$}{\color{black}\Verb$"$}\Verb$ $\Verb$ $\Verb$ ${\color{Tan}\Verb$-$}{\color{Tan}\Verb$>$}{\color{Tan}\Verb$ $}{\color{Tan}\Verb$E$}{\color{Tan}\Verb$n$}{\color{Tan}\Verb$t$}{\color{Tan}\Verb$e$}{\color{Tan}\Verb$r$}{\color{Tan}\Verb$[$}{\color{Tan}\Verb$A$}{\color{Tan}\Verb$p$}{\color{Tan}\Verb$p$}{\color{Tan}\Verb$l$}{\color{Tan}\Verb$y$}{\color{Tan}\Verb$[$}{\color{Tan}\Verb$"$}{\color{Tan}\Verb$a$}{\color{Tan}\Verb$n$}{\color{Tan}\Verb$y$}{\color{Tan}\Verb$t$}{\color{Tan}\Verb$h$}{\color{Tan}\Verb$i$}{\color{Tan}\Verb$n$}{\color{Tan}\Verb$g$}{\color{Tan}\Verb$"$}{\color{Tan}\Verb$]$}{\color{Tan}\Verb$,$}{\color{blue}\Verb$e$}{\color{Tan}\Verb$]$}{\color{Tan}\\}{\color{Tan}\Verb$ $}\Verb$ $\Verb$ $\Verb$ $\Verb$ $\Verb$ $\Verb$|$\Verb$ ${\color{black}\Verb$'$}{\color{black}\Verb$"$}{\color{black}\Verb$'$}\Verb$ $\Verb$ ${\color{red}\Verb$u$}{\color{red}\Verb$n$}{\color{red}\Verb$t$}{\color{red}\Verb$i$}{\color{red}\Verb$l$}{\color{green}\Verb$($}{\color{green}\Verb$'$}{\color{green}\Verb$"$}{\color{green}\Verb$'$}{\color{green}\Verb$ $}{\color{green}\Verb$)$}{\color{Aquamarine}\Verb$:$}{\color{Aquamarine}\Verb$s$}\Verb$ $\Verb$ $\Verb$ $\Verb$ $\Verb$ ${\color{Tan}\Verb$-$}{\color{Tan}\Verb$>$}{\color{Tan}\Verb$ $}{\color{Tan}\Verb$A$}{\color{Tan}\Verb$p$}{\color{Tan}\Verb$p$}{\color{Tan}\Verb$l$}{\color{Tan}\Verb$y$}{\color{Tan}\Verb$[$}{\color{Tan}\Verb$"$}{\color{Tan}\Verb$t$}{\color{Tan}\Verb$o$}{\color{Tan}\Verb$k$}{\color{Tan}\Verb$e$}{\color{Tan}\Verb$n$}{\color{Tan}\Verb$"$}{\color{Tan}\Verb$ $}{\color{Tan}\Verb$,$}{\color{Tan}\Verb$q$}{\color{Tan}\Verb$u$}{\color{Tan}\Verb$o$}{\color{Tan}\Verb$t$}{\color{Tan}\Verb$e$}{\color{Tan}\Verb$($}{\color{Aquamarine}\Verb$s$}{\color{Tan}\Verb$)$}{\color{Tan}\Verb$]$}{\color{Tan}\\}{\color{Tan}\Verb$ $}\Verb$ $\Verb$ $\Verb$ $\Verb$ $\Verb$ $\Verb$|$\Verb$ ${\color{black}\Verb$'$}{\color{black}\Verb$\$}{\color{black}\Verb$'$}{\color{black}\Verb$'$}\Verb$ ${\color{red}\Verb$u$}{\color{red}\Verb$n$}{\color{red}\Verb$t$}{\color{red}\Verb$i$}{\color{red}\Verb$l$}{\color{green}\Verb$($}{\color{green}\Verb$'$}{\color{green}\Verb$\$}{\color{green}\Verb$'$}{\color{green}\Verb$'$}{\color{green}\Verb$)$}{\color{Aquamarine}\Verb$:$}{\color{Aquamarine}\Verb$s$}\Verb$ $\Verb$ $\Verb$ $\Verb$ $\Verb$ ${\color{Tan}\Verb$-$}{\color{Tan}\Verb$>$}{\color{Tan}\Verb$ $}{\color{Tan}\Verb$A$}{\color{Tan}\Verb$p$}{\color{Tan}\Verb$p$}{\color{Tan}\Verb$l$}{\color{Tan}\Verb$y$}{\color{Tan}\Verb$[$}{\color{Tan}\Verb$"$}{\color{Tan}\Verb$s$}{\color{Tan}\Verb$e$}{\color{Tan}\Verb$q$}{\color{Tan}\Verb$"$}{\color{Tan}\Verb$ $}{\color{Tan}\Verb$ $}{\color{Tan}\Verb$ $}{\color{Tan}\Verb$,$}{\color{Tan}\Verb$q$}{\color{Tan}\Verb$u$}{\color{Tan}\Verb$o$}{\color{Tan}\Verb$t$}{\color{Tan}\Verb$e$}{\color{Tan}\Verb$($}{\color{Aquamarine}\Verb$s$}{\color{Tan}\Verb$)$}{\color{Tan}\Verb$]$}{\color{Tan}\\}{\color{Tan}\Verb$ $}\Verb$ $\Verb$ $\Verb$ $\Verb$ $\Verb$ $\Verb$|$\Verb$ ${\color{black}\Verb$'$}{\color{black}\Verb$#$}{\color{black}\Verb$'$}\Verb$ $\Verb$ ${\color{red}\Verb$l$}{\color{red}\Verb$i$}{\color{red}\Verb$n$}{\color{red}\Verb$e$}{\color{Aquamarine}\Verb$:$}{\color{Aquamarine}\Verb$s$}\Verb$ $\Verb$ $\Verb$ $\Verb$ $\Verb$ $\Verb$ $\Verb$ $\Verb$ $\Verb$ $\Verb$ $\Verb$ $\Verb$ ${\color{Tan}\Verb$-$}{\color{Tan}\Verb$>$}{\color{Tan}\Verb$ $}{\color{Tan}\Verb$C$}{\color{Tan}\Verb$o$}{\color{Tan}\Verb$m$}{\color{Tan}\Verb$m$}{\color{Tan}\Verb$e$}{\color{Tan}\Verb$n$}{\color{Tan}\Verb$t$}{\color{Tan}\Verb$[$}{\color{Aquamarine}\Verb$s$}{\color{Tan}\Verb$]$}{\color{Tan}\\}{\color{Tan}\Verb$ $}\Verb$ $\Verb$ $\Verb$ $\Verb$ $\Verb$ $\Verb$|$\Verb$ ${\color{black}\Verb$'$}{\color{black}\Verb$($}{\color{black}\Verb$'$}\Verb$ ${\color{red}\Verb$e$}{\color{red}\Verb$x$}{\color{red}\Verb$p$}{\color{red}\Verb$r$}{\color{red}\Verb$e$}{\color{red}\Verb$s$}{\color{red}\Verb$s$}{\color{red}\Verb$i$}{\color{red}\Verb$o$}{\color{red}\Verb$n$}{\color{Aquamarine}\Verb$:$}{\color{Aquamarine}\Verb$x$}\Verb$ ${\color{black}\Verb$"$}{\color{black}\Verb$)$}{\color{black}\Verb$[$}{\color{black}\Verb$]$}{\color{black}\Verb$"$}\Verb$ ${\color{Tan}\Verb${$}{\color{Aquamarine}\Verb$x$}{\color{Tan}\Verb$}$}{\color{blue}\Verb$=$}{\color{blue}\Verb$>$}{\color{red}\Verb$c$}{\color{red}\Verb$o$}{\color{red}\Verb$l$}{\color{red}\Verb$l$}{\color{red}\Verb$e$}{\color{red}\Verb$c$}{\color{red}\Verb$t$}\Verb$ $\\\Verb$ $\Verb$ $\Verb$ $\Verb$ $\Verb$ $\Verb$ $\Verb$|$\Verb$ ${\color{black}\Verb$'$}{\color{black}\Verb$($}{\color{black}\Verb$'$}\Verb$ ${\color{red}\Verb$e$}{\color{red}\Verb$x$}{\color{red}\Verb$p$}{\color{red}\Verb$r$}{\color{red}\Verb$e$}{\color{red}\Verb$s$}{\color{red}\Verb$s$}{\color{red}\Verb$i$}{\color{red}\Verb$o$}{\color{red}\Verb$n$}{\color{Aquamarine}\Verb$:$}{\color{Aquamarine}\Verb$x$}\Verb$ ${\color{black}\Verb$"$}{\color{black}\Verb$)$}{\color{black}\Verb$"$}\Verb$ $\Verb$ $\Verb$ ${\color{Tan}\Verb$-$}{\color{Tan}\Verb$>$}{\color{Tan}\Verb$ $}{\color{Aquamarine}\Verb$x$}
}\end{spacing}

\end{grambox}

\chapter{  Implementation } \label{regreg}
In this chapter we describe main techniques used in amethyst parser generator. We introduce novel notion of structured grammars and formalism of relativized regular expressions that enables us to produce effective top-down parsers for wide family of languages.

A top-down parsing implementation can be viewed as bunch of mutually recursive functions recognizing individual rules in grammar description. 
Top-down parsers are easy to implement and furn fast for simple grammars.

But naively implemented parser of the following rule:\\
\verb$R="aa" R | "a" R$ \\
on \extext{aaaaaaaaaaaaaaaaaaa...} can take exponential time. 

Incorporating left recursion also causes problems. A naive parser of \\
\verb$L=L a$ \\
would call L infinitely many times. 

In natural language processing we typically want to enumerate possible interpretations of ambiguous grammar.

Frost \cite{frost} gave $O(n^4)$ algorithm that outputs compact representation of all parses \cite{tomita} and handles left recursion as recursive descend. 
Parsing expression grammars allow unlimited lookahead. Okhotin \cite{bool} suggest to extend context free grammars with lookahead to class of {\it boolean grammars}. Again his algorithm for boolean grammars had complexity $O(n^4)$. Both these algorithms were improved by variant of Valiant algorithm \cite{valiant} to obtain complexity $O(M(n) \log n)$ where $M(n)$ is time of matrix multiplication. When boolean grammars are restricted to {\it unambiguous boolean grammars} there exists $O(n^2)$ algorithm. 

For programming languages ambiguity is undesirable. One of approaches are parsing expression grammars defined by Ford \cite{ford}. A parsing expression grammars (\peg{} for short) can be viewed as a top-down parser that places three additional constraints. First is that rules are deterministic. Second is restricting choice operator \verb$|$ to {\it ordered choice} operator \verb$/$. Once an alternative of ordered choice succeeds then choice succeeds then we do not backtrack if something fails later. Third is that iteration is greedy and does not backtrack. 

This definition without backtracking introduced problem of {\it prefix hiding}, an expression \verb$"a"/"ab"$    does not match string \extext{ab}.

Seaton in his Katahdin language \cite{katahdin} uses different {\it longest choice} operator to partially solve this problem. The longest choice tries all alternatives and deterministically chooses the longest match. However this does not eliminate the prefix hiding completely. Parser of:\\
\texttt{"\textvisiblespace"* "\textvisiblespace{}foo"} (\verb$*$ is iteration operator) still does not match \extext{\textvisiblespace{}foo}. 

We take another approach. Programming languages use only two types of recursion: iteration and nested recursion. By making this information explicit we can generate linear time parsers that are equivalent to the fully backtracking ones.

We present new formalism of relativized regular expressions  \regreg{}. Our formalism relaxes determinism of \peg{} grammars. As in \peg{} we support arbitrary lookaheads. Previous results can be easily derived using  \regreg{} formalism.

Although \regreg{} seems stronger than  \peg{} we show that \peg{} and  \regreg{} are equivalent.

\section{Structured grammars}

We devise approach to describe programming languages which we call {\it structured grammars}. We build on an analogy with structured programming languages. 

As programs used arbitrary {\it goto} constructs, grammars use arbitrary forms of recursion.
To make programs more readable, programming languages was extended by adding structured control flow constructs
making it easier for developers to read the code on a local basis without spending hours to understand
the whole context. We seek similar goals with introduction of structured grammars.

Assume we are given a grammar for the fully-backtracking top-down parser. We say it is {\it structured grammar} if it satisfies the following conditions:

\begin{enumerate}[label*=\arabic*.]
\item {\it Transparency of semantic actions}. We can imagine that parser is augmented by an oracle that may decide that alternative will eventually fail. The parser should display same output regardless if we tried that alternative and failed or used the hint from the oracle. Lookaheads form important case. We always revert actions made by lookaheads.

\item Recursion is restricted to {\it iterative} and {\it nested recursion}.\\
\begin{enumerate}
\item Iterative:
 For example, arguments of function in C are lists of \verb$expressions$ separated by "\verb$,$". We typically use iteration \verb$*$ operator. Iteration can be also described by left recursive or by right recursive rules. When possible iteration should be written in way that is associative.\\
\item Nested recursion:
What is not iteration can be described by {\it \texttt{start}} and {\it \texttt{end}} delimiters.  
We require user to annotate this concept by operator \verb$nested(${\it \texttt{start,middle,end}}\verb$)$. 
\end{enumerate}

Simplest example are properly parenthised expressions. They can be described as:\\
\verb$exp = nested('(',(| exp |),')')$\\
We show two less trivial examples in structured grammar formalism. A while loop in \verb$C$ is matched by:\\
\verb$'while' exp nested('{',(| stmts |),'}')$ \\
Python uses indentation to describe nesting. We use a semantic predicate to find where we end. We match python while loops in amethyst as:\\
\verb$ nested((|   '\n' ' '*:x 'while' exp |),(| stmts |),$\\
\verb$        (| &('\n' ' '*:y &{x.size>y.size}) |))$

A nesting should satisfy three natural conditions.
\begin{enumerate}[label*=\arabic*.]
\item Position of {\it\texttt{end}} delimiter is determined by position {\it\texttt{start}} delimiter.
\item When \verb$nested$ starts in smaller position it should end in strictly larger position.
\item When both \verb$nested(${\it \texttt{start,mid1,end}}\verb$)$ and  \verb$nested(${\it \texttt{start,mid2,end}}\verb$)$ match string then their end positions should agree.
\end{enumerate}
\end{enumerate}

Note that programming languages implicitly follow this convention. Other types of recursion are undesirable because user can not reason about them locally.

One of reasons is that programming languages were described as deterministic context free grammars. Thus they can be written by deterministic push-down automaton. We can model push/pop pair by calling \verb$nested$. Indeed if we did not include lookaheads our class would be equivalent to class of deterministic context free grammars. 

Structured grammars offer additional advantages. 
For example, we can use the structure information to semiautomatically construct error correction tool.

For equivalence with top-down parser our parsing algorithm needs condition 2.1. Without condition 2.2 a parser would be quadratic instead linear time. Condition 2.3 is design guideline which is not needed in our algorithm.

\newpage
\section{\peg{} and \regreg{} operators} 

In this chapter we use \peg{} operators as originally defined by Ford \cite{ford}.

\vskip -0.3em \noindent\begin{tabular}{| l | p{12cm} | }
\hline
\verb$'s'     $& Match string.\\
\verb$ r      $& Rule application.\\
\verb$e1 e2   $& Sequencing.\\
\verb$e1/e2   $& Ordered choice.\\
\verb$ e*   e+$& Iteration.\\
\verb$&e   ~e $& Positive and negative lookahead.\\
\verb${a} &{a}$& Semantic action and predicate.\\
\hline
\end{tabular} \vskip -0.3em

We relax determinism of \peg{} to \regreg{} expressions. We can describe every structured grammar by \regreg{} rules with linear time guarantee.
A \regreg{} expressions mostly use the same operators as \peg{}. Difference is that operators do backtracking except of \verb$nested$ which behaves deterministically. 

\vskip -0.3em \noindent\begin{tabular}{| l | p{9cm} |}
\hline
\verb$nested(start,mid,end)$& Nested operator.\\
\verb$e1|e2 $& Priorized choice.\\
\verb$e* e+ $& Backtracking iteration.\\
\verb$e1[e2]$& Enter operator.\\
\hline
\end{tabular} \vskip -0.3em

\subsection{Simple algorithm}\label{simplealg}
We will describe our parser in functional programming style pseudocode in continuation passing style \cite{contpass}.
We denote lambda as: \\
\verb$\lambda(arguments){body}$ and call it with \verb$call$ method.

We start with simple implementation and will progressively add more details. 

A \regreg{} parser behaves mostly as a top-down parser.  We use the function \verb$match(e,s,cont)$ where \verb$e$ is expression we match, \verb$s$ is current position and \verb$cont$ is a continuation \cite{continuation} represented as lambda.
\begin{prolbox}
\begin{verbatim}
match( r  ,s,cont) = match(body(r),s,cont)
match('c' ,s,cont) = if s.head=='c'              ;cont.call(s.tail)
                     else                        ;fail

match(e f ,s,cont) = match(e,s,\(s2){ match(f,s2,cont) }
match(e|f ,s,cont) = if match(e,s,cont)          ;success
                     else                        ;match(f,s,cont)

match(~e  ,s,cont) = if match(e,s,\(s2){success});fail
                     else                        ;cont.call(s)
match(e*  ,s,cont) =
  cont2 <-    \(s2){ if match(e,s2,cont2)        ;success
                     else                        ;cont.call(s2) 
                   }
  cont2.call
\end{verbatim}
\end{prolbox}
Pseudocode above describe naive top-down parser. For \regreg{} class we restrict recursion and add \verb$nested$ operator:
\begin{prolbox}
\begin{verbatim}
match(nested(st,mi,en) ,s,cont) =
  s3 <- match((st mi en),s,\(s2){success})
  if s3    ; cont.call(s3)
  else     ; fail
\end{verbatim}
\end{prolbox}

\subsection{Equivalence with top-down parsers and \peg{}}

We prove that for structured grammars  \regreg{} parser finds same derivation as fully backtracking one. As top-down parser does not directly support left recursion we do not consider left recursion in this section. 

An implementation of the fully backtracking parser is same as the implementation of \regreg{} parser in Section \ref{simplealg} except of \verb$nested$:
\begin{prolbox}
\vskip -0.5em
\begin{verbatim}
match(nested(st,mi,en), s, cont) = match((st mi en), s, cont)
\end{verbatim}
\vskip -0.5em
\end{prolbox}
\vskip -0.5em

For sake of proof we transform rewrite implementation of \verb$nested$ in \regreg parser to equivalent one. In \verb$nested$ we only consider the first alternative in the way the following pseudocode suggests:
\begin{prolbox}
\begin{verbatim}
match(nested(st,mi,en), s, cont) = first <- true
  match(s, (st mi en), \(s2){
    if first ; first <- false
             ; cont.call(s2)
    else     ; fail 
  })
\end{verbatim}
\end{prolbox}
\vskip -0.5em

An equivalence with top-down parser can be proved by easy induction on the nesting level. 
\vskip -0.5em
\begin{enumerate}
\vskip -0.5em
\item When expression contain no nesting we have identical implementation. 
\vskip -0.5em
\item Assume we proved proposition for nesting level $\ell-1$. We prove level $\ell$ by second induction on the number of \verb$nested$ calls in the continuation of level $\ell-1$.
\vskip -0.5em
\begin{enumerate}
\item For continuation that does not call \verb$nested$ we use same argument as in 1.
\vskip -0.5em
\item Assume we have continuation that calls \verb$nested$ $n$ times. Consider first time we call \verb$nested$. If this call fails it, by induction, also fails in the fully backtracking parser and we are done.

Otherwise  \regreg{} and the fully backtracking parser first try lexicographically smallest alternative in the recursion tree. If a continuation succeeds a derivation is same by induction.

If a continuation fails we use assumption 2.1. of structured grammars.
Our parser does not try alternatives further. A backtracking parser enumerates all derivations. As every derivation ends in same position and continuation will always fail. Thus the backtracking parser behaves like \regreg{} parser.
\end{enumerate}

\end{enumerate}

Like not every C program is structured program not every  \regreg{} grammar is structured one. We can use \verb$nested$ with empty {\it\texttt{start}} and {\it\texttt{end}} to implement \peg{} operators. This gives us inclusion \peg{} $\subseteq$ \regreg{}. An opposite inclusion is true but not very enlightening. As there are only finitely many pairs $(e,cont)$ we can for each pair write a \peg{} rule that emulates  \regreg{} algorithm.

For linear time guarantee we still require every recursion except left and right recursion to be annotated by \verb$nested$.

\section{Relativized regular machines}
To better understand languages recognized by relativized regular expressions we introduce the {\it relativized regular machines} that are similar to nondeterministic finite state machines \cite{nfa}. We use this formalism as an inspiration for effective low-level implementation of parsers.

It is easy to see that a continuation corresponds to syntactic right congruence class. We use representation that unifies identical expressions and continuations. This can be viewed as NFA state minimization\footnote{NFA minimization is NP-hard in general case. Our approach is a good heuristic.}. 

A {\it relativized regular machine} is similar to nondeterministic finite state machine. A machine can be described by triple 
$M=(S,t,a)$ where 

\begin{tabular}{l}
S is set of states, \\
$t: (S,N,S) \to (M,S)$ set of transitions and \\
$a \subset S$ a set of accepting states.\\
\end{tabular} \vskip -0.3em

We have elementary machines that match single character. 

Transitions from state $s$ are done in the following way. We put $(M_i,s_i)=t((s,i,r_{i-1}))$ then recursively call machine $M_i$ and if it succeeds we move to its end position and set state to $s_i$. Based of accepting state $r_i$ this choice reaches we choose a next choice. 

\newpage
\section{Effective implementation} \label{implement}
An implementation above runs in linear time but constant factor is quite high. For better constant factor our parser generator applies various optimizations. We use a low-level representation that is suitable for these optimizations.

In this section we describe parser that does not consider semantic actions. Semantic actions will be added in the next section. 

Representation of expressions is similar to syntax tree. We use similar technique as compact representation of derivations  in Tomita algorithm \cite{tomita}:

\begin{enumerate}
\item All nodes are immutable.

\item We represent all identical subtrees by single object. When we are asked to construct a node optimizer first tries to simplify node by algebraic identities. If after simplification we obtain node identical to previously constructed node we return previously constructed node.
\end{enumerate}

We will again use function \verb$match(e,Args[ ... ] ) -> Result[ ... ]$.\\
We will extend several times what \verb$Args$ and \verb$Result$ objects contain. Initially we define the following fields:

\begin{tabular}{l l}
\verb$Args.s$ &is starting position of string,\\
\verb$Result.s$  &is end position of string,  \\
\verb$Args.cont$ &is a continuation.\\
\end{tabular} \vskip -0.3em

Objects \verb$Args$ and \verb$Result$ have method \verb$change$ that creates new object with appropriately changed fields.
\vskip -2em
\subsection{Sequencing}
\vskip -1em

We represent sequencing operator \verb$head tail$ by object with pattern \verb$Seq[head tail]$. Representing sequencing in this way allows \verb$tail$ parts to be shared.
Implementation is straightforward.

\begin{prolbox}
\begin{verbatim}
match( Seq[head tail],a) = match(head,a.change(
  cont:\(a2){
    match(tail,a.change(cont: a.cont))
  }))
\end{verbatim}
\end{prolbox}

\vskip -2em
\subsection{Choice and lookaheads}
\vskip -1em

Inspired by relativized regular machines we model choice and lookaheads by more general \verb$Switch$ operator. First we need add field \verb$Result.state$. This state will be used to pass information from rules to the \verb$Switch$ operator.

The \verb$Switch$ operator satisfies the following pattern:\\
\verb$Switch[ head alt:{state=>tail} merge ]$\\
\verb$Switch$ operator first matches a head. Then it looks what end state head reached and matches tail entry corresponding to that state. Finally it computes final state from states of children by merge method. 

For simplicity in this paper we use only two states  \verb$success$ and \verb$fail$. We use identity function as a merge method. We also add \verb$success$ and \verb$fail$ rules with obvious implementation:
\begin{verbatim}
match(Rule[success]) = Result[state: success]
match(Rule[fail   ]) = Result[state: fail   ]
\end{verbatim}

This is quite general operator and we illustrate its uses on several examples.\\
The choice operator backtracks until success state was reached. An implementation is:

\begin{verbatim}
e1|e2  ->    Switch[ e1 {success: success
                            fail: e2}]
\end{verbatim}

Lookaheads can be modeled like:\\
\begin{verbatim}
~e   ->    Switch[ Seq[e success ]
                  {success:   fail,
                   fail:      empty} ]
&e   ->    Switch[ Seq[e success ]
                  {success:   empty,
                   fail:      fail } ]
\end{verbatim}

A \verb$Switch$  makes optimizations easy.  
Switches can be easily composed. To compose switches \verb$A$ and \verb$B$ a simplest way is to use states that are pairs (state from \verb$A$,state from \verb$B$). We need to define merge method to compute final state. We can represent these pairs compactly as bit vector.
Another optimization is predication. When we know first character we can simplify expression:

\begin{verbatim}
Switch[ Result[ first_character ]
                  { 'a': expressions that can start by a,
                    'b': expressions that can start by b, 
                    ...   
\end{verbatim}

For choice \verb$e1|e2$ we can, based on the result of the partial match of e1, simplify matching of \verb$e2$. For example, consider expression:

\begin{verbatim}
  (a|b) c (d|f)
| (b|c) c f
\end{verbatim}
on string \extext{bcd}. \\
When first alternative matches \extext{d} then we know that second alternative will not match. Last choice could pass state to inform first choice about this condition.

An implementation of \verb$Switch$ is the following. We hide technical details to merge method. For details see our implementation \cite{implementation}. 

\begin{prolbox}
\begin{verbatim}
  match_memo(e,a) = 
  if memo[e,a]; memo[e,a]
  else        ; memo[e,a] <- match(e,a)

  match(Switch[ head alt merge ],a) =
    r  <- match_memo(head,a)
    r2 <- match_memo(alt[r.state],a)
    merge(r,r2)
\end{verbatim}
\end{prolbox}

\subsection{Iteration}

We use low-level repeat-until and \verb$Stop$ operators from previous chapter to represent iteration.

Repeat-until can terminate if and only if we encountered corresponding \verb$Stop$ in current iteration. We add \verb$stops$ field to \verb$Args$ to collect encountered stops.

This allows to describe normal iteration \verb$e*$ and eager iteration \verb$e*?$ as \verb$(e|Stop)**$ and \verb$(Stop|e)**$ respectively.  Repeat-until is equivalent to right-recursion. For example, we can flip between rules\\
\verb$ R =  a R | b      | c R | d       $\\
and\\
\verb$ R = (a   | b Stop | c   | d Stop)*$.

Except of stop condition the implementation is nearly identical to implementation of \verb$*$ operator from Section \ref{simplealg}.
\begin{prolbox}
\begin{verbatim}
match(Stop[st]   ,a) = a.cont.call(  a.change(stops: a.stops+st))
match(Many[st e] ,a) = 
  cont2 <- \(a2){ 
    if a2.stops & st ; a.cont.call( a2.change(stops:a2.stops-st))
    else             ; match(e,a2.change(cont:cont2 ))
  }
  cont2.call(a)
\end{verbatim}
\end{prolbox}

\subsection{Rule call}
Rule call only affects scope of variables. When no semantic actions are present we can directly move expression to separate rule and back.

\begin{prolbox}
\begin{verbatim}
match(Rule[ e ], a) = match(e ,a)
\end{verbatim}
\end{prolbox}

For \verb$nested$ we use similar implementation as before. 

\begin{prolbox}
\begin{verbatim}
  match(Nested[st mi en],a) =
    r = match_memo(Seq[st mi en],Args[s:a.s,cont:\(m){success}])
    if r.state==success ; a.cont.call(a.change(s:r.s))
    else                ; fail
\end{verbatim}
\end{prolbox}

\newpage
\section{Semantic actions} \label{semact}
We add semantic actions as was explained in the previous chapter.

While semantic actions are easy to add they complicate other parts of the parser.

We add the following fields:\\
\verb$Args.closure$ closure for semantic actions.\\
\verb$Args.returned$ the result of last expression.\\
\verb$Result.returned$ returned result.

We model semantic act as a function that modifies arguments. For simplicity we model variable binding by semantic act.

\begin{prolbox}
\begin{verbatim}
  match( Act[ f ] ,a ) = a2 <- f.call(a.closure)
    a.cont.call(a.change(a2))
\end{verbatim}
\end{prolbox}

Now we are ready to add enter operator. 

\begin{prolbox}
\begin{verbatim}
  match( Enter[e1 e2], a) =
    match(e1,a.change(cont: \(a2){
       match(a2.change(s:a2.returned),cont:\(a3){
         a.cont.call(a3.change(s:a.s))
       }
    }
\end{verbatim}
\end{prolbox}

Semantic actions in rule invocation have shared scope. We use closure object to achieve this. A rule invocation becomes:

\begin{prolbox}
\begin{verbatim}
match( Rule[ e ] ,a ) = match(e,
                              a.change(closure:new_closure,
         cont:\(a2){ a.cont.call(a.change(s:a.s,
                                 returned:a.returned))}
     )
\end{verbatim}
\end{prolbox}

We also add semantic predicates from the previous chapter This complicates memoization and we, for simplicity, disable memoization when semantic predicate is present.

Support parametrized rules and lambdas is bit technical to add. For parametrized rule we first model arguments by semantic act bound to argument variables. Then we add field consisting of pairs (argument variable,parameter variable) and we initialize new closure according to pairs. For lambda we bind (expression,closure) pair to corresponding variable. We disable memoization when parametrized rule is present for same reasons as with {\it semantic predicate}.

Memoization becomes more technical. A simplest way how to get linear time complexity is to use two pass parser which in first pass run parser from Section \ref{implement} and second time we just constructs parse tree. We refine this idea and run both phases in parallel. We use functor \verb$forget_semantic_actions$:

\begin{prolbox}
\begin{verbatim}
match_memo_state(e,a) =
  if (has_predicate(e) | has_predicate(a)) ; match(e,a)
  else ; e2 <- forget_semantic_actions(e)
         a2 <- forget_semantic_actions(a)
         if    memo[e2,a2]  ; memo[e2,a2]
         else               ; memo[e2,a2] <- match(e,a)

\end{verbatim}
\end{prolbox}

A simple implementation of \verb$Switch$ can be:

\begin{prolbox}
\begin{verbatim}
match(Switch[ head alts merge ], a)
  r  <- match_memo_state(head,a)
  r2 <- match_memo_state(alts[r.state],a)
  if r2.state==fail
    fail(r2)
  else merge(match(head,a),
             match(alts[r.state],a))
\end{verbatim}
\end{prolbox}

Sometimes \verb$Switch$ knows that the result is not needed. Then we can directly call expression simplified by \verb$forget_semantic_action$. This always happens for lookaheads. 

\newpage
\subsection{Time complexity}
Ford \cite{ford} rewrites iteration to recursion for linear time complexity. However most implementations naively use a loop. 

It is possible to construct test cases where arbitrary (say k) number of loops are nested together and each fails at the end of input.  This leads to
time complexity at least $n^k$ for arbitrary k. This can be seen on the following expression:\\
\begin{verbatim}
( ( ( ( 'a' )* 'b'
    / 'a' )* 'c'
  / 'a' )* 'd'
/ 'a' )* 'e'
\end{verbatim}
on \extext{aaaaaaaaaaaaaaaaaaaaaaa...}\\

We memoize continuations precisely for this reason.

For parser from Section \ref{implement} there are only finitely many expressions and continuations. Thus there are only $O(n)$ memoization pairs \verb$(e,a)$.

With semantic actions we sometimes need to recalculate the result. For a given pair (\verb$nested,position$) we need to recalculate result of every \verb$(e,a)$ pair at most once.
For general \regreg{} expressions time complexity $O(n^2)$ follows. 

For structured grammars this behavior can not happen. We do not have to recalculate when the result state is \verb$fail$ or we match in lookaheads. What is left is that we could have two invocations of same \verb$nested$ expression with two different positions that recalculates same \verb$(e,a)$ pair. But this would mean that both invocations will be accepted with same end position which is in contradiction with condition 2.2 of structured grammars. Consequently the parser of structured grammars runs in linear time

With semantic predicates we can not give any complexity guarantee. To integrate them correctly we disable memoization when continuation contains semantic predicate.

\section{Memory consumption of  \regreg{} parsers} \label{pegmemory}
Mizushima et al  \cite{memo} propose way to decrease the memory usage. We describe similar but simpler approach. 

The parser implementation maintain set of live branches in a list \verb$live$. The list is maintained in the following way:

\begin{itemize}
\item When parser descends into choice operator then its branches are added to \verb$live$ list.
\item When parser descends into branch, then it is removed from \verb$live$ list.
\item When parser encounters cut then branches that were cut are removed from \verb$live$ list.
\end{itemize}

When \verb$live$ list is empty we know that subsequent parsing can not return to position smaller than current. We can safely delete all \verb$memo$ entries with smaller position.
One can observe that \verb$live$ list is not  needed.  The implementation can be further simplified by only keeping
track of the size of the list in a counter \verb$alternatives$.

The parser then deletes stale entries from memo table lazily.  It keeps
track of the rightmost position where \verb$alternatives$ was zero.  At a time
table expansion is needed, all earlier entries are deleted. This avoids the
need for the expansion if the table after deletion is at most half full.

Note that if we want to incorporate destructive semantic actions we can in same way defer their evaluation until \verb$alternatives$ is zero.

For practical grammars this extension gives nearly constant memory usage. However we can construct examples where this approach does not help, for example in expression:\\
\verb$ exp* 'x' | exp* 'y' $ \\
we need to keep memoization entries until end is reached.

\subsection*{Memoization in general setting}

The memoization is viewed as alternative to dynamic programming.
A naive memoization can have big memory consumption. We show that with few simple trick we can obtain better performance with order of magnitude smaller memory usage.
Memoization strategy and automatic memoization were unsurprisingly developed in context of context-free parsing \cite{memoization}. 
We use memoization strategy that applies to reducing memoization memory consumption in general. 

First step to reduce memory usage is to save values for parameters in the hash table. This gives memory consumption proportional to number of saved values.

When we do not ask question: "What functions must be memoized?" but right one "What parameters must be memoized?" then solution is surprisingly simple: \\
We have only memoize those parameters that took at least say 512 cycles to compute. 

There are three additional improvements:
\begin{itemize}
\item We can count only cycles that were not memoized by son rule. 
\item We keep additional small (say 512 element) directly mapped write-back cache. We use this when we notice that values before our threshold are typically rarely reused after say 100 steps. 
\item Some functions never reach our threshold so we can use separate table to save unnecessary lookups. We are more worried that these lookups thrash cache than direct cost.
\end{itemize}

This has same asymptotic time complexity as when we memoize everything because rule is memoized at most once and when we do not memoize time complexity is constant.

There is technical problem how measure number of cycles. While x86 offers timestamp counter and core2 it takes about 24 cycles. We currently use simple estimate by counting number of functions that we called as we typically have constant overhead. Alternative way is first run a version that gathers profiling data and then use estimates from that version.

We illustrate our ideas on the following memoized version of Fibonacci numbers in \verb$C$ language:
\vskip -1.8em\begin{spacing}{0.8}
{\small
\begin{exambox}
\Verb$t$\Verb$y$\Verb$p$\Verb$e$\Verb$d$\Verb$e$\Verb$f$\Verb$ $\Verb$s$\Verb$t$\Verb$r$\Verb$u$\Verb$c$\Verb$t$\Verb$ $\Verb${$\Verb$l$\Verb$o$\Verb$n$\Verb$g$\Verb$ $\Verb$t$\Verb$i$\Verb$m$\Verb$e$\Verb$;$\Verb$l$\Verb$o$\Verb$n$\Verb$g$\Verb$ $\Verb$s$\Verb$a$\Verb$v$\Verb$e$\Verb$d$\Verb$;$\Verb$l$\Verb$o$\Verb$n$\Verb$g$\Verb$ $\Verb$r$\Verb$e$\Verb$s$\Verb$u$\Verb$l$\Verb$t$\Verb$;$\Verb$}$\Verb$ $\Verb$t$\Verb$i$\Verb$m$\Verb$e$\Verb$_$\Verb$s$\Verb$t$\Verb$r$\Verb$u$\Verb$c$\Verb$t$\Verb$;$\\\Verb$t$\Verb$i$\Verb$m$\Verb$e$\Verb$_$\Verb$s$\Verb$t$\Verb$r$\Verb$u$\Verb$c$\Verb$t$\Verb$ $\Verb$t$\Verb$i$\Verb$m$\Verb$e$\Verb$s$\Verb$t$\Verb$a$\Verb$m$\Verb$p$\Verb$;$\\\Verb$s$\Verb$t$\Verb$r$\Verb$u$\Verb$c$\Verb$t$\Verb${$\Verb$i$\Verb$n$\Verb$t$\Verb$ $\Verb$k$\Verb$e$\Verb$y$\Verb$;$\Verb$l$\Verb$o$\Verb$n$\Verb$g$\Verb$ $\Verb$v$\Verb$a$\Verb$l$\Verb$u$\Verb$e$\Verb$;$\Verb$}$\Verb$ $\Verb$c$\Verb$a$\Verb$c$\Verb$h$\Verb$e$\Verb$[$\Verb$5$\Verb$1$\Verb$2$\Verb$]$\Verb$;$\\\Verb$l$\Verb$o$\Verb$n$\Verb$g$\Verb$ $\Verb$ $\Verb$m$\Verb$e$\Verb$m$\Verb$o$\Verb$[$\Verb$1$\Verb$0$\Verb$0$\Verb$0$\Verb$0$\Verb$0$\Verb$0$\Verb$]$\Verb$;$\\\Verb$t$\Verb$i$\Verb$m$\Verb$e$\Verb$_$\Verb$s$\Verb$t$\Verb$r$\Verb$u$\Verb$c$\Verb$t$\Verb$ $\Verb$m$\Verb$e$\Verb$m$\Verb$o$\Verb$i$\Verb$z$\Verb$e$\Verb$_$\Verb$s$\Verb$t$\Verb$a$\Verb$r$\Verb$t$\Verb$($\Verb$i$\Verb$n$\Verb$t$\Verb$ $\Verb$k$\Verb$e$\Verb$y$\Verb$)$\Verb${$\\\Verb$ $\Verb$ $\Verb$i$\Verb$f$\Verb$ $\Verb$($\Verb$c$\Verb$a$\Verb$c$\Verb$h$\Verb$e$\Verb$[$\Verb$k$\Verb$e$\Verb$y$\Verb$
\end{exambox}
}
\end{spacing}\vskip -0.4em

\newpage
\section{From \regreg{} back to $\mathrm{REG}$}
We establish a $reg$ functor. We use it to analyze \regreg{} expressions. 

A $reg$ functor assigns to each relativized regular expression $e$ a regular expression $reg(e)$. A $reg(e)$ satisfies approximation condition that if $e$ accepts $s$ then $reg(e)$ accepts $s$ but converse is not necessary true.

We can extract useful information testing if the intersection with a suitable regular language is empty. 

\vskip -0.3em \noindent\verb$  empty(e)           = reg(e    )$ $\mathtt{\cap}$\verb$ reg( ''   )$\\
         \verb$  first_char('c',e)  = reg(e    )$ $\mathtt{\cap}$\verb$ reg('c' .*)$\\
         \verb$  overlap(e1,e2)     = reg(e1 .*)$ $\mathtt{\cap}$\verb$ reg(e2  .*)$\\

If \verb$overlap(e1,e2)$ does not match anything then we can freely flip between \verb$e1|e2$ and \verb$e2|e1$. Also note that if this occurs then choice is deterministic and we do not have to backtrack if first alternative happens.

Mizushima \cite{memo} also transforms grammar to more deterministic one. We use stronger analysis. Using \verb$overlap$ we can determine where we can insert return states that inform \verb$Switch$ that next alternatives can not occur\footnote{We can also consider continuations for better results}. 

While bounds \verb$minsize(e)$, \verb$maxsize(e)$ on minimal and maximal sizes of string that matches \verb$e$ can be discovered by intersecting with suitable languages it is faster to compute them by dataflow analysis.

Functor $reg$ can be defined in the following way:\\
\verb$reg( 'c'                  ) = c$\\
\verb$reg( r                    ) = reg(r)$\\
\verb$reg( a* )                   = reg(a)*$\\
\verb$reg( nested(start,mid,end)) = reg(start) .* reg(end)$\\
\verb$reg( a   b )                = reg(a)   reg(b)$\\
\verb$reg( a | b )                = reg(a) | reg(b)$\\
\verb$reg(&a   b )                = reg(a) $ $\mathtt{\cap}$\verb$ reg(b)$

We use rough approximation of middle of \verb$nested$. In typical case inside nesting could be practically anything so trying to improve this approximation leads only to larger expressions without any new insights.

We shall remark that better result can be obtained by first using relativized regular machine and then converting to regular machine. This gives two advantages:\\
First is that \verb$Switch$ describes also lookaheads and we can describe intersection by lookahead.\\
Second is that we can use facts:\\
If \verb$A$ is unambiguous then \texttt{A B $\cap$ A C = A ( B $\cap$ C)}.\\
If \verb$A$ is unambiguous then \texttt{A B | A C = A ( B | C)}.

As there only finitely many $(continuation,cuts,stops)$ triples size of our machine is finite.

\newpage
\section{Problems of left recursion}\label{leftrec}
Left recursion handling deserves topic of its own. Various approaches were suggested and various counterexamples found.

In \peg{}  implementing left recursion correctly is an impossible task. Consider rule:

\begin{verbatim}
L = &( L 'cd'   ) 'abc' # a -> abc -> abcbc -> ab
  | &( L 'bcd'  ) 'ab'  #       ^               |
  | L 'bc'              #       |               V
  | L 'cb'              #     abcbcb     <-   abcb
  | 'a'
\end{verbatim}

On \extext{abcbcbcd}. 

It creates infinite cycle in the recursion. This problem is more fundamental as there is a paradox:

\begin{verbatim}
  L = ~L
\end{verbatim}

We reject such self references and raise an error when lookahead refers to possibly indirectly left recursive rule. Note that in boolean grammars same problem was recognized \cite{bool}.

Left recursion can be handled by recursive descend/ascend. A rule:\\
\verb$ L = L 'bc' | L 'c' | 'ab' | 'a'$ \\
on \extext{abc}
is recognized as \extext{(a(bc))} by recursive descend parser but as \extext{((ab)c)} by recursive ascend one.
All previous approaches in \peg{} and context-free bottom-up parser used a recursive ascend variant of left recursion.
A simplest algorithm is attributed to Paull \cite{aho}. It consist of rewriting direct left recursion to equivalent rule: \\
\verb$L = L a | b | L c | d        $ \\
\verb$L =      (b |       d) (a|c)*$. 

An indirect left recursion is removed by inlining and thus reducing to direct recursion case.

In 1965 Kuno \cite{kuno} suggested to limit recursion depth by $n$. It was rejected in \peg{} setting as in presence of semantic predicates some recursive rules need more than $n$ calls. Also it was not clear how handle infinite streams. But it was rejected prematurely. 

Using $reg$ functor (or simple dataflow) we can for each expression compute lower bound on minimal length of a string that matches that expression. Using this information we can easily estimate minimal size of current continuation. When this bound exceeds the length of our string we can fail.

 For infinite streams we can guess bound by guessing initially 1 and doubling bound when recursion could continue.  We do not use this approach as it has an exponential complexity in the worst case. 

Note that same technique can improve to Frost's algorithm \cite{frost}.

In packrat setting Ford used Paull algorithm to remove direct left recursion. He rejected to support left recursion with the following reason \cite{ford}: 

``At least until left recursion in TDPL is studied further, utilizing such a feature would amount
to opening a syntactic Pandora's Box, which clearly defeats the pragmatic purpose for which
the simple left recursion transformation is provided.''

Warth, Douglass, Millstein \cite{ometa} attempted to add runtime detection of left recursion. With bit of imagination it could be interpreted as doing Paull algorithm at runtime. However this approach has several flaws.

One discovered by Tratt \cite{tratt} is that seed growing introduces ambiguity of direct left recursion when right recursive alternative is also present.

A revised algorithm of Tratt still contains a flaw. Tratt at certain times forbids expansion of right recursion.

Tratt approach fails to handle right-recursive lookahead as the following counterexample shows.

\begin{verbatim}
  L = L 'a'
    | ~('b' L) 'b'
    | 'c'
\end{verbatim}

Third issue was discovered by Peter Goodman \cite{goodman}. Warth algorithm does not handle the following grammar.

\begin{verbatim}
A = A 'a'  / B
B = B 'b'  / A  / C
C = C 'c'  / B  / 'd'
\end{verbatim}

Medeiros in unpublished paper \cite{medeir} devised a revised version of seed growing algorithm.

One of possible advantages of seed growing could be support of higher order parametrized rules. In amethyst parser most of higher order rules are inlined making this point a moot one.

\subsection{Left recursion in \regreg{} parser}

We combine two techniques. First we just rewrite recursion by Paull algorithm. A second technique is that continuation passing style does implicit finite state machine minimization. This is simpler and leads to smaller grammars than Moore's left corner transform heuristic \cite{moore}.

We handle left recursion inside iteration by unrolling one level. 

With some bookkeeping we can transform left recursion to recursive descend. Idea is that each alternative returns its derivation and we choose a lexicographically smallest in recursion tree. This can be done in $O(1)$ time using dynamic lowest common ancestor \cite{lca}.

\newpage
\section{State handling} \label{handlingstate}
We show how to implement techniques for state handling that we described in section \ref{tamingstate}. 

We use a simple memoizing top-down \peg{} parser implemented in Ruby as an example. Implementation in amethyst uses similar ideas but intermixed with handling of other features.

\vskip -1.8em\begin{spacing}{0.8}
{\small
\begin{exambox}
\Verb$c$\Verb$l$\Verb$a$\Verb$s$\Verb$s$\Verb$ $\Verb$M$\Verb$a$\Verb$t$\Verb$c$\Verb$h$\\\Verb$ $\Verb$ $\Verb$[$\Verb$"$\Verb$s$\Verb$r$\Verb$c$\Verb$"$\Verb$,$\Verb$"$\Verb$p$\Verb$o$\Verb$s$\Verb$"$\Verb$,$\Verb$"$\Verb$r$\Verb$e$\Verb$s$\Verb$u$\Verb$l$\Verb$t$\Verb$"$\Verb$,$\Verb$"$\Verb$c$\Verb$o$\Verb$n$\Verb$t$\Verb$e$\Verb$x$\Verb$t$\Verb$u$\Verb$a$\Verb$l$\Verb$_$\Verb$a$\Verb$r$\Verb$g$\Verb$u$\Verb$m$\Verb$e$\Verb$n$\Verb$t$\Verb$s$\Verb$"$\Verb$,$\\\Verb$ $\Verb$ $\Verb$ $\Verb$"$\Verb$c$\Verb$o$\Verb$n$\Verb$t$\Verb$e$\Verb$x$\Verb$t$\Verb$u$\Verb$a$\Verb$l$\Verb$_$\Verb$r$\Verb$e$\Verb$t$\Verb$u$\Verb$r$\Verb$n$\Verb$s$\Verb$"$\Verb$,$\Verb$"$\Verb$l$\Verb$o$\Verb$c$\Verb$a$\Verb$l$\Verb$s$\Verb$"$\Verb$]$\Verb$.$\Verb$e$\Verb$a$\Verb$c$\Verb$h$\Verb${$\Verb$|$\Verb$n$\Verb$a$\Verb$m$\Verb$e$\Verb$|$\\\Verb$ $\Verb$ $\Verb$ $\Verb$ $\Verb$e$\Verb$v$\Verb$a$\Verb$l$\Verb$ $\Verb$"$\\\Verb$ $\Verb$ $\Verb$ $\Verb$ $\Verb$ $\Verb$ $\Verb$d$\Verb$e$\Verb$f$\Verb$ $\Verb$#$\Verb${$\Verb$n$\Verb$a$\Verb$m$\Verb$e$\Verb$}$\Verb$ $\Verb$($\Verb$ $\Verb$)$\Verb$;$\Verb$@$\Verb$h$\Verb$a$\Verb$s$\Verb$h$\Verb$[$\Verb$\$\Verb$"$\Verb$#$\Verb${$\Verb$n$\Verb$a$\Verb$m$\Verb$e$\Verb$}$\Verb$\$\Verb$"$\Verb$]$\Verb$ $\Verb$ $\Verb$;$\Verb$ $\Verb$e$\Verb$n$\Verb$d$\\\Verb$ $\Verb$ $\Verb$ $\Verb$ $\Verb$ $\Verb$ $\Verb$d$\Verb$e$\Verb$f$\Verb$ $\Verb$#$\Verb${$\Verb$n$\Verb$a$\Verb$m$\Verb$e$\Verb$}$\Verb$=$\Verb$($\Verb$v$\Verb$)$\Verb$;$\Verb$@$\Verb$h$\Verb$a$\Verb$s$\Verb$h$\Verb$[$\Verb$\$\Verb$"$\Verb$#$\Verb${$\Verb$n$\Verb$a$\Verb$m$\Verb$e$\Verb$}$\Verb$\$\Verb$"$\Verb$]$\Verb$=$\Verb$v$\Verb$;$\Verb$ $\Verb$e$\Verb$n$\Verb$d$\\\Verb$ $\Verb$ $\Verb$ $\Verb$ $\Verb$"$\\\Verb$ $\Verb$ $\Verb$}$\\\Verb$ $\Verb$ $\Verb$d$\Verb$e$\Verb$f$\Verb$ $\Verb$t$\Verb$i$\Verb$m$\Verb$e$\Verb$s$\Verb$t$\Verb$a$\Verb$m$\Verb$p$\Verb$ $\Verb$;$\Verb$ $\Verb$d$\Verb$e$\Verb$e$\Verb$p$\Verb$_$\Verb$c$\Verb$l$\Verb$o$\Verb$n$\Verb$e$\Verb$($\Verb$@$\Verb$h$\Verb$a$\Verb$s$\Verb$h$\Verb$)$\Verb$;$\Verb$ $\Verb$e$\Verb$n$\Verb$d$\\\Verb$ $\Verb$ $\Verb$d$\Verb$e$\Verb$f$\Verb$ $\Verb$r$\Verb$e$\Verb$v$\Verb$e$\Verb$r$\Verb$t$\Verb$($\Verb$t$\Verb$s$\Verb$)$\Verb$;$\Verb$ $\Verb$@$\Verb$h$\Verb$a$\Verb$s$\Verb$h$\Verb$=$\Verb$t$\Verb$s$\Verb$ $\Verb$ $\Verb$ $\Verb$ $\Verb$ $\Verb$ $\Verb$ $\Verb$ $\Verb$ $\Verb$;$\Verb$ $\Verb$e$\Verb$n$\Verb$d$\\\Verb$ $\Verb$ $\Verb$d$\Verb$e$\Verb$f$\Verb$ $\Verb$m$\Verb$e$\Verb$m$\Verb$o$\Verb$_$\Verb$i$\Verb$d$\Verb$;$\Verb$ $\Verb$[$\Verb$s$\Verb$r$\Verb$c$\Verb$,$\Verb$p$\Verb$o$\Verb$s$\Verb$,$\Verb$d$\Verb$e$\Verb$e$\Verb$p$\Verb$_$\Verb$c$\Verb$l$\Verb$o$\Verb$n$\Verb$e$\Verb$($\Verb$c$\Verb$o$\Verb$n$\Verb$t$\Verb$e$\Verb$x$\Verb$t$\Verb$u$\Verb$a$\Verb$l$\Verb$_$\Verb$a$\Verb$r$\Verb$g$\Verb$u$\Verb$m$\Verb$e$\Verb$n$\Verb$t$\Verb$s$\Verb$)$\Verb$]$\Verb$;$\Verb$ $\Verb$ $\Verb$e$\Verb$n$\Verb$d$\\\Verb$ $\Verb$ $\Verb$a$\Verb$t$\Verb$t$\Verb$r$\Verb$_$\Verb$a$\Verb$c$\Verb$c$\Verb$e$\Verb$s$\Verb$s$\Verb$o$\Verb$r$\Verb$ $\Verb$:$\Verb$m$\Verb$e$\Verb$m$\Verb$o$\Verb$i$\Verb$z$\Verb$e$\Verb$d$\Verb$ $\Verb$#$\Verb$w$\Verb$e$\Verb$ $\Verb$d$\Verb$o$\Verb$n$\Verb$'$\Verb$t$\Verb$ $\Verb$r$\Verb$e$\Verb$v$\Verb$e$\Verb$r$\Verb$t$\Verb$ $\Verb$m$\Verb$e$\Verb$m$\Verb$o$\Verb$ $\Verb$t$\Verb$a$\Verb$b$\Verb$l$\Verb$e$
\end{exambox}
}
\end{spacing}\vskip -0.4em

One could instead of deep cloning track what changes we did and revert them. This has same time complexity as without persistence because every revert will be payed by corresponding addition. 

\vskip -1.8em\begin{spacing}{0.8}
{\small
\begin{exambox}
\Verb$d$\Verb$e$\Verb$f$\Verb$ $\Verb$m$\Verb$a$\Verb$t$\Verb$c$\Verb$h$\Verb$($\Verb$e$\Verb$x$\Verb$p$\Verb$)$\\\Verb$ $\Verb$ $\Verb$c$\Verb$a$\Verb$s$\Verb$e$\Verb$ $\Verb$e$\Verb$x$\Verb$p$\\\Verb$ $\Verb$ $\Verb$w$\Verb$h$\Verb$e$\Verb$n$\Verb$ $\Verb$C$\Verb$a$\Verb$l$\Verb$l$\Verb$ $\Verb$ $\Verb$ $\Verb$ $\Verb$ $\Verb$i$\Verb$d$\Verb$=$\Verb$m$\Verb$e$\Verb$m$\Verb$o$\Verb$_$\Verb$i$\Verb$d$\Verb$ $\Verb$+$\Verb$ $\Verb$e$\Verb$x$\Verb$p$\Verb$.$\Verb$n$\Verb$a$\Verb$m$\Verb$e$\\\Verb$ $\Verb$ $\Verb$ $\Verb$ $\Verb$ $\Verb$ $\Verb$ $\Verb$ $\Verb$ $\Verb$ $\Verb$ $\Verb$ $\Verb$ $\Verb$ $\Verb$ $\Verb$ $\Verb$i$\Verb$f$\Verb$ $\Verb$!$\Verb$m$\Verb$e$\Verb$m$\Verb$o$\Verb$i$\Verb$z$\Verb$e$\Verb$d$\Verb$[$\Verb$i$\Verb$d$\Verb$]$\\\Verb$ $\Verb$ $\Verb$ $\Verb$ $\Verb$ $\Verb$ $\Verb$ $\Verb$ $\Verb$ $\Verb$ $\Verb$ $\Verb$ $\Verb$ $\Verb$ $\Verb$ $\Verb$ $\Verb$ $\Verb$ $\Verb$t$\Verb$s$\Verb$ $\Verb$=$\Verb$t$\Verb$i$\Verb$m$\Verb$e$\Verb$s$\Verb$t$\Verb$a$\Verb$m$\Verb$p$\\\Verb$ $\Verb$ $\Verb$ $\Verb$ $\Verb$ $\Verb$ $\Verb$ $\Verb$ $\Verb$ $\Verb$ $\Verb$ $\Verb$ $\Verb$ $\Verb$ $\Verb$ $\Verb$ $\Verb$ $\Verb$ $\Verb$l$\Verb$o$\Verb$c$\Verb$a$\Verb$l$\Verb$s$\Verb$,$\Verb$ $\Verb$c$\Verb$o$\Verb$n$\Verb$t$\Verb$e$\Verb$x$\Verb$t$\Verb$u$\Verb$a$\Verb$l$\Verb$_$\Verb$r$\Verb$e$\Verb$t$\Verb$u$\Verb$r$\Verb$n$\Verb$s$\Verb$ $\Verb$=$\Verb$ $\Verb${$\Verb$}$\Verb$ $\Verb$,$\Verb$ $\Verb${$\Verb$}$\\\Verb$ $\Verb$ $\Verb$ $\Verb$ $\Verb$ $\Verb$ $\Verb$ $\Verb$ $\Verb$ $\Verb$ $\Verb$ $\Verb$ $\Verb$ $\Verb$ $\Verb$ $\Verb$ $\Verb$ $\Verb$ $\Verb$r$\Verb$=$\Verb$m$\Verb$a$\Verb$t$\Verb$c$\Verb$h$\Verb$($\Verb+$+\Verb$r$\Verb$u$\Verb$l$\Verb$e$\Verb$s$\Verb$[$\Verb$e$\Verb$x$\Verb$p$\Verb$.$\Verb$n$\Verb$a$\Verb$m$\Verb$e$\Verb$]$\Verb$)$\\\Verb$ $\Verb$ $\Verb$ $\Verb$ $\Verb$ $\Verb$ $\Verb$ $\Verb$ $\Verb$ $\Verb$ $\Verb$ $\Verb$ $\Verb$ $\Verb$ $\Verb$ $\Verb$ $\Verb$ $\Verb$ $\Verb$m$\Verb$e$\Verb$m$\Verb$o$\Verb$i$\Verb$z$\Verb$e$\Verb$d$\Verb$[$\Verb$i$\Verb$d$\Verb$]$\Verb$=$\Verb$d$\Verb$e$\Verb$e$\Verb$p$\Verb$_$\Verb$f$\Verb$r$\Verb$e$\Verb$e$\Verb$z$\Verb$e$\Verb$($\Verb$c$\Verb$l$\Verb$o$\Verb$n$\Verb$e$\Verb$)$\\\Verb$ $\Verb$ $\Verb$ $\Verb$ $\Verb$ $\Verb$ $\Verb$ $\Verb$ $\Verb$ $\Verb$ $\Verb$ $\Verb$ $\Verb$ $\Verb$ $\Verb$ $\Verb$ $\Verb$ $\Verb$ $\Verb$r$\Verb$e$\Verb$v$\Verb$e$\Verb$r$\Verb$t$\Verb$($\Verb$t$\Verb$s$\Verb$)$\\\Verb$ $\Verb$ $\Verb$ $\Verb$ $\Verb$ $\Verb$ $\Verb$ $\Verb$ $\Verb$ $\Verb$ $\Verb$ $\Verb$ $\Verb$ $\Verb$ $\Verb$ $\Verb$ $\Verb$e$\Verb$n$\Verb$d$\\\Verb$ $\Verb$ $\Verb$ $\Verb$ $\Verb$ $\Verb$ $\Verb$ $\Verb$ $\Verb$ $\Verb$ $\Verb$ $\Verb$ $\Verb$ $\Verb$ $\Verb$ $\Verb$ $\Verb$c$\Verb$=$\Verb$m$\Verb$e$\Verb$m$\Verb$o$\Verb$i$\Verb$z$\Verb$e$\Verb$d$\Verb$[$\Verb$i$\Verb$d$\Verb$]$\\\Verb$ $\Verb$ $\Verb$ $\Verb$ $\Verb$ $\Verb$ $\Verb$ $\Verb$ $\Verb$ $\Verb$ $\Verb$ $\Verb$ $\Verb$ $\Verb$ $\Verb$ $\Verb$ $\Verb$r$\Verb$e$\Verb$s$\Verb$u$\Verb$l$\Verb$t$\Verb$,$\Verb$p$\Verb$o$\Verb$s$\Verb$=$\Verb$c$\Verb$.$\Verb$r$\Verb$e$\Verb$s$\Verb$u$\Verb$l$\Verb$t$\Verb$,$\Verb$c$\Verb$.$\Verb$p$\Verb$o$\Verb$s$\\\Verb$ $\Verb$ $\Verb$ $\Verb$ $\Verb$ $\Verb$ $\Verb$ $\Verb$ $\Verb$ $\Verb$ $\Verb$ $\Verb$ $\Verb$ $\Verb$ $\Verb$ $\Verb$ $\Verb$c$\Verb$.$\Verb$c$\Verb$o$\Verb$n$\Verb$t$\Verb$e$\Verb$x$\Verb$t$\Verb$u$\Verb$a$\Verb$l$\Verb$_$\Verb$r$\Verb$e$\Verb$t$\Verb$u$\Verb$r$\Verb$n$\Verb$s$\Verb$.$\Verb$e$\Verb$a$\Verb$c$\Verb$h$\Verb${$\Verb$|$\Verb$k$\Verb$,$\Verb$v$\Verb$|$\Verb$ $\\\Verb$ $\Verb$ $\Verb$ $\Verb$ $\Verb$ $\Verb$ $\Verb$ $\Verb$ $\Verb$ $\Verb$ $\Verb$ $\Verb$ $\Verb$ $\Verb$ $\Verb$ $\Verb$ $\Verb$ $\Verb$ $\Verb$c$\Verb$o$\Verb$n$\Verb$t$\Verb$e$\Verb$x$\Verb$t$\Verb$u$\Verb$a$\Verb$l$\Verb$_$\Verb$r$\Verb$e$\Verb$t$\Verb$u$\Verb$r$\Verb$n$\Verb$s$\Verb$[$\Verb$k$\Verb$]$\Verb$ $\Verb$<$\Verb$<$\Verb$ $\Verb$v$\\\Verb$ $\Verb$ $\Verb$ $\Verb$ $\Verb$ $\Verb$ $\Verb$ $\Verb$ $\Verb$ $\Verb$ $\Verb$ $\Verb$ $\Verb$ $\Verb$ $\Verb$ $\Verb$ $\Verb$}$\\\Verb$ $\Verb$ $\Verb$w$\Verb$h$\Verb$e$\Verb$n$\Verb$ $\Verb$C$\Verb$h$\Verb$a$\Verb$r$\Verb$;$\Verb$ $\Verb$ $\Verb$ $\Verb$ $\Verb$i$\Verb$f$\Verb$ $\Verb$s$\Verb$r$\Verb$c$\Verb$[$\Verb$p$\Verb$o$\Verb$s$\Verb$]$\Verb$=$\Verb$=$\Verb$e$\Verb$x$\Verb$p$\Verb$.$\Verb$c$\Verb$h$\Verb$a$\Verb$r$\Verb$ $\Verb$ $\Verb$ $\Verb$;$\Verb$ $\Verb$ $\Verb$r$\Verb$e$\Verb$s$\Verb$u$\Verb$l$\Verb$t$\Verb$=$\Verb$e$\Verb$x$\Verb$p$\Verb$.$\Verb$c$\Verb$h$\Verb$a$\Verb$r$\Verb$;$\Verb$ $\Verb$s$\Verb$r$\Verb$c$\Verb$+$\Verb$=$\Verb$1$\\\Verb$ $\Verb$ $\Verb$ $\Verb$ $\Verb$ $\Verb$ $\Verb$ $\Verb$ $\Verb$ $\Verb$ $\Verb$ $\Verb$ $\Verb$ $\Verb$ $\Verb$ $\Verb$ $\Verb$e$\Verb$l$\Verb$s$\Verb$e$\Verb$ $\Verb$ $\Verb$ $\Verb$ $\Verb$ $\Verb$ $\Verb$ $\Verb$ $\Verb$ $\Verb$ $\Verb$ $\Verb$ $\Verb$ $\Verb$ $\Verb$ $\Verb$ $\Verb$ $\Verb$ $\Verb$ $\Verb$ $\Verb$;$\Verb$ $\Verb$ $\Verb$r$\Verb$e$\Verb$s$\Verb$u$\Verb$l$\Verb$t$\Verb$=$\Verb$:$\Verb$f$\Verb$a$\Verb$i$\Verb$l$\\\Verb$ $\Verb$ $\Verb$ $\Verb$ $\Verb$ $\Verb$ $\Verb$ $\Verb$ $\Verb$ $\Verb$ $\Verb$ $\Verb$ $\Verb$ $\Verb$ $\Verb$ $\Verb$ $\Verb$e$\Verb$n$\Verb$d$\\\Verb$ $\Verb$ $\Verb$w$\Verb$h$\Verb$e$\Verb$n$\Verb$ $\Verb$S$\Verb$e$\Verb$q$\Verb$ $\Verb$ $\Verb$ $\Verb$ $\Verb$ $\Verb$ $\Verb$e$\Verb$x$\Verb$p$\Verb$.$\Verb$e$\Verb$a$\Verb$c$\Verb$h$\Verb${$\Verb$|$\Verb$s$\Verb$e$\Verb$q$\Verb$|$\\\Verb$ $\Verb$ $\Verb$ $\Verb$ $\Verb$ $\Verb$ $\Verb$ $\Verb$ $\Verb$ $\Verb$ $\Verb$ $\Verb$ $\Verb$ $\Verb$ $\Verb$ $\Verb$ $\Verb$ $\Verb$ $\Verb$i$\Verb$f$\Verb$ $\Verb$m$\Verb$a$\Verb$t$\Verb$c$\Verb$h$\Verb$($\Verb$s$\Verb$e$\Verb$q$\Verb$)$\Verb$ $\Verb$=$\Verb$=$\Verb$ $\Verb$:$\Verb$f$\Verb$a$\Verb$i$\Verb$l$\Verb$;$\Verb$ $\Verb$ $\Verb$r$\Verb$e$\Verb$t$\Verb$u$\Verb$r$\Verb$n$\Verb$ $\Verb$:$\Verb$f$\Verb$a$\Verb$i$\Verb$l$\Verb$ $\\\Verb$ $\Verb$ $\Verb$ $\Verb$ $\Verb$ $\Verb$ $\Verb$ $\Verb$ $\Verb$ $\Verb$ $\Verb$ $\Verb$ $\Verb$ $\Verb$ $\Verb$ $\Verb$ $\Verb$ $\Verb$ $\Verb$e$\Verb$n$\Verb$d$\\\Verb$ $\Verb$ $\Verb$ $\Verb$ $\Verb$ $\Verb$ $\Verb$ $\Verb$ $\Verb$ $\Verb$ $\Verb$ $\Verb$ $\Verb$ $\Verb$ $\Verb$ $\Verb$ $\Verb$}$\\\Verb$ $\Verb$ $\Verb$w$\Verb$h$\Verb$e$\Verb$n$\Verb$ $\Verb$O$\Verb$r$\Verb$ $\Verb$ $\Verb$ $\Verb$ $\Verb$ $\Verb$ $\Verb$ $\Verb$e$\Verb$x$\Verb$p$\Verb$.$\Verb$e$\Verb$a$\Verb$c$\Verb$h$\Verb${$\Verb$|$\Verb$a$\Verb$l$\Verb$t$\Verb$|$\\\Verb$ $\Verb$ $\Verb$ $\Verb$ $\Verb$ $\Verb$ $\Verb$ $\Verb$ $\Verb$ $\Verb$ $\Verb$ $\Verb$ $\Verb$ $\Verb$ $\Verb$ $\Verb$ $\Verb$ $\Verb$ $\Verb$t$\Verb$s$\Verb$ $\Verb$=$\Verb$ $\Verb$t$\Verb$i$\Verb$m$\Verb$e$\Verb$s$\Verb$t$\Verb$a$\Verb$m$\Verb$p$\\\Verb$ $\Verb$ $\Verb$ $\Verb$ $\Verb$ $\Verb$ $\Verb$ $\Verb$ $\Verb$ $\Verb$ $\Verb$ $\Verb$ $\Verb$ $\Verb$ $\Verb$ $\Verb$ $\Verb$ $\Verb$ $\Verb$i$\Verb$f$\Verb$ $\Verb$m$\Verb$a$\Verb$t$\Verb$c$\Verb$h$\Verb$($\Verb$a$\Verb$l$\Verb$t$\Verb$)$\Verb$ $\Verb$!$\Verb$=$\Verb$ $\Verb$:$\Verb$f$\Verb$a$\Verb$i$\Verb$l$\Verb$;$\Verb$ $\Verb$ $\Verb$r$\Verb$e$\Verb$t$\Verb$u$\Verb$r$\Verb$n$\Verb$ $\Verb$r$\Verb$e$\Verb$s$\Verb$u$\Verb$l$\Verb$t$\\\Verb$ $\Verb$ $\Verb$ $\Verb$ $\Verb$ $\Verb$ $\Verb$ $\Verb$ $\Verb$ $\Verb$ $\Verb$ $\Verb$ $\Verb$ $\Verb$ $\Verb$ $\Verb$ $\Verb$ $\Verb$ $\Verb$e$\Verb$l$\Verb$s$\Verb$e$\Verb$ $\Verb$ $\Verb$ $\Verb$ $\Verb$ $\Verb$ $\Verb$ $\Verb$ $\Verb$ $\Verb$ $\Verb$ $\Verb$ $\Verb$ $\Verb$ $\Verb$ $\Verb$ $\Verb$ $\Verb$ $\Verb$;$\Verb$ $\Verb$ $\Verb$r$\Verb$e$\Verb$v$\Verb$e$\Verb$r$\Verb$t$\Verb$($\Verb$t$\Verb$s$\Verb$)$\\\Verb$ $\Verb$ $\Verb$ $\Verb$ $\Verb$ $\Verb$ $\Verb$ $\Verb$ $\Verb$ $\Verb$ $\Verb$ $\Verb$ $\Verb$ $\Verb$ $\Verb$ $\Verb$ $\Verb$ $\Verb$ $\Verb$e$\Verb$n$\Verb$d$\\\Verb$ $\Verb$ $\Verb$ $\Verb$ $\Verb$ $\Verb$ $\Verb$ $\Verb$ $\Verb$ $\Verb$ $\Verb$ $\Verb$ $\Verb$ $\Verb$ $\Verb$ $\Verb$ $\Verb$}$\\\Verb$ $\Verb$ $\Verb$ $\Verb$ $\Verb$ $\Verb$ $\Verb$ $\Verb$ $\Verb$ $\Verb$ $\Verb$ $\Verb$ $\Verb$ $\Verb$ $\Verb$ $\Verb$ $\Verb$r$\Verb$e$\Verb$t$\Verb$u$\Verb$r$\Verb$n$\Verb$ $\Verb$:$\Verb$f$\Verb$a$\Verb$i$\Verb$l$\\\Verb$ $\Verb$ $\Verb$w$\Verb$h$\Verb$e$\Verb$n$\Verb$ $\Verb$A$\Verb$c$\Verb$t$\Verb$ $\Verb$ $\Verb$ $\Verb$ $\Verb$ $\Verb$ $\Verb$r$\Verb$e$\Verb$s$\Verb$u$\Verb$l$\Verb$t$\Verb$=$\Verb$e$\Verb$x$\Verb$p$\Verb$.$\Verb$c$\Verb$a$\Verb$l$\Verb$l$\Verb$($\Verb$s$\Verb$e$\Verb$l$\Verb$f$\Verb$)$\\\Verb$ $\Verb$ $\Verb$w$\Verb$h$\Verb$e$\Verb$n$\Verb$ $\Verb$B$\Verb$i$\Verb$n$\Verb$d$\Verb$ $\Verb$ $\Verb$ $\Verb$ $\Verb$ $\Verb$l$\Verb$o$\Verb$c$\Verb$a$\Verb$l$\Verb$s$\Verb$[$\Verb$e$\Verb$x$\Verb$p$\Verb$.$\Verb$n$\Verb$a$\Verb$m$\Verb$e$\Verb$]$\Verb$=$\Verb$r$\Verb$e$\Verb$s$\Verb$u$\Verb$l$\Verb$t$\\\Verb$ $\Verb$ $\Verb$e$\Verb$n$\Verb$d$\\\Verb$ $\Verb$ $\Verb$r$\Verb$e$\Verb$t$\Verb$u$\Verb$r$\Verb$n$\Verb$ $\Verb$r$\Verb$e$\Verb$s$\Verb$u$\Verb$l$\Verb$t$\\\Verb$e$\Verb$n$\Verb$d$
\end{exambox}
}
\end{spacing}\vskip -0.4em

\chapter{Dynamic parsing} \label{dynparsing}
Normal parser processes files in batch fashion. Amethyst allows dynamic parsing where the user is allowed to add and delete characters from string and query current parser output. 

Editors and IDE try to maintain syntax highlighting and error detection often in ad-hoc way.
Syntax highlighting typically uses regular expressions to determine meaning of text edited. This yields several problems, one is that regular expression can be confused with certain inputs. Other is updating regular expression for new versions of grammar. Dynamic parsing solves these problems in a robust way 

In this chapter we develop a generic way how transform memoizing top-down  parser to dynamic one. The update operation of the dynamic parser has the worst case time complexity $O(r \ln n)$ where $r$ is number of rules that need recomputed. For typical workloads running time is $O(r)$.  Our techniques can be applied to packrat parsers, parsing algorithm of Frost, and \regreg{} parser.

Main idea of our approach is to annotate memoized rule with an interval of input used to calculate it. We use a balanced tree to detect if this interval changed or not. After receiving an update to the input we update version of corresponding element. When we need to recalculate memoized rule we check if there was a change in its interval and recalculate it as necessary. 

\section{Interface to memoizing top-down parsers}
All three algorithms (\regreg{}, \peg{}, Frost's algorithm, ...) allows separation of memoization into independent module. In dynamic parsing this module serve as intermediary interface between user (IDE, editor, ...) and parser.

Our parser can be easily generalized to matching arrays of arbitrary type with no modications in algorithm.

A user interface consists of following four methods:

\noindent\begin{tabular}{ l  l}
\verb$chr(p)  $ & Character at position p\\
\verb$ins(p,c) $ & Insert character c at position p\\
\verb$del(p)   $ & Delete character at position p\\
\verb$parse    $ & Return result of parser\\
\end{tabular}

Interface with parser is more interesting. A parser can access string only by pointers that always point to same character regardless of modifications.

\noindent\begin{tabular}{ l  l}
\verb$char(ptr)$            & Value of current character.\\
\verb$next(ptr)$            & Next character.\\
\verb$get_memo(rule,ptr)$ & Returns memoized value.\\
\verb$set_memo(rule,ptr,value)$ & Memoizies given value.\\
\end{tabular}

We assume that parser calls \verb$get_memo$ calls on entering rule and \verb$set_memo$ on exiting rule. Parser does not have to call \verb$set_memo$ if it decides not to memoize a rule.
\section{Data structure}
We present data structures for $O(r \ln n)$ time bound. 

Our structure is a balanced tree. We maintain several properties:

\noindent\begin{tabular}{ l  l}
\verb$value$& value of character\\
\verb$sons $& number of sons in subtree\\
\verb$ts   $& timestamp\\
\verb$maxts$& maximal timestamp of son nodes\\
\verb$memoized$&Memoization entries starting at this position.\\
\end{tabular}

We use timestamp that increases after call of each \verb$parse$. Each insertion/deletion gets assigned this timestamp

Our implementation needs several auxiliary methods:

\noindent\begin{tabular}{ l  l}
\verb$timestamp(first,last)$& maximal timestamp in interval specified by first and last.\\
\verb$ index(ptr)$& position of character in current string.\\
\verb$rindex(n)$ & pointer to n-th character in current string.\\
\end{tabular}

Writing balanced tree that supports \verb$chr,ins,del,timestamp,index,rindex$ methods in $O(\ln n)$ while maintaining properties above is typical homework exercise. Note that queries that we made exhibit spatial locality.  Thus a splay tree \cite{splaytree} looks like good candidate to obtain $O(1)$ running time in practice. 

Implementing our data structure as tree with node for each character is unpractical. Observe that order in which we modify string between calls to parse method is not important.
We can modify our data structure to rope data structure \cite{rope} with property that leaf substrings have same timestamp. When splitting substring we need also update table entries. As character can participate in at most $O(\ln n)$ splits amortized $O(\ln n)$ time complexity still holds.

There is a technical problem with the deletion. We need to save somewhere that deletion occurred. We keep nodes that contain no character for this purpose. Luckily we can always merge two empty adjacent nodes into one. It is easy to see that number of empty nodes will be at most the number of nonempty ones.
\newpage
\section{Algorithm}
We will present two implementations. First implementation is an extension of amethyst that overwrites '.' operator. Second is in pseudocode that serves as overview of the effective \verb$C$ implementation from \verb$lib/dynamic$  subdirectory of the amethyst project.
\vskip -1.8em\begin{spacing}{0.8}
{\small
\begin{exambox}
\Verb$a$\Verb$m$\Verb$e$\Verb$t$\Verb$h$\Verb$y$\Verb$s$\Verb$t$\Verb$ $\Verb$D$\Verb$y$\Verb$n$\Verb$a$\Verb$m$\Verb$i$\Verb$c$\Verb$ $\Verb${$\\\Verb$ $\Verb$ ${\color{red}\Verb$i$}{\color{red}\Verb$n$}{\color{red}\Verb$i$}{\color{red}\Verb$t$}\Verb$ $\Verb$ $\Verb$=$\Verb$ ${\color{Tan}\Verb${$}{\color{Tan}\Verb$ $}{\color{Aquamarine}\Verb$@$}{\color{Aquamarine}\Verb$@$}{\color{Aquamarine}\Verb$m$}{\color{Aquamarine}\Verb$e$}{\color{Aquamarine}\Verb$m$}{\color{Aquamarine}\Verb$o$}{\color{Aquamarine}\Verb$i$}{\color{Aquamarine}\Verb$z$}{\color{Aquamarine}\Verb$e$}{\color{Aquamarine}\Verb$d$}{\color{Tan}\Verb$=$}{\color{Tan}\Verb${$}{\color{Tan}\Verb$}$}{\color{Tan}\Verb$;$}{\color{Tan}\Verb$ $}{\color{Aquamarine}\Verb$@$}{\color{Aquamarine}\Verb$@$}{\color{Aquamarine}\Verb$r$}{\color{Aquamarine}\Verb$i$}{\color{Aquamarine}\Verb$g$}{\color{Aquamarine}\Verb$h$}{\color{Aquamarine}\Verb$t$}{\color{Aquamarine}\Verb$m$}{\color{Aquamarine}\Verb$o$}{\color{Aquamarine}\Verb$s$}{\color{Aquamarine}\Verb$t$}{\color{Tan}\Verb$=$}{\color{Tan}\Verb$0$}{\color{Tan}\Verb$}$}\\\Verb$ $\Verb$ ${\color{red}\Verb$i$}{\color{red}\Verb$f$}{\color{green}\Verb$($}{\color{green}\Verb$x$}{\color{green}\Verb$)$}\Verb$ $\Verb$=$\Verb$ ${\color{Violet}\Verb$&$}{\color{Violet}\Verb${$}{\color{Tan}\Verb$x$}{\color{Violet}\Verb$}$}\\\\\Verb$ $\Verb$ ${\color{red}\Verb$m$}{\color{red}\Verb$e$}{\color{red}\Verb$m$}{\color{red}\Verb$o$}{\color{green}\Verb$($}{\color{green}\Verb$r$}{\color{green}\Verb$u$}{\color{green}\Verb$l$}{\color{green}\Verb$e$}{\color{green}\Verb$)$}\Verb$ $\Verb$=$\Verb$ $\\\Verb$ $\Verb$ $\Verb$ $\Verb$ $\Verb$ $\Verb$ ${\color{Tan}\Verb${$}{\color{Tan}\Verb$i$}{\color{Tan}\Verb$d$}{\color{Tan}\Verb$($}{\color{Tan}\Verb$r$}{\color{Tan}\Verb$u$}{\color{Tan}\Verb$l$}{\color{Tan}\Verb$e$}{\color{Tan}\Verb$,$}{\color{Tan}\Verb$s$}{\color{Tan}\Verb$r$}{\color{Tan}\Verb$c$}{\color{Tan}\Verb$,$}{\color{Tan}\Verb$p$}{\color{Tan}\Verb$o$}{\color{Tan}\Verb$s$}{\color{Tan}\Verb$)$}{\color{Tan}\Verb$}$}{\color{Aquamarine}\Verb$:$}{\color{Aquamarine}\Verb$i$}{\color{Aquamarine}\Verb$d$}\Verb$ $\\\Verb$ $\Verb$ $\Verb$ $\Verb$ $\Verb$ $\Verb$ ${\color{Tan}\Verb${$}{\color{Tan}\Verb$p$}{\color{Tan}\Verb$o$}{\color{Tan}\Verb$s$}{\color{Tan}\Verb$}$}{\color{Aquamarine}\Verb$:$}{\color{Aquamarine}\Verb$o$}{\color{Aquamarine}\Verb$l$}{\color{Aquamarine}\Verb$d$}{\color{Aquamarine}\Verb$p$}{\color{Aquamarine}\Verb$o$}{\color{Aquamarine}\Verb$s$}\Verb$ $\\\Verb$ $\Verb$ $\Verb$ $\Verb$ $\Verb$ $\Verb$ ${\color{Tan}\Verb${$}{\color{Aquamarine}\Verb$@$}{\color{Aquamarine}\Verb$@$}{\color{Aquamarine}\Verb$r$}{\color{Aquamarine}\Verb$i$}{\color{Aquamarine}\Verb$g$}{\color{Aquamarine}\Verb$h$}{\color{Aquamarine}\Verb$t$}{\color{Aquamarine}\Verb$m$}{\color{Aquamarine}\Verb$o$}{\color{Aquamarine}\Verb$s$}{\color{Aquamarine}\Verb$t$}{\color{Tan}\Verb$}$}{\color{Aquamarine}\Verb$:$}{\color{Aquamarine}\Verb$o$}{\color{Aquamarine}\Verb$l$}{\color{Aquamarine}\Verb$d$}{\color{Aquamarine}\Verb$r$}{\color{Aquamarine}\Verb$i$}{\color{Aquamarine}\Verb$g$}{\color{Aquamarine}\Verb$h$}{\color{Aquamarine}\Verb$t$}\Verb$ $\\\Verb$ $\Verb$ $\Verb$ $\Verb$ $\Verb$ $\Verb$ ${\color{Tan}\Verb${$}{\color{Tan}\Verb$p$}{\color{Tan}\Verb$o$}{\color{Tan}\Verb$s$}{\color{Tan}\Verb$}$}{\color{Aquamarine}\Verb$:$}{\color{Aquamarine}\Verb$@$}{\color{Aquamarine}\Verb$@$}{\color{Aquamarine}\Verb$r$}{\color{Aquamarine}\Verb$i$}{\color{Aquamarine}\Verb$g$}{\color{Aquamarine}\Verb$h$}{\color{Aquamarine}\Verb$t$}{\color{Aquamarine}\Verb$m$}{\color{Aquamarine}\Verb$o$}{\color{Aquamarine}\Verb$s$}{\color{Aquamarine}\Verb$t$}\\\Verb$ $\Verb$ $\Verb$ $\Verb$ $\Verb$ $\Verb$ $\Verb$($\Verb$ ${\color{red}\Verb$i$}{\color{red}\Verb$f$}{\color{green}\Verb$($}{\color{green}\Verb$!$}{\color{green}\Verb$m$}{\color{green}\Verb$e$}{\color{green}\Verb$m$}{\color{green}\Verb$o$}{\color{green}\Verb$i$}{\color{green}\Verb$z$}{\color{green}\Verb$e$}{\color{green}\Verb$d$}{\color{green}\Verb$[$}{\color{Aquamarine}\Verb$i$}{\color{Aquamarine}\Verb$d$}{\color{green}\Verb$]$}{\color{green}\Verb$)$}\Verb$ $\\\Verb$ $\Verb$ $\Verb$ $\Verb$ $\Verb$ $\Verb$ $\Verb$ $\Verb$ $\Verb$ $\Verb$ $\Verb$(${\color{red}\Verb$a$}{\color{red}\Verb$p$}{\color{red}\Verb$p$}{\color{red}\Verb$l$}{\color{red}\Verb$y$}{\color{green}\Verb$($}{\color{green}\Verb$r$}{\color{green}\Verb$u$}{\color{green}\Verb$l$}{\color{green}\Verb$e$}{\color{green}\Verb$)$}\Verb$ $\Verb$|$\Verb$ ${\color{Tan}\Verb${$}{\color{Tan}\Verb$"$}{\color{Tan}\Verb$f$}{\color{Tan}\Verb$a$}{\color{Tan}\Verb$i$}{\color{Tan}\Verb$l$}{\color{Tan}\Verb$e$}{\color{Tan}\Verb$d$}{\color{Tan}\Verb$"$}{\color{Tan}\Verb$}$}\Verb$)${\color{Aquamarine}\Verb$:$}{\color{Aquamarine}\Verb$r$}{\color{Aquamarine}\Verb$e$}{\color{Aquamarine}\Verb$s$}{\color{Aquamarine}\Verb$u$}{\color{Aquamarine}\Verb$l$}{\color{Aquamarine}\Verb$t$}\Verb$ $\Verb$ $\Verb$ $\Verb$ $\Verb$ $\Verb$ $\Verb$ $\Verb$ $\Verb$ $\Verb$ $\Verb$ $\Verb$ $\Verb$ $\Verb$ $\Verb$ $\Verb$ $\Verb$ $\Verb$ $\\\Verb$ $\Verb$ $\Verb$ $\Verb$ $\Verb$ $\Verb$ $\Verb$ $\Verb$ $\Verb$ $\Verb$ ${\color{Tan}\Verb${$}{\color{Tan}\Verb$ $}{\color{Tan}\Verb$m$}{\color{Tan}\Verb$e$}{\color{Tan}\Verb$m$}{\color{Tan}\Verb$o$}{\color{Tan}\Verb$i$}{\color{Tan}\Verb$z$}{\color{Tan}\Verb$e$}{\color{Tan}\Verb$d$}{\color{Tan}\Verb$[$}{\color{Aquamarine}\Verb$i$}{\color{Aquamarine}\Verb$d$}{\color{Tan}\Verb$]$}{\color{Tan}\Verb$=$}{\color{Tan}\Verb$M$}{\color{Tan}\Verb$e$}{\color{Tan}\Verb$m$}{\color{Tan}\Verb$o$}{\color{Tan}\Verb$[$}{\color{Tan}\Verb$p$}{\color{Tan}\Verb$o$}{\color{Tan}\Verb$s$}{\color{Tan}\Verb$-$}{\color{Aquamarine}\Verb$o$}{\color{Aquamarine}\Verb$l$}{\color{Aquamarine}\Verb$d$}{\color{Aquamarine}\Verb$p$}{\color{Aquamarine}\Verb$o$}{\color{Aquamarine}\Verb$s$}{\color{Tan}\Verb$,$}{\color{Aquamarine}\Verb$@$}{\color{Aquamarine}\Verb$@$}{\color{Aquamarine}\Verb$r$}{\color{Aquamarine}\Verb$i$}{\color{Aquamarine}\Verb$g$}{\color{Aquamarine}\Verb$h$}{\color{Aquamarine}\Verb$t$}{\color{Aquamarine}\Verb$m$}{\color{Aquamarine}\Verb$o$}{\color{Aquamarine}\Verb$s$}{\color{Aquamarine}\Verb$t$}{\color{Tan}\Verb$-$}{\color{Aquamarine}\Verb$o$}{\color{Aquamarine}\Verb$l$}{\color{Aquamarine}\Verb$d$}{\color{Aquamarine}\Verb$p$}{\color{Aquamarine}\Verb$o$}{\color{Aquamarine}\Verb$s$}{\color{Tan}\Verb$,$}{\color{Aquamarine}\Verb$r$}{\color{Aquamarine}\Verb$e$}{\color{Aquamarine}\Verb$s$}{\color{Aquamarine}\Verb$u$}{\color{Aquamarine}\Verb$l$}{\color{Aquamarine}\Verb$t$}{\color{Tan}\Verb$]$}{\color{Tan}\Verb$ $}{\color{Tan}\Verb$}$}\\\Verb$ $\Verb$ $\Verb$ $\Verb$ $\Verb$ $\Verb$ $\Verb$ $\Verb$ $\Verb$|$\Verb$ ${\color{Tan}\Verb$-$}{\color{Tan}\Verb$>$}{\color{Tan}\Verb$ $}{\color{Tan}\Verb$n$}{\color{Tan}\Verb$i$}{\color{Tan}\Verb$l$}{\color{Tan}\\}{\color{Tan}\Verb$ $}\Verb$ $\Verb$ $\Verb$ $\Verb$ $\Verb$ $\Verb$)$\\\Verb$ $\Verb$ $\Verb$ $\Verb$ $\Verb$ $\Verb$ ${\color{Tan}\Verb${$}{\color{Tan}\Verb$ $}{\color{Tan}\Verb$p$}{\color{Tan}\Verb$o$}{\color{Tan}\Verb$s$}{\color{Tan}\Verb$=$}{\color{Aquamarine}\Verb$o$}{\color{Aquamarine}\Verb$l$}{\color{Aquamarine}\Verb$d$}{\color{Aquamarine}\Verb$p$}{\color{Aquamarine}\Verb$o$}{\color{Aquamarine}\Verb$s$}{\color{Tan}\Verb$+$}{\color{Tan}\Verb$m$}{\color{Tan}\Verb$e$}{\color{Tan}\Verb$m$}{\color{Tan}\Verb$o$}{\color{Tan}\Verb$i$}{\color{Tan}\Verb$z$}{\color{Tan}\Verb$e$}{\color{Tan}\Verb$d$}{\color{Tan}\Verb$[$}{\color{Aquamarine}\Verb$i$}{\color{Aquamarine}\Verb$d$}{\color{Tan}\Verb$]$}{\color{Tan}\Verb$.$}{\color{Tan}\Verb$a$}{\color{Tan}\Verb$d$}{\color{Tan}\Verb$v$}{\color{Tan}\Verb$a$}{\color{Tan}\Verb$n$}{\color{Tan}\Verb$c$}{\color{Tan}\Verb$e$}{\color{Tan}\\}{\color{Tan}\Verb$ $}{\color{Tan}\Verb$ $}{\color{Tan}\Verb$ $}{\color{Tan}\Verb$ $}{\color{Tan}\Verb$ $}{\color{Tan}\Verb$ $}{\color{Tan}\Verb$ $}{\color{Tan}\Verb$ $}{\color{Aquamarine}\Verb$@$}{\color{Aquamarine}\Verb$@$}{\color{Aquamarine}\Verb$r$}{\color{Aquamarine}\Verb$i$}{\color{Aquamarine}\Verb$g$}{\color{Aquamarine}\Verb$h$}{\color{Aquamarine}\Verb$t$}{\color{Aquamarine}\Verb$m$}{\color{Aquamarine}\Verb$o$}{\color{Aquamarine}\Verb$s$}{\color{Aquamarine}\Verb$t$}{\color{Tan}\Verb$=$}{\color{Tan}\Verb$m$}{\color{Tan}\Verb$a$}{\color{Tan}\Verb$x$}{\color{Tan}\Verb$($}{\color{Aquamarine}\Verb$o$}{\color{Aquamarine}\Verb$l$}{\color{Aquamarine}\Verb$d$}{\color{Aquamarine}\Verb$r$}{\color{Aquamarine}\Verb$i$}{\color{Aquamarine}\Verb$g$}{\color{Aquamarine}\Verb$h$}{\color{Aquamarine}\Verb$t$}{\color{Tan}\Verb$,$}{\color{Tan}\Verb$p$}{\color{Tan}\Verb$o$}{\color{Tan}\Verb$s$}{\color{Tan}\Verb$+$}{\color{Tan}\Verb$m$}{\color{Tan}\Verb$e$}{\color{Tan}\Verb$m$}{\color{Tan}\Verb$o$}{\color{Tan}\Verb$i$}{\color{Tan}\Verb$z$}{\color{Tan}\Verb$e$}{\color{Tan}\Verb$d$}{\color{Tan}\Verb$[$}{\color{Aquamarine}\Verb$i$}{\color{Aquamarine}\Verb$d$}{\color{Tan}\Verb$]$}{\color{Tan}\Verb$.$}{\color{Tan}\Verb$r$}{\color{Tan}\Verb$m$}{\color{Tan}\Verb$_$}{\color{Tan}\Verb$a$}{\color{Tan}\Verb$d$}{\color{Tan}\Verb$v$}{\color{Tan}\Verb$a$}{\color{Tan}\Verb$n$}{\color{Tan}\Verb$c$}{\color{Tan}\Verb$e$}{\color{Tan}\Verb$)$}{\color{Tan}\\}{\color{Tan}\Verb$ $}{\color{Tan}\Verb$ $}{\color{Tan}\Verb$ $}{\color{Tan}\Verb$ $}{\color{Tan}\Verb$ $}{\color{Tan}\Verb$ $}{\color{Tan}\Verb$}$}\\\Verb$ $\Verb$ $\Verb$ $\Verb$ $\Verb$ $\Verb$ $\Verb$($\Verb$ ${\color{red}\Verb$i$}{\color{red}\Verb$f$}{\color{green}\Verb$($}{\color{green}\Verb$m$}{\color{green}\Verb$e$}{\color{green}\Verb$m$}{\color{green}\Verb$o$}{\color{green}\Verb$i$}{\color{green}\Verb$z$}{\color{green}\Verb$e$}{\color{green}\Verb$d$}{\color{green}\Verb$[$}{\color{Aquamarine}\Verb$i$}{\color{Aquamarine}\Verb$d$}{\color{green}\Verb$]$}{\color{green}\Verb$.$}{\color{green}\Verb$r$}{\color{green}\Verb$e$}{\color{green}\Verb$s$}{\color{green}\Verb$u$}{\color{green}\Verb$l$}{\color{green}\Verb$t$}{\color{green}\Verb$=$}{\color{green}\Verb$=$}{\color{green}\Verb$"$}{\color{green}\Verb$f$}{\color{green}\Verb$a$}{\color{green}\Verb$i$}{\color{green}\Verb$l$}{\color{green}\Verb$e$}{\color{green}\Verb$d$}{\color{green}\Verb$"$}{\color{green}\Verb$)$}\Verb$ $\Verb$ $\Verb$ ${\color{red}\Verb$f$}{\color{red}\Verb$a$}{\color{red}\Verb$i$}{\color{red}\Verb$l$}\\\Verb$ $\Verb$ $\Verb$ $\Verb$ $\Verb$ $\Verb$ $\Verb$ $\Verb$ $\Verb$|$\Verb$ $\Verb$ $\Verb$ $\Verb$ $\Verb$ $\Verb$ $\Verb$ $\Verb$ $\Verb$ $\Verb$ $\Verb$ $\Verb$ $\Verb$ $\Verb$ $\Verb$ $\Verb$ $\Verb$ $\Verb$ $\Verb$ $\Verb$ $\Verb$ $\Verb$ $\Verb$ $\Verb$ $\Verb$ $\Verb$ $\Verb$ $\Verb$ $\Verb$ $\Verb$ $\Verb$ $\Verb$ $\Verb$ $\Verb$ $\Verb$ ${\color{Tan}\Verb$-$}{\color{Tan}\Verb$>$}{\color{Tan}\Verb$ $}{\color{Tan}\Verb$m$}{\color{Tan}\Verb$e$}{\color{Tan}\Verb$m$}{\color{Tan}\Verb$o$}{\color{Tan}\Verb$i$}{\color{Tan}\Verb$z$}{\color{Tan}\Verb$e$}{\color{Tan}\Verb$d$}{\color{Tan}\Verb$[$}{\color{Aquamarine}\Verb$i$}{\color{Aquamarine}\Verb$d$}{\color{Tan}\Verb$]$}{\color{Tan}\Verb$.$}{\color{Tan}\Verb$r$}{\color{Tan}\Verb$e$}{\color{Tan}\Verb$s$}{\color{Tan}\Verb$u$}{\color{Tan}\Verb$l$}{\color{Tan}\Verb$t$}{\color{Tan}\\}{\color{Tan}\Verb$ $}\Verb$ $\Verb$ $\Verb$ $\Verb$ $\Verb$ $\Verb$)$\\\\\Verb$ $\Verb$ ${\color{red}\Verb$a$}{\color{red}\Verb$n$}{\color{red}\Verb$y$}{\color{red}\Verb$t$}{\color{red}\Verb$h$}{\color{red}\Verb$i$}{\color{red}\Verb$n$}{\color{red}\Verb$g$}\Verb$ $\Verb$=$\Verb$ ${\color{red}\Verb$i$}{\color{red}\Verb$f$}{\color{green}\Verb$($}{\color{green}\Verb$p$}{\color{green}\Verb$o$}{\color{green}\Verb$s$}{\color{green}\Verb$>$}{\color{green}\Verb$=$}{\color{green}\Verb$l$}{\color{green}\Verb$e$}{\color{green}\Verb$n$}{\color{green}\Verb$)$}\Verb$ ${\color{red}\Verb$f$}{\color{red}\Verb$a$}{\color{red}\Verb$i$}{\color{red}\Verb$l$}\\\Verb$ $\Verb$ $\Verb$ $\Verb$ $\Verb$ $\Verb$ $\Verb$ $\Verb$ $\Verb$ $\Verb$ $\Verb$ $\Verb$|$\Verb$ ${\color{Tan}\Verb${$}{\color{Aquamarine}\Verb$@$}{\color{Aquamarine}\Verb$@$}{\color{Aquamarine}\Verb$r$}{\color{Aquamarine}\Verb$i$}{\color{Aquamarine}\Verb$g$}{\color{Aquamarine}\Verb$h$}{\color{Aquamarine}\Verb$t$}{\color{Aquamarine}\Verb$m$}{\color{Aquamarine}\Verb$o$}{\color{Aquamarine}\Verb$s$}{\color{Aquamarine}\Verb$t$}{\color{Tan}\Verb$=$}{\color{Tan}\Verb$m$}{\color{Tan}\Verb$a$}{\color{Tan}\Verb$x$}{\color{Tan}\Verb$($}{\color{Aquamarine}\Verb$@$}{\color{Aquamarine}\Verb$@$}{\color{Aquamarine}\Verb$r$}{\color{Aquamarine}\Verb$i$}{\color{Aquamarine}\Verb$g$}{\color{Aquamarine}\Verb$h$}{\color{Aquamarine}\Verb$t$}{\color{Aquamarine}\Verb$m$}{\color{Aquamarine}\Verb$o$}{\color{Aquamarine}\Verb$s$}{\color{Aquamarine}\Verb$t$}{\color{Tan}\Verb$,$}{\color{Tan}\Verb$p$}{\color{Tan}\Verb$o$}{\color{Tan}\Verb$s$}{\color{Tan}\Verb$)$}{\color{Tan}\Verb$;$}{\color{Tan}\Verb$p$}{\color{Tan}\Verb$o$}{\color{Tan}\Verb$s$}{\color{Tan}\Verb$=$}{\color{Tan}\Verb$p$}{\color{Tan}\Verb$o$}{\color{Tan}\Verb$s$}{\color{Tan}\Verb$+$}{\color{Tan}\Verb$1$}{\color{Tan}\Verb$}$}\Verb$ ${\color{Tan}\Verb$-$}{\color{Tan}\Verb$>$}{\color{Tan}\Verb$ $}{\color{Tan}\Verb$s$}{\color{Tan}\Verb$r$}{\color{Tan}\Verb$c$}{\color{Tan}\Verb$[$}{\color{Tan}\Verb$p$}{\color{Tan}\Verb$o$}{\color{Tan}\Verb$s$}{\color{Tan}\Verb$-$}{\color{Tan}\Verb$1$}{\color{Tan}\Verb$]$}{\color{Tan}\\}{\color{Tan}\Verb$ $}\Verb$ ${\color{gray}\Verb$#$}{\color{gray}\Verb$s$}{\color{gray}\Verb$e$}{\color{gray}\Verb$q$}{\color{gray}\Verb$ $}{\color{gray}\Verb$i$}{\color{gray}\Verb$s$}{\color{gray}\Verb$ $}{\color{gray}\Verb$a$}{\color{gray}\Verb$n$}{\color{gray}\Verb$a$}{\color{gray}\Verb$l$}{\color{gray}\Verb$o$}{\color{gray}\Verb$g$}{\color{gray}\Verb$o$}{\color{gray}\Verb$u$}{\color{gray}\Verb$s$}{\color{gray}\\}\Verb$}$\\\Verb$c$\Verb$l$\Verb$a$\Verb$s$\Verb$s$\Verb$ $\Verb$M$\Verb$e$\Verb$m$\Verb$o$\\\Verb$ $\Verb$ $\Verb$a$\Verb$t$\Verb$t$\Verb$r$\Verb$_$\Verb$a$\Verb$c$\Verb$c$\Verb$e$\Verb$s$\Verb$s$\Verb$o$\Verb$r$\Verb$ $\Verb$:$\Verb$a$\Verb$d$\Verb$v$\Verb$a$\Verb$n$\Verb$c$\Verb$e$\Verb$,$\Verb$:$\Verb$r$\Verb$a$\Verb$d$\Verb$v$\Verb$a$\Verb$n$\Verb$c$\Verb$e$\Verb$,$\Verb$:$\Verb$r$\Verb$e$\Verb$s$\Verb$u$\Verb$l$\Verb$t$\\\Verb$ $\Verb$ $\Verb$d$\Verb$e$\Verb$f$\Verb$ $\Verb$s$\Verb$e$\Verb$l$\Verb$f$\Verb$.$\Verb$[$\Verb$]$\Verb$($\Verb$a$\Verb$d$\Verb$v$\Verb$a$\Verb$n$\Verb$c$\Verb$e$\Verb$,$\Verb$r$\Verb$a$\Verb$d$\Verb$v$\Verb$a$\Verb$n$\Verb$c$\Verb$e$\Verb$,$\Verb$r$\Verb$e$\Verb$s$\Verb$u$\Verb$l$\Verb$t$\Verb$)$\\\Verb$ $\Verb$ $\Verb$ $\Verb$ $\Verb$m$\Verb$=$\Verb$M$\Verb$e$\Verb$m$\Verb$o$\Verb$.$\Verb$n$\Verb$e$\Verb$w$\\\Verb$ $\Verb$ $\Verb$ $\Verb$ $\Verb$m$\Verb$.$\Verb$a$\Verb$d$\Verb$v$\Verb$a$\Verb$n$\Verb$c$\Verb$e$\Verb$,$\Verb$m$\Verb$.$\Verb$r$\Verb$a$\Verb$d$\Verb$v$\Verb$a$\Verb$n$\Verb$c$\Verb$e$\Verb$,$\Verb$m$\Verb$.$\Verb$r$\Verb$e$\Verb$s$\Verb$u$\Verb$l$\Verb$t$\Verb$=$\Verb$a$\Verb$d$\Verb$v$\Verb$a$\Verb$n$\Verb$c$\Verb$e$\Verb$,$\Verb$r$\Verb$a$\Verb$d$\Verb$v$\Verb$a$\Verb$n$\Verb$c$\Verb$e$\Verb$,$\Verb$r$\Verb$e$\Verb$s$\Verb$u$\Verb$l$\Verb$t$\\\Verb$ $\Verb$ $\Verb$ $\Verb$ $\Verb$m$\\\Verb$ $\Verb$ $\Verb$e$\Verb$n$\Verb$d$\\\Verb$e$\Verb$n$\Verb$d$
\end{exambox}
}
\end{spacing}\vskip -0.4em

Our algorithm maintains stack that mirrors a call stack of parser. For each rule we find a rightmost position that can affect result of rule. 

\begin{verbatim}
stack_struct stack 
stack_push(rule,ptr){
  stack.push
  stack.top.rule=rule
  stack.top.ptr =ptr
  stack.top.last=ptr
}
stack_pop(rule,ptr){
  last=stack.top.last
  stack.pop
  update_last(last)
}


update_last(ptr){
  if (index(ptr)>index(stack.top.last))
    stack.top.last=ptr
}
\end{verbatim}

There is technical issue that saving rightmost position as pointer is unwieldy. Instead we represent rightmost position as number of characters from starting position. With this improvement we for example do not have to worry what happens if rightmost position is deleted.

\begin{verbatim}
char(ptr){
  update_last(ptr)
  return ptr.value;
}
get_memo(rule,ptr){
  if (ptr.memoized[rule])
    last=rindex(ptr.index+ptr.memoized[rule].advance)
    if(timestamp(ptr,last)==ptr.memoized[rule].saved){
      update_last(last)
      return ptr.memoized[rule].value
    }
  }
  stack_push(rule,ptr)
  return nil
}
set_memo(rule,ptr,value){
  while (stack.top.rule!=rule ||
         stack.top.ptr!=ptr)
    stack_pop //parser decided not to memoize
  ptr.memoized[rule].value   = value
  ptr.memoized[rule].advance = stack.top.last.index
                             - ptr.index
  ptr.memoized[rule].saved   = timestamp(ptr,stack.top.last)
  stack_pop
}
\end{verbatim}
\newpage

\chapter{Peridot}
We use peridot as an example of using amethyst in language design. 

\section{Basic concepts}
Peridot does not differ much from mainstream dynamic programming languages. We assume that reader is familiar with concepts like class, method, dynamic dispatching. 

Currently Peridot has classes for integers, arrays and strings with basic methods and operators. For their description see Peridot documentation.

As in Ruby variables are defined by assignment.
\section{Peridot grammar in amethyst}
We use a parts of Peridot grammar to illustrate use of amethyst. Entire grammar can be found in Appendix \ref{append_per}.

We try to design an operator precedence that avoids pitfalls of \verb$C$ language that expressions \verb$1 + 1<<2 == 5$ and \verb$1&2 == 5&2$ are false.

\begin{grambox}
\begin{spacing}{0.8}
{\small
{\color{red}\Verb$b$}{\color{red}\Verb$i$}{\color{red}\Verb$n$}{\color{red}\Verb$a$}{\color{red}\Verb$r$}{\color{red}\Verb$y$}{\color{red}\Verb$_$}{\color{red}\Verb$o$}{\color{red}\Verb$p$}{\color{green}\Verb$($}{\color{green}\Verb$e$}{\color{green}\Verb$x$}{\color{green}\Verb$p$}{\color{green}\Verb$,$}{\color{green}\Verb$o$}{\color{green}\Verb$p$}{\color{green}\Verb$e$}{\color{green}\Verb$r$}{\color{green}\Verb$)$}\Verb$ $\Verb$=$\Verb$ ${\color{red}\Verb$a$}{\color{red}\Verb$p$}{\color{red}\Verb$p$}{\color{red}\Verb$l$}{\color{red}\Verb$y$}{\color{green}\Verb$($}{\color{green}\Verb$e$}{\color{green}\Verb$x$}{\color{green}\Verb$p$}{\color{green}\Verb$)$}{\color{Aquamarine}\Verb$:$}{\color{Aquamarine}\Verb$a$}\Verb$ $\Verb$(${\color{black}\Verb$"$}{\color{black}\Verb$"$}\Verb$ ${\color{red}\Verb$a$}{\color{red}\Verb$p$}{\color{red}\Verb$p$}{\color{red}\Verb$l$}{\color{red}\Verb$y$}{\color{green}\Verb$($}{\color{green}\Verb$o$}{\color{green}\Verb$p$}{\color{green}\Verb$e$}{\color{green}\Verb$r$}{\color{green}\Verb$)$}{\color{Aquamarine}\Verb$:$}{\color{Aquamarine}\Verb$o$}{\color{Aquamarine}\Verb$p$}\Verb$ $\\\Verb$ $\Verb$ $\Verb$ $\Verb$ $\Verb$ $\Verb$ $\Verb$ $\Verb$ $\Verb$ $\Verb$ $\Verb$ $\Verb$ $\Verb$ $\Verb$ $\Verb$ $\Verb$ $\Verb$ $\Verb$ $\Verb$ $\Verb$ $\Verb$ $\Verb$ ${\color{red}\Verb$a$}{\color{red}\Verb$p$}{\color{red}\Verb$p$}{\color{red}\Verb$l$}{\color{red}\Verb$y$}{\color{green}\Verb$($}{\color{green}\Verb$e$}{\color{green}\Verb$x$}{\color{green}\Verb$p$}{\color{green}\Verb$)$}{\color{Aquamarine}\Verb$:$}{\color{Aquamarine}\Verb$b$}\Verb$ ${\color{Tan}\Verb${$}{\color{Tan}\Verb$c$}{\color{Tan}\Verb$a$}{\color{Tan}\Verb$l$}{\color{Tan}\Verb$l$}{\color{Tan}\Verb$($}{\color{Aquamarine}\Verb$o$}{\color{Aquamarine}\Verb$p$}{\color{Tan}\Verb$,$}{\color{Aquamarine}\Verb$a$}{\color{Tan}\Verb$,$}{\color{Aquamarine}\Verb$b$}{\color{Tan}\Verb$)$}{\color{Tan}\Verb$}$}{\color{Aquamarine}\Verb$:$}{\color{Aquamarine}\Verb$a$}\Verb$)${\color{black}\Verb$*$}\Verb$ ${\color{Tan}\Verb$-$}{\color{Tan}\Verb$>$}{\color{Tan}\Verb$ $}{\color{Aquamarine}\Verb$a$}{\color{Tan}\\}{\color{Tan}\\}{\color{red}\Verb$e$}{\color{red}\Verb$x$}{\color{red}\Verb$p$}{\color{red}\Verb$r$}{\color{red}\Verb$_$}{\color{red}\Verb$o$}{\color{red}\Verb$r$}{\color{red}\Verb$_$}{\color{red}\Verb$l$}\Verb$ $\Verb$ $\Verb$=$\Verb$ ${\color{red}\Verb$b$}{\color{red}\Verb$i$}{\color{red}\Verb$n$}{\color{red}\Verb$a$}{\color{red}\Verb$r$}{\color{red}\Verb$y$}{\color{red}\Verb$_$}{\color{red}\Verb$o$}{\color{red}\Verb$p$}{\color{green}\Verb$($}{\color{green}\Verb$'$}{\color{green}\Verb$e$}{\color{green}\Verb$x$}{\color{green}\Verb$p$}{\color{green}\Verb$r$}{\color{green}\Verb$_$}{\color{green}\Verb$a$}{\color{green}\Verb$n$}{\color{green}\Verb$d$}{\color{green}\Verb$_$}{\color{green}\Verb$l$}{\color{green}\Verb$'$}{\color{green}\Verb$,$}{\color{Violet}\Verb$($}{\color{Violet}\Verb$|$}{\color{green}\Verb$ $}{\color{black}\Verb$"$}{\color{black}\Verb$|$}{\color{black}\Verb$|$}{\color{black}\Verb$"$}{\color{green}\Verb$ $}{\color{Violet}\Verb$|$}{\color{Violet}\Verb$)$}{\color{green}\Verb$)$}\\{\color{red}\Verb$e$}{\color{red}\Verb$x$}{\color{red}\Verb$p$}{\color{red}\Verb$r$}{\color{red}\Verb$_$}{\color{red}\Verb$a$}{\color{red}\Verb$n$}{\color{red}\Verb$d$}{\color{red}\Verb$_$}{\color{red}\Verb$l$}\Verb$ $\Verb$=$\Verb$ ${\color{red}\Verb$b$}{\color{red}\Verb$i$}{\color{red}\Verb$n$}{\color{red}\Verb$a$}{\color{red}\Verb$r$}{\color{red}\Verb$y$}{\color{red}\Verb$_$}{\color{red}\Verb$o$}{\color{red}\Verb$p$}{\color{green}\Verb$($}{\color{green}\Verb$'$}{\color{green}\Verb$e$}{\color{green}\Verb$x$}{\color{green}\Verb$p$}{\color{green}\Verb$r$}{\color{green}\Verb$_$}{\color{green}\Verb$c$}{\color{green}\Verb$m$}{\color{green}\Verb$p$}{\color{green}\Verb$'$}{\color{green}\Verb$ $}{\color{green}\Verb$ $}{\color{green}\Verb$,$}{\color{Violet}\Verb$($}{\color{Violet}\Verb$|$}{\color{green}\Verb$ $}{\color{black}\Verb$"$}{\color{black}\Verb$&$}{\color{black}\Verb$&$}{\color{black}\Verb$"$}{\color{green}\Verb$ $}{\color{Violet}\Verb$|$}{\color{Violet}\Verb$)$}{\color{green}\Verb$)$}\\\\{\color{red}\Verb$e$}{\color{red}\Verb$x$}{\color{red}\Verb$p$}{\color{red}\Verb$r$}{\color{red}\Verb$_$}{\color{red}\Verb$c$}{\color{red}\Verb$m$}{\color{red}\Verb$p$}\Verb$ $\Verb$=$\Verb$ ${\color{red}\Verb$b$}{\color{red}\Verb$i$}{\color{red}\Verb$n$}{\color{red}\Verb$a$}{\color{red}\Verb$r$}{\color{red}\Verb$y$}{\color{red}\Verb$_$}{\color{red}\Verb$o$}{\color{red}\Verb$p$}{\color{green}\Verb$($}{\color{green}\Verb$'$}{\color{green}\Verb$e$}{\color{green}\Verb$x$}{\color{green}\Verb$p$}{\color{green}\Verb$r$}{\color{green}\Verb$_$}{\color{green}\Verb$a$}{\color{green}\Verb$r$}{\color{green}\Verb$1$}{\color{green}\Verb$'$}{\color{green}\Verb$,$}{\color{Violet}\Verb$($}{\color{Violet}\Verb$|$}{\color{green}\Verb$ $}{\color{black}\Verb$'$}{\color{black}\Verb$<$}{\color{black}\Verb$'$}{\color{green}\Verb$ $}{\color{green}\Verb$|$}{\color{black}\Verb$'$}{\color{black}\Verb$<$}{\color{black}\Verb$=$}{\color{black}\Verb$'$}{\color{green}\Verb$|$}{\color{black}\Verb$'$}{\color{black}\Verb$<$}{\color{black}\Verb$=$}{\color{black}\Verb$>$}{\color{black}\Verb$'$}{\color{green}\\}{\color{green}\Verb$ $}{\color{green}\Verb$ $}{\color{green}\Verb$ $}{\color{green}\Verb$ $}{\color{green}\Verb$ $}{\color{green}\Verb$ $}{\color{green}\Verb$ $}{\color{green}\Verb$ $}{\color{green}\Verb$ $}{\color{green}\Verb$ $}{\color{green}\Verb$ $}{\color{green}\Verb$ $}{\color{green}\Verb$ $}{\color{green}\Verb$ $}{\color{green}\Verb$ $}{\color{green}\Verb$ $}{\color{green}\Verb$ $}{\color{green}\Verb$ $}{\color{green}\Verb$ $}{\color{green}\Verb$ $}{\color{green}\Verb$ $}{\color{green}\Verb$ $}{\color{green}\Verb$ $}{\color{green}\Verb$ $}{\color{green}\Verb$ $}{\color{green}\Verb$ $}{\color{green}\Verb$ $}{\color{green}\Verb$ $}{\color{green}\Verb$|$}{\color{black}\Verb$'$}{\color{black}\Verb$>$}{\color{black}\Verb$=$}{\color{black}\Verb$'$}{\color{green}\Verb$|$}{\color{green}\Verb$ $}{\color{black}\Verb$'$}{\color{black}\Verb$>$}{\color{black}\Verb$'$}{\color{green}\Verb$ $}{\color{green}\Verb$|$}{\color{black}\Verb$'$}{\color{black}\Verb$=$}{\color{black}\Verb$=$}{\color{black}\Verb$'$}{\color{green}\Verb$|$}{\color{black}\Verb$'$}{\color{black}\Verb$!$}{\color{black}\Verb$=$}{\color{black}\Verb$'$}{\color{green}\Verb$ $}{\color{Violet}\Verb$|$}{\color{Violet}\Verb$)$}{\color{green}\Verb$)$}\\\\{\color{red}\Verb$e$}{\color{red}\Verb$x$}{\color{red}\Verb$p$}{\color{red}\Verb$r$}{\color{red}\Verb$_$}{\color{red}\Verb$a$}{\color{red}\Verb$r$}{\color{red}\Verb$1$}\Verb$ $\Verb$=$\Verb$ ${\color{red}\Verb$b$}{\color{red}\Verb$i$}{\color{red}\Verb$n$}{\color{red}\Verb$a$}{\color{red}\Verb$r$}{\color{red}\Verb$y$}{\color{red}\Verb$_$}{\color{red}\Verb$o$}{\color{red}\Verb$p$}{\color{green}\Verb$($}{\color{green}\Verb$'$}{\color{green}\Verb$e$}{\color{green}\Verb$x$}{\color{green}\Verb$p$}{\color{green}\Verb$r$}{\color{green}\Verb$_$}{\color{green}\Verb$a$}{\color{green}\Verb$r$}{\color{green}\Verb$2$}{\color{green}\Verb$'$}{\color{green}\Verb$,$}{\color{Violet}\Verb$($}{\color{Violet}\Verb$|$}{\color{green}\Verb$ $}{\color{black}\Verb$'$}{\color{black}\Verb$+$}{\color{black}\Verb$'$}{\color{green}\Verb$ $}{\color{green}\Verb$|$}{\color{black}\Verb$'$}{\color{black}\Verb$-$}{\color{black}\Verb$'$}{\color{green}\Verb$ $}{\color{green}\Verb$ $}{\color{green}\Verb$ $}{\color{green}\Verb$ $}{\color{green}\Verb$ $}{\color{green}\Verb$ $}{\color{green}\Verb$ $}{\color{Violet}\Verb$|$}{\color{Violet}\Verb$)$}{\color{green}\Verb$)$}\\{\color{red}\Verb$e$}{\color{red}\Verb$x$}{\color{red}\Verb$p$}{\color{red}\Verb$r$}{\color{red}\Verb$_$}{\color{red}\Verb$a$}{\color{red}\Verb$r$}{\color{red}\Verb$2$}\Verb$ $\Verb$=$\Verb$ ${\color{red}\Verb$b$}{\color{red}\Verb$i$}{\color{red}\Verb$n$}{\color{red}\Verb$a$}{\color{red}\Verb$r$}{\color{red}\Verb$y$}{\color{red}\Verb$_$}{\color{red}\Verb$o$}{\color{red}\Verb$p$}{\color{green}\Verb$($}{\color{green}\Verb$'$}{\color{green}\Verb$e$}{\color{green}\Verb$x$}{\color{green}\Verb$p$}{\color{green}\Verb$r$}{\color{green}\Verb$_$}{\color{green}\Verb$o$}{\color{green}\Verb$r$}{\color{green}\Verb$'$}{\color{green}\Verb$ $}{\color{green}\Verb$,$}{\color{Violet}\Verb$($}{\color{Violet}\Verb$|$}{\color{green}\Verb$ $}{\color{black}\Verb$'$}{\color{black}\Verb$*$}{\color{black}\Verb$'$}{\color{green}\Verb$ $}{\color{green}\Verb$|$}{\color{black}\Verb$'$}{\color{black}\Verb$/$}{\color{black}\Verb$'$}{\color{green}\Verb$ $}{\color{green}\Verb$|$}{\color{black}\Verb$'$}{\color{black}\Verb$
}\end{spacing}

\end{grambox}

\newpage
\section{Functional programming style}
Peridot supports several functional programming features.

Lambdas are supported with same syntax as in amethyst.

Ruby extensively uses the block passing style. You use it even for loops:

\vskip -1.8em\begin{spacing}{0.8}
{\small
\begin{exambox}
\Verb$4$\Verb$.$\Verb$t$\Verb$i$\Verb$m$\Verb$e$\Verb$s$\Verb${$\Verb$|$\Verb$i$\Verb$|$\\\Verb$ $\Verb$ $\Verb$p$\Verb$u$\Verb$t$\Verb$s$\Verb$ $\Verb$i$\\\Verb$}$\\\Verb$#$\Verb$e$\Verb$q$\Verb$u$\Verb$i$\Verb$v$\Verb$a$\Verb$l$\Verb$e$\Verb$n$\Verb$t$\Verb$ $\Verb$c$\Verb$o$\Verb$d$\Verb$e$\\\Verb$4$\Verb$.$\Verb$t$\Verb$i$\Verb$m$\Verb$e$\Verb$s$\Verb$($\Verb$&$\Verb$p$\Verb$r$\Verb$o$\Verb$c$\Verb${$\Verb$|$\Verb$i$\Verb$|$\Verb$ $\Verb$p$\Verb$u$\Verb$t$\Verb$s$\Verb$ $\Verb$i$\Verb$}$\Verb$)$
\end{exambox}
}
\end{spacing}\vskip -0.4em

This construction can be viewed as a case of the continuation passing style \cite{steele}.

We want have better support for continuation passing style.

In Peridot a yield keyword returns a (result, continuation) pair. The block syntax \verb$a(b){block}$ is a shortcut for:

\vskip -1.8em\begin{spacing}{0.8}
{\small
\begin{exambox}
\Verb$c$\Verb$o$\Verb$n$\Verb$t$\Verb$ $\Verb$=$\Verb$ $\Verb$a$\Verb$($\Verb$b$\Verb$)$\\\Verb$ $\Verb$ $\Verb$w$\Verb$h$\Verb$i$\Verb$l$\Verb$e$\Verb$ $\Verb$($\Verb$c$\Verb$o$\Verb$n$\Verb$t$\Verb$.$\Verb$i$\Verb$s$\Verb$_$\Verb$a$\Verb$?$\Verb$($\Verb$C$\Verb$o$\Verb$n$\Verb$t$\Verb$i$\Verb$n$\Verb$u$\Verb$a$\Verb$t$\Verb$a$\Verb$t$\Verb$i$\Verb$o$\Verb$n$\Verb$)$\Verb$)$\\\Verb$ $\Verb$ $\Verb$ $\Verb$ $\Verb$r$\Verb$ $\Verb$ $\Verb$ $\Verb$ $\Verb$=$\Verb$ $\Verb$b$\Verb$l$\Verb$o$\Verb$c$\Verb$k$\Verb$.$\Verb$c$\Verb$a$\Verb$l$\Verb$l$\Verb$($\Verb$c$\Verb$o$\Verb$n$\Verb$t$\Verb$.$\Verb$r$\Verb$e$\Verb$s$\Verb$u$\Verb$l$\Verb$t$\Verb$)$\\\Verb$ $\Verb$ $\Verb$ $\Verb$ $\Verb$c$\Verb$o$\Verb$n$\Verb$t$\Verb$ $\Verb$=$\Verb$ $\Verb$c$\Verb$o$\Verb$n$\Verb$t$\Verb$.$\Verb$c$\Verb$a$\Verb$l$\Verb$l$\Verb$($\Verb$r$\Verb$)$\\\Verb$ $\Verb$ $\Verb$e$\Verb$n$\Verb$d$
\end{exambox}
}
\end{spacing}\vskip -0.4em

When no block is specified you can use returned \verb$Continuation$ object is similar way as \verb$Enumerator$ object in Ruby. A main difference is that in \verb$Continuation$ object communication goes in both directions. 

As hypothetical example consider a binary search tree that implements a generic binary search \verb$bsearch$ method that takes a block with supplied comparison method. For example guess a number game can be done as:

\vskip -1.8em\begin{spacing}{0.8}
{\small
\begin{exambox}
\Verb$t$\Verb$=$\Verb$T$\Verb$r$\Verb$e$\Verb$e$\Verb$.$\Verb$n$\Verb$e$\Verb$w$\\\Verb$t$\Verb$.$\Verb$b$\Verb$s$\Verb$e$\Verb$a$\Verb$r$\Verb$c$\Verb$h$\Verb${$\Verb$|$\Verb$x$\Verb$|$\Verb$ $\\\Verb$ $\Verb$ $\Verb$p$\Verb$u$\Verb$t$\Verb$s$\Verb$ $\Verb$"$\Verb$i$\Verb$s$\Verb$"$\Verb$+$\Verb$x$\Verb$+$\Verb$"$\Verb$m$\Verb$o$\Verb$r$\Verb$e$\Verb$/$\Verb$l$\Verb$e$\Verb$s$\Verb$s$\Verb$/$\Verb$e$\Verb$q$\Verb$u$\Verb$a$\Verb$l$\Verb$ $\Verb$t$\Verb$o$\Verb$ $\Verb$y$\Verb$o$\Verb$u$\Verb$r$\Verb$ $\Verb$n$\Verb$u$\Verb$m$\Verb$b$\Verb$e$\Verb$r$\Verb$?$\Verb$"$\\\Verb$ $\Verb$ $\Verb$g$\Verb$e$\Verb$t$\Verb$s$\\\Verb$}$
\end{exambox}
}
\end{spacing}\vskip -0.4em

We can change numbers passed to block by map method:

\vskip -1.8em\begin{spacing}{0.8}
{\small
\begin{exambox}
\Verb$t$\Verb$=$\Verb$T$\Verb$r$\Verb$e$\Verb$e$\Verb$.$\Verb$n$\Verb$e$\Verb$w$\\\Verb$c$\Verb$=$\Verb$t$\Verb$.$\Verb$b$\Verb$s$\Verb$e$\Verb$a$\Verb$r$\Verb$c$\Verb$h$\\\Verb$c$\Verb$=$\Verb$c$\Verb$.$\Verb$m$\Verb$a$\Verb$p$\Verb${$\Verb$|$\Verb$x$\Verb$|$\Verb$ $\Verb$r$\Verb$o$\Verb$m$\Verb$a$\Verb$n$\Verb$_$\Verb$n$\Verb$u$\Verb$m$\Verb$e$\Verb$r$\Verb$a$\Verb$l$\Verb$($\Verb$x$\Verb$)$\Verb$ $\Verb$}$\Verb$ $\\\Verb$c$\Verb$.$\Verb$c$\Verb$a$\Verb$l$\Verb$l$\Verb${$\Verb$|$\Verb$x$\Verb$|$\Verb$ $\\\Verb$ $\Verb$ $\Verb$p$\Verb$u$\Verb$t$\Verb$s$\Verb$ $\Verb$"$\Verb$i$\Verb$s$\Verb$"$\Verb$+$\Verb$x$\Verb$+$\Verb$"$\Verb$m$\Verb$o$\Verb$r$\Verb$e$\Verb$/$\Verb$l$\Verb$e$\Verb$s$\Verb$s$\Verb$/$\Verb$e$\Verb$q$\Verb$u$\Verb$a$\Verb$l$\Verb$ $\Verb$t$\Verb$o$\Verb$ $\Verb$y$\Verb$o$\Verb$u$\Verb$r$\Verb$ $\Verb$n$\Verb$u$\Verb$m$\Verb$b$\Verb$e$\Verb$r$\Verb$?$\Verb$"$\\\Verb$ $\Verb$ $\Verb$g$\Verb$e$\Verb$t$\Verb$s$\\\Verb$}$
\end{exambox}
}
\end{spacing}\vskip -0.4em

And values returned back. For example reversing direction can be done by changing third line of previous example to:

\vskip -1.8em\begin{spacing}{0.8}
{\small
\begin{exambox}
\Verb$c$\Verb$=$\Verb$c$\Verb$.$\Verb$r$\Verb$e$\Verb$v$\Verb$_$\Verb$m$\Verb$a$\Verb$p$\Verb${$\Verb$|$\Verb$x$\Verb$|$\\\Verb$ $\Verb$ $\Verb$i$\Verb$f$\Verb$ $\Verb$x$\Verb$=$\Verb$=$\Verb$"$\Verb$m$\Verb$o$\Verb$r$\Verb$e$\Verb$"$\\\Verb$ $\Verb$ $\Verb$ $\Verb$ $\Verb$"$\Verb$l$\Verb$e$\Verb$s$\Verb$s$\Verb$"$\\\Verb$ $\Verb$ $\Verb$e$\Verb$l$\Verb$s$\Verb$e$\\\Verb$ $\Verb$ $\Verb$ $\Verb$ $\Verb$i$\Verb$f$\Verb$ $\Verb$x$\Verb$=$\Verb$=$\Verb$"$\Verb$l$\Verb$e$\Verb$s$\Verb$s$\Verb$"$\\\Verb$ $\Verb$ $\Verb$ $\Verb$ $\Verb$ $\Verb$ $\Verb$"$\Verb$m$\Verb$o$\Verb$r$\Verb$e$\Verb$"$\\\Verb$ $\Verb$ $\Verb$ $\Verb$ $\Verb$e$\Verb$l$\Verb$s$\Verb$e$\\\Verb$ $\Verb$ $\Verb$ $\Verb$ $\Verb$ $\Verb$ $\Verb$"$\Verb$e$\Verb$q$\Verb$u$\Verb$a$\Verb$l$\Verb$"$\\\Verb$ $\Verb$ $\Verb$ $\Verb$ $\Verb$e$\Verb$n$\Verb$d$\\\Verb$ $\Verb$ $\Verb$e$\Verb$n$\Verb$d$\\\Verb$}$
\end{exambox}
}
\end{spacing}\vskip -0.4em

\appendix
\chapter{Amethyst syntax summary} \label{syntsummary}
In this section we recapitulate semantic of amethyst in systematic manner.

\subsection*{Rules}

\vskip 0.2em\noindent\begin{tabular}{| l || p{8.5cm} |}
\hline
Pattern     &Description \\
\hline
{\color{red}\Verb$r$}{\color{red}\Verb$u$}{\color{red}\Verb$l$}{\color{red}\Verb$e$}\Verb$ $\Verb$ $\Verb$ $\Verb$ $\Verb$ $\Verb$ $\Verb$ $\Verb$ $\Verb$ $\Verb$ $\Verb$ $\Verb$=$\Verb$ ${\color{red}\Verb$e$}{\color{red}\Verb$x$}{\color{red}\Verb$p$} & Rule definition.\\
{\color{red}\Verb$r$}{\color{red}\Verb$u$}{\color{red}\Verb$l$}{\color{red}\Verb$e$} & Rule call.\\
{\color{red}\Verb$r$}{\color{red}\Verb$u$}{\color{red}\Verb$l$}{\color{red}\Verb$e$}{\color{green}\Verb$($}{\color{green}\Verb$v$}{\color{green}\Verb$1$}{\color{green}\Verb$,$}{\color{green}\Verb$v$}{\color{green}\Verb$2$}{\color{green}\Verb$.$}{\color{green}\Verb$.$}{\color{green}\Verb$.$}{\color{green}\Verb$)$}\Verb$ $\Verb$=$\Verb$ ${\color{red}\Verb$e$}{\color{red}\Verb$x$}{\color{red}\Verb$p$} & Parametrized definition.\\
{\color{red}\Verb$r$}{\color{red}\Verb$u$}{\color{red}\Verb$l$}{\color{red}\Verb$e$}{\color{green}\Verb$($}{\color{green}\Verb$c$}{\color{green}\Verb$1$}{\color{green}\Verb$,$}{\color{green}\Verb$c$}{\color{green}\Verb$2$}{\color{green}\Verb$.$}{\color{green}\Verb$.$}{\color{green}\Verb$.$}{\color{green}\Verb$)$} & Parametrized call.\\
{\color{red}\Verb$G$}{\color{red}\Verb$r$}{\color{red}\Verb$a$}{\color{red}\Verb$m$}{\color{red}\Verb$m$}{\color{red}\Verb$a$}{\color{red}\Verb$r$}{\color{red}\Verb$:$}{\color{red}\Verb$:$}{\color{red}\Verb$r$}{\color{red}\Verb$u$}{\color{red}\Verb$l$}{\color{red}\Verb$e$} & Call rule from given Grammar.\\
\Verb$($\Verb$|$\Verb$ ${\color{blue}\Verb$e$}\Verb$ $\Verb$|$\Verb$)$ & Lambda\\
\hline
\end{tabular}\vskip 0.2em

\subsection*{Semantic actions an predicates}

\vskip 0.2em\noindent\begin{tabular}{| l | l || p{8.2cm} |}
\hline
Pattern     &Shortcut&Description \\
\hline
{\color{Tan}\Verb${$}{\color{Tan}\Verb$c$}{\color{Tan}\Verb$}$}   & core &Semantic action.\\
{\color{Violet}\Verb$&$}{\color{Violet}\Verb${$}{\color{Tan}\Verb$c$}{\color{Violet}\Verb$}$} &{\color{red}\Verb$c$}{\color{red}\Verb$o$}{\color{red}\Verb$r$}{\color{red}\Verb$e$} &Semantic predicate.\\
{\color{Violet}\Verb$~$}{\color{Violet}\Verb${$}{\color{Tan}\Verb$c$}{\color{Violet}\Verb$}$} &{\color{Violet}\Verb$&$}{\color{Violet}\Verb${$}{\color{Tan}\Verb$!$}{\color{Tan}\Verb$c$}{\color{Violet}\Verb$}$} &Negative semantic predicate.\\
{\color{Tan}\Verb$-$}{\color{Tan}\Verb$>$}{\color{Tan}\Verb$ $}{\color{Tan}\Verb$c$}{\color{Tan}\Verb$ $}{\color{Tan}\Verb$n$}{\color{Tan}\Verb$e$}{\color{Tan}\Verb$w$}{\color{Tan}\Verb$l$}{\color{Tan}\Verb$i$}{\color{Tan}\Verb$n$}{\color{Tan}\Verb$e$} &{\color{Tan}\Verb${$}{\color{Tan}\Verb$c$}{\color{Tan}\Verb$}$}   & Alternative syntax of semantic action. \\
\hline
\end{tabular}\vskip 0.2em

In semantic actions and predicates we do following substitutions.

\vskip 0.2em\noindent\begin{tabular}{| l | l | | p{7.1cm} |}
\hline
Pattern     &Shortcut&Description \\
\hline
{\color{Aquamarine}\Verb$@$}{\color{Aquamarine}\Verb$m$}{\color{Aquamarine}\Verb$e$}{\color{Aquamarine}\Verb$t$}{\color{Aquamarine}\Verb$h$}{\color{Aquamarine}\Verb$o$}{\color{Aquamarine}\Verb$d$}       &{\color{red}\Verb$s$}{\color{red}\Verb$r$}{\color{red}\Verb$c$}\Verb$.${\color{red}\Verb$m$}{\color{red}\Verb$e$}{\color{red}\Verb$t$}{\color{red}\Verb$h$}{\color{red}\Verb$o$}{\color{red}\Verb$d$}  & method call (parameters are allowed)\\
{\color{Aquamarine}\Verb$@$}{\color{Aquamarine}\Verb$C$}{\color{Aquamarine}\Verb$l$}{\color{Aquamarine}\Verb$a$}{\color{Aquamarine}\Verb$s$}{\color{Aquamarine}\Verb$s$}       &{\color{red}\Verb$C$}{\color{red}\Verb$l$}{\color{red}\Verb$a$}{\color{red}\Verb$s$}{\color{red}\Verb$s$}\Verb$.${\color{red}\Verb$c$}{\color{red}\Verb$r$}{\color{red}\Verb$e$}{\color{red}\Verb$a$}{\color{red}\Verb$t$}{\color{red}\Verb$e$}{\color{green}\Verb$($}{\color{green}\Verb$.$}{\color{green}\Verb$.$}{\color{green}\Verb$.$}{\color{green}\Verb$)$}  & Construct object\\
{\color{Aquamarine}\Verb$@$}{\color{Aquamarine}\Verb$>$}{\color{Aquamarine}\Verb$n$}{\color{Aquamarine}\Verb$a$}{\color{Aquamarine}\Verb$m$}{\color{Aquamarine}\Verb$e$}       &{\color{red}\Verb$a$}{\color{red}\Verb$r$}{\color{red}\Verb$g$}{\color{red}\Verb$u$}{\color{red}\Verb$m$}{\color{red}\Verb$e$}{\color{red}\Verb$n$}{\color{red}\Verb$t$}{\color{red}\Verb$s$}{\color{blue}\Verb$[$}{\color{black}\Verb$"$}{\color{black}\Verb$n$}{\color{black}\Verb$a$}{\color{black}\Verb$m$}{\color{black}\Verb$e$}{\color{black}\Verb$"$}{\color{blue}\Verb$]$}  & Contextual argument\\
{\color{Aquamarine}\Verb$@$}{\color{Aquamarine}\Verb$<$}{\color{Aquamarine}\Verb$n$}{\color{Aquamarine}\Verb$a$}{\color{Aquamarine}\Verb$m$}{\color{Aquamarine}\Verb$e$}       &{\color{red}\Verb$r$}{\color{red}\Verb$e$}{\color{red}\Verb$t$}{\color{red}\Verb$u$}{\color{red}\Verb$r$}{\color{red}\Verb$n$}{\color{red}\Verb$s$}{\color{blue}\Verb$[$}{\color{black}\Verb$"$}{\color{black}\Verb$n$}{\color{black}\Verb$a$}{\color{black}\Verb$m$}{\color{black}\Verb$e$}{\color{black}\Verb$"$}{\color{blue}\Verb$]$}  & Contextual return\\
\Verb$($\Verb$|${\color{blue}\Verb$e$}\Verb$|$\Verb$)$       & core                    & Lambda\\
\hline
\end{tabular}\vskip 0.2em

\subsection*{Variable binding}

\vskip 0.2em\noindent\begin{tabular}{| l | l || p{8.6cm} |}
\hline
Pattern     &Shortcut&Description \\
\hline
{\color{blue}\Verb$e$}{\color{Aquamarine}\Verb$:$}{\color{Aquamarine}\Verb$v$} & core & variable binding \\
{\color{blue}\Verb$e$}{\color{Aquamarine}\Verb$:$}{\color{Tan}\Verb${$}{\color{Tan}\Verb$c$}{\color{Tan}\Verb$}$} &{\color{blue}\Verb$e$}{\color{Aquamarine}\Verb$:$}{\color{Aquamarine}\Verb$i$}{\color{Aquamarine}\Verb$t$}\Verb$ ${\color{Tan}\Verb${$}{\color{Tan}\Verb$c$}{\color{Tan}\Verb$}$} & For conversions etc.\\
{\color{blue}\Verb$e$}{\color{Aquamarine}\Verb$:$}{\color{Aquamarine}\Verb$[$}{\color{Aquamarine}\Verb$c$}{\color{Aquamarine}\Verb$]$} &{\color{blue}\Verb$e$}{\color{Aquamarine}\Verb$:$}{\color{Tan}\Verb${$}{\color{Tan}\Verb$v$}{\color{Tan}\Verb$ $}{\color{Tan}\Verb$<$}{\color{Tan}\Verb$<$}{\color{Tan}\Verb$ $}{\color{Aquamarine}\Verb$i$}{\color{Aquamarine}\Verb$t$}{\color{Tan}\Verb$}$} &Append result to array  {\color{red}\Verb$c$}.\\
\hline
\end{tabular}\vskip 0.2em

\subsection*{Sequencing and choice}

\vskip 0.2em\noindent\begin{tabular}{| l | l || p{8cm} |}
\hline
Operation & Expansion & Description \\
\hline
{\color{blue}\Verb$e$}{\color{black}\Verb$?$}  & {\color{blue}\Verb$e$}\Verb$|${\color{Tan}\Verb${$}{\color{Tan}\Verb$n$}{\color{Tan}\Verb$i$}{\color{Tan}\Verb$l$}{\color{Tan}\Verb$}$} & Make {\color{blue}\Verb$e$} optional.  \\
{\color{red}\Verb$C$}{\color{red}\Verb$u$}{\color{red}\Verb$t$}  & auxiliary & Like ! in prolog\\
{\color{Violet}\Verb$~$}{\color{blue}\Verb$e$}  &{\color{blue}\Verb$e$}\Verb$ ${\color{red}\Verb$C$}{\color{red}\Verb$u$}{\color{red}\Verb$t$}\Verb$ ${\color{red}\Verb$f$}{\color{red}\Verb$a$}{\color{red}\Verb$i$}{\color{red}\Verb$l$}{\color{red}\Verb$s$}\Verb$ $\Verb$|$\Verb$ ${\color{Tan}\Verb${$}{\color{Tan}\Verb$n$}{\color{Tan}\Verb$i$}{\color{Tan}\Verb$l$}{\color{Tan}\Verb$}$} & Negative lookahead. \\
{\color{blue}\Verb$e$}{\color{blue}\Verb$1$}\Verb$ ${\color{Violet}\Verb$&$}{\color{Violet}\Verb$ $}{\color{blue}\Verb$e$}{\color{blue}\Verb$2$}  &{\color{Violet}\Verb$~$}{\color{Violet}\Verb$~$}{\color{blue}\Verb$e$}{\color{blue}\Verb$1$}\Verb$ ${\color{blue}\Verb$e$}{\color{blue}\Verb$2$} & Positive lookahead. \\
\hline
\end{tabular}\vskip 0.2em

\subsection*{Iteration}

\vskip 0.2em\noindent\begin{tabular}{| l | l || p{10.9cm} |}
\hline
Pattern     &Shortcut&Description \\
\hline
{\color{blue}\Verb$e$}{\color{black}\Verb$*$}{\color{black}\Verb$*$}         & auxiliary & repeat-until\\
{\color{red}\Verb$S$}{\color{red}\Verb$t$}{\color{red}\Verb$o$}{\color{red}\Verb$p$}         & auxiliary & Stop iteration\\
{\color{blue}\Verb$e$}{\color{black}\Verb$*$}         &{\color{blue}\Verb$e$}{\color{black}\Verb$*$}{\color{black}\Verb$*$}          & When {\color{blue}\Verb$e$} contains \verb$Stop$,\\
{\color{blue}\Verb$e$}{\color{black}\Verb$*$}         &\Verb$(${\color{blue}\Verb$e$}\Verb$|${\color{red}\Verb$S$}{\color{red}\Verb$t$}{\color{red}\Verb$o$}{\color{red}\Verb$p$}\Verb$)${\color{black}\Verb$*$}{\color{black}\Verb$*$}          & otherwise. \\
\Verb$b$\Verb$r$\Verb$e$\Verb$a$\Verb$k$         &{\color{red}\Verb$C$}{\color{red}\Verb$u$}{\color{red}\Verb$t$}\Verb$ ${\color{red}\Verb$S$}{\color{red}\Verb$t$}{\color{red}\Verb$o$}{\color{red}\Verb$p$}          & Possible expansion.\\
\hline
\end{tabular}\vskip 0.2em

\subsection*{Object orientated constructs}

\vskip 0.2em\noindent\begin{tabular}{| l | l || p{9.8cm} |}
\hline
Pattern     &Shortcut&Description \\
\hline
{\color{blue}\Verb$e$}{\color{blue}\Verb$1$}{\color{blue}\Verb$[$}{\color{blue}\Verb$e$}{\color{blue}\Verb$2$}{\color{blue}\Verb$]$}   &core                    & Enter operator \\
{\color{blue}\Verb$[$}{\color{blue}\Verb$e$}{\color{blue}\Verb$]$}      &\Verb$.${\color{blue}\Verb$[$}{\color{blue}\Verb$e$}{\color{blue}\Verb$]$}              & We can omit leading . .\\
{\color{blue}\Verb$e$}{\color{blue}\Verb$1$}\Verb$=$\Verb$>${\color{blue}\Verb$e$}{\color{blue}\Verb$2$}   &{\color{blue}\Verb$e$}{\color{blue}\Verb$1$}{\color{Aquamarine}\Verb$:$}{\color{Aquamarine}\Verb$a$}\Verb$ ${\color{Tan}\Verb${$}{\color{Tan}\Verb$[$}{\color{Aquamarine}\Verb$a$}{\color{Tan}\Verb$]$}{\color{Tan}\Verb$}$}{\color{blue}\Verb$[$}{\color{blue}\Verb$e$}{\color{blue}\Verb$2$}{\color{blue}\Verb$]$}  & Pass operator\\
{\color{red}\Verb$C$}{\color{red}\Verb$l$}{\color{red}\Verb$a$}{\color{red}\Verb$s$}{\color{red}\Verb$s$}    &{\color{red}\Verb$m$}{\color{red}\Verb$e$}{\color{red}\Verb$m$}{\color{red}\Verb$b$}{\color{red}\Verb$e$}{\color{red}\Verb$r$}{\color{green}\Verb$($}{\color{green}\Verb$C$}{\color{green}\Verb$l$}{\color{green}\Verb$a$}{\color{green}\Verb$s$}{\color{green}\Verb$s$}{\color{green}\Verb$)$}    & Test class membership\\
\hline
\end{tabular}

\chapter{Standard prologue} \label{append_prolog}
\begin{prolbox}
\vskip -1.8em\begin{spacing}{0.8}
{\small
\begin{exambox}
\Verb$a$\Verb$m$\Verb$e$\Verb$t$\Verb$h$\Verb$y$\Verb$s$\Verb$t$\Verb$ $\Verb$A$\Verb$m$\Verb$e$\Verb$t$\Verb$h$\Verb$y$\Verb$s$\Verb$t$\Verb$ $\Verb$<$\Verb$ $\Verb$A$\Verb$m$\Verb$e$\Verb$t$\Verb$h$\Verb$y$\Verb$s$\Verb$t$\Verb$C$\Verb$o$\Verb$r$\Verb$e$\Verb$ $\Verb${$\\\Verb$ $\Verb$ ${\color{red}\Verb$s$}{\color{red}\Verb$p$}{\color{red}\Verb$a$}{\color{red}\Verb$c$}{\color{red}\Verb$e$}\Verb$ $\Verb$=$\Verb$ $\Verb$<$\Verb$\$\Verb$s$\Verb$\$\Verb$t$\Verb$\$\Verb$r$\Verb$\$\Verb$n$\Verb$\$\Verb$f$\Verb$>$\\\Verb$ $\Verb$ ${\color{red}\Verb$s$}{\color{red}\Verb$p$}{\color{red}\Verb$a$}{\color{red}\Verb$c$}{\color{red}\Verb$e$}{\color{red}\Verb$s$}\Verb$=$\Verb$ ${\color{red}\Verb$s$}{\color{red}\Verb$p$}{\color{red}\Verb$a$}{\color{red}\Verb$c$}{\color{red}\Verb$e$}{\color{black}\Verb$*$}\\\Verb$ $\Verb$ ${\color{red}\Verb$t$}{\color{red}\Verb$o$}{\color{red}\Verb$k$}{\color{red}\Verb$e$}{\color{red}\Verb$n$}{\color{green}\Verb$($}{\color{green}\Verb$s$}{\color{green}\Verb$)$}\Verb$ $\Verb$=$\Verb$ ${\color{red}\Verb$s$}{\color{red}\Verb$p$}{\color{red}\Verb$a$}{\color{red}\Verb$c$}{\color{red}\Verb$e$}{\color{red}\Verb$s$}\Verb$ ${\color{red}\Verb$s$}{\color{red}\Verb$e$}{\color{red}\Verb$q$}{\color{green}\Verb$($}{\color{green}\Verb$s$}{\color{green}\Verb$)$}\\\Verb$ $\Verb$ ${\color{red}\Verb$_$}\Verb$ $\Verb$ $\Verb$ $\Verb$ $\Verb$ $\Verb$=$\Verb$ ${\color{red}\Verb$s$}{\color{red}\Verb$p$}{\color{red}\Verb$a$}{\color{red}\Verb$c$}{\color{red}\Verb$e$}\\\\\Verb$ $\Verb$ ${\color{red}\Verb$l$}{\color{red}\Verb$o$}{\color{red}\Verb$w$}{\color{red}\Verb$e$}{\color{red}\Verb$r$}\Verb$ $\Verb$ $\Verb$=$\Verb$ $\Verb$<$\Verb$a$\Verb$-$\Verb$z$\Verb$>$\\\Verb$ $\Verb$ ${\color{red}\Verb$u$}{\color{red}\Verb$p$}{\color{red}\Verb$p$}{\color{red}\Verb$e$}{\color{red}\Verb$r$}\Verb$ $\Verb$ $\Verb$=$\Verb$ $\Verb$<$\Verb$A$\Verb$-$\Verb$Z$\Verb$>$\\\Verb$ $\Verb$ ${\color{red}\Verb$a$}{\color{red}\Verb$l$}{\color{red}\Verb$p$}{\color{red}\Verb$h$}{\color{red}\Verb$a$}\Verb$ $\Verb$ $\Verb$=$\Verb$ ${\color{red}\Verb$l$}{\color{red}\Verb$o$}{\color{red}\Verb$w$}{\color{red}\Verb$e$}{\color{red}\Verb$r$}\Verb$ $\Verb$|$\Verb$ ${\color{red}\Verb$u$}{\color{red}\Verb$p$}{\color{red}\Verb$p$}{\color{red}\Verb$e$}{\color{red}\Verb$r$}\\\Verb$ $\Verb$ ${\color{red}\Verb$a$}{\color{red}\Verb$l$}{\color{red}\Verb$n$}{\color{red}\Verb$u$}{\color{red}\Verb$m$}\Verb$ $\Verb$ $\Verb$=$\Verb$ ${\color{red}\Verb$a$}{\color{red}\Verb$l$}{\color{red}\Verb$p$}{\color{red}\Verb$h$}{\color{red}\Verb$a$}\Verb$ $\Verb$|$\Verb$ ${\color{red}\Verb$d$}{\color{red}\Verb$i$}{\color{red}\Verb$g$}{\color{red}\Verb$i$}{\color{red}\Verb$t$}\\\Verb$ $\Verb$ ${\color{red}\Verb$d$}{\color{red}\Verb$i$}{\color{red}\Verb$g$}{\color{red}\Verb$i$}{\color{red}\Verb$t$}\Verb$ $\Verb$ $\Verb$=$\Verb$ $\Verb$<$\Verb$0$\Verb$-$\Verb$9$\Verb$>$\\\Verb$ $\Verb$ ${\color{red}\Verb$x$}{\color{red}\Verb$d$}{\color{red}\Verb$i$}{\color{red}\Verb$g$}{\color{red}\Verb$i$}{\color{red}\Verb$t$}\Verb$ $\Verb$=$\Verb$ $\Verb$<$\Verb$0$\Verb$-$\Verb$9$\Verb$a$\Verb$-$\Verb$f$\Verb$A$\Verb$-$\Verb$F$\Verb$>$\\\Verb$ $\Verb$ ${\color{red}\Verb$w$}{\color{red}\Verb$o$}{\color{red}\Verb$r$}{\color{red}\Verb$d$}\Verb$ $\Verb$ $\Verb$ $\Verb$=$\Verb$ ${\color{red}\Verb$a$}{\color{red}\Verb$l$}{\color{red}\Verb$p$}{\color{red}\Verb$h$}{\color{red}\Verb$a$}\Verb$ $\Verb$|$\Verb$ ${\color{black}\Verb$'$}{\color{black}\Verb$_$}{\color{black}\Verb$'$}\\\\\Verb$ $\Verb$ ${\color{red}\Verb$n$}{\color{red}\Verb$e$}{\color{red}\Verb$w$}{\color{red}\Verb$l$}{\color{red}\Verb$i$}{\color{red}\Verb$n$}{\color{red}\Verb$e$}\Verb$ $\Verb$=$\Verb$ $\Verb$ ${\color{black}\Verb$'$}{\color{black}\Verb$\$}{\color{black}\Verb$r$}{\color{black}\Verb$\$}{\color{black}\Verb$n$}{\color{black}\Verb$'$}\Verb$ $\Verb$|$\Verb$ ${\color{black}\Verb$'$}{\color{black}\Verb$\$}{\color{black}\Verb$r$}{\color{black}\Verb$'$}\Verb$ $\Verb$|$\Verb$ ${\color{black}\Verb$'$}{\color{black}\Verb$\$}{\color{black}\Verb$n$}{\color{black}\Verb$'$}\\\Verb$ $\Verb$ ${\color{red}\Verb$l$}{\color{red}\Verb$i$}{\color{red}\Verb$n$}{\color{red}\Verb$e$}\Verb$ $\Verb$=$\Verb$ $\Verb$(${\color{red}\Verb$n$}{\color{red}\Verb$e$}{\color{red}\Verb$w$}{\color{red}\Verb$l$}{\color{red}\Verb$i$}{\color{red}\Verb$n$}{\color{red}\Verb$e$}\Verb$ $\Verb$b$\Verb$r$\Verb$e$\Verb$a$\Verb$k$\Verb$ $\Verb$|$\Verb$ $\Verb$.$\Verb$)${\color{black}\Verb$*$}{\color{Aquamarine}\Verb$:$}{\color{Tan}\Verb${$}{\color{Aquamarine}\Verb$i$}{\color{Aquamarine}\Verb$t$}{\color{Tan}\Verb$*$}{\color{Tan}\Verb$"$}{\color{Tan}\Verb$"$}{\color{Tan}\Verb$}$}\\\\\Verb$ $\Verb$ ${\color{red}\Verb$e$}{\color{red}\Verb$m$}{\color{red}\Verb$p$}{\color{red}\Verb$t$}{\color{red}\Verb$y$}\Verb$ $\Verb$=$\Verb$ ${\color{Tan}\Verb$-$}{\color{Tan}\Verb$>$}{\color{Tan}\Verb$ $}{\color{Tan}\Verb$n$}{\color{Tan}\Verb$i$}{\color{Tan}\Verb$l$}{\color{Tan}\\}{\color{Tan}\Verb$ $}\Verb$ ${\color{red}\Verb$e$}{\color{red}\Verb$o$}{\color{red}\Verb$f$}\Verb$=$\Verb$ ${\color{Violet}\Verb$~$}{\color{Violet}\Verb$.$}\\\\\Verb$ $\Verb$ ${\color{red}\Verb$s$}{\color{red}\Verb$e$}{\color{red}\Verb$q$}{\color{green}\Verb$($}{\color{green}\Verb$s$}{\color{green}\Verb$)$}\Verb$ $\Verb$=$\Verb$ ${\color{red}\Verb$_$}{\color{red}\Verb$s$}{\color{red}\Verb$e$}{\color{red}\Verb$q$}{\color{green}\Verb$($}{\color{green}\Verb$s$}{\color{green}\Verb$)$}\Verb$ ${\color{Tan}\Verb${$}{\color{Tan}\Verb$s$}{\color{Tan}\Verb$}$}\\\\\Verb$ $\Verb$ ${\color{red}\Verb$i$}{\color{red}\Verb$n$}{\color{red}\Verb$t$}\Verb$ $\Verb$=$\Verb$ $\Verb$(${\color{black}\Verb$'$}{\color{black}\Verb$-$}{\color{black}\Verb$'$}\Verb$|${\color{Tan}\Verb${$}{\color{Tan}\Verb$"$}{\color{Tan}\Verb$"$}{\color{Tan}\Verb$}$}\Verb$)${\color{Aquamarine}\Verb$:$}{\color{Aquamarine}\Verb$s$}\Verb$ $\Verb$($\Verb$ ${\color{black}\Verb$'$}{\color{black}\Verb$0$}{\color{black}\Verb$x$}{\color{black}\Verb$'$}\Verb$ $\Verb$<$\Verb$0$\Verb$-$\Verb$9$\Verb$a$\Verb$-$\Verb$f$\Verb$A$\Verb$-$\Verb$F$\Verb$>${\color{black}\Verb$+$}\Verb$ $\Verb$|$\Verb$ ${\color{black}\Verb$'$}{\color{black}\Verb$0$}{\color{black}\Verb$b$}{\color{black}\Verb$'$}\Verb$ $\Verb$<$\Verb$0$\Verb$1$\Verb$>${\color{black}\Verb$+$}\Verb$ $\\\Verb$ $\Verb$ $\Verb$ $\Verb$ $\Verb$ $\Verb$ $\Verb$ $\Verb$ $\Verb$ $\Verb$ $\Verb$ $\Verb$ $\Verb$ $\Verb$ $\Verb$ $\Verb$ $\Verb$ $\Verb$ $\Verb$ $\Verb$ $\Verb$ $\Verb$|$\Verb$ ${\color{black}\Verb$'$}{\color{black}\Verb$0$}{\color{black}\Verb$o$}{\color{black}\Verb$'$}\Verb$ $\Verb$<$\Verb$0$\Verb$-$\Verb$7$\Verb$>${\color{black}\Verb$+$}\Verb$ $\Verb$|$\Verb$ $\Verb$<$\Verb$0$\Verb$-$\Verb$9$\Verb$>${\color{black}\Verb$+$}\Verb$)$\Verb$[$\Verb$]${\color{Aquamarine}\Verb$:$}{\color{Aquamarine}\Verb$n$}\Verb$ ${\color{Tan}\Verb${$}{\color{Tan}\Verb$($}{\color{Aquamarine}\Verb$s$}{\color{Tan}\Verb$+$}{\color{Aquamarine}\Verb$n$}{\color{Tan}\Verb$*$}{\color{Tan}\Verb$"$}{\color{Tan}\Verb$"$}{\color{Tan}\Verb$)$}{\color{Tan}\Verb$.$}{\color{Tan}\Verb$t$}{\color{Tan}\Verb$o$}{\color{Tan}\Verb$_$}{\color{Tan}\Verb$i$}{\color{Tan}\Verb$}$}\\\Verb$ $\Verb$ ${\color{red}\Verb$n$}{\color{red}\Verb$u$}{\color{red}\Verb$m$}{\color{red}\Verb$b$}{\color{red}\Verb$e$}{\color{red}\Verb$r$}\Verb$ $\Verb$=$\Verb$ ${\color{red}\Verb$i$}{\color{red}\Verb$n$}{\color{red}\Verb$t$}\\\\\\\Verb$ $\Verb$ ${\color{red}\Verb$f$}{\color{red}\Verb$i$}{\color{red}\Verb$n$}{\color{red}\Verb$d$}{\color{green}\Verb$($}{\color{green}\Verb$e$}{\color{green}\Verb$x$}{\color{green}\Verb$p$}{\color{green}\Verb$)$}\Verb$ $\Verb$ $\Verb$ $\Verb$ $\Verb$=$\Verb$ $\Verb$(${\color{red}\Verb$a$}{\color{red}\Verb$p$}{\color{red}\Verb$p$}{\color{red}\Verb$l$}{\color{red}\Verb$y$}{\color{green}\Verb$($}{\color{green}\Verb$e$}{\color{green}\Verb$x$}{\color{green}\Verb$p$}{\color{green}\Verb$)$}{\color{Aquamarine}\Verb$:$}{\color{blue}\Verb$e$}\Verb$ $\Verb$b$\Verb$r$\Verb$e$\Verb$a$\Verb$k$\Verb$ $\Verb$|$\Verb$ $\Verb$.$\Verb$)${\color{black}\Verb$*$}\Verb$ $\Verb$.${\color{black}\Verb$*$}\Verb$ ${\color{Tan}\Verb$-$}{\color{Tan}\Verb$>$}{\color{Tan}\Verb$ $}{\color{blue}\Verb$e$}{\color{Tan}\\}{\color{Tan}\Verb$ $}\Verb$ ${\color{red}\Verb$r$}{\color{red}\Verb$e$}{\color{red}\Verb$p$}{\color{red}\Verb$l$}{\color{red}\Verb$a$}{\color{red}\Verb$c$}{\color{red}\Verb$e$}{\color{green}\Verb$($}{\color{green}\Verb$e$}{\color{green}\Verb$x$}{\color{green}\Verb$p$}{\color{green}\Verb$)$}\Verb$ $\Verb$=$\Verb$ $\Verb$(${\color{red}\Verb$a$}{\color{red}\Verb$p$}{\color{red}\Verb$p$}{\color{red}\Verb$l$}{\color{red}\Verb$y$}{\color{green}\Verb$($}{\color{green}\Verb$e$}{\color{green}\Verb$x$}{\color{green}\Verb$p$}{\color{green}\Verb$)$}\Verb$ $\Verb$ $\Verb$ $\Verb$ $\Verb$ $\Verb$ $\Verb$ $\Verb$ $\Verb$ $\Verb$|$\Verb$ $\Verb$.$\Verb$)${\color{black}\Verb$*$}{\color{Aquamarine}\Verb$:$}{\color{Tan}\Verb${$}{\color{Aquamarine}\Verb$i$}{\color{Aquamarine}\Verb$t$}{\color{Tan}\Verb$*$}{\color{Tan}\Verb$"$}{\color{Tan}\Verb$"$}{\color{Tan}\Verb$}$}\\\\\Verb$ $\Verb$ ${\color{red}\Verb$u$}{\color{red}\Verb$n$}{\color{red}\Verb$t$}{\color{red}\Verb$i$}{\color{red}\Verb$l$}{\color{green}\Verb$($}{\color{blue}\Verb$e$}{\color{green}\Verb$)$}\Verb$ $\Verb$ $\Verb$ $\Verb$ $\Verb$ $\Verb$=$\Verb$ $\Verb$($\Verb$ ${\color{red}\Verb$s$}{\color{red}\Verb$e$}{\color{red}\Verb$q$}{\color{green}\Verb$($}{\color{blue}\Verb$e$}{\color{green}\Verb$)$}\Verb$ $\Verb$b$\Verb$r$\Verb$e$\Verb$a$\Verb$k$\Verb$ $\\\Verb$ $\Verb$ $\Verb$ $\Verb$ $\Verb$ $\Verb$ $\Verb$ $\Verb$ $\Verb$ $\Verb$ $\Verb$ $\Verb$ $\Verb$ $\Verb$ $\Verb$ $\Verb$|$\Verb$ $\Verb$(${\color{black}\Verb$'$}{\color{black}\Verb$\$}{\color{black}\Verb$\$}{\color{black}\Verb$'$}{\color{Aquamarine}\Verb$:$}{\color{Aquamarine}\Verb$[$}{\color{Aquamarine}\Verb$x$}{\color{Aquamarine}\Verb$]$}\Verb$)${\color{black}\Verb$?$}\Verb$ $\Verb$.${\color{Aquamarine}\Verb$:$}{\color{Aquamarine}\Verb$[$}{\color{Aquamarine}\Verb$x$}{\color{Aquamarine}\Verb$]$}\\\Verb$ $\Verb$ $\Verb$ $\Verb$ $\Verb$ $\Verb$ $\Verb$ $\Verb$ $\Verb$ $\Verb$ $\Verb$ $\Verb$ $\Verb$ $\Verb$ $\Verb$ $\Verb$)${\color{black}\Verb$*$}\Verb$ ${\color{Tan}\Verb$-$}{\color{Tan}\Verb$>$}{\color{Tan}\Verb$ $}{\color{Aquamarine}\Verb$x$}{\color{Tan}\Verb$.$}{\color{Tan}\Verb$j$}{\color{Tan}\Verb$o$}{\color{Tan}\Verb$i$}{\color{Tan}\Verb$n$}{\color{Tan}\\}{\color{Tan}\\}\Verb$ $\Verb$ ${\color{red}\Verb$l$}{\color{red}\Verb$i$}{\color{red}\Verb$s$}{\color{red}\Verb$t$}{\color{red}\Verb$O$}{\color{red}\Verb$f$}{\color{green}\Verb$($}{\color{green}\Verb$r$}{\color{green}\Verb$u$}{\color{green}\Verb$l$}{\color{green}\Verb$e$}{\color{green}\Verb$,$}{\color{green}\Verb$d$}{\color{green}\Verb$e$}{\color{green}\Verb$l$}{\color{green}\Verb$)$}\Verb$ $\Verb$=$\Verb$ ${\color{red}\Verb$a$}{\color{red}\Verb$p$}{\color{red}\Verb$p$}{\color{red}\Verb$l$}{\color{red}\Verb$y$}{\color{green}\Verb$($}{\color{green}\Verb$r$}{\color{green}\Verb$u$}{\color{green}\Verb$l$}{\color{green}\Verb$e$}{\color{green}\Verb$)$}{\color{Aquamarine}\Verb$:$}{\color{Aquamarine}\Verb$[$}{\color{Aquamarine}\Verb$f$}{\color{Aquamarine}\Verb$]$}\Verb$ $\Verb$(${\color{red}\Verb$s$}{\color{red}\Verb$e$}{\color{red}\Verb$q$}{\color{green}\Verb$($}{\color{green}\Verb$d$}{\color{green}\Verb$e$}{\color{green}\Verb$l$}{\color{green}\Verb$)$}\Verb$ ${\color{red}\Verb$a$}{\color{red}\Verb$p$}{\color{red}\Verb$p$}{\color{red}\Verb$l$}{\color{red}\Verb$y$}{\color{green}\Verb$($}{\color{green}\Verb$r$}{\color{green}\Verb$u$}{\color{green}\Verb$l$}{\color{green}\Verb$e$}{\color{green}\Verb$)$}\Verb$)${\color{black}\Verb$*$}{\color{Aquamarine}\Verb$:$}{\color{Aquamarine}\Verb$[$}{\color{Aquamarine}\Verb$*$}{\color{Aquamarine}\Verb$f$}{\color{Aquamarine}\Verb$]$}\Verb$ ${\color{Tan}\Verb$-$}{\color{Tan}\Verb$>$}{\color{Aquamarine}\Verb$f$}{\color{Tan}\\}{\color{Tan}\Verb$ $}\Verb$ $\Verb$ $\Verb$ $\Verb$ $\Verb$ $\Verb$ $\Verb$ $\Verb$ $\Verb$ $\Verb$ $\Verb$ $\Verb$ $\Verb$ $\Verb$ $\Verb$ $\Verb$ $\Verb$ $\Verb$ $\Verb$|$\Verb$ ${\color{red}\Verb$e$}{\color{red}\Verb$m$}{\color{red}\Verb$p$}{\color{red}\Verb$t$}{\color{red}\Verb$y$}\Verb$ ${\color{Tan}\Verb$-$}{\color{Tan}\Verb$>$}{\color{Tan}\Verb$ $}{\color{Tan}\Verb$[$}{\color{Tan}\Verb$]$}{\color{Tan}\\}{\color{Tan}\\}\Verb$ $\Verb$ ${\color{red}\Verb$r$}{\color{red}\Verb$e$}{\color{red}\Verb$v$}{\color{red}\Verb$e$}{\color{red}\Verb$r$}{\color{red}\Verb$s$}{\color{red}\Verb$e$}{\color{green}\Verb$($}{\color{green}\Verb$l$}{\color{green}\Verb$)$}\Verb$ $\Verb$=$\Verb$ ${\color{Tan}\Verb${$}{\color{Aquamarine}\Verb$@$}{\color{Aquamarine}\Verb$@$}{\color{Aquamarine}\Verb$r$}{\color{Aquamarine}\Verb$e$}{\color{Aquamarine}\Verb$v$}{\color{Tan}\Verb$|$}{\color{Tan}\Verb$|$}{\color{Tan}\Verb$=$}{\color{Tan}\Verb$H$}{\color{Tan}\Verb$a$}{\color{Tan}\Verb$s$}{\color{Tan}\Verb$h$}{\color{Tan}\Verb$.$}{\color{Tan}\Verb$n$}{\color{Tan}\Verb$e$}{\color{Tan}\Verb$w$}{\color{Tan}\Verb${$}{\color{Tan}\Verb$|$}{\color{Tan}\Verb$h$}{\color{Tan}\Verb$,$}{\color{Tan}\Verb$k$}{\color{Tan}\Verb$|$}{\color{Tan}\Verb$ $}{\color{Tan}\Verb$h$}{\color{Tan}\Verb$[$}{\color{Tan}\Verb$k$}{\color{Tan}\Verb$]$}{\color{Tan}\Verb$=$}{\color{Tan}\Verb$k$}{\color{Tan}\Verb$.$}{\color{Tan}\Verb$r$}{\color{Tan}\Verb$e$}{\color{Tan}\Verb$v$}{\color{Tan}\Verb$e$}{\color{Tan}\Verb$r$}{\color{Tan}\Verb$s$}{\color{Tan}\Verb$e$}{\color{Tan}\Verb$ $}{\color{Tan}\Verb$}$}{\color{Tan}\Verb$}$}\Verb$ $\\\Verb$ $\Verb$ $\Verb$ $\Verb$ $\Verb$ $\Verb$ $\Verb$ $\Verb$ $\Verb$ $\Verb$ $\Verb$ $\Verb$ $\Verb$ $\Verb$ $\Verb$ $\Verb$ ${\color{red}\Verb$_$}{\color{red}\Verb$r$}{\color{red}\Verb$e$}{\color{red}\Verb$v$}{\color{red}\Verb$e$}{\color{red}\Verb$r$}{\color{red}\Verb$s$}{\color{red}\Verb$e$}{\color{green}\Verb$($}{\color{Aquamarine}\Verb$@$}{\color{Aquamarine}\Verb$@$}{\color{Aquamarine}\Verb$r$}{\color{Aquamarine}\Verb$e$}{\color{Aquamarine}\Verb$v$}{\color{green}\Verb$[$}{\color{Aquamarine}\Verb$@$}{\color{Aquamarine}\Verb$s$}{\color{Aquamarine}\Verb$e$}{\color{Aquamarine}\Verb$l$}{\color{Aquamarine}\Verb$f$}{\color{green}\Verb$]$}{\color{green}\Verb$)$}\Verb$ $\\\Verb$ $\Verb$ $\Verb$ $\Verb$ $\Verb$ $\Verb$ $\Verb$ $\Verb$ $\Verb$ $\Verb$ $\Verb$ $\Verb$ $\Verb$ $\Verb$ $\Verb$ $\Verb$ ${\color{red}\Verb$a$}{\color{red}\Verb$p$}{\color{red}\Verb$p$}{\color{red}\Verb$l$}{\color{red}\Verb$y$}{\color{green}\Verb$($}{\color{green}\Verb$l$}{\color{green}\Verb$)$}{\color{Aquamarine}\Verb$:$}{\color{Aquamarine}\Verb$r$}{\color{Aquamarine}\Verb$e$}{\color{Aquamarine}\Verb$v$}\Verb$ $\\\Verb$ $\Verb$ $\Verb$ $\Verb$ $\Verb$ $\Verb$ $\Verb$ $\Verb$ $\Verb$ $\Verb$ $\Verb$ $\Verb$ $\Verb$ $\Verb$ $\Verb$ $\Verb$ ${\color{red}\Verb$_$}{\color{red}\Verb$r$}{\color{red}\Verb$e$}{\color{red}\Verb$v$}{\color{red}\Verb$e$}{\color{red}\Verb$r$}{\color{red}\Verb$s$}{\color{red}\Verb$e$}{\color{green}\Verb$($}{\color{Aquamarine}\Verb$@$}{\color{Aquamarine}\Verb$@$}{\color{Aquamarine}\Verb$r$}{\color{Aquamarine}\Verb$e$}{\color{Aquamarine}\Verb$v$}{\color{green}\Verb$[$}{\color{Aquamarine}\Verb$@$}{\color{Aquamarine}\Verb$s$}{\color{Aquamarine}\Verb$e$}{\color{Aquamarine}\Verb$l$}{\color{Aquamarine}\Verb$f$}{\color{green}\Verb$]$}{\color{green}\Verb$)$}\Verb$ $\\\Verb$ $\Verb$ $\Verb$ $\Verb$ $\Verb$ $\Verb$ $\Verb$ $\Verb$ $\Verb$ $\Verb$ $\Verb$ $\Verb$ $\Verb$ $\Verb$ $\Verb$ ${\color{Tan}\Verb${$}{\color{Aquamarine}\Verb$r$}{\color{Aquamarine}\Verb$e$}{\color{Aquamarine}\Verb$v$}{\color{Tan}\Verb$}$}\Verb$ $\\\Verb$ $\Verb$ $\\\Verb$ $\Verb$ ${\color{red}\Verb$f$}{\color{red}\Verb$a$}{\color{red}\Verb$i$}{\color{red}\Verb$l$}{\color{red}\Verb$s$}\Verb$ $\Verb$=$\Verb$ ${\color{Violet}\Verb$&$}{\color{Violet}\Verb${$}{\color{Tan}\Verb$f$}{\color{Tan}\Verb$a$}{\color{Tan}\Verb$l$}{\color{Tan}\Verb$s$}{\color{Tan}\Verb$e$}{\color{Violet}\Verb$}$}\\\\\Verb$ $\Verb$ ${\color{red}\Verb$c$}{\color{red}\Verb$h$}{\color{red}\Verb$a$}{\color{red}\Verb$r$}\Verb$=$\Verb$ $\Verb$.${\color{Aquamarine}\Verb$:$}{\color{Aquamarine}\Verb$c$}\Verb$ ${\color{Violet}\Verb$&$}{\color{Violet}\Verb${$}{\color{Aquamarine}\Verb$c$}{\color{Tan}\Verb$.$}{\color{Tan}\Verb$i$}{\color{Tan}\Verb$s$}{\color{Tan}\Verb$_$}{\color{Tan}\Verb$a$}{\color{Tan}\Verb$?$}{\color{Tan}\Verb$ $}{\color{Tan}\Verb$S$}{\color{Tan}\Verb$t$}{\color{Tan}\Verb$r$}{\color{Tan}\Verb$i$}{\color{Tan}\Verb$n$}{\color{Tan}\Verb$g$}{\color{Tan}\Verb$ $}{\color{Violet}\Verb$}$}\Verb$ ${\color{Tan}\Verb$-$}{\color{Tan}\Verb$>$}{\color{Tan}\Verb$ $}{\color{Aquamarine}\Verb$c$}{\color{Tan}\\}{\color{Tan}\\}\Verb$ $\Verb$ ${\color{red}\Verb$m$}{\color{red}\Verb$e$}{\color{red}\Verb$m$}{\color{red}\Verb$b$}{\color{red}\Verb$e$}{\color{red}\Verb$r$}{\color{green}\Verb$($}{\color{green}\Verb$x$}{\color{green}\Verb$)$}\Verb$ $\Verb$ $\Verb$ $\Verb$ $\Verb$ $\Verb$=$\Verb$ $\Verb$.${\color{Aquamarine}\Verb$:$}{\color{Aquamarine}\Verb$a$}\Verb$ ${\color{Violet}\Verb$&$}{\color{Violet}\Verb${$}{\color{Tan}\Verb$x$}{\color{Tan}\Verb$ $}{\color{Tan}\Verb$=$}{\color{Tan}\Verb$=$}{\color{Tan}\Verb$=$}{\color{Tan}\Verb$ $}{\color{Aquamarine}\Verb$a$}{\color{Violet}\Verb$}$}\Verb$ ${\color{Tan}\Verb${$}{\color{Aquamarine}\Verb$a$}{\color{Tan}\Verb$}$}\\\Verb$ $\Verb$ ${\color{red}\Verb$t$}{\color{red}\Verb$r$}{\color{red}\Verb$u$}{\color{red}\Verb$e$}\Verb$ $\Verb$ $\Verb$ $\Verb$ $\Verb$ $\Verb$ $\Verb$ $\Verb$ $\Verb$ $\Verb$ $\Verb$=$\Verb$ ${\color{red}\Verb$m$}{\color{red}\Verb$e$}{\color{red}\Verb$m$}{\color{red}\Verb$b$}{\color{red}\Verb$e$}{\color{red}\Verb$r$}{\color{green}\Verb$($}{\color{green}\Verb$t$}{\color{green}\Verb$r$}{\color{green}\Verb$u$}{\color{green}\Verb$e$}{\color{green}\Verb$)$}\Verb$ $\\\Verb$ $\Verb$ ${\color{red}\Verb$f$}{\color{red}\Verb$a$}{\color{red}\Verb$l$}{\color{red}\Verb$s$}{\color{red}\Verb$e$}\Verb$ $\Verb$ $\Verb$ $\Verb$ $\Verb$ $\Verb$ $\Verb$ $\Verb$ $\Verb$ $\Verb$=$\Verb$ ${\color{red}\Verb$m$}{\color{red}\Verb$e$}{\color{red}\Verb$m$}{\color{red}\Verb$b$}{\color{red}\Verb$e$}{\color{red}\Verb$r$}{\color{green}\Verb$($}{\color{green}\Verb$f$}{\color{green}\Verb$a$}{\color{green}\Verb$l$}{\color{green}\Verb$s$}{\color{green}\Verb$e$}{\color{green}\Verb$)$}\Verb$ $\Verb$ $\\\Verb$ $\Verb$ ${\color{red}\Verb$n$}{\color{red}\Verb$i$}{\color{red}\Verb$l$}\Verb$ $\Verb$ $\Verb$ $\Verb$ $\Verb$ $\Verb$ $\Verb$ $\Verb$ $\Verb$ $\Verb$ $\Verb$ $\Verb$=$\Verb$ ${\color{red}\Verb$m$}{\color{red}\Verb$e$}{\color{red}\Verb$m$}{\color{red}\Verb$b$}{\color{red}\Verb$e$}{\color{red}\Verb$r$}{\color{green}\Verb$($}{\color{green}\Verb$n$}{\color{green}\Verb$i$}{\color{green}\Verb$l$}{\color{green}\Verb$)$}\Verb$ $\\\Verb$ $\Verb$ ${\color{red}\Verb$c$}{\color{red}\Verb$l$}{\color{red}\Verb$a$}{\color{red}\Verb$s$}{\color{green}\Verb$($}{\color{green}\Verb$c$}{\color{green}\Verb$l$}{\color{green}\Verb$s$}{\color{green}\Verb$)$}\Verb$ $\Verb$ $\Verb$ $\Verb$ $\Verb$ $\Verb$=$\Verb$ ${\color{red}\Verb$m$}{\color{red}\Verb$e$}{\color{red}\Verb$m$}{\color{red}\Verb$b$}{\color{red}\Verb$e$}{\color{red}\Verb$r$}{\color{green}\Verb$($}{\color{green}\Verb$c$}{\color{green}\Verb$l$}{\color{green}\Verb$s$}{\color{green}\Verb$)$}\\\Verb$ $\Verb$ ${\color{red}\Verb$r$}{\color{red}\Verb$a$}{\color{red}\Verb$n$}{\color{red}\Verb$g$}{\color{red}\Verb$e$}{\color{red}\Verb$_$}{\color{red}\Verb$i$}{\color{red}\Verb$n$}{\color{green}\Verb$($}{\color{green}\Verb$a$}{\color{green}\Verb$,$}{\color{green}\Verb$b$}{\color{green}\Verb$)$}\Verb$ $\Verb$=$\Verb$ ${\color{red}\Verb$m$}{\color{red}\Verb$e$}{\color{red}\Verb$m$}{\color{red}\Verb$b$}{\color{red}\Verb$e$}{\color{red}\Verb$r$}{\color{green}\Verb$($}{\color{green}\Verb$a$}{\color{green}\Verb$.$}{\color{green}\Verb$.$}{\color{green}\Verb$ $}{\color{green}\Verb$b$}{\color{green}\Verb$)$}\\\Verb$ $\Verb$ ${\color{red}\Verb$r$}{\color{red}\Verb$a$}{\color{red}\Verb$n$}{\color{red}\Verb$g$}{\color{red}\Verb$e$}{\color{red}\Verb$_$}{\color{red}\Verb$e$}{\color{red}\Verb$x$}{\color{green}\Verb$($}{\color{green}\Verb$a$}{\color{green}\Verb$,$}{\color{green}\Verb$b$}{\color{green}\Verb$)$}\Verb$ $\Verb$=$\Verb$ ${\color{red}\Verb$m$}{\color{red}\Verb$e$}{\color{red}\Verb$m$}{\color{red}\Verb$b$}{\color{red}\Verb$e$}{\color{red}\Verb$r$}{\color{green}\Verb$($}{\color{green}\Verb$a$}{\color{green}\Verb$.$}{\color{green}\Verb$.$}{\color{green}\Verb$.$}{\color{green}\Verb$b$}{\color{green}\Verb$)$}\\\Verb$ $\Verb$ ${\color{red}\Verb$r$}{\color{red}\Verb$e$}{\color{red}\Verb$g$}{\color{red}\Verb$c$}{\color{red}\Verb$h$}{\color{green}\Verb$($}{\color{green}\Verb$r$}{\color{green}\Verb$e$}{\color{green}\Verb$g$}{\color{green}\Verb$e$}{\color{green}\Verb$x$}{\color{green}\Verb$)$}\Verb$ $\Verb$ $\Verb$=$\Verb$ ${\color{red}\Verb$m$}{\color{red}\Verb$e$}{\color{red}\Verb$m$}{\color{red}\Verb$b$}{\color{red}\Verb$e$}{\color{red}\Verb$r$}{\color{green}\Verb$($}{\color{green}\Verb$r$}{\color{green}\Verb$e$}{\color{green}\Verb$g$}{\color{green}\Verb$e$}{\color{green}\Verb$x$}{\color{green}\Verb$)$}\\\\\Verb$ $\Verb$ ${\color{red}\Verb$p$}{\color{red}\Verb$a$}{\color{red}\Verb$r$}{\color{red}\Verb$s$}{\color{red}\Verb$e$}{\color{green}\Verb$($}{\color{green}\Verb$r$}{\color{green}\Verb$u$}{\color{green}\Verb$l$}{\color{green}\Verb$e$}{\color{green}\Verb$,$}{\color{green}\Verb$o$}{\color{green}\Verb$b$}{\color{green}\Verb$j$}{\color{green}\Verb$,$}{\color{green}\Verb$a$}{\color{green}\Verb$)$}\Verb$ $\Verb$=$\Verb$ ${\color{Tan}\Verb${$}{\color{Tan}\Verb$o$}{\color{Tan}\Verb$b$}{\color{Tan}\Verb$j$}{\color{Tan}\Verb$}$}{\color{blue}\Verb$[$}{\color{Tan}\Verb${$}{\color{Tan}\Verb$a$}{\color{Tan}\Verb$p$}{\color{Tan}\Verb$p$}{\color{Tan}\Verb$l$}{\color{Tan}\Verb$y$}{\color{Tan}\Verb$($}{\color{Tan}\Verb$r$}{\color{Tan}\Verb$u$}{\color{Tan}\Verb$l$}{\color{Tan}\Verb$e$}{\color{Tan}\Verb$,$}{\color{Tan}\Verb$*$}{\color{Tan}\Verb$a$}{\color{Tan}\Verb$)$}{\color{Tan}\Verb$}$}{\color{Aquamarine}\Verb$:$}{\color{Aquamarine}\Verb$r$}\Verb$ ${\color{Tan}\Verb${$}{\color{Tan}\Verb$s$}{\color{Tan}\Verb$e$}{\color{Tan}\Verb$l$}{\color{Tan}\Verb$f$}{\color{Tan}\Verb$.$}{\color{Tan}\Verb$p$}{\color{Tan}\Verb$r$}{\color{Tan}\Verb$o$}{\color{Tan}\Verb$f$}{\color{Tan}\Verb$_$}{\color{Tan}\Verb$r$}{\color{Tan}\Verb$e$}{\color{Tan}\Verb$p$}{\color{Tan}\Verb$o$}{\color{Tan}\Verb$r$}{\color{Tan}\Verb$t$}{\color{Tan}\Verb$;$}{\color{Aquamarine}\Verb$r$}{\color{Tan}\Verb$}$}{\color{blue}\Verb$]$}\\\\\Verb$ $\Verb$ ${\color{red}\Verb$n$}{\color{red}\Verb$e$}{\color{red}\Verb$s$}{\color{red}\Verb$t$}{\color{red}\Verb$e$}{\color{red}\Verb$d$}{\color{green}\Verb$($}{\color{green}\Verb$s$}{\color{green}\Verb$t$}{\color{green}\Verb$a$}{\color{green}\Verb$r$}{\color{green}\Verb$t$}{\color{green}\Verb$,$}{\color{green}\Verb$m$}{\color{green}\Verb$i$}{\color{green}\Verb$d$}{\color{green}\Verb$,$}{\color{green}\Verb$e$}{\color{green}\Verb$n$}{\color{green}\Verb$d$}{\color{green}\Verb$)$}\Verb$ $\Verb$=$\Verb$ ${\color{red}\Verb$s$}{\color{red}\Verb$e$}{\color{red}\Verb$q$}{\color{green}\Verb$($}{\color{green}\Verb$s$}{\color{green}\Verb$t$}{\color{green}\Verb$a$}{\color{green}\Verb$r$}{\color{green}\Verb$t$}{\color{green}\Verb$)$}\Verb$ ${\color{red}\Verb$a$}{\color{red}\Verb$p$}{\color{red}\Verb$p$}{\color{red}\Verb$l$}{\color{red}\Verb$y$}{\color{green}\Verb$($}{\color{green}\Verb$m$}{\color{green}\Verb$i$}{\color{green}\Verb$d$}{\color{green}\Verb$)$}\Verb$ ${\color{red}\Verb$s$}{\color{red}\Verb$e$}{\color{red}\Verb$q$}{\color{green}\Verb$($}{\color{green}\Verb$e$}{\color{green}\Verb$n$}{\color{green}\Verb$d$}{\color{green}\Verb$)$}\\\Verb$}$
\end{exambox}
}
\end{spacing}\vskip -0.4em

\end{prolbox}

\chapter{Peridot grammar} \label{append_per}
\vskip -1.8em\begin{spacing}{0.8}
{\small
\begin{exambox}
\Verb$c$\Verb$l$\Verb$a$\Verb$s$\Verb$s$\Verb$ $\Verb$P$\Verb$e$\Verb$r$\Verb$i$\Verb$d$\Verb$o$\Verb$t$\Verb$_$\Verb$p$\Verb$a$\Verb$r$\Verb$s$\Verb$e$\Verb$r$\Verb$ $\Verb$<$\Verb$ $\Verb$A$\Verb$m$\Verb$e$\Verb$t$\Verb$h$\Verb$y$\Verb$s$\Verb$t$\\\Verb$ $\Verb$ $\Verb$d$\Verb$e$\Verb$f$\Verb$ $\Verb$c$\Verb$a$\Verb$l$\Verb$l$\Verb$($\Verb$n$\Verb$a$\Verb$m$\Verb$e$\Verb$,$\Verb$*$\Verb$a$\Verb$r$\Verb$g$\Verb$s$\Verb$)$\\\Verb$ $\Verb$ $\Verb$ $\Verb$ $\Verb$C$\Verb$a$\Verb$l$\Verb$l$\Verb$[$\Verb${$\Verb$:$\Verb$n$\Verb$a$\Verb$m$\Verb$e$\Verb$=$\Verb$>$\Verb$l$\Verb$e$\Verb$t$\Verb$e$\Verb$r$\Verb$i$\Verb$z$\Verb$e$\Verb$($\Verb$n$\Verb$a$\Verb$m$\Verb$e$\Verb$)$\Verb$,$\Verb$:$\Verb$a$\Verb$r$\Verb$y$\Verb$=$\Verb$>$\Verb$a$\Verb$r$\Verb$g$\Verb$s$\Verb$}$\Verb$]$\\\Verb$ $\Verb$ $\Verb$e$\Verb$n$\Verb$d$\\\Verb$e$\Verb$n$\Verb$d$\\\Verb$a$\Verb$m$\Verb$e$\Verb$t$\Verb$h$\Verb$y$\Verb$s$\Verb$t$\Verb$ $\Verb$P$\Verb$e$\Verb$r$\Verb$i$\Verb$d$\Verb$o$\Verb$t$\Verb$_$\Verb$p$\Verb$a$\Verb$r$\Verb$s$\Verb$e$\Verb$r$\Verb${$\\\Verb$ $\Verb$ ${\color{red}\Verb$r$}{\color{red}\Verb$o$}{\color{red}\Verb$o$}{\color{red}\Verb$t$}\Verb$ $\Verb$=$\Verb$ $\Verb$(${\color{red}\Verb$b$}{\color{red}\Verb$o$}{\color{red}\Verb$d$}{\color{red}\Verb$y$}\Verb$ $\Verb$|$\Verb$ ${\color{red}\Verb$d$}{\color{red}\Verb$e$}{\color{red}\Verb$f$}{\color{red}\Verb$i$}\Verb$ $\Verb$|$\Verb$ ${\color{red}\Verb$s$}{\color{red}\Verb$e$}{\color{red}\Verb$q$}{\color{red}\Verb$u$}{\color{red}\Verb$e$}{\color{red}\Verb$n$}{\color{red}\Verb$c$}{\color{red}\Verb$e$}\Verb$)${\color{black}\Verb$*$}{\color{Aquamarine}\Verb$:$}{\color{Aquamarine}\Verb$a$}\Verb$ $\Verb$.${\color{black}\Verb$*$}\Verb$ ${\color{Tan}\Verb${$}{\color{Aquamarine}\Verb$a$}{\color{Tan}\Verb$}$}\\\Verb$ $\Verb$ $\\\Verb$ $\Verb$ ${\color{red}\Verb$b$}{\color{red}\Verb$o$}{\color{red}\Verb$d$}{\color{red}\Verb$y$}\Verb$ $\Verb$=$\Verb$ ${\color{black}\Verb$"$}{\color{black}\Verb$c$}{\color{black}\Verb$l$}{\color{black}\Verb$a$}{\color{black}\Verb$s$}{\color{black}\Verb$s$}{\color{black}\Verb$"$}\Verb$ ${\color{black}\Verb$"$}{\color{black}\Verb$"$}\Verb$ ${\color{red}\Verb$n$}{\color{red}\Verb$a$}{\color{red}\Verb$m$}{\color{red}\Verb$e$}{\color{Aquamarine}\Verb$:$}{\color{Aquamarine}\Verb$n$}{\color{Aquamarine}\Verb$a$}{\color{Aquamarine}\Verb$m$}{\color{Aquamarine}\Verb$e$}\Verb$ ${\color{red}\Verb$d$}{\color{red}\Verb$e$}{\color{red}\Verb$f$}{\color{red}\Verb$i$}{\color{black}\Verb$*$}{\color{Aquamarine}\Verb$:$}{\color{Aquamarine}\Verb$a$}{\color{Aquamarine}\Verb$r$}{\color{Aquamarine}\Verb$y$}\Verb$ ${\color{black}\Verb$"$}{\color{black}\Verb$e$}{\color{black}\Verb$n$}{\color{black}\Verb$d$}{\color{black}\Verb$"$}\Verb$ ${\color{Aquamarine}\Verb$@$}{\color{Aquamarine}\Verb$K$}{\color{Aquamarine}\Verb$l$}{\color{Aquamarine}\Verb$a$}{\color{Aquamarine}\Verb$s$}{\color{Aquamarine}\Verb$s$}\\\\\Verb$ $\Verb$ ${\color{red}\Verb$d$}{\color{red}\Verb$e$}{\color{red}\Verb$f$}{\color{red}\Verb$a$}{\color{red}\Verb$r$}{\color{red}\Verb$g$}\Verb$=$\Verb$ $\Verb$<$\Verb$^$\Verb$,$\Verb$)$\Verb$>${\color{black}\Verb$+$}{\color{Aquamarine}\Verb$:$}{\color{Tan}\Verb${$}{\color{Aquamarine}\Verb$i$}{\color{Aquamarine}\Verb$t$}{\color{Tan}\Verb$.$}{\color{Tan}\Verb$j$}{\color{Tan}\Verb$o$}{\color{Tan}\Verb$i$}{\color{Tan}\Verb$n$}{\color{Tan}\Verb$}$}\\\Verb$ $\Verb$ $\\\Verb$ $\Verb$ ${\color{red}\Verb$d$}{\color{red}\Verb$e$}{\color{red}\Verb$f$}{\color{red}\Verb$i$}\Verb$ $\Verb$=$\Verb$ ${\color{black}\Verb$"$}{\color{black}\Verb$d$}{\color{black}\Verb$e$}{\color{black}\Verb$f$}{\color{black}\Verb$"$}\Verb$ ${\color{black}\Verb$"$}{\color{black}\Verb$"$}\Verb$ ${\color{red}\Verb$d$}{\color{red}\Verb$e$}{\color{red}\Verb$f$}{\color{red}\Verb$n$}{\color{red}\Verb$a$}{\color{red}\Verb$m$}{\color{red}\Verb$e$}{\color{Aquamarine}\Verb$:$}{\color{Aquamarine}\Verb$n$}{\color{Aquamarine}\Verb$a$}{\color{Aquamarine}\Verb$m$}{\color{Aquamarine}\Verb$e$}\Verb$ ${\color{Tan}\Verb${$}{\color{Tan}\Verb$[$}{\color{Tan}\Verb$"$}{\color{Tan}\Verb$o$}{\color{Tan}\Verb$b$}{\color{Tan}\Verb$j$}{\color{Tan}\Verb$ $}{\color{Tan}\Verb$s$}{\color{Tan}\Verb$e$}{\color{Tan}\Verb$l$}{\color{Tan}\Verb$f$}{\color{Tan}\Verb$"$}{\color{Tan}\Verb$]$}{\color{Tan}\Verb$}$}{\color{Aquamarine}\Verb$:$}{\color{Aquamarine}\Verb$a$}{\color{Aquamarine}\Verb$r$}{\color{Aquamarine}\Verb$g$}{\color{Aquamarine}\Verb$s$}\Verb$ $\Verb$ $\\\Verb$ $\Verb$ $\Verb$ $\Verb$ $\Verb$ $\Verb$ $\Verb$ $\Verb$ $\Verb$ ${\color{black}\Verb$'$}{\color{black}\Verb$($}{\color{black}\Verb$'$}\Verb$ ${\color{red}\Verb$l$}{\color{red}\Verb$i$}{\color{red}\Verb$s$}{\color{red}\Verb$t$}{\color{red}\Verb$O$}{\color{red}\Verb$f$}{\color{green}\Verb$($}{\color{green}\Verb$'$}{\color{green}\Verb$d$}{\color{green}\Verb$e$}{\color{green}\Verb$f$}{\color{green}\Verb$a$}{\color{green}\Verb$r$}{\color{green}\Verb$g$}{\color{green}\Verb$'$}{\color{green}\Verb$,$}{\color{green}\Verb$'$}{\color{green}\Verb$,$}{\color{green}\Verb$'$}{\color{green}\Verb$)$}{\color{Aquamarine}\Verb$:$}{\color{Aquamarine}\Verb$[$}{\color{Aquamarine}\Verb$a$}{\color{Aquamarine}\Verb$r$}{\color{Aquamarine}\Verb$g$}{\color{Aquamarine}\Verb$s$}{\color{Aquamarine}\Verb$]$}\Verb$ ${\color{black}\Verb$'$}{\color{black}\Verb$)$}{\color{black}\Verb$'$}\Verb$ $\Verb$ $\\\Verb$ $\Verb$ $\Verb$ $\Verb$ $\Verb$ $\Verb$ $\Verb$ $\Verb$ $\Verb$ ${\color{red}\Verb$s$}{\color{red}\Verb$e$}{\color{red}\Verb$q$}{\color{red}\Verb$u$}{\color{red}\Verb$e$}{\color{red}\Verb$n$}{\color{red}\Verb$c$}{\color{red}\Verb$e$}{\color{Aquamarine}\Verb$:$}{\color{Aquamarine}\Verb$[$}{\color{Aquamarine}\Verb$a$}{\color{Aquamarine}\Verb$r$}{\color{Aquamarine}\Verb$y$}{\color{Aquamarine}\Verb$]$}\Verb$ ${\color{black}\Verb$"$}{\color{black}\Verb$e$}{\color{black}\Verb$n$}{\color{black}\Verb$d$}{\color{black}\Verb$"$}\Verb$ ${\color{Aquamarine}\Verb$@$}{\color{Aquamarine}\Verb$D$}{\color{Aquamarine}\Verb$e$}{\color{Aquamarine}\Verb$f$}\\\Verb$ $\Verb$ ${\color{red}\Verb$n$}{\color{red}\Verb$a$}{\color{red}\Verb$m$}{\color{red}\Verb$e$}\Verb$ $\Verb$=$\Verb$ $\Verb$<$\Verb$a$\Verb$-$\Verb$z$\Verb$A$\Verb$-$\Verb$Z$\Verb$_$\Verb$>${\color{Aquamarine}\Verb$:$}{\color{Aquamarine}\Verb$s$}\Verb$ $\Verb$<$\Verb$a$\Verb$-$\Verb$z$\Verb$A$\Verb$-$\Verb$Z$\Verb$0$\Verb$-$\Verb$9$\Verb$_$\Verb$>${\color{black}\Verb$*$}{\color{Aquamarine}\Verb$:$}{\color{Tan}\Verb${$}{\color{Aquamarine}\Verb$s$}{\color{Tan}\Verb$+$}{\color{Aquamarine}\Verb$i$}{\color{Aquamarine}\Verb$t$}{\color{Tan}\Verb$*$}{\color{Tan}\Verb$"$}{\color{Tan}\Verb$"$}{\color{Tan}\Verb$}$}\\\Verb$ $\Verb$ ${\color{red}\Verb$d$}{\color{red}\Verb$e$}{\color{red}\Verb$f$}{\color{red}\Verb$n$}{\color{red}\Verb$a$}{\color{red}\Verb$m$}{\color{red}\Verb$e$}\Verb$ $\Verb$=$\Verb$ $\Verb$($\Verb$<$\Verb$^$\Verb$ $\Verb$\$\Verb$t$\Verb$\$\Verb$r$\Verb$\$\Verb$n$\Verb$($\Verb$)$\Verb$>$\Verb$)${\color{black}\Verb$*$}{\color{Aquamarine}\Verb$:$}{\color{Aquamarine}\Verb$x$}\Verb$ ${\color{Tan}\Verb${$}{\color{Tan}\Verb$l$}{\color{Tan}\Verb$e$}{\color{Tan}\Verb$t$}{\color{Tan}\Verb$e$}{\color{Tan}\Verb$r$}{\color{Tan}\Verb$i$}{\color{Tan}\Verb$z$}{\color{Tan}\Verb$e$}{\color{Tan}\Verb$($}{\color{Aquamarine}\Verb$x$}{\color{Tan}\Verb$*$}{\color{Tan}\Verb$"$}{\color{Tan}\Verb$"$}{\color{Tan}\Verb$)$}{\color{Tan}\Verb$}$}\\\Verb$ $\Verb$ $\\\Verb$ $\Verb$ ${\color{red}\Verb$a$}{\color{red}\Verb$t$}{\color{red}\Verb$o$}{\color{red}\Verb$m$}\Verb$ $\Verb$=$\Verb$ ${\color{black}\Verb$"$}{\color{black}\Verb$"$}\\\Verb$ $\Verb$ $\Verb$ $\Verb$ $\Verb$ $\Verb$ $\Verb$ $\Verb$($\Verb$ ${\color{red}\Verb$n$}{\color{red}\Verb$u$}{\color{red}\Verb$m$}{\color{red}\Verb$b$}{\color{red}\Verb$e$}{\color{red}\Verb$r$}{\color{Aquamarine}\Verb$:$}{\color{Aquamarine}\Verb$n$}\Verb$ $\Verb$ $\Verb$ $\Verb$ $\Verb$ $\Verb$ $\Verb$ $\Verb$ $\Verb$ ${\color{Tan}\Verb$-$}{\color{Tan}\Verb$>$}{\color{Tan}\Verb$ $}{\color{Tan}\Verb$C$}{\color{Tan}\Verb$C$}{\color{Tan}\Verb$o$}{\color{Tan}\Verb$d$}{\color{Tan}\Verb$e$}{\color{Tan}\Verb$[$}{\color{Tan}\Verb$"$}{\color{Tan}\Verb$I$}{\color{Tan}\Verb$n$}{\color{Tan}\Verb$t$}{\color{Tan}\Verb$($}{\color{Tan}\Verb$#$}{\color{Tan}\Verb${$}{\color{Aquamarine}\Verb$n$}{\color{Tan}\Verb$}$}{\color{Tan}\Verb$)$}{\color{Tan}\Verb$"$}{\color{Tan}\Verb$]$}{\color{Tan}\\}{\color{Tan}\Verb$ $}\Verb$ $\Verb$ $\Verb$ $\Verb$ $\Verb$ $\Verb$ $\Verb$|$\Verb$ ${\color{black}\Verb$'$}{\color{black}\Verb$"$}{\color{black}\Verb$'$}\Verb$ ${\color{red}\Verb$u$}{\color{red}\Verb$n$}{\color{red}\Verb$t$}{\color{red}\Verb$i$}{\color{red}\Verb$l$}{\color{green}\Verb$($}{\color{green}\Verb$'$}{\color{green}\Verb$"$}{\color{green}\Verb$'$}{\color{green}\Verb$)$}{\color{Aquamarine}\Verb$:$}{\color{Aquamarine}\Verb$s$}\Verb$ ${\color{Tan}\Verb$-$}{\color{Tan}\Verb$>$}{\color{Tan}\Verb$ $}{\color{Tan}\Verb$C$}{\color{Tan}\Verb$C$}{\color{Tan}\Verb$o$}{\color{Tan}\Verb$d$}{\color{Tan}\Verb$e$}{\color{Tan}\Verb$[$}{\color{Tan}\Verb$"$}{\color{Tan}\Verb$S$}{\color{Tan}\Verb$t$}{\color{Tan}\Verb$r$}{\color{Tan}\Verb$($}{\color{Tan}\Verb$#$}{\color{Tan}\Verb${$}{\color{Aquamarine}\Verb$s$}{\color{Tan}\Verb$.$}{\color{Tan}\Verb$i$}{\color{Tan}\Verb$n$}{\color{Tan}\Verb$s$}{\color{Tan}\Verb$p$}{\color{Tan}\Verb$e$}{\color{Tan}\Verb$c$}{\color{Tan}\Verb$t$}{\color{Tan}\Verb$}$}{\color{Tan}\Verb$)$}{\color{Tan}\Verb$"$}{\color{Tan}\Verb$]$}{\color{Tan}\\}{\color{Tan}\Verb$ $}\Verb$ $\Verb$ $\Verb$ $\Verb$ $\Verb$ $\Verb$ $\Verb$|$\Verb$ ${\color{black}\Verb$'$}{\color{black}\Verb$($}{\color{black}\Verb$|$}{\color{black}\Verb$'$}\Verb$ ${\color{red}\Verb$e$}{\color{red}\Verb$x$}{\color{red}\Verb$p$}{\color{red}\Verb$r$}{\color{Aquamarine}\Verb$:$}{\color{blue}\Verb$e$}\Verb$ ${\color{black}\Verb$"$}{\color{black}\Verb$|$}{\color{black}\Verb$)$}{\color{black}\Verb$"$}\Verb$ ${\color{Tan}\Verb$-$}{\color{Tan}\Verb$>$}{\color{Tan}\Verb$ $}{\color{Tan}\Verb$L$}{\color{Tan}\Verb$a$}{\color{Tan}\Verb$m$}{\color{Tan}\Verb$b$}{\color{Tan}\Verb$d$}{\color{Tan}\Verb$a$}{\color{Tan}\Verb$[$}{\color{blue}\Verb$e$}{\color{Tan}\Verb$]$}{\color{Tan}\\}{\color{Tan}\Verb$ $}\Verb$ $\Verb$ $\Verb$ $\Verb$ $\Verb$ $\Verb$ $\Verb$|$\Verb$ ${\color{black}\Verb$'$}{\color{black}\Verb$($}{\color{black}\Verb$'$}\Verb$ ${\color{red}\Verb$e$}{\color{red}\Verb$x$}{\color{red}\Verb$p$}{\color{red}\Verb$r$}{\color{Aquamarine}\Verb$:$}{\color{blue}\Verb$e$}\Verb$ ${\color{black}\Verb$"$}{\color{black}\Verb$)$}{\color{black}\Verb$"$}\Verb$ ${\color{Tan}\Verb${$}{\color{blue}\Verb$e$}{\color{Tan}\Verb$}$}\\\Verb$ $\Verb$ $\Verb$ $\Verb$ $\Verb$ $\Verb$ $\Verb$ $\Verb$|$\Verb$ ${\color{black}\Verb$'$}{\color{black}\Verb$i$}{\color{black}\Verb$f$}{\color{black}\Verb$'$}\Verb$ ${\color{black}\Verb$"$}{\color{black}\Verb$($}{\color{black}\Verb$"$}\Verb$ ${\color{red}\Verb$e$}{\color{red}\Verb$x$}{\color{red}\Verb$p$}{\color{red}\Verb$r$}{\color{Aquamarine}\Verb$:$}{\color{Aquamarine}\Verb$e$}{\color{Aquamarine}\Verb$x$}{\color{Aquamarine}\Verb$p$}{\color{Aquamarine}\Verb$r$}\Verb$ ${\color{black}\Verb$"$}{\color{black}\Verb$)$}{\color{black}\Verb$"$}\Verb$ ${\color{red}\Verb$b$}{\color{red}\Verb$l$}{\color{red}\Verb$o$}{\color{red}\Verb$c$}{\color{red}\Verb$k$}{\color{Aquamarine}\Verb$:$}{\color{Aquamarine}\Verb$b$}{\color{Aquamarine}\Verb$l$}{\color{Aquamarine}\Verb$o$}{\color{Aquamarine}\Verb$c$}{\color{Aquamarine}\Verb$k$}\Verb$ $\\\Verb$ $\Verb$ $\Verb$ $\Verb$ $\Verb$ $\Verb$ $\Verb$ $\Verb$ $\Verb$ $\Verb$ $\Verb$ $\Verb$ $\Verb$ ${\color{Tan}\Verb$-$}{\color{Tan}\Verb$>$}{\color{Tan}\Verb$ $}{\color{Tan}\Verb$I$}{\color{Tan}\Verb$f$}{\color{Tan}\Verb$[$}{\color{Tan}\Verb${$}{\color{Tan}\Verb$:$}{\color{Tan}\Verb$e$}{\color{Tan}\Verb$x$}{\color{Tan}\Verb$p$}{\color{Tan}\Verb$r$}{\color{Tan}\Verb$=$}{\color{Tan}\Verb$>$}{\color{Aquamarine}\Verb$e$}{\color{Aquamarine}\Verb$x$}{\color{Aquamarine}\Verb$p$}{\color{Aquamarine}\Verb$r$}{\color{Tan}\Verb$,$}{\color{Tan}\Verb$:$}{\color{Tan}\Verb$b$}{\color{Tan}\Verb$l$}{\color{Tan}\Verb$o$}{\color{Tan}\Verb$c$}{\color{Tan}\Verb$k$}{\color{Tan}\Verb$=$}{\color{Tan}\Verb$>$}{\color{Aquamarine}\Verb$b$}{\color{Aquamarine}\Verb$l$}{\color{Aquamarine}\Verb$o$}{\color{Aquamarine}\Verb$c$}{\color{Aquamarine}\Verb$k$}{\color{Tan}\Verb$}$}{\color{Tan}\Verb$]$}{\color{Tan}\\}{\color{Tan}\Verb$ $}\Verb$ $\Verb$ $\Verb$ $\Verb$ $\Verb$ $\Verb$ $\Verb$|$\Verb$ ${\color{black}\Verb$'$}{\color{black}\Verb${$}{\color{black}\Verb$'$}\Verb$ $\Verb$(${\color{black}\Verb$'$}{\color{black}\Verb$}$}{\color{black}\Verb$'$}\Verb$ $\Verb$b$\Verb$r$\Verb$e$\Verb$a$\Verb$k$\Verb$ $\Verb$|$\Verb$ ${\color{Violet}\Verb$&$}{\color{Violet}\Verb$'$}{\color{black}\Verb${$}{\color{Violet}\Verb$'$}\Verb$ ${\color{red}\Verb$a$}{\color{red}\Verb$t$}{\color{red}\Verb$o$}{\color{red}\Verb$m$}{\color{Aquamarine}\Verb$:$}{\color{Tan}\Verb${$}{\color{Tan}\Verb$'$}{\color{Tan}\Verb${$}{\color{Tan}\Verb$'$}{\color{Tan}\Verb$+$}{\color{Aquamarine}\Verb$i$}{\color{Aquamarine}\Verb$t$}{\color{Tan}\Verb$[$}{\color{Tan}\Verb$0$}{\color{Tan}\Verb$]$}{\color{Tan}\Verb$+$}{\color{Tan}\Verb$'$}{\color{Tan}\Verb$}$}{\color{Tan}\Verb$'$}{\color{Tan}\Verb$}$}{\color{Aquamarine}\Verb$:$}{\color{Aquamarine}\Verb$[$}{\color{Aquamarine}\Verb$s$}{\color{Aquamarine}\Verb$]$}\Verb$ $\Verb$|$\Verb$ $\Verb$.${\color{Aquamarine}\Verb$:$}{\color{Aquamarine}\Verb$[$}{\color{Aquamarine}\Verb$s$}{\color{Aquamarine}\Verb$]$}\Verb$)${\color{black}\Verb$*$}\Verb$ $\\\Verb$ $\Verb$ $\Verb$ $\Verb$ $\Verb$ $\Verb$ $\Verb$ $\Verb$ $\Verb$ $\Verb$ $\Verb$ $\Verb$ $\Verb$ ${\color{Tan}\Verb$-$}{\color{Tan}\Verb$>$}{\color{Tan}\Verb$ $}{\color{Tan}\Verb$C$}{\color{Tan}\Verb$C$}{\color{Tan}\Verb$o$}{\color{Tan}\Verb$d$}{\color{Tan}\Verb$e$}{\color{Tan}\Verb$[$}{\color{Aquamarine}\Verb$s$}{\color{Tan}\Verb$*$}{\color{Tan}\Verb$"$}{\color{Tan}\Verb$"$}{\color{Tan}\Verb$]$}{\color{Tan}\\}{\color{Tan}\Verb$ $}\Verb$ $\Verb$ $\Verb$ $\Verb$ $\Verb$ $\Verb$ $\Verb$|$\Verb$ ${\color{black}\Verb$'$}{\color{black}\Verb$y$}{\color{black}\Verb$i$}{\color{black}\Verb$e$}{\color{black}\Verb$l$}{\color{black}\Verb$d$}{\color{black}\Verb$'$}\Verb$ ${\color{red}\Verb$a$}{\color{red}\Verb$t$}{\color{red}\Verb$o$}{\color{red}\Verb$m$}{\color{Aquamarine}\Verb$:$}{\color{Aquamarine}\Verb$a$}\Verb$ $\Verb$ $\Verb$ ${\color{Tan}\Verb$-$}{\color{Tan}\Verb$>$}{\color{Tan}\Verb$ $}{\color{Tan}\Verb$Y$}{\color{Tan}\Verb$i$}{\color{Tan}\Verb$e$}{\color{Tan}\Verb$l$}{\color{Tan}\Verb$d$}{\color{Tan}\Verb$[$}{\color{Aquamarine}\Verb$a$}{\color{Tan}\Verb$]$}{\color{Tan}\Verb$ $}{\color{Tan}\\}{\color{Tan}\Verb$ $}\Verb$ $\Verb$ $\Verb$ $\Verb$ $\Verb$ $\Verb$ $\Verb$|$\Verb$ ${\color{red}\Verb$m$}{\color{red}\Verb$e$}{\color{red}\Verb$t$}{\color{red}\Verb$h$}{\color{red}\Verb$o$}{\color{red}\Verb$d$}{\color{green}\Verb$($}{\color{green}\Verb$"$}{\color{green}\Verb$s$}{\color{green}\Verb$e$}{\color{green}\Verb$l$}{\color{green}\Verb$f$}{\color{green}\Verb$"$}{\color{green}\Verb$)$}\\\Verb$ $\Verb$ $\Verb$ $\Verb$ $\Verb$ $\Verb$ $\Verb$ $\Verb$|$\Verb$ ${\color{red}\Verb$l$}{\color{red}\Verb$o$}{\color{red}\Verb$c$}{\color{red}\Verb$a$}{\color{red}\Verb$l$}\\\Verb$ $\Verb$ $\Verb$ $\Verb$ $\Verb$ $\Verb$ $\Verb$ $\Verb$)$\\\\\Verb$ $\Verb$ ${\color{red}\Verb$l$}{\color{red}\Verb$o$}{\color{red}\Verb$c$}{\color{red}\Verb$a$}{\color{red}\Verb$l$}\Verb$ $\Verb$=$\Verb$ $\Verb$ ${\color{Violet}\Verb$~$}{\color{Violet}\Verb$'$}{\color{black}\Verb$e$}{\color{black}\Verb$n$}{\color{black}\Verb$d$}{\color{Violet}\Verb$'$}\Verb$ ${\color{red}\Verb$n$}{\color{red}\Verb$a$}{\color{red}\Verb$m$}{\color{red}\Verb$e$}{\color{Aquamarine}\Verb$:$}{\color{Aquamarine}\Verb$n$}{\color{Aquamarine}\Verb$a$}{\color{Aquamarine}\Verb$m$}{\color{Aquamarine}\Verb$e$}\Verb$ ${\color{Aquamarine}\Verb$@$}{\color{Aquamarine}\Verb$V$}{\color{Aquamarine}\Verb$a$}{\color{Aquamarine}\Verb$r$}\\\\\Verb$ $\Verb$ ${\color{red}\Verb$a$}{\color{red}\Verb$r$}{\color{red}\Verb$g$}{\color{red}\Verb$s$}\Verb$ $\Verb$=$\Verb$ ${\color{red}\Verb$l$}{\color{red}\Verb$i$}{\color{red}\Verb$s$}{\color{red}\Verb$t$}{\color{red}\Verb$O$}{\color{red}\Verb$f$}{\color{green}\Verb$($}{\color{green}\Verb$'$}{\color{green}\Verb$e$}{\color{green}\Verb$x$}{\color{green}\Verb$p$}{\color{green}\Verb$r$}{\color{green}\Verb$'$}{\color{green}\Verb$,$}{\color{green}\Verb$'$}{\color{green}\Verb$,$}{\color{green}\Verb$'$}{\color{green}\Verb$)$}\\\Verb$ $\Verb$ ${\color{red}\Verb$m$}{\color{red}\Verb$e$}{\color{red}\Verb$t$}{\color{red}\Verb$h$}{\color{red}\Verb$o$}{\color{red}\Verb$d$}{\color{green}\Verb$($}{\color{green}\Verb$o$}{\color{green}\Verb$b$}{\color{green}\Verb$j$}{\color{green}\Verb$)$}\Verb$ $\Verb$=$\Verb$ ${\color{red}\Verb$n$}{\color{red}\Verb$a$}{\color{red}\Verb$m$}{\color{red}\Verb$e$}{\color{Aquamarine}\Verb$:$}{\color{Aquamarine}\Verb$n$}{\color{Aquamarine}\Verb$a$}{\color{Aquamarine}\Verb$m$}{\color{Aquamarine}\Verb$e$}\Verb$ ${\color{black}\Verb$'$}{\color{black}\Verb$($}{\color{black}\Verb$'$}\Verb$ $\Verb$(${\color{red}\Verb$a$}{\color{red}\Verb$r$}{\color{red}\Verb$g$}{\color{red}\Verb$s$}\Verb$ $\Verb$|$\Verb$ ${\color{Tan}\Verb${$}{\color{Tan}\Verb$[$}{\color{Tan}\Verb$]$}{\color{Tan}\Verb$}$}\Verb$)${\color{Aquamarine}\Verb$:$}{\color{Aquamarine}\Verb$a$}{\color{Aquamarine}\Verb$r$}{\color{Aquamarine}\Verb$g$}\Verb$ ${\color{black}\Verb$"$}{\color{black}\Verb$)$}{\color{black}\Verb$"$}\Verb$ $\\\Verb$ $\Verb$ $\Verb$ $\Verb$ $\Verb$ $\Verb$ $\Verb$ $\Verb$ $\Verb$ $\Verb$ $\Verb$ $\Verb$ $\Verb$ $\Verb$ $\Verb$ $\Verb$ $\Verb$ $\Verb$ $\Verb$ $\Verb$ $\Verb$($\Verb$ ${\color{red}\Verb$b$}{\color{red}\Verb$l$}{\color{red}\Verb$o$}{\color{red}\Verb$c$}{\color{red}\Verb$k$}{\color{Aquamarine}\Verb$:$}{\color{Aquamarine}\Verb$b$}\Verb$ $\Verb$ ${\color{Tan}\Verb$-$}{\color{Tan}\Verb$>$}{\color{Tan}\Verb$ $}{\color{Tan}\Verb$I$}{\color{Tan}\Verb$t$}{\color{Tan}\Verb$e$}{\color{Tan}\Verb$r$}{\color{Tan}\Verb$a$}{\color{Tan}\Verb$t$}{\color{Tan}\Verb$e$}{\color{Tan}\Verb$[$}{\color{Tan}\Verb$c$}{\color{Tan}\Verb$a$}{\color{Tan}\Verb$l$}{\color{Tan}\Verb$l$}{\color{Tan}\Verb$($}{\color{Aquamarine}\Verb$n$}{\color{Aquamarine}\Verb$a$}{\color{Aquamarine}\Verb$m$}{\color{Aquamarine}\Verb$e$}{\color{Tan}\Verb$,$}{\color{Tan}\Verb$o$}{\color{Tan}\Verb$b$}{\color{Tan}\Verb$j$}{\color{Tan}\Verb$,$}{\color{Tan}\Verb$*$}{\color{Aquamarine}\Verb$a$}{\color{Aquamarine}\Verb$r$}{\color{Aquamarine}\Verb$g$}{\color{Tan}\Verb$)$}{\color{Tan}\Verb$,$}{\color{Aquamarine}\Verb$b$}{\color{Tan}\Verb$]$}{\color{Tan}\\}{\color{Tan}\Verb$ $}\Verb$ $\Verb$ $\Verb$ $\Verb$ $\Verb$ $\Verb$ $\Verb$ $\Verb$ $\Verb$ $\Verb$ $\Verb$ $\Verb$ $\Verb$ $\Verb$ $\Verb$ $\Verb$ $\Verb$ $\Verb$ $\Verb$ $\Verb$|$\Verb$ $\Verb$ $\Verb$ $\Verb$ $\Verb$ $\Verb$ $\Verb$ $\Verb$ $\Verb$ $\Verb$ ${\color{Tan}\Verb$-$}{\color{Tan}\Verb$>$}{\color{Tan}\Verb$ $}{\color{Tan}\Verb$c$}{\color{Tan}\Verb$a$}{\color{Tan}\Verb$l$}{\color{Tan}\Verb$l$}{\color{Tan}\Verb$($}{\color{Aquamarine}\Verb$n$}{\color{Aquamarine}\Verb$a$}{\color{Aquamarine}\Verb$m$}{\color{Aquamarine}\Verb$e$}{\color{Tan}\Verb$,$}{\color{Tan}\Verb$o$}{\color{Tan}\Verb$b$}{\color{Tan}\Verb$j$}{\color{Tan}\Verb$,$}{\color{Tan}\Verb$*$}{\color{Aquamarine}\Verb$a$}{\color{Aquamarine}\Verb$r$}{\color{Aquamarine}\Verb$g$}{\color{Tan}\Verb$)$}{\color{Tan}\\}{\color{Tan}\Verb$ $}\Verb$ $\Verb$ $\Verb$ $\Verb$ $\Verb$ $\Verb$ $\Verb$ $\Verb$ $\Verb$ $\Verb$ $\Verb$ $\Verb$ $\Verb$ $\Verb$ $\Verb$ $\Verb$ $\Verb$ $\Verb$ $\Verb$ $\Verb$)$\\\Verb$ $\Verb$ $\\\Verb$ $\Verb$ ${\color{red}\Verb$e$}{\color{red}\Verb$x$}{\color{red}\Verb$p$}{\color{red}\Verb$r$}{\color{red}\Verb$_$}{\color{red}\Verb$p$}{\color{red}\Verb$o$}{\color{red}\Verb$s$}{\color{red}\Verb$t$}{\color{red}\Verb$f$}{\color{red}\Verb$i$}{\color{red}\Verb$x$}{\color{red}\Verb$e$}{\color{red}\Verb$d$}\Verb$ $\Verb$=$\Verb$ ${\color{red}\Verb$a$}{\color{red}\Verb$t$}{\color{red}\Verb$o$}{\color{red}\Verb$m$}{\color{Aquamarine}\Verb$:$}{\color{Aquamarine}\Verb$a$}\Verb$ $\Verb$($\Verb$ $\\\Verb$ $\Verb$ $\Verb$ $\Verb$ $\Verb$ $\Verb$ $\Verb$ $\Verb$($\Verb$ ${\color{black}\Verb$'$}{\color{black}\Verb$[$}{\color{black}\Verb$'$}\Verb$ ${\color{red}\Verb$a$}{\color{red}\Verb$r$}{\color{red}\Verb$g$}{\color{red}\Verb$s$}{\color{Aquamarine}\Verb$:$}{\color{Aquamarine}\Verb$a$}{\color{Aquamarine}\Verb$r$}{\color{Aquamarine}\Verb$g$}\Verb$ ${\color{black}\Verb$"$}{\color{black}\Verb$]$}{\color{black}\Verb$"$}\Verb$ ${\color{black}\Verb$"$}{\color{black}\Verb$=$}{\color{black}\Verb$"$}\Verb$ ${\color{red}\Verb$e$}{\color{red}\Verb$x$}{\color{red}\Verb$p$}{\color{red}\Verb$r$}{\color{Aquamarine}\Verb$:$}{\color{Aquamarine}\Verb$a$}{\color{Aquamarine}\Verb$r$}{\color{Aquamarine}\Verb$g$}{\color{Aquamarine}\Verb$2$}\Verb$ ${\color{Tan}\Verb$-$}{\color{Tan}\Verb$>$}{\color{Tan}\Verb$ $}{\color{Tan}\Verb$c$}{\color{Tan}\Verb$a$}{\color{Tan}\Verb$l$}{\color{Tan}\Verb$l$}{\color{Tan}\Verb$($}{\color{Tan}\Verb$"$}{\color{Tan}\Verb$[$}{\color{Tan}\Verb$]$}{\color{Tan}\Verb$=$}{\color{Tan}\Verb$"$}{\color{Tan}\Verb$,$}{\color{Aquamarine}\Verb$a$}{\color{Tan}\Verb$,$}{\color{Tan}\Verb$*$}{\color{Aquamarine}\Verb$a$}{\color{Aquamarine}\Verb$r$}{\color{Aquamarine}\Verb$g$}{\color{Tan}\Verb$,$}{\color{Aquamarine}\Verb$a$}{\color{Aquamarine}\Verb$r$}{\color{Aquamarine}\Verb$g$}{\color{Aquamarine}\Verb$2$}{\color{Tan}\Verb$)$}{\color{Tan}\\}{\color{Tan}\Verb$ $}\Verb$ $\Verb$ $\Verb$ $\Verb$ $\Verb$ $\Verb$ $\Verb$|$\Verb$ ${\color{black}\Verb$'$}{\color{black}\Verb$[$}{\color{black}\Verb$'$}\Verb$ ${\color{red}\Verb$a$}{\color{red}\Verb$r$}{\color{red}\Verb$g$}{\color{red}\Verb$s$}{\color{Aquamarine}\Verb$:$}{\color{Aquamarine}\Verb$a$}{\color{Aquamarine}\Verb$r$}{\color{Aquamarine}\Verb$g$}\Verb$ ${\color{black}\Verb$"$}{\color{black}\Verb$]$}{\color{black}\Verb$"$}\Verb$ $\Verb$ $\Verb$ $\Verb$ $\Verb$ $\Verb$ $\Verb$ $\Verb$ $\Verb$ $\Verb$ $\Verb$ $\Verb$ $\Verb$ $\Verb$ $\Verb$ ${\color{Tan}\Verb$-$}{\color{Tan}\Verb$>$}{\color{Tan}\Verb$ $}{\color{Tan}\Verb$c$}{\color{Tan}\Verb$a$}{\color{Tan}\Verb$l$}{\color{Tan}\Verb$l$}{\color{Tan}\Verb$($}{\color{Tan}\Verb$"$}{\color{Tan}\Verb$[$}{\color{Tan}\Verb$]$}{\color{Tan}\Verb$"$}{\color{Tan}\Verb$,$}{\color{Aquamarine}\Verb$a$}{\color{Tan}\Verb$,$}{\color{Tan}\Verb$*$}{\color{Aquamarine}\Verb$a$}{\color{Aquamarine}\Verb$r$}{\color{Aquamarine}\Verb$g$}{\color{Tan}\Verb$)$}{\color{Tan}\\}{\color{Tan}\Verb$ $}\Verb$ $\Verb$ $\Verb$ $\Verb$ $\Verb$ $\Verb$ $\Verb$|$\Verb$ ${\color{black}\Verb$'$}{\color{black}\Verb$.$}{\color{black}\Verb$'$}\Verb$ ${\color{red}\Verb$m$}{\color{red}\Verb$e$}{\color{red}\Verb$t$}{\color{red}\Verb$h$}{\color{red}\Verb$o$}{\color{red}\Verb$d$}{\color{green}\Verb$($}{\color{Aquamarine}\Verb$a$}{\color{green}\Verb$)$}\\\Verb$ $\Verb$ $\Verb$ $\Verb$ $\Verb$ $\Verb$ $\Verb$ $\Verb$|$\Verb$ ${\color{black}\Verb$'$}{\color{black}\Verb$.$}{\color{black}\Verb$'$}\Verb$ ${\color{red}\Verb$n$}{\color{red}\Verb$a$}{\color{red}\Verb$m$}{\color{red}\Verb$e$}{\color{Aquamarine}\Verb$:$}{\color{Aquamarine}\Verb$n$}{\color{Aquamarine}\Verb$a$}{\color{Aquamarine}\Verb$m$}{\color{Aquamarine}\Verb$e$}\Verb$ $\Verb$ $\Verb$ $\Verb$ $\Verb$ $\Verb$ $\Verb$ $\Verb$ $\Verb$ $\Verb$ $\Verb$ $\Verb$ $\Verb$ $\Verb$ $\Verb$ $\Verb$ $\Verb$ $\Verb$ ${\color{Tan}\Verb$-$}{\color{Tan}\Verb$>$}{\color{Tan}\Verb$ $}{\color{Tan}\Verb$c$}{\color{Tan}\Verb$a$}{\color{Tan}\Verb$l$}{\color{Tan}\Verb$l$}{\color{Tan}\Verb$($}{\color{Aquamarine}\Verb$n$}{\color{Aquamarine}\Verb$a$}{\color{Aquamarine}\Verb$m$}{\color{Aquamarine}\Verb$e$}{\color{Tan}\Verb$,$}{\color{Aquamarine}\Verb$a$}{\color{Tan}\Verb$)$}{\color{Tan}\\}{\color{Tan}\Verb$ $}\Verb$ $\Verb$ $\Verb$ $\Verb$ $\Verb$ $\Verb$ $\Verb$)${\color{Aquamarine}\Verb$:$}{\color{Aquamarine}\Verb$a$}\\\Verb$ $\Verb$ $\Verb$ $\Verb$ $\Verb$ $\Verb$ $\Verb$)${\color{black}\Verb$*$}\Verb$ ${\color{Tan}\Verb${$}{\color{Aquamarine}\Verb$a$}{\color{Tan}\Verb$}$}\\\Verb$ $\Verb$ $\Verb$ $\Verb$ $\Verb$ $\Verb$ $\Verb$ $\Verb$ $\Verb$ $\Verb$ $\Verb$ $\Verb$ $\Verb$ $\Verb$ $\Verb$ $\\\Verb$ $\Verb$ ${\color{red}\Verb$e$}{\color{red}\Verb$x$}{\color{red}\Verb$p$}{\color{red}\Verb$r$}\Verb$ $\Verb$=$\Verb$ ${\color{red}\Verb$e$}{\color{red}\Verb$x$}{\color{red}\Verb$p$}{\color{red}\Verb$r$}{\color{red}\Verb$_$}{\color{red}\Verb$a$}{\color{red}\Verb$s$}{\color{red}\Verb$s$}\\\\\Verb$ $\Verb$ $\\\Verb$ $\Verb$ ${\color{red}\Verb$b$}{\color{red}\Verb$i$}{\color{red}\Verb$n$}{\color{red}\Verb$a$}{\color{red}\Verb$r$}{\color{red}\Verb$y$}{\color{red}\Verb$_$}{\color{red}\Verb$o$}{\color{red}\Verb$p$}{\color{green}\Verb$($}{\color{green}\Verb$e$}{\color{green}\Verb$x$}{\color{green}\Verb$p$}{\color{green}\Verb$,$}{\color{green}\Verb$o$}{\color{green}\Verb$p$}{\color{green}\Verb$e$}{\color{green}\Verb$r$}{\color{green}\Verb$)$}\Verb$ $\Verb$=$\Verb$ ${\color{red}\Verb$a$}{\color{red}\Verb$p$}{\color{red}\Verb$p$}{\color{red}\Verb$l$}{\color{red}\Verb$y$}{\color{green}\Verb$($}{\color{green}\Verb$e$}{\color{green}\Verb$x$}{\color{green}\Verb$p$}{\color{green}\Verb$)$}{\color{Aquamarine}\Verb$:$}{\color{Aquamarine}\Verb$a$}\Verb$ $\\\Verb$ $\Verb$ $\Verb$ $\Verb$ $\Verb$ $\Verb$ $\Verb$ $\Verb$ $\Verb$ $\Verb$ $\Verb$ $\Verb$(${\color{black}\Verb$"$}{\color{black}\Verb$"$}\Verb$ ${\color{red}\Verb$a$}{\color{red}\Verb$p$}{\color{red}\Verb$p$}{\color{red}\Verb$l$}{\color{red}\Verb$y$}{\color{green}\Verb$($}{\color{green}\Verb$o$}{\color{green}\Verb$p$}{\color{green}\Verb$e$}{\color{green}\Verb$r$}{\color{green}\Verb$)$}{\color{Aquamarine}\Verb$:$}{\color{Aquamarine}\Verb$o$}{\color{Aquamarine}\Verb$p$}\Verb$ ${\color{red}\Verb$a$}{\color{red}\Verb$p$}{\color{red}\Verb$p$}{\color{red}\Verb$l$}{\color{red}\Verb$y$}{\color{green}\Verb$($}{\color{green}\Verb$e$}{\color{green}\Verb$x$}{\color{green}\Verb$p$}{\color{green}\Verb$)$}{\color{Aquamarine}\Verb$:$}{\color{Aquamarine}\Verb$b$}\Verb$ ${\color{Tan}\Verb${$}{\color{Tan}\Verb$c$}{\color{Tan}\Verb$a$}{\color{Tan}\Verb$l$}{\color{Tan}\Verb$l$}{\color{Tan}\Verb$($}{\color{Aquamarine}\Verb$o$}{\color{Aquamarine}\Verb$p$}{\color{Tan}\Verb$,$}{\color{Aquamarine}\Verb$a$}{\color{Tan}\Verb$,$}{\color{Aquamarine}\Verb$b$}{\color{Tan}\Verb$)$}{\color{Tan}\Verb$}$}{\color{Aquamarine}\Verb$:$}{\color{Aquamarine}\Verb$a$}\Verb$)${\color{black}\Verb$*$}\Verb$ ${\color{Tan}\Verb$-$}{\color{Tan}\Verb$>$}{\color{Tan}\Verb$ $}{\color{Aquamarine}\Verb$a$}{\color{Tan}\Verb$ $}{\color{Tan}\\}{\color{Tan}\Verb$ $}\Verb$ $\\\Verb$ $\Verb$ ${\color{red}\Verb$e$}{\color{red}\Verb$x$}{\color{red}\Verb$p$}{\color{red}\Verb$r$}{\color{red}\Verb$_$}{\color{red}\Verb$a$}{\color{red}\Verb$s$}{\color{red}\Verb$s$}\Verb$ $\Verb$=$\Verb$ ${\color{black}\Verb$"$}{\color{black}\Verb$"$}\Verb$ ${\color{red}\Verb$n$}{\color{red}\Verb$a$}{\color{red}\Verb$m$}{\color{red}\Verb$e$}{\color{Aquamarine}\Verb$:$}{\color{Aquamarine}\Verb$n$}{\color{Aquamarine}\Verb$a$}{\color{Aquamarine}\Verb$m$}{\color{Aquamarine}\Verb$e$}\Verb$ ${\color{black}\Verb$'$}{\color{black}\Verb$=$}{\color{black}\Verb$'$}\Verb$ ${\color{red}\Verb$e$}{\color{red}\Verb$x$}{\color{red}\Verb$p$}{\color{red}\Verb$r$}{\color{Aquamarine}\Verb$:$}{\color{Aquamarine}\Verb$e$}{\color{Aquamarine}\Verb$x$}{\color{Aquamarine}\Verb$p$}{\color{Aquamarine}\Verb$r$}\Verb$ ${\color{Aquamarine}\Verb$@$}{\color{Aquamarine}\Verb$A$}{\color{Aquamarine}\Verb$s$}{\color{Aquamarine}\Verb$s$}{\color{Aquamarine}\Verb$i$}{\color{Aquamarine}\Verb$g$}{\color{Aquamarine}\Verb$n$}\\\Verb$ $\Verb$ $\Verb$ $\Verb$ $\Verb$ $\Verb$ $\Verb$ $\Verb$ $\Verb$ $\Verb$ $\Verb$ $\Verb$|$\Verb$ ${\color{red}\Verb$e$}{\color{red}\Verb$x$}{\color{red}\Verb$p$}{\color{red}\Verb$r$}{\color{red}\Verb$_$}{\color{red}\Verb$o$}{\color{red}\Verb$r$}{\color{red}\Verb$_$}{\color{red}\Verb$l$}
\end{exambox}
}
\end{spacing}\vskip -0.4em

\begin{exambox}
\begin{spacing}{0.8}
{\small
{\color{red}\Verb$e$}{\color{red}\Verb$x$}{\color{red}\Verb$p$}{\color{red}\Verb$r$}{\color{red}\Verb$_$}{\color{red}\Verb$o$}{\color{red}\Verb$r$}{\color{red}\Verb$_$}{\color{red}\Verb$l$}\Verb$ $\Verb$ $\Verb$=$\Verb$ ${\color{red}\Verb$b$}{\color{red}\Verb$i$}{\color{red}\Verb$n$}{\color{red}\Verb$a$}{\color{red}\Verb$r$}{\color{red}\Verb$y$}{\color{red}\Verb$_$}{\color{red}\Verb$o$}{\color{red}\Verb$p$}{\color{green}\Verb$($}{\color{green}\Verb$'$}{\color{green}\Verb$e$}{\color{green}\Verb$x$}{\color{green}\Verb$p$}{\color{green}\Verb$r$}{\color{green}\Verb$_$}{\color{green}\Verb$a$}{\color{green}\Verb$n$}{\color{green}\Verb$d$}{\color{green}\Verb$_$}{\color{green}\Verb$l$}{\color{green}\Verb$'$}{\color{green}\Verb$,$}{\color{Violet}\Verb$($}{\color{Violet}\Verb$|$}{\color{green}\Verb$ $}{\color{black}\Verb$"$}{\color{black}\Verb$|$}{\color{black}\Verb$|$}{\color{black}\Verb$"$}{\color{green}\Verb$ $}{\color{Violet}\Verb$|$}{\color{Violet}\Verb$)$}{\color{green}\Verb$)$}\\\Verb$ $\Verb$ ${\color{red}\Verb$e$}{\color{red}\Verb$x$}{\color{red}\Verb$p$}{\color{red}\Verb$r$}{\color{red}\Verb$_$}{\color{red}\Verb$a$}{\color{red}\Verb$n$}{\color{red}\Verb$d$}{\color{red}\Verb$_$}{\color{red}\Verb$l$}\Verb$ $\Verb$=$\Verb$ ${\color{red}\Verb$b$}{\color{red}\Verb$i$}{\color{red}\Verb$n$}{\color{red}\Verb$a$}{\color{red}\Verb$r$}{\color{red}\Verb$y$}{\color{red}\Verb$_$}{\color{red}\Verb$o$}{\color{red}\Verb$p$}{\color{green}\Verb$($}{\color{green}\Verb$'$}{\color{green}\Verb$e$}{\color{green}\Verb$x$}{\color{green}\Verb$p$}{\color{green}\Verb$r$}{\color{green}\Verb$_$}{\color{green}\Verb$c$}{\color{green}\Verb$m$}{\color{green}\Verb$p$}{\color{green}\Verb$'$}{\color{green}\Verb$ $}{\color{green}\Verb$ $}{\color{green}\Verb$,$}{\color{Violet}\Verb$($}{\color{Violet}\Verb$|$}{\color{green}\Verb$ $}{\color{black}\Verb$"$}{\color{black}\Verb$&$}{\color{black}\Verb$&$}{\color{black}\Verb$"$}{\color{green}\Verb$ $}{\color{Violet}\Verb$|$}{\color{Violet}\Verb$)$}{\color{green}\Verb$)$}\\\\\Verb$ $\Verb$ ${\color{red}\Verb$e$}{\color{red}\Verb$x$}{\color{red}\Verb$p$}{\color{red}\Verb$r$}{\color{red}\Verb$_$}{\color{red}\Verb$c$}{\color{red}\Verb$m$}{\color{red}\Verb$p$}\Verb$ $\Verb$=$\Verb$ ${\color{red}\Verb$b$}{\color{red}\Verb$i$}{\color{red}\Verb$n$}{\color{red}\Verb$a$}{\color{red}\Verb$r$}{\color{red}\Verb$y$}{\color{red}\Verb$_$}{\color{red}\Verb$o$}{\color{red}\Verb$p$}{\color{green}\Verb$($}{\color{green}\Verb$'$}{\color{green}\Verb$e$}{\color{green}\Verb$x$}{\color{green}\Verb$p$}{\color{green}\Verb$r$}{\color{green}\Verb$_$}{\color{green}\Verb$a$}{\color{green}\Verb$r$}{\color{green}\Verb$1$}{\color{green}\Verb$'$}{\color{green}\Verb$,$}{\color{Violet}\Verb$($}{\color{Violet}\Verb$|$}{\color{green}\Verb$ $}{\color{black}\Verb$"$}{\color{black}\Verb$<$}{\color{black}\Verb$"$}{\color{green}\Verb$ $}{\color{green}\Verb$|$}{\color{black}\Verb$"$}{\color{black}\Verb$<$}{\color{black}\Verb$=$}{\color{black}\Verb$"$}{\color{green}\Verb$|$}{\color{black}\Verb$"$}{\color{black}\Verb$<$}{\color{black}\Verb$=$}{\color{black}\Verb$>$}{\color{black}\Verb$"$}{\color{green}\Verb$|$}{\color{green}\\}{\color{green}\Verb$ $}{\color{green}\Verb$ $}{\color{green}\Verb$ $}{\color{green}\Verb$ $}{\color{green}\Verb$ $}{\color{green}\Verb$ $}{\color{green}\Verb$ $}{\color{green}\Verb$ $}{\color{green}\Verb$ $}{\color{green}\Verb$ $}{\color{green}\Verb$ $}{\color{green}\Verb$ $}{\color{green}\Verb$ $}{\color{green}\Verb$ $}{\color{green}\Verb$ $}{\color{green}\Verb$ $}{\color{green}\Verb$ $}{\color{green}\Verb$ $}{\color{green}\Verb$ $}{\color{green}\Verb$ $}{\color{green}\Verb$ $}{\color{green}\Verb$ $}{\color{green}\Verb$ $}{\color{green}\Verb$ $}{\color{green}\Verb$ $}{\color{green}\Verb$ $}{\color{green}\Verb$ $}{\color{green}\Verb$ $}{\color{green}\Verb$ $}{\color{green}\Verb$ $}{\color{green}\Verb$ $}{\color{green}\Verb$ $}{\color{black}\Verb$"$}{\color{black}\Verb$>$}{\color{black}\Verb$=$}{\color{black}\Verb$"$}{\color{green}\Verb$|$}{\color{black}\Verb$"$}{\color{black}\Verb$>$}{\color{black}\Verb$"$}{\color{green}\Verb$ $}{\color{green}\Verb$|$}{\color{black}\Verb$"$}{\color{black}\Verb$=$}{\color{black}\Verb$=$}{\color{black}\Verb$"$}{\color{green}\Verb$|$}{\color{black}\Verb$"$}{\color{black}\Verb$!$}{\color{black}\Verb$=$}{\color{black}\Verb$"$}{\color{green}\Verb$ $}{\color{Violet}\Verb$|$}{\color{Violet}\Verb$)$}{\color{green}\Verb$)$}\\\Verb$ $\Verb$ ${\color{red}\Verb$e$}{\color{red}\Verb$x$}{\color{red}\Verb$p$}{\color{red}\Verb$r$}{\color{red}\Verb$_$}{\color{red}\Verb$a$}{\color{red}\Verb$r$}{\color{red}\Verb$1$}\Verb$ $\Verb$=$\Verb$ ${\color{red}\Verb$e$}{\color{red}\Verb$x$}{\color{red}\Verb$p$}{\color{red}\Verb$r$}{\color{red}\Verb$_$}{\color{red}\Verb$a$}{\color{red}\Verb$r$}{\color{red}\Verb$2$}{\color{Aquamarine}\Verb$:$}{\color{Aquamarine}\Verb$a$}\Verb$ $\Verb$($\Verb$(${\color{black}\Verb$"$}{\color{black}\Verb$+$}{\color{black}\Verb$"$}\Verb$ ${\color{red}\Verb$e$}{\color{red}\Verb$x$}{\color{red}\Verb$p$}{\color{red}\Verb$r$}{\color{red}\Verb$_$}{\color{red}\Verb$a$}{\color{red}\Verb$r$}{\color{red}\Verb$2$}\\\Verb$ $\Verb$ $\Verb$ $\Verb$ $\Verb$ $\Verb$ $\Verb$ $\Verb$ $\Verb$ $\Verb$ $\Verb$ $\Verb$ $\Verb$ $\Verb$ $\Verb$ $\Verb$ $\Verb$ $\Verb$ $\Verb$ $\Verb$ $\Verb$ $\Verb$ $\Verb$ $\Verb$ $\Verb$ $\Verb$|${\color{black}\Verb$"$}{\color{black}\Verb$-$}{\color{black}\Verb$"$}\Verb$ ${\color{red}\Verb$e$}{\color{red}\Verb$x$}{\color{red}\Verb$p$}{\color{red}\Verb$r$}{\color{red}\Verb$_$}{\color{red}\Verb$a$}{\color{red}\Verb$r$}{\color{red}\Verb$2$}{\color{Aquamarine}\Verb$:$}{\color{Tan}\Verb${$}{\color{Tan}\Verb$c$}{\color{Tan}\Verb$a$}{\color{Tan}\Verb$l$}{\color{Tan}\Verb$l$}{\color{Tan}\Verb$($}{\color{Tan}\Verb$'$}{\color{Tan}\Verb$-$}{\color{Tan}\Verb$'$}{\color{Tan}\Verb$,$}{\color{Aquamarine}\Verb$i$}{\color{Aquamarine}\Verb$t$}{\color{Tan}\Verb$)$}{\color{Tan}\Verb$}$}\Verb$)${\color{Aquamarine}\Verb$:$}{\color{Aquamarine}\Verb$b$}\Verb$ $\\\Verb$ $\Verb$ $\Verb$ $\Verb$ $\Verb$ $\Verb$ $\Verb$ $\Verb$ $\Verb$ $\Verb$ $\Verb$ $\Verb$ $\Verb$ $\Verb$ $\Verb$ $\Verb$ $\Verb$ $\Verb$ $\Verb$ $\Verb$ $\Verb$ $\Verb$ $\Verb$ $\Verb$ ${\color{Tan}\Verb${$}{\color{Tan}\Verb$c$}{\color{Tan}\Verb$a$}{\color{Tan}\Verb$l$}{\color{Tan}\Verb$l$}{\color{Tan}\Verb$($}{\color{Tan}\Verb$'$}{\color{Tan}\Verb$+$}{\color{Tan}\Verb$'$}{\color{Tan}\Verb$,$}{\color{Aquamarine}\Verb$a$}{\color{Tan}\Verb$,$}{\color{Aquamarine}\Verb$b$}{\color{Tan}\Verb$)$}{\color{Tan}\Verb$}$}{\color{Aquamarine}\Verb$:$}{\color{Aquamarine}\Verb$a$}\Verb$ $\Verb$ $\Verb$ $\Verb$)${\color{black}\Verb$*$}\Verb$ ${\color{Tan}\Verb$-$}{\color{Tan}\Verb$>$}{\color{Tan}\Verb$ $}{\color{Aquamarine}\Verb$a$}{\color{Tan}\\}{\color{Tan}\Verb$ $}\Verb$ ${\color{red}\Verb$e$}{\color{red}\Verb$x$}{\color{red}\Verb$p$}{\color{red}\Verb$r$}{\color{red}\Verb$_$}{\color{red}\Verb$a$}{\color{red}\Verb$r$}{\color{red}\Verb$2$}\Verb$ $\Verb$=$\Verb$ ${\color{red}\Verb$b$}{\color{red}\Verb$i$}{\color{red}\Verb$n$}{\color{red}\Verb$a$}{\color{red}\Verb$r$}{\color{red}\Verb$y$}{\color{red}\Verb$_$}{\color{red}\Verb$o$}{\color{red}\Verb$p$}{\color{green}\Verb$($}{\color{green}\Verb$'$}{\color{green}\Verb$e$}{\color{green}\Verb$x$}{\color{green}\Verb$p$}{\color{green}\Verb$r$}{\color{green}\Verb$_$}{\color{green}\Verb$o$}{\color{green}\Verb$r$}{\color{green}\Verb$'$}{\color{green}\Verb$ $}{\color{green}\Verb$,$}{\color{Violet}\Verb$($}{\color{Violet}\Verb$|$}{\color{green}\Verb$ $}{\color{black}\Verb$'$}{\color{black}\Verb$*$}{\color{black}\Verb$'$}{\color{green}\Verb$ $}{\color{green}\Verb$|$}{\color{black}\Verb$'$}{\color{black}\Verb$/$}{\color{black}\Verb$'$}{\color{green}\Verb$ $}{\color{green}\Verb$|$}{\color{black}\Verb$'$}{\color{black}\Verb$
}\end{spacing}

\end{exambox}




\def\bibname{Bibliography}




\openright

\begin{thebibliography}{99}
\addcontentsline{toc}{chapter}{\bibname}

\bibitem{aho} Aho V., Sethi R., Ullman J. D. \emph{Compilers: Principles, Techniques, and Tools.} Addison-Wesley Publishing Company, Reading, Massachusetts. 1986.
\bibitem{scope} Backus J. W., Bauer F. L., Green J., Katz, C., McCarthy J., Perlis A. J., Rutishauser H., Samelson K. et al. , Naur Peter ed. \emph{ Report on the Algorithmic Language ALGOL 60}. Copenhagen. , May 1960
\bibitem{example} \verb$http://kam.mff.cuni.cz/~ondra/amethyst/parser_highlight.ame.html$.
\bibitem{lowlevel} \verb$http://kam.mff.cuni.cz/~ondra/benchmark_string$.
\bibitem{implementation} \verb$http://kam.mff.cuni.cz/~ondra/regreg$.
\bibitem{rope} Boehm H-J., Atkinson R., Plass M. \emph{ Ropes: an Alternative to Strings}, Software—Practice  Experience (New York, NY, USA: John Wiley  Sons, Inc.) 25 (12): 1315–1330 and computation, Addison-Wesley, Reading (1979).
\bibitem{combinators} Burge W. H. \emph{ Recursive Programming Techniques,} The Systems programming series. Addison-Wesley. 1975
\bibitem{contextfree} Chomsky N. \emph{ Three models for the description of language}, Information Theory, IEEE Transactions 2 (3): 113–124, September 1956
\bibitem{lambdakalkul} Church A \emph{ A set of postulates for the foundation of logic}, Annals of Mathematics, Series 2, 33:346–366 (1932)
\bibitem{lca} Cole R.,  Ramesh H. \emph{Dynamic LCA queries on trees,} Proceeding SODA '99 Proceedings of the tenth annual ACM-SIAM symposium on Discrete algorithms
\bibitem{persistent} { Driscoll  J. R., Sarnak N., Sleator D. D., Tarjan R. E., } \emph{Making Data Structures Persistent}, Journal of Computer and System Sciences, Vol. 38, No. 1, 1989
\bibitem{ford} Ford Bryan, \emph{Packrat Parsing: a Practical Linear-Time Algorithm with Backtracking}, Master's thesis, MIT, September 2002
\bibitem{frost}Frost R., Hafiz R., Callaghan P.\emph{ Modular and Efficient Top-Down Parsing for Ambiguous Left-Recursive Grammars,} 10th International Workshop on Parsing Technologies (IWPT), ACL-SIGPARSE (Prague), June 2007
\bibitem{goodman} Goodman P,  \emph{ Re: [PEG] Res: Problem w/ nullable left recursion and trailing rules in "Packrat Parsers Can Support Left Recursion"} \verb$http://www.mail-archive.com/peg@lists.csail.mit.edu/msg00185.html$, 2008
\bibitem{contpass} Greif I. \emph{ Semantics of Communicating Parallel Processes. Ph.D. thesis,} Technical Report MAC-TR-154, Project MAC, MIT (Cambridge), September 1975
\bibitem{rats} Grimm R. \emph{ Better extensibility through modular syntax,} In Proceedings of the ACM SIGPLAN 2006 Conference on Programming Language Design and Implementation (PLDI ’06), pp. 38–51, June 2006
\bibitem{rightcongruence}  Hopcroft J. E., Ullman  J. D. \emph{ Introduction to automata theory}, languages
\bibitem{john} Johnson M. \emph{Memoization in top-down parsing,} Comput. Linguist., 21(3): 405-417, 1995.
\bibitem{regexp} {  Kleene S. C. }  \emph{ Representation of Events in Nerve Nets and Finite Automata}, In Shannon, Claude E.; McCarthy, John. Automata Studies. Princeton University Press. pp. 3–42, 1956
\bibitem{dataflow} Kildall G. \emph{ A Unified Approach to Global Program Optimization,} Proceedings of the 1st Annual ACM SIGACT-SIGPLAN Symposium on Principles of Programming Languages: 194–206., 1973
\bibitem{kuno} Kuno S, \emph{The predictive analyzer and a path elimination technique,} Comm. ACM 8(7) 453–462, 1965.
\bibitem{finstatemachine} McCulloch W. S., Pitts E. \emph{ A logical calculus of the ideas imminent in nervous activity}, Bulletin of Mathematical Biophysics: 541–544, 1943
\bibitem{oreg} Medeiros Se., Mascarenhas F., Ierusalimschy R., \emph{From Regular Expressions to Parsing Expression Grammars,}  SBLP, September 2011.
\bibitem{medeir} Medeiros S.\emph{ Left Recursion in PEGs} \\ \verb$http://www.lua.inf.puc-rio.br/~sergio/leftpeglist.pdf$, 2010
\bibitem{memo} Mizushima K., Maeda A., Yamaguchi Y.: \emph{Packrat parsers can handle practical grammars in mostly constant space}, Proceedings of the 9th ACM SIGPLAN-SIGSOFT Workshop on Program Analysis for Software Tools and Engineering, PASTE'10, Toronto, Ontario, Canada, June 5-6, 2010 (S. Lerner, A. Rountev, Eds.), ACM, 2010.
\bibitem{moore} Moore C. \emph{Removing left recursion from context-free grammars}, In Proc. 1st North American chapter of the Association for Computational Linguistics conference, pages 249–255, 2000.
\bibitem{memoization} Norvig P. \emph{ Techniques for Automatic Memoization with Applications to Context-Free Parsing,} Computational Linguistics, Vol. 17 No. 1, pp. 91–98, March 1991
\bibitem{boolean}Okhotin A. \emph{ Boolean grammars }, Information and Computation, Volume 194, Issue 1, 10 October 2004
\bibitem{bool} Okhotin A.  \emph{LR parsing for Boolean grammars}, International Journal of Foundations of Computer Science 17:3 (2006), 629--664.
\bibitem{nfa} Rabin M. O., Scott D. \emph{Finite automata and their decision problems}, IBM J. Res. Dev. vol 3, 1959
\bibitem{continuation} Reynolds J. C.\emph{ Definitional interpreters for higher order programming languages}, In ACM Conference Proceedings, 1972.
\bibitem{radz1} Redziejowsky R. \emph{BITES instead of FIRST for Parsing Expression Grammar}, Fundamenta Informaticae 109, 3 (2011) 323-337.
\bibitem{katahdin} Seaton G. S. \emph{A Programming Language Where the Syntax and Semantics Are Mutable at Runtime}, University of Bristol, Master thesis, \verb$http://www.chrisseaton.com/katahdin/katahdin.pdf$
\bibitem{incrementaldf} Ryder B. G. \emph{ Incremental data flow analysis}, In Conference Record 10th Annual ACM Symposium on Principles of Programming Languages (Austin, TX): 167-176, ACM, New York, Jan. 1983
\bibitem{tanter} Tanter E. \emph{Contextual values}, Proceeding DLS '08 Proceedings of the 2008 symposium on Dynamic languages
\bibitem{tomita} Tomita M. \emph{An efficient augmented-context-free parsing algorithm}, Comput. Linguist. 13, 31--46, 1987
\bibitem{splaytree} Sleator D. D., Tarjan R. E. \emph{Self-Adjusting Binary Search Trees}, Journal of the ACM (Association for Computing Machinery) 32, 1985
\bibitem{reftrans} Søndergaard, Harald, Sestoft \emph{Referential transparency, definiteness and unfoldability}, Acta Informatica 27 (6): 505–517, 1990
\bibitem{steele} Sussman G. J., Steele G. L. Jr. \emph{ Scheme: An Interpreter for Extended Lambda Calculus}, MIT AI Lab. AI Lab Memo AIM-349. December 1975. 
\bibitem{tratt} Tratt  L., \emph{Direct Left-recursive Parsing Expression Grammars}, Technical report EIS-10-01, Middlesex University, October 2010
\bibitem{valiant} Valiant L. G. \emph{General context-free recognition in less than cubic time}, Journal of Computer and System Sciences 10, 1975
\bibitem{ometa}  Warth A. \emph{Experimenting with Programming Languages}, PhD dissertation, University of California, Los Angeles, 2009 
Alessandro Warth and Ian Piumarta, \emph{ OMeta: an Object-Oriented Language for Pattern Matching}, Dynamic Language Symposium 2007, October 2007.

\end{thebibliography}
\end{document}